\documentclass[useAMS,usegraphicx,usenatbib]{mn2e}

\usepackage{txfonts,amssymb,hyperref,lscape} 
\usepackage{subfigure} 

\newcommand{\EBV}{\hbox{$E(B\!-\!V)$}}
\newcommand{\apropto}{\,\rlap{\raise 0.4ex\hbox{$\propto$}}{\lower
    0.5ex\hbox{$\sim$}}\,}

\newcommand{\bspsmall}{\vspace{0.5cm}\small\noindent This paper has been typeset
from a \TeX/\LaTeX\ file prepared by the author.\normalsize}

\defcitealias{Wild:2006p405}{WHP06}

\def\AlIII{Al\,{\sc iii}}
\def\CI{C\,{\sc i}}
\def\CIIf{C\,{\sc ii}$^\ast$}
\def\CIV{C\,{\sc iv}}
\def\CaII{Ca\,{\sc ii}}

\def\CrII{Cr\,{\sc ii}}

\def\MgI{Mg\,{\sc i}}
\def\MgII{Mg\,{\sc ii}}
\def\MnII{Mn\,{\sc ii}}

\def\FeI{Fe\,{\sc i}}
\def\FeII{Fe\,{\sc ii}}
\def\SiII{Si\,{\sc ii}}
\def\SiIIf{Si\,{\sc ii}$^\ast$}
\def\TiII{Ti\,{\sc ii}}
\def\ZnII{Zn\,{\sc ii}}
\def\HI{H\,{\sc i}}

\def\Lya{\rm{Ly}\alpha}
\def\zabs{z_{\rm abs}}

\def\UVESpopler{{\sc uves popler}}
\def\VPFIT{{\sc vpfit}}

\title[\CaII\ absorber environments]{Dust depletion, chemical uniformity and environment of \CaII\ H\&K quasar absorbers}

\author[B.~J.~Zych et al.]
       {Berkeley~J.~Zych,$^{1,2}$\thanks{E-mail:~\href{mailto:bjz@ast.cam.ac.uk}{bjz@ast.cam.ac.uk} (BJZ)}
         Michael~T.~Murphy,$^2$ Paul~C.~Hewett$^1$ and Jason~X.~Prochaska$^3$\\
         $^1$Institute of Astronomy, University of Cambridge, Madingley Road, Cambridge CB3 0HA\\
         $^2$Centre for Astrophysics \& Supercomputing, Swinburne University of Technology, P.O. Box 218, Victoria 3122, Australia\\
	 $^3$Department of Astronomy and Astrophysics, UCO/Lick Observatory, University of California, 1156 High Street, Santa Cruz, CA 95064}
\begin{document}

\date{Accepted ---. Received ---; in original form ---}
\pagerange{\pageref{firstpage}--\pageref{lastpage}}
\pubyear{2008}

\maketitle

\label{firstpage}

\begin{abstract}
\CaII$\,\lambda\lambda3934,3969$ absorbers, which are likely to be a
subset of damped Lyman $\alpha$ systems, are the most dusty quasar
absorbers known, with an order of magnitude more extinction in
$E(B\!-\!V)$ than other absorption systems. There is also
evidence that \CaII\ absorbers trace galaxies with more ongoing
star-formation than the average quasar absorber. Despite this,
relatively little is known in detail about these unusual absorption
systems. Here we present the first high resolution spectroscopic study
of 19 \CaII\ quasar absorbers, in the range $0.6 \leq z_{\rm abs} \leq
1.2$, with \CaII$\,\lambda3934$ equivalent widths, $W_0^{3934} \geq
0.2\,\rm{\AA}$. Their general elemental depletion patterns are found to
be similar to measurements in the warm halo phase of the Milky Way
(MW) and Magellanic Clouds interstellar medium. Dust depletions and
$\alpha$-enrichments profiles of sub-samples of 7 and 3 absorbers,
respectively, are measured using a combination of Voigt profile
fitting and apparent optical depth techniques. Deviations in ${\rm
  [Cr/Zn]} \sim 0.3 \pm 0.1\,\rm{dex}$ and in ${\rm [Si/Fe]}
\gtrsim0.8 \pm 0.1\,\rm{dex}$ are detected across the profile of one
absorber, which we attribute to differential dust depletion. The
remaining absorbers have $<0.3\,\rm{dex}$ ($3\sigma$ limit) variation
in [Cr/Zn], much like the general DLA population, though the dustiest
\CaII\ absorbers, those with $W_0^{3934} > 0.7\,\rm{\AA}$, remain
relatively unprobed in our sample. A limit on electron densities in
\CaII\ absorbers, $n_e < 0.1\,\rm{cm}^{-3}$, is derived using the
ratio of neutral and singly ionised species and assuming a MW-like
radiation field. These electron densities may imply hydrogen densities
sufficient for the presence of molecular hydrogen in the absorbers.
The \CaII\ absorber sample comprises a wide range of velocity
widths, $\Delta v_{90} = 50 - 470\,\rm{km}\,\rm{s}^{-1}$, and velocity
structures, thus a range of physical models for their origin, from
simple discs to galactic outflows and mergers, would be required to
explain the observations.
\end{abstract}


\section{Introduction}
Quasar absorption line systems are understood to be probes of galaxies
not subject to the luminosity bias of traditional surveys, being
instead biased by their gas content and galaxy cross-section. Despite
this, our exact understanding of how absorbers and galaxies are
related is uncertain due to the difficulty of interpreting absorption
line characteristics in the context of the associated galaxy. Damped
Lyman $\alpha$ systems (DLAs), defined by virtue of having neutral
hydrogen column densities $N_{\rm H\textsc{i}}\geq
2\times10^{20}\,\rm{atoms}\,\rm{cm}^{-2}$, are some of the best
studied absorption systems. In particular, it was postulated that DLAs
may be the reservoirs of neutral gas in the Universe from which stars
form because the amount of neutral gas in DLAs at $z\sim 3$ is similar
to the amount of gas locked up in stars today
\citep*{Wolfe:1993p281,Wolfe:2005p66}. Attempts to measure the
star-formation associated with DLAs through a variety of direct and
indirect techniques
\citep*[e.g.][]{Wolfe:2003p215,Chen:2005p1631,Wolfe:2006p28} have met
with limited success, generally finding low star-formation rates.

Recently, \cite{Wild:2005p407} used the statistical power of the Sloan
Digital Sky Survey (SDSS) to study the rare, so-called strong
\CaII\ absorbers (those with rest-frame equivalent widths, $W_0^{3934}
\geq 0.2\,\rm{\AA}$). \cite{Wild:2005p407} postulated that
\CaII\ absorbers were an unusual sub-class of DLAs because lines of
sight through our own Galaxy with similar strengths of
\CaII\ absorbers have neutral hydrogen column densities consistent
with DLAs. Furthermore, \cite{Wild:2005p407} found that the
\MgII\ equivalent widths of \CaII\ absorbers are large ($W_0^{2796}
\gtrsim 1.0\,\rm{\AA}$), thus they have a high probability of being
DLAs \citep{Rao:2000p1664}.  The observed sample of \CaII\ absorbers
has also been found to be much more dusty than either the general DLA
population or strong \MgII\ absorbers, with average $\EBV \sim
0.1\,\rm{mag}$ \citep*[][henceforth
  \citetalias{Wild:2006p405}]{Wild:2006p405}, compared to $\EBV \simeq
0.013\,\rm{mag}$ for \MgII\ absorbers (\citetalias{Wild:2006p405};
\citealp{York:2006p89}) and $\EBV < 0.02\,\rm{mag}$ for DLAs
\citep{Murphy:2004p99,Vladilo:2008p701}. The low levels of dust in
DLAs are confirmed by studies of radio-selected quasars, which find
little evidence for optical quasar drop-outs due to dusty DLAs
(\citealp{Ellison:2001p393};
\citealp*{Ellison:2005p1345}). Nonetheless, the presence of dust
detected by \citetalias{Wild:2006p405} means that \CaII\ absorbers are
significantly under-represented in existing samples of DLAs, a
conclusion consistent with those of \citet{Nestor_etal_2008a}, who
find that only $\sim$10 per cent of DLAs possess very strong,
$W_0^{3934} \geq 0.5\,\rm{\AA}$, \CaII\ absorption. The suggested
number of optical quasar drop-outs due to dusty \CaII\ absorbers is
still consistent with the statistics of the \citet{Ellison:2001p393}
radio-selected quasar sample. \citet{Nestor_etal_2008a} also find
indirect indications from photo-ionisation modeling, constrained by
their observations, that \CaII\ absorbers host molecular hydrogen.

In addition to these indirect indicators of star-formation activity,
such as dust and molecular hydrogen, direct studies at both
intermediate and low redshift have shown \CaII\ absorbers to be
associated more strongly with star formation than other absorbers.  At
$z\sim1$ \CaII\ absorbers are found to have mean {\em in situ} star
formation rates (SFRs) from $0.11-0.48\,\rm{M}_\odot\,\rm{yr}^{-1}$
compared to just $0.11-0.14\,\rm{M}_\odot\,\rm{yr}^{-1}$ for DLAs
\citep*{Wild:2007p15}.  These SFRs are as measured inside the
3\arcsec\ diameter SDSS fibre centred on each quasar and thus do not
account for star-formation in the absorber host galaxy if it falls
outside this radius.  At lower redshifts, $z\lesssim 0.4$, direct
measurements from the host galaxies of \CaII\ absorbers give SFRs in
the range $0.3-30\,\rm{M}_\odot\,\rm{yr}^{-1}$, though it is likely
the work is biased towards $L^\ast$ galaxies \citep{Zych:2007p392}.

Thus, there is now much evidence to suggest \CaII\ absorbers are
associated with star-formation, at least to some extent, both with
direct measurements of the star formation rate and through signatures
generally associated with star formation such as dust and molecular
hydrogen. However, few details of why \CaII\ absorbers might trace
star-formation more readily than other absorbers are understood. Nor
do we know much about the physical environment of the absorbers
themselves (rather than their host galaxies). In this paper we present
the first results of a high resolution study of the absorbing gas
using observations from the Ultra-Violet Echelle Spectrograph (UVES)
at the VLT and the HIgh Resolution Echelle Spectrograph (HIRES) at
Keck. With these data we are able to study the dust depletion and
chemical uniformity ($\alpha$ enhancement) of individual absorbers as
well as constrain the physical environment giving rise to the
absorbing gas based on the velocity structure of the absorber.

The paper is structured as follows: in Section~\ref{sec:data} we
present the sample selection, observations, data reduction and
analysis techniques used in the rest of the
paper. Section~\ref{sec:results} contains the results of our analysis,
divided into dust depletion, enrichment and environment
sub-sections. Section~\ref{sec:discussion} places these results into
context with more general results from the literature and finally,
Section~\ref{sec:conclusions} summarizes the main points from the
paper.


\section{Sample selection, observations, data reduction and analysis}\label{sec:data}

\subsection{Selection}
The \CaII$\,\lambda\lambda3934,3969$ absorption lines are relatively
weak and thus detection of the \MgII$\,\lambda\lambda2796,2803$
absorption lines at the same redshift are required to confirm that the
absorption system is real. In practice no real \CaII\ absorbers are
detected with $W_0^{2796} < 0.5\,\rm{\AA}$, thus this can be used as a
constraint to weed out false-positive detections
\citep{Wild:2005p407,Wild:2006p405}. Absorbers were identified by
searching for features with rest frame equivalent width,
$W_0^{3934}\geq 0.2\,\rm{\AA}$ at $4\,\sigma$ significance in the
sample of all 58835 SDSS DR4 quasars classified by having `specClass'
$=3$ or $4$ in the SDSS. To confirm a detection as real, it was also
required that the \CaII\,$\lambda3969$ line equivalent width had a
significance $>1\,\sigma$ and that \MgII$\,\lambda2796$ was detected
at $>6\,\sigma$ significance with $W_0^{2796}\geq 0.5\,\rm{\AA}$. This
requirement implies a minimum velocity width $\sim
50\,\rm{km}\,\rm{s}^{-1}$ for our absorbers, which being at the lower
limit of the velocity width distribution for DLAs will not strongly
bias any comparison between \CaII\ absorbers and DLAs (See Section
\ref{sec:disc_env}). Absorbers separated by less than
$500\,\rm{km}\,\rm{s}^{-1}$ were classified as a single absorption
system. For the purposes of this project the sample was restricted to
$\zabs>0.73$ so that \ZnII\,$\lambda2026$ was above 3500\,\AA, below
which the efficiency of UVES (and HIRES) begins to drop more
rapidly. Finally, suitable candidates for follow-up high resolution
observations required the Zn and Cr transitions to lie outside the
$\Lya$ forest.


Targets with $g$-band magnitudes $\leq 18.5\,\rm{mag}$ were selected
to achieve signal-to-noise ratio (SNR) $\sim10\,{\rm per}\,{\rm pix}$
in the continuum at the wavelength of \ZnII$\,\lambda\lambda2026,2062$
and \CrII$\,\lambda2056,2062,2066$. Declination was
restricted to allow observations from the VLT at Paranal in Chile
(except for the one target observed with Keck in Hawaii) and R.A. was
scheduling dependant.

These SDSS-selected absorbers were complemented by a search through
the ESO/UVES science archive for any absorbers which also exhibited
strong \CaII\ absorption at the limits defined above.

\subsection{Observations}
The new UVES data were taken in service mode during the period 2007
May 18 $-$ July 25 as part of ESO observing programme 79.A-0656. The
programme was not completed, so we only have partial data on some
absorbers, therefore we will highlight in each section which absorbers
have data which contributed to the respective analyses. We observed
J0334$-$0711 on 2008 January 12 for one $2700\,\rm{s}$ exposure with
the C1 decker on Keck/HIRESb \citep{Vogt:1994p1796}. The observations
are summarised in Table~\ref{tab:obs}. The emission redshifts are as
given by SDSS and the absorber redshifts are as measured from the SDSS
\CaII\ detections. For all UVES observations the slit width was
$1.2\arcsec$, which was comparable to the seeing in all cases. The
data were binned, $2\times2$, to improve the SNR. All observations
were taken with the slit at the parallactic angle to compensate for
atmospheric dispersion.

To supplement these newly observed data we include 12 \CaII\ absorbers
detected in quasar spectra from the ESO/UVES science archive (See
Table \ref{tab:arXiv}). Equivalent width measurements of
\CaII$\,\lambda3934$ and \MgII$\,\lambda2796$ for all the absorbers
are available in Table \ref{tab:EWs}.

\begin{table*}
  \caption{A summary of the observations taken with UVES during ESO
    Period 79 and HIRES for this work. We define the resolving power,
    $R\equiv\Delta\lambda/\lambda$, where $\lambda$ is the UVES
    dichroic central wavelength measured in $nm$.}
  \label{tab:obs}
  \begin{tabular}{llcccccc}
    \hline
    Object & SDSS name & $g$-band mag. & $z_{\rm em}$ & $z_{\rm abs}$ & Dichroic & R & Total exposure time [$s$]\\[1.0ex]
    \hline
    J0334$-$0711$^1$ & SDSSJ033438.28$-$071149.0 & 17.25 & 0.635 & 0.59760 & $-$ & 54000 & 2700\\
    J0846$+$0529 & SDSSJ084650.44$+$052946.0 & 17.65 & 1.052 & 0.74294 & 390 & 54256 & 2930\\
                 &                           &       &       &       & 564 & 54256 & 1465\\
                 &                           &       &       &       & 760 & 54256 & 1465\\
    J0953$+$0801 & SDSSJ095352.69$+$080103.6 & 17.89 & 1.720 & 1.02316 & 390 & 54240 & 2930\\
                 &                           &       &       &       & 437 & 54240 & 2930\\
                 &                           &       &       &       & 580 & 54240 & 2930\\
                 &                           &       &       &       & 760 & 54240 & 2930\\
    J1005$+$1157 & SDSSJ100523.73$+$115712.4 & 18.20 & 1.657 & 0.83460 & 390 & 54259 & 2930\\
                 &                           &       &       &       & 564 & 54259 & 2930\\
    J1129$+$0204 & SDSSJ112932.71$+$020422.7 & 17.65 & 1.193 & 0.96497 & 390 & 54260 & 2930\\
                 &                           &       &       &       & 437 & 54256 & 2930\\
                 &                           &       &       &       & 564 & 54260 & 2930\\
                 &                           &       &       &       & 760 & 54256 & 2930\\
    J1203$+$1028 & SDSSJ120342.24$+$102831.7 & 17.88 & 1.888 & 0.74630 & 390 & 54280 & 9805\\
                 &                           &       &       &       & 564 & 54313 & 4395\\
                 &                           &       &       &       & 760 & 54277 & 5411\\
    J1430$+$0149 & SDSSJ143040.83$+$014939.9 & 17.79 & 2.113 & 1.24180 & 437 & 54252 & 12147\\
                 &                           &       &       &       & 760 & 54252 & 12146\\
    \hline
  \end{tabular}

\medskip
$^1$ HIRESb target which used the C1 decker with FWHM
$\sim$$7\,\rm{km}\,\rm{s}^{-1}$ spectral resolution
\end{table*}

\begin{table*}
 \caption{A summary of spectra taken from the ESO/UVES Science Archive
   for this work, with an indication of how the data is
   used. Magnitudes are SDSS $g$-band values unless otherwise
   referenced. B1950 names are included for convenience.}
 \label{tab:arXiv}
 \begin{tabular}{lcclcccc}
	\hline
	Object & $\alpha$ (J2000) & $\delta$ (J2000) & B1950 & magnitude & $z_{\rm em}$ & $z_{\rm abs}$ & Notes$^1$\\[1.0ex]
	\hline
	J0004$-$4157 & 00\,04\,48.07 & -41\,57\,27.7 & Q0002$-$4214 & 17.2\,V$^2$ & 2.760 & 0.83663 & V\\
	J0256$+$0110 & 02\,56\,07.24 & +01\,10\,38.6 & Q0253$+$0058 & 18.95\,g & 1.348 & 0.72578    & V\\
	J0407$-$4410 & 04\,07\,18.08 & -44\,10\,14.6 & Q0405$-$4418 & 17.6\,B$^3$ & 3.000 & 0.81841 & V\\
	J0517$-$4410 & 05\,17\,07.82 & -44\,10\,55.4 & Q0515$-$4414 & 14.9\,V$^4$ & 1.710 & 1.14955 & V, D, A\\
	J0830$+$2410 & 08\,30\,52.09 & +24\,10\,59.8 & Q0827$+$243 & 17.4\,g & 0.940 & 0.52477      & V\\
	J1028$-$0100 & 10\,28\,37.02 & -01\,00\,27.6 & Q1026$-$0045 & 17.2\,g & 1.531 & 0.63214     & V\\
	J1107$+$0048 & 11\,07\,29.04 & +00\,48\,11.2 & Q1104$-$0104 & 17.7\,g & 1.392 & 0.74030   & V, D\\
	J1130$-$1449 & 11\,30\,07.05 & -14\,49\,27.4 & Q1127$-$145 & 16.9\,V$^2$& 1.184 & 0.31272   & V\\
	J1211$+$1030 & 12\,11\,40.60 & +10\,30\,02.0 & Q1209$+$1046 & 18.1\,g & 2.192 & 0.62962     & V\\
	J1232$-$0224 & 12\,32\,00.01 & -02\,24\,04.3 & Q1229$-$021 & 17.2\,g & 1.043 & 0.39544      & V\\
	J1323$-$0021 & 13\,23\,23.79 & -00\,21\,55.3 & Q1320$-$0006 & 18.5\,g & 1.388 & 0.71608     & V\\
	J2328$+$0022 & 23\,28\,20.37 & +00\,22\,38.2 & Q2325$+$0006 & 18.0\,g & 1.308 & 0.65200     & V\\
	\hline 
 \end{tabular}
 
 \medskip
 $^1$ Letters denote the following: (V)elocity structure analysis, (D)epletion profile analysis, (A)lpha enhancement analysis
 
 $^2$ \citet*{Rao:2006p610}
 $^3$ \citet{Maza:1993p51}
 $^4$ \citet{Reimers:1998p96}
\end{table*}

\subsection{Data reduction}
The UVES data were reduced using the new UVES Common Pipeline Language
(CPL) pipeline from ESO. The Keck/HIRESb data were reduced using the
HIRedux software
package\footnote{\href{http://www.ucolick.org/~xavier/HIRedux/index.html}{http://www.ucolick.org/$\sim$xavier/HIRedux/index.html}}
bundled within the XIDL software
package\footnote{\href{http://www.ucolick.org/~xavier/IDL/index.html}{http://www.ucolick.org/$\sim$xavier/IDL/index.html}}.
The algorithms calibrate, sky subtract and optimally extract the
echelle orders of the spectrum \citep*{Bernstein_etal_2008a}.

The individual orders and exposures were then air$-$vacuum and
heliocentric corrected and combined using
\UVESpopler.\footnote{\UVESpopler\ is available
  from\\ \href{http://astronomy.swin.edu.au/~mmurphy/UVES_popler.html}{http://www.astronomy.swin.edu.au/$\sim$mmurphy/UVES\_popler.html}}
During combination all spectra were rebinned to a
$2.5\,\rm{km}\,\rm{s}^{-1}$ pixel scale. Both instruments have FWHM
(Full-Width Half-Maximum) $\sim$$7\,\rm{km}\,\rm{s}^{-1}$ spectral
resolution, thereby giving $\sim$3 pixel sampling for unresolved
features.

\subsection{Analysis}
The analysis of the final spectra involves two main
approaches. Firstly, abundance profile analysis based on Voigt profile
fits to the various ionic transitions\footnote{Transition wavelengths
  and oscillator strengths were taken from \\\citet{Morton:2003p205}.}
in the absorber and secondly a comparison of the velocity structure of
these absorbers to data and models in the literature. The first
approach allows us to make a quantitative analysis of chemical and
dust-depletion uniformity and, in principle, chemical enrichment
history via models of relative abundances of $\alpha$ and Fe-peak
elements. Unfortunately, the wavelength coverage and SNR of the
spectra are not sufficient for a full decomposition of the
star-formation history
\citep[e.g.][]{DessaugesZavadsky:2007p1485}. Nonetheless
non-uniformity in the $\alpha$ enrichment and dust depletion of the
absorbers will provide clues as to the physical environment of the
absorbing gas (See Sections~\ref{sec:res_chem} \&
\ref{sec:disc_chem}). Furthermore, the electron density of the
absorbing cloud can be constrained from measured limits from the
$N\left({\rm Fe \textsc{i}}\right)/N\left({\rm Fe \textsc{ii}}\right)$
and $N\left({\rm Mg \textsc{i}}\right)/N\left({\rm Mg
  \textsc{ii}}\right)$ ratios. The second approach involves study of
the velocity profiles and widths of each absorber, allowing
qualitative assessment of the physical models able to explain the
velocity structure of the gas, such as simple galaxy discs, haloes,
inflows or outflows.

\subsubsection{Column densities and equivalent widths}\label{sec:vpfit}
We used \VPFIT\footnote{\VPFIT\ is available from
  \href{http://www.ast.cam.ac.uk/~rfc/vpfit.html}{http://www.ast.cam.ac.uk/$\sim$rfc/vpfit.html}.}
to fit multiple velocity component Voigt profiles to each unsaturated
metal ion transition of interest in each absorber. The transitions of
interest for our analysis were \FeII, \MgI, \SiII, \ZnII, \CrII,
\MnII, \TiII\ and \AlIII. An initial guess for the fit was constructed
from \FeII\ as multiple \FeII\ transitions at a variety of wavelengths
were usually observed, which allows \VPFIT\ to constrain the velocity
structure more accurately. For the remaining transitions fitted, the
redshifts and Doppler parameters, $b$, of individual components were
tied to \FeII\ (or \MgI$\,\lambda2852$ when \FeII\ was unavailable),
whilst column densities for each component were allowed to vary
freely. At the resolution and SNR of the spectra we saw no evidence
for these transitions requiring independent redshifts and $b$
parameters.

The total column densities were reported directly from \VPFIT, which
allows for a more accurate estimate of the formal error because it
alleviates the problem of blending between neighbouring components,
which produces degeneracies between column densities and $b$
parameters for those components (see \VPFIT\ documentation). Upper
limits were taken at $3\sigma$ using the error reported from \VPFIT.

Both the \MgI$\,\lambda2026$ and \ZnII$\,\lambda2026$ and
\ZnII$\,\lambda2062$ and \CrII$\,\lambda2062$ lines are usually
blended due to the velocity width of the absorption system. Despite
this blending the fits to these lines are not degenerate;
\VPFIT\ simultaneously fits these lines, along with unblended lines
such as \CrII$\,\lambda2056$ and \MgI$\,\lambda2852$, using the
predetermined velocity structure. The spectra are always of sufficient
wavelength coverage with high enough SNR for a non-degenerate fit to
be reached. For example the \ZnII\ velocity structure may appear
uncertain if the absorption profile is sufficiently wide to blend
together the \ZnII$\,\lambda2026$ and \MgI$\,\lambda2026$
profiles\footnote{Note that whilst \CrII$\,\lambda2026$ also
  contributes to the flux decrement it is generally too weak to
  confuse the deblending process.} and also the \ZnII$\,\lambda2062$
and \CrII$\,\lambda2062$ profiles. However, since the \CrII\ velocity
structure is independently determined by \CrII$\,\lambda2056$ and/or
\CrII$\,\lambda2066$, and since fitting the \MgI$\,\lambda2852$
profile determines the \MgI$\,\lambda2026$ profile, the
\ZnII\ velocity structure can be recovered reliably.

When all transitions of a given ionic species are saturated
(e.g. \MgII) the column density of the species is unconstrained by
Voigt profile fitting. Instead, for \MgII$\,\lambda2796,2803$ we
measure the equivalent width of the profile, which gives an indication
of the velocity spread in the absorber. The equivalent widths were
calculated by summing across each pixel, $i$, in the profile
\begin{equation}
  W = \sum_i \left(1-f_i\right)\Delta\lambda_i
\end{equation}
where $f_i$ is the normalised flux and $\Delta\lambda_i$ is the width
of the $i$th pixel in rest-frame wavelength space. The associated
errors are the formal errors in the sum and thus do not account for
errors associated with continuum fitting.

\subsubsection{Measuring velocity profile uniformity}\label{sec:anal_AOD}
The apparent optical depth (AOD) method, as first described by
\citet{Savage:1991p1362}, offered an alternative way of measuring the
column densities of velocity components in an absorber to more
traditional Voigt profile analyses. The reader is referred to
\citet{Prochaska:2003p110} for a discussion of the merits of each
technique. \cite{Prochaska:2003p110} conducted a velocity bin based
analysis of the chemical uniformity {\em within} DLAs by calculating
the column density in each velocity bin using the AOD method. Here we
will use the same technique to measure variation across the velocity
profile of the absorber in both dust depletion (via Zn and Cr) and
chemical uniformity (via Fe and Si) where possible. In particular we
hope to show whether the dust depletion profiles are relatively
uniform, or whether the dust is concentrated in one or two velocity
components as this may indicate the presence of molecular hydrogen in
the absorber \citep[e.g.][]{Ledoux:2003p1351}.

\cite{Prochaska:2003p110} use unsaturated, unblended transitions such
as Fe and Si and convert the flux pixels to optical depth pixels,
binning the result in velocity space. These bins are not components in
the traditional sense, but just velocity bins across the profile, as
such they may contain one or more real velocity
components. Nonetheless, if there is any non-uniformity across the
absorption profile, it will still be reflected in the constructed
velocity bins. Unlike \cite{Prochaska:2003p110} we will be applying
this technique to blended transitions such as \ZnII\ and \CrII. We
circumvent the problem of line blending by binning the Voigt profile
fits to the data, rather than the spectra themselves. The error in
each pixel is still drawn from the actual spectral error array. This
technique utilises the fact that we have multiple transitions of each
blended element, which a straight-forward analysis of the data cannot
account for. Using the error array corresponding to the flux in each
pixel, rather than adapting the \VPFIT\ code to return the error
associated with each pixel in the fit, will not underestimate the real
error too greatly. This assertion can be verified by applying the
technique to transitions such as \FeII\ and \SiII\ which do not suffer
from blending. A further advantage of using the Voigt profile fits to
the spectra is that taking account of multiple transitions at
different wavelengths makes the subsequent analysis more resilient to
changes in bin size. That is, the effect of Root Mean Square (RMS)
fluctuations in the data of an individual transition is reduced.

We will use $20\,\rm{km}\,\rm{s}^{-1}$ bins for our analysis,
corresponding to eight $2.5\,\rm{km}\,\rm{s}^{-1}$ pixels per velocity
bin; this gives reasonable SNR in each bin, whilst still providing
enough bins to discuss the absorbers' depletion properties. In
practice the size of the bin does not alter our conclusions as the
component-to-component dust depletions are resilient to changes in bin
size; binning mainly effects the error bars in our analysis. We also
performed the same analysis with the velocity bins offset by half a
bin from the initial analysis to test whether the results were
resilient to shifts in velocity space. We found this made no
difference to our conclusions.

\subsubsection{Electron densities}\label{sec:anal_ne}
The electron density of the gas can be constrained by the relative
column densities of the neutral and singly ionised transitions of an
element \citep[e.g.][]{Prochaska:2006p737}. For instance, under steady
state conditions and photoionization equilibrium the balance of
photoionisation and recombination for \MgI\ and \MgII\ gives
\begin{equation}
  n_e = \frac{N\left({\rm Mg \textsc{i}}\right)}{N\left({\rm Mg
      \textsc{ii}}\right)} \frac{\Gamma\left({\rm
      Mg\textsc{i}}\right)}{\alpha\left({\rm Mg\textsc{ii},T}\right)}
  \label{eqn:ne}
\end{equation}
where $N$ are the measured column densities of the named transitions,
$\alpha\left({\rm Mg\textsc{ii}},T\right)$ is the total recombination
coefficient (radiative and dielectric) dependent on temperature, T,
and $\Gamma\left({\rm Mg\textsc{i}}\right) = {\rm Const.} \times
\sigma_{\rm ph}\left({\rm Mg\textsc{i}}\right) \times G/G_0$
represents the photoionisation rate of \MgI. $\sigma_{\rm
  ph}\left({\rm Mg\textsc{i}}\right)$ is cross-section to ionisation
integrated over the incident radiation field and $G_0=2.72\times
10^{-3}\,\rm{erg}\,\rm{cm}^{-2}\,\rm{s}^{-1}$ is $1.7\times$ Habing's
constant \citep{Habing:1968p421,Gondhalekar:1980p272}, where the
factor of 1.7 is required to normalise $G$ to the value of the Milky
Way UV radiation field strength. A similar relationship can be derived
for \FeI\ \& \FeII. Thus observations of the ratio of the column
densities of a neutral and singly ionised species can constrain the
electron density, $n_e$, for a given strength of radiation field,
$G/G_0$. See \citet{Prochaska:2006p737} Section 4.2 \& Appendix for
details of the calculations. For our analysis we use the
\FeI$\,\lambda2484$, \FeII$\,\lambda\lambda2260,2374$,
\MgI$\,\lambda2852$ and \MgII$\,\lambda2803$ transitions.  Note that
for the ionisation of \FeI, H{\sc i} charge transfer can dominate the
transition, dependent on the exact ratio of $n_{\mathrm H{\textsc i}}$
to $n_e$ \citep{Kingdon_Ferland_1999a}. If H{\sc i} charge transfer does
dominate then equation~\ref{eqn:ne} will underestimate the true
electron density. The results and conclusions drawn from observed
limits on electron densities are dominated by measurements of
\MgI\ and \MgII\ (See Fig.~\ref{fig:ne}). Thus, the increased
uncertainty in measurements of $n_e$ based on Fe due to H{\sc i}
charge transfer will not effect any conclusions.

\subsubsection{Constraints on absorber environment from velocity profiles}\label{sec:anal_env}
There have been extensive studies of the velocity profile structure in
both DLAs and \MgII\ absorbers. Early work focused on modeling DLAs as
rotating discs of gas extending out from the associated galaxy
\citep{Steidel:2002p103}. Later, more complex models including hot and
cold rotational components, haloes and spherical accretion
\citep{Prochaska:1997p73,Charlton:1998p1325} or outflows, such as
super-bubbles expanding out into the Inter-Galactic Medium (IGM)
\citep{Bond:2001p1149} were called upon to explain the increasingly
complex \MgII\ absorption profiles being observed and their sometimes
highly symmetric velocity profiles. Given the paucity of strong
\CaII\ absorbers we do not have a statistical sample of absorbers
required for a quantitative analysis, therefore we will restrict
ourselves to a qualitative comparison with models from the
literature. It will, nonetheless, be possible to place constraints on
the physical environment of these absorbers such as whether they can
be reproduced using simple disc models, outflows, etc.

\section{Results}\label{sec:results}
Voigt profile fits for each absorber are shown in Appendix
\ref{adx:fits}, whilst the unfitted \MgII\ and \CaII\ profiles are
presented in Appendix \ref{adx:MgII}.

The measured column densities for all the absorbers in the sample are
given in Table~\ref{tab:abundances}, whilst measured equivalent widths
of \MgII\ and \CaII\ and the velocity width, as measured by $\Delta
v_{90}$ \citep{Prochaska:1997p73}, are given in
Table~\ref{tab:EWs}. The \CaII\ transition in J1203$+$1028 is
contaminated by sky absorption, but because the absorption features
are not flat bottomed (i.e. not all the flux is absorbed) we are able
to make a correction to the $W_0^{3934}$ measurement for the sky
absorption as follows. The first step was to find exposures with
uncontaminated sky absorption of similar air-mass, date and time to
the J1203$+$1028 exposures. For each uncontaminated sky exposure the
equivalent width of absorption due to the sky, $W^{\rm sky}$, was
measured. We then apply a correction to the matching exposure from
J1203$+$1028 such that, $W^{3934}_{\rm real} = W^{3934}_{\rm obs} -
W^{\rm sky}$, where $W^{3934}_{\rm obs}$ is measured across the same
wavelength range as $W^{\rm sky}$. The errors are propagated formally,
so will not include any systematic error such as that due to continuum
fitting.

\begin{table*}
\begin{minipage}{1.0\textwidth}
\caption{The ionic total column densities measured for each absorber
  in this work. Upper limits are quoted at 3$\sigma$ significance. Systematic
  errors due to continuum level and component fitting are estimated at
  $0.03\,\rm{dex}$. Quoted errors include random and systematic
  contributions added in quadrature.}
\label{tab:abundances}
\begin{tabular}{lcccccccccc}
\hline
&\multicolumn{9}{c}{$\log N$ [${\rm atom}\,{\rm cm}^{-2}$]}\\
\raisebox{1.0ex}{Object} & ${\rm Ca\textsc{ii}}$ & ${\rm Fe\textsc{ii}}$ & ${\rm Mg\textsc{i}}$ & ${\rm Zn\textsc{ii}}$ & ${\rm Cr\textsc{ii}}$ & ${\rm Si\textsc{ii}}$ & ${\rm Al\textsc{iii}}$ & ${\rm Mn\textsc{ii}}$ & ${\rm Ti\textsc{ii}}$\\
\hline
J0334 & $-$ & 14.94 $\pm$ 0.04 & 12.84 $\pm$ 0.07 & 12.58 $\pm$ 0.09 &13.19 $\pm$ 0.08 &$-$ &$-$ & 12.65 $\pm$ 0.04 & 11.9 $\pm$ 0.1\\
J0517 & 12.74 $\pm$ 0.03 & 14.31 $\pm$ 0.03$^1$ & 12.74 $\pm$ 0.03 & 12.22 $\pm$ 0.04 & 12.53 $\pm$ 0.06 & 14.77 $\pm$ 0.03$^2$ & 12.76 $\pm$ 0.03 & 12.21 $\pm$ 0.03 & $<$10.6\\
J0846 & 13.06 $\pm$ 0.06 & 15.21 $\pm$ 0.04 & 12.96 $\pm$ 0.05 & $<$12.8 & $<$13.4 & $-$ & $-$ & 13.21 $\pm$ 0.06 & 13.00 $\pm$ 0.05\\
J0953 & 13.57 $\pm$ 0.04 & 15.09 $\pm$ 0.04 & 13.24 $\pm$ 0.03 & 12.25 $\pm$ 0.06 & 13.17 $\pm$ 0.04 & 15.61 $\pm$ 0.07 & 15.31 $\pm$ 0.03 & 13.0 $\pm$ 0.7 & 12.1 $\pm$ 0.1\\
J1005  & $-$ & 15.30 $\pm$ 0.05 & 13.10 $\pm$ 0.03 & $<$12.8 & $<$13.3 & $-$ & $<$13.5 & 13.03 $\pm$ 0.05 & $<$12.4\\
J1107 & $-$ & 15.52 $\pm$ 0.03 & 13.00 $\pm$ 0.07 & 13.23 $\pm$ 0.04 & 13.83 $\pm$ 0.09 & $-$ & $-$ & 13.37 $\pm$ 0.04 & 13.12 $\pm$ 0.03\\
J1129 & 13.11 $\pm$ 0.04 & 15.35 $\pm$ 0.03 & 13.14 $\pm$ 0.09$^3$ & 12.80 $\pm$ 0.05 & 13.66 $\pm$ 0.04 & 16.0 $\pm$ 0.5 & 13.81 $\pm$ 0.09 & 13.13 $\pm$ 0.04 & 12.79 $\pm$ 0.04\\
J1203  & 13.05 $\pm$ 0.03 & 14.83 $\pm$ 0.03  & 12.99 $\pm$ 0.03 &  12.63 $\pm$ 0.06 & 13.38 $\pm$ 0.05 & $-$ & $-$ & 12.77 $\pm$ 0.05 & $<$11.2\\
J1430  & 12.85 $\pm$ 0.04 & $-$ & 13.07 $\pm$ 0.05 & 12.94 $\pm$ 0.04 & 13.45 $\pm$ 0.04 & 15.75 $\pm$ 0.03 & 13.72 $\pm$ 0.03 & $-$ & 12.79 $\pm$ 0.04\\
\hline
\end{tabular}

\medskip
$^1$ \FeII\ measured over full $\sim800\,\rm{km}\,\rm{s}^{-1}$ range of
absorber. $N_{\rm Fe\textsc{ii}} = 14.25 \pm 0.03$ over
$\sim100\,\rm{km}\,\rm{s}^{-1}$ region which other elements could be
measured.

$^2$ \SiII\ constrained by simultaneous fit to
\SiII$\,\lambda1526,1808$ and intervening Ly$\alpha$ forest.  

$^3$ \MgI$\,\lambda 2852$ not covered by spectrum. \MgI\ column
density derived from residual from \ZnII\ and \CrII\ profile fits.
\end{minipage}
\end{table*}

\begin{table}
\caption{Equivalent width and $\Delta v_{90}$ measurements from high
  resolution spectra for each absorber in this work. The 1$\sigma$
  error in $\Delta v_{90}$ is $\lesssim10\,\rm{km}\,\rm{s}^{-1}$,
  which is dominated by uncertainty in continuum fitting. Where no
  high resolution data exists the measurement from SDSS spectra are
  given in parentheses when available.}
\label{tab:EWs}
\begin{tabular}{lccc}
\hline
&\multicolumn{2}{c}{Rest-frame equivalent width [\AA]}\\
\raisebox{1.0ex}{Object} & ${\rm Mg\textsc{ii}}\,\lambda2796$ & ${\rm Ca\textsc{ii}}\,\lambda3934$ & \raisebox{1.0ex}{$\Delta v_{90}$ [${\rm km}\,{\rm s}^{-1}$]}\\
\hline
J0004$-$4157 & 4.424 $\pm$ 0.003 & 0.966 $\pm$ 0.005 & 400\\
J0256$+$0110 & 3.27 $\pm$ 0.01   & 0.39 $\pm$ 0.04$^1$ & 350\\
J0334$-$0711 & 3.49--4.65$^2$ & (0.31 $\pm$ 0.05) & 410\\
J0407$-$4410 & 1.668 $\pm$ 0.002 & 0.282 $\pm$ 0.009 & 200\\
J0517$-$4410 & 2.336 $\pm$ 0.002 & 0.346 $\pm$ 0.002 & 470\\
J0830$+$2410 & $-$ & 0.284 $\pm$ 0.007 & 190\\
J0846$+$0529 & 2.22 $\pm$ 0.03 & 0.50 $\pm$ 0.03 & 160\\
J0953$+$0801 & 0.468 $\pm$ 0.008$^3$ & 0.34 $\pm$ 0.02 & 50\\
J1005$+$1157 & 2.58 $\pm$ 0.02 & (0.78 $\pm$ 0.08) & 210\\
J1028$-$0100 & (1.47 $\pm$ 0.08) & 0.29 $\pm$ 0.02 & 100\\
J1107$+$0048 & 2.766 $\pm$ 0.009 & 0.34 $\pm$ 0.01$^1$ & 190\\
J1129$+$0204 & 2.11 $\pm$ 0.01 & 0.69 $\pm$ 0.03 & 170\\
J1130$-$1449 & 1.794 $\pm$ 0.007 & 0.374 $\pm$ 0.005 & 120\\
J1203$+$1028 & 2.58 $\pm$ 0.02 & 0.630 $\pm$ 0.006$^4$ & 270\\
J1211$+$1030 & $-$ & 0.24 $\pm$ 0.01 & 330\\
J1232$-$0224 & 2.071 $\pm$ 0.005 & 0.203 $\pm$ 0.007 & 120\\
J1323$-$0021 & 2.156 $\pm$ 0.009 & 0.82 $\pm$ 0.02$^1$ & 120\\
J1430$+$0149 & 2.898 $\pm$ 0.009 & 0.30 $\pm$ 0.01 & 220\\
J2328$+$0022 & (1.98 $\pm$ 0.07) & 0.250 $\pm$ 0.008 & 200\\
\hline
\end{tabular}

\medskip
$^1$ Reported measurement from \cite{Nestor_etal_2008a}

$^2$ Velocity structure too complex to separate \MgII$\,\lambda2796$ and \MgII$\,\lambda2803$. Therefore we measured the total equivalent width over both lines and assumed doublet ratios of 1:1 and 2:1 to calculate the range of possible $W_0^{2796}$.

$^3$ $W_0^{2796} = 0.88 \pm 0.05\,\rm{\AA}$ when measured in the SDSS DR4 spectrum of this quasar, thus it is included in the sample.

$^4$ Corrected for sky absorption contamination, as discussed in Section~\ref{sec:results}, the associated error only includes the formal statistical error and is therefore an underestimate.

\end{table}

\subsection{Depletions}
\subsubsection{Total depletions}\label{sec:tot_dep}
\begin{figure}
  \includegraphics[scale=0.5,trim=20 0 0 0,angle=0,clip=true]{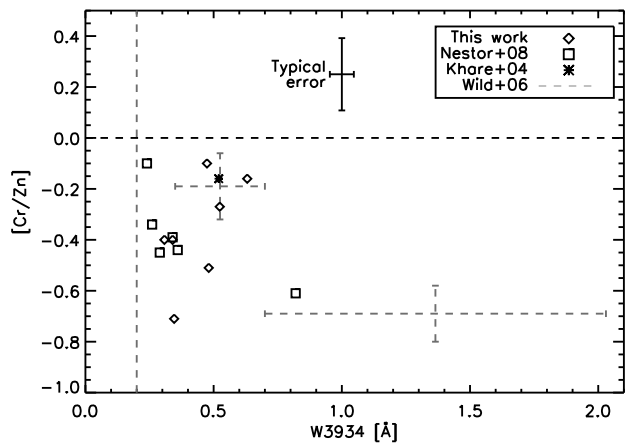}
\caption{[Cr/Zn] vs. $W^{3934}_{\rm r}$ for each of the absorbers. The
  black diamonds show [Cr/Zn] measured in the UVES or HIRES spectra
  vs. $W^{3934}_{\rm r}$ measured from the SDSS spectra. An indication
  of the typical error is shown at the centre-top of the plot. The
  dashed grey regions show the average relation found by
  \citetalias{Wild:2006p405}. Other measurements of [Cr/Zn] for
  \CaII\ absorbers in the literature are shown as black squares and
  asterisks \citep{Nestor_etal_2008a,Khare:2004p86}. The horizontal
  dashed black line shows the level of solar depletion. The vertical
  dashed grey line shows the $W_0^{3934}$ equivalent width limit of
  our sample selection.}
\label{fig:depletion}
\end{figure}

\begin{figure}
  \includegraphics[scale=0.5,trim=20 0 0 0,angle=0,clip=true]{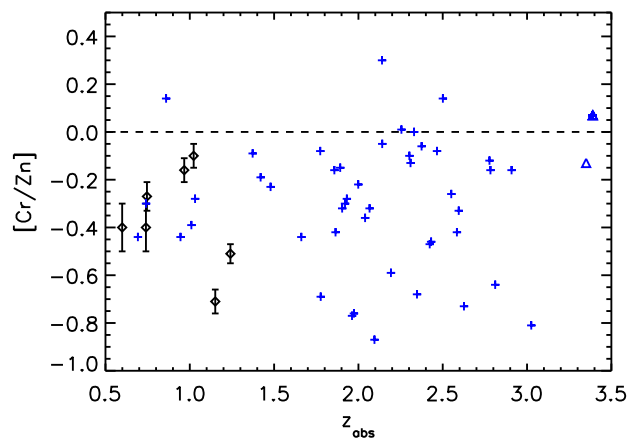}
\caption{[Cr/Zn] vs. $z_{\rm abs}$ for the DLAs from
  \citet{Akerman:2005p81} and each of the absorbers. The black
  diamonds show [Cr/Zn] measured in the UVES or HIRES spectra. The
  blue crosses show the DLA measurements of [Cr/Zn]
  whilst the blue triangles show upper limits to the depletion. The
  dashed black line shows the level of solar depletion.}
\label{fig:DLA_depletion}
\end{figure}

The total column densities derived for \ZnII\ and \CrII\ in
Section~\ref{sec:results}\ were used to calculate the depletion level
in each of the absorbers. In high $N_{\rm H{\textsc i}}$ systems
self-shielding ensures that the metallic elements considered here are
predominantly in the lowest ionization state with an ionization
potential less than $13.6\,\rm{eV}$. Thus we may assume that the
ratio of $N_{\rm Cr\textsc{ii}}/N_{\rm Zn\textsc{ii}} \sim N_{\rm
  Cr}/N_{\rm Zn}$, i.e.~no correction for ionization is
required. Measured [Cr/Zn] ratios are given in
Table~\ref{tab:abundance_ratios}. Fig.~\ref{fig:depletion}\ shows the
depletion in each of the systems versus $W^{3934}_0$ for each
absorber. Included in the plot are the DLAs with known \ZnII,
\CrII\ and \CaII\ absorption from \cite{Nestor_etal_2008a} and
\cite{Khare:2004p86}. For comparison we show the mean values of
     [Cr/Zn] vs.~$W^{3934}_0$ measured statistically by
     \citetalias{Wild:2006p405} from SDSS quasar spectra with
     \CaII\ absorbers. The high $W_0^{3934}$ regime remains relatively
     unexplored for measurements from high resolution
     spectroscopy. Similarly, Fig.~\ref{fig:DLA_depletion} shows
     [Cr/Zn] vs.~redshift for the \CaII\ absorbers with the
     compilation of measurements of [Cr/Zn] in DLAs given by
     \cite{Akerman:2005p81} included for comparison. We note that the
     measurements from the observed \CaII\ absorber sample seem
     consistent with those observed in other DLAs, being neither
     particularly dusty nor dust-free. Performing a Kolmogorov-Smirnov
     (KS) test of the measured [Cr/Zn] ratios indicates an 87 per cent
     likelihood that the \CaII\ and DLA samples are drawn from the
     same underlying population.

\begin{table*}
\begin{minipage}{0.7\textwidth}
\caption{A table of abundance ratios in the absorber sample, relative
  to solar, traditionally used for assessing dust-depletion and
  chemical enrichment history. Solar abundances are taken from
  \citet{Lodders:2003p1174}.}
\label{tab:abundance_ratios}
\begin{tabular}{lccccccc}
	\hline
	Object & [Cr/Zn] & [Mn/Fe] & [Ti/Fe] & [Zn/Fe] & [Si/Fe] & [Si/Zn] & [Si/Ti]\\
	\hline
	J0334$-$0711 & $-0.4 \pm 0.1$ & $-0.33 \pm 0.06$ & $-0.5 \pm 0.1$ & 0.5 $\pm$ 0.1 & $-$ & $-$ & $-$\\
	J0517$-$4410$^1$ & $-0.71 \pm 0.07$ & $-0.08 \pm 0.04$ & $<-0.70$ & 0.81 $\pm$ 0.05 & 0.44 $\pm$ 0.04 & $-0.36 \pm 0.05$ & $>1.14$\\
	J0846$+$0529 & $-$ & $-0.02 \pm 0.07$ & $0.35 \pm 0.06$ & $<0.62$ & $-$ & $-$ & $-$\\
	J0953$+$0801 & $-0.10 \pm 0.08$ & $-0.2 \pm 0.7$ & $-0.5 \pm 0.1$ & 0.00 $\pm$ 0.07 & 0.45 $\pm$ 0.08 & 0.45 $\pm$ 0.09 & 0.9 $\pm$ 0.1\\
	J1005$+$1157 & $-$ & $-0.30 \pm 0.06$ & $<-0.39$ & $<0.32$ & $-$ & $-$ & $-$\\
	J1107$+$0048 & $-0.4 \pm 0.1$ & $-0.17 \pm 0.06$ & 0.17 $\pm$ 0.05 & 0.53 $\pm$ 0.06 & $-$ & $-$ & $-$\\
	J1129$+$0204 & $-0.16 \pm 0.06$ & $-0.26 \pm 0.05$ & $-0.01 \pm 0.05$ & 0.29 $\pm$ 0.06 & 0.5 $\pm$ 0.5 &  0.2 $\pm$ 0.5 & 0.5 $\pm$ 0.5\\
	J1203$+$1028 & $-0.27 \pm 0.06$ & $-0.08 \pm 0.06$ & $<-1.08$ & 0.64 $\pm$ 0.07 & $-$ & $-$ & $-$\\
	J1430$+$0149 & $-0.51 \pm 0.06$ & $-$ & $-$ & $-$ & $-$ & $-0.11 \pm 0.05$ & 0.34 $\pm$ 0.05\\
	\hline
\end{tabular}

\medskip
$^1$ Abundance ratios for this absorber based only on regions in
velocity space where both transitions are detectable; in particular
the \FeII\ transitions cover a much broader region of velocity space
than most other transitions.

\end{minipage}
\end{table*}


Table~\ref{tab:abundance_ratios} also shows several abundance ratios
sensitive to dust depletion. It is evident that the depletion
properties of these absorbers differ slightly from local measurements
of the MW, LMC and SMC
\citep{Savage:1996p64,Welty:1999p1542,Welty:2001p1556}; whilst the
      [Zn/Fe] ratios in these \CaII\ absorbers are similar to the warm
      halo phase in either the MW, LMC or SMC, the [Si/Zn] ratios are
      $\sim0.5\,\rm{dex}$ higher than typically observed locally.

\subsubsection{Depletion uniformity}\label{sec:res_dep_prof}
Using the apparent optical depth method detailed in
Section~\ref{sec:anal_AOD}, we calculate the variation in [Cr/Zn]
across the velocity profile in each absorber for which both \ZnII\ and
\CrII\ are detectable. The results are shown in
Fig.~\ref{fig:dep_prof}.

\setcounter{subfigure}{0}
\begin{figure*}
  \centerline{
  \begin{tabular}{cc}
  \subfigure[tight][The depletion profile across the $z=0.59760$ absorber towards J0334$-$0711.]{ 
    \includegraphics[scale=0.5,trim=0  0 0 0,angle=0,clip=true]{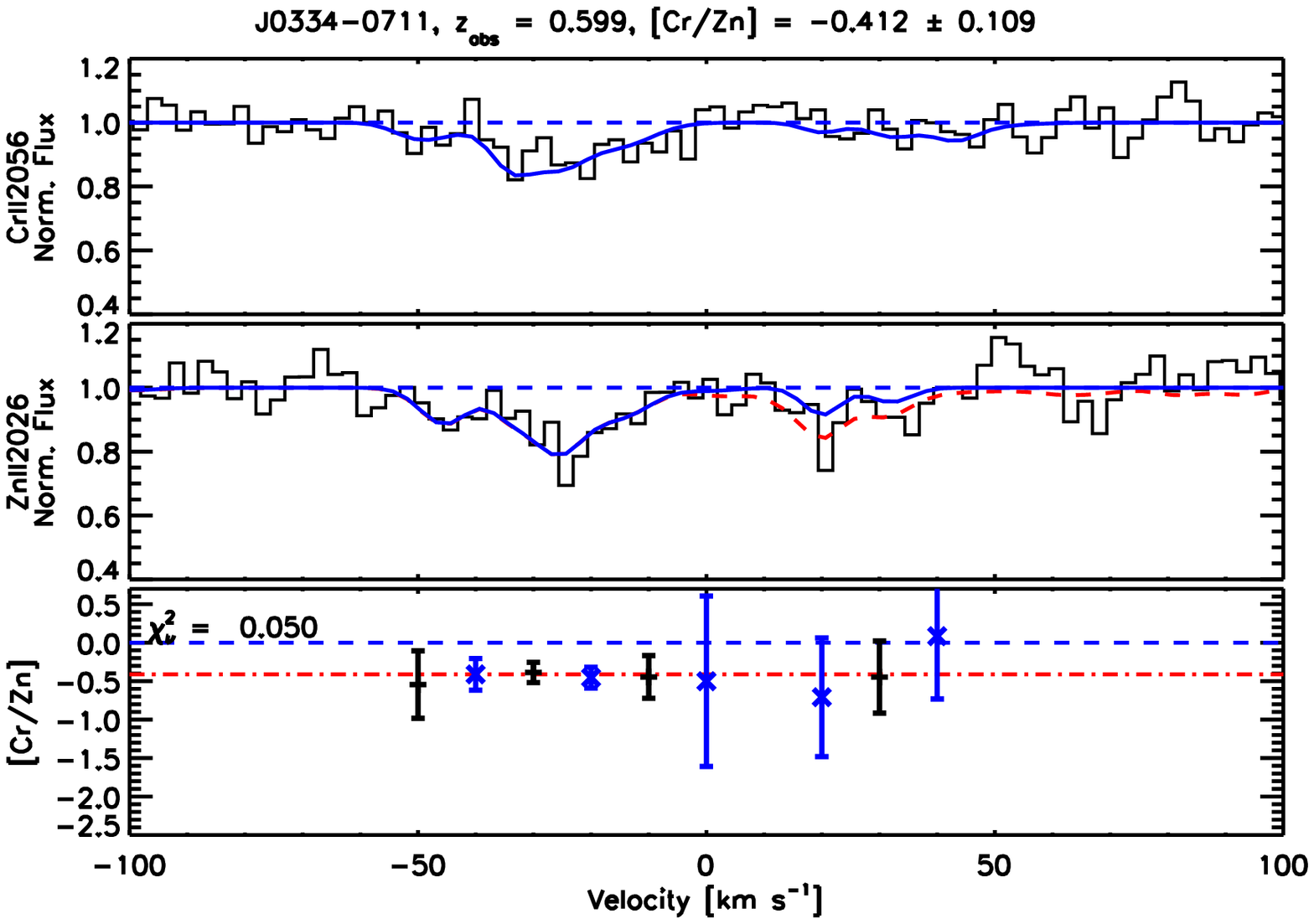}
    \label{fig:dep_prof_J0334}
  } & 
  \subfigure[tight][The depletion profile across the $z=1.02316$ absorber towards J0953$+$0801. The velocity profile for this absorber is too narrow to measure variation in $\rm{[Cr/Zn]}$.]{
    \includegraphics[scale=0.5,trim=0 0 0 0,angle=0,clip=true]{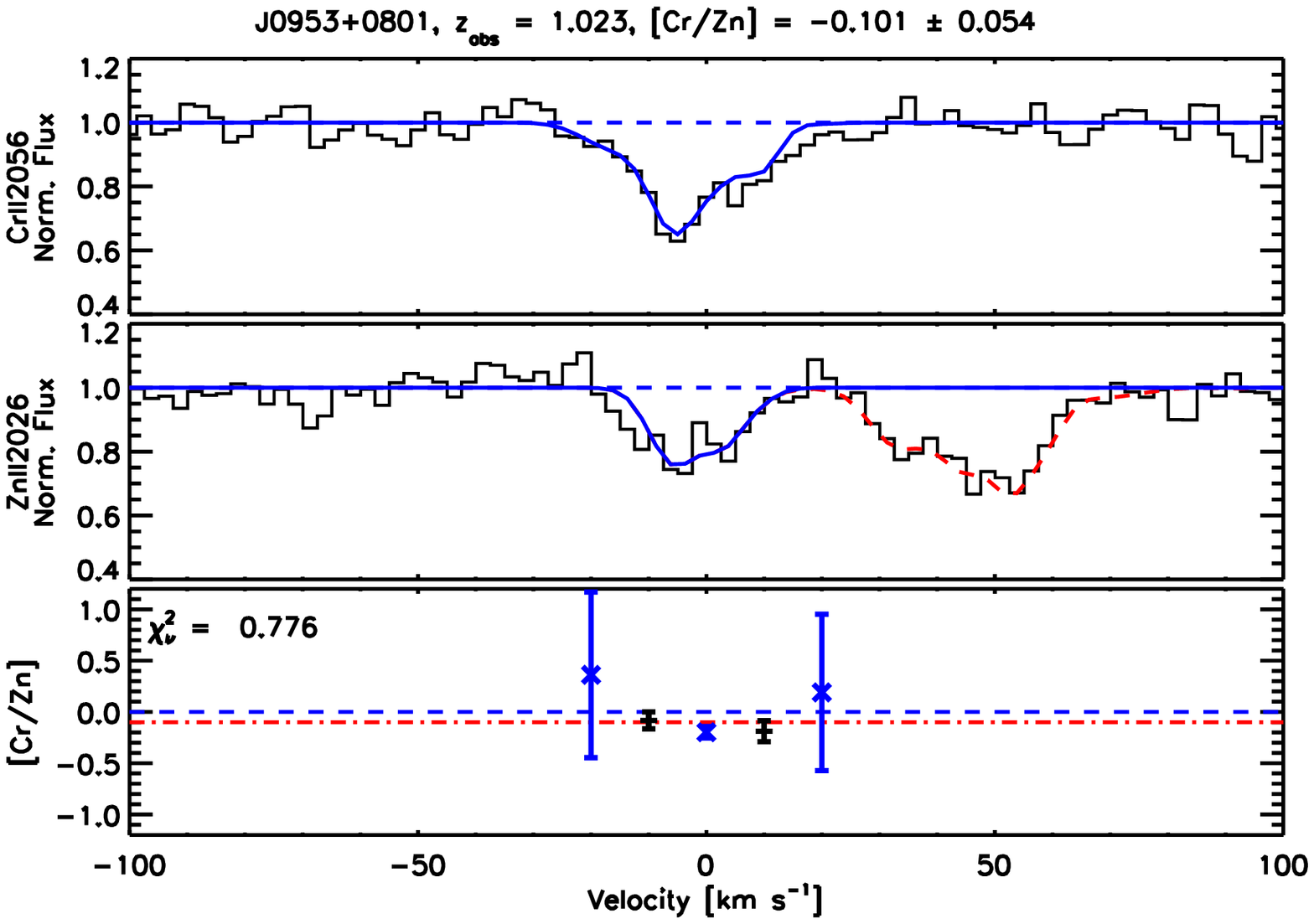}
    \label{fig:dep_prof_J0953}
  } \\
  \subfigure[tight][The depletion profile across the $z=1.14955$ absorber towards J0517$-$4410. There is evidence for significant variation, $\sim0.3\,\rm{dex} \left(3\sigma\right)$, in the absorber's $\rm{[Cr/Zn]}$ profile, consistent with findings of previous works (e.g. \citealt{Quast:2008p443}).]{
    \includegraphics[scale=0.5,trim=0 0 0 0,angle=0,clip=true]{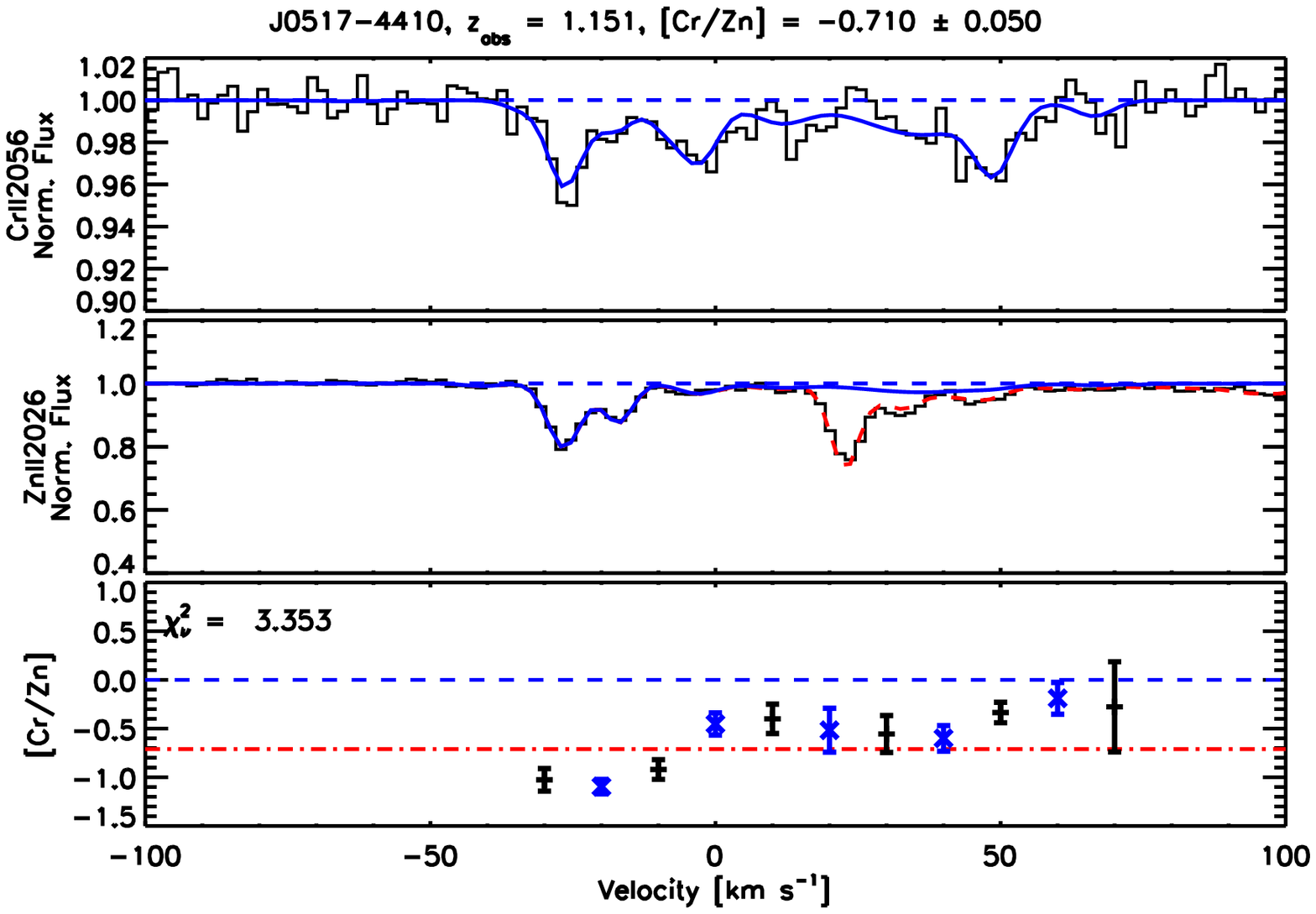}
    \label{fig:dep_prof_J0517}
  } &
  \subfigure[tight][The depletion profile across the $z=0.96497$ absorber towards J1129$+$0204. \MgI$\,\lambda2852$ was not observed so the profile fits for \ZnII\ and \CrII\ are less certain, making it difficult to assess profile uniformity.]{
    \includegraphics[scale=0.5,trim=0 0 0 0,angle=0,clip=true]{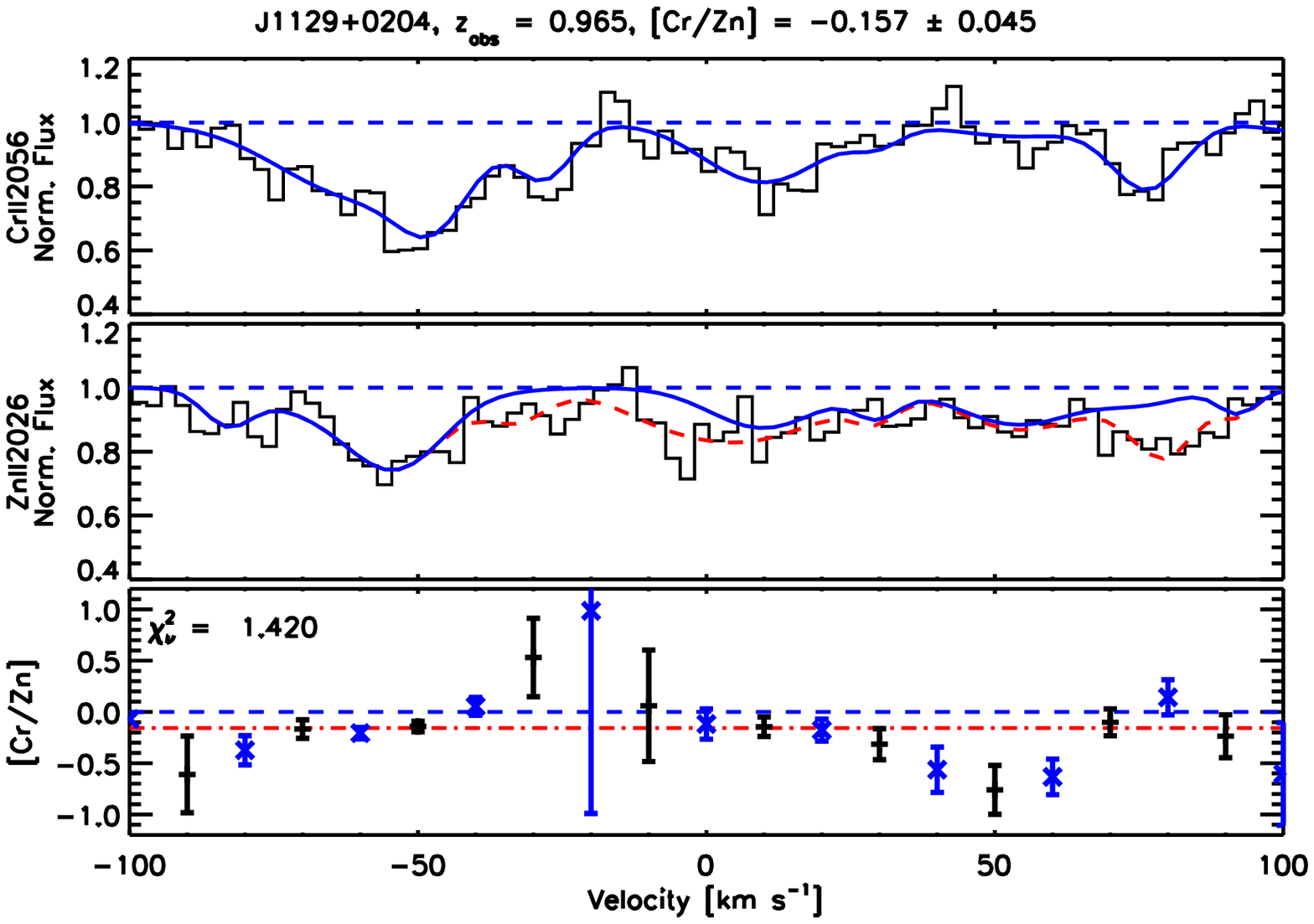}
    \label{fig:dep_prof_J1129}
  }
  \end{tabular}
}
\caption{Each top panel shows the \CrII$\,\lambda2056$ transition, each
  middle panel shows the blended \ZnII$\,\lambda2026$ and
  \MgI$\,\lambda2026$ transitions. The flux is shown as black
  histograms, the continuum as a dashed line (blue) and the fit to the
  \CrII\ or \ZnII\ transition is shown as a solid line (blue). The
  dashed line (red) through the data shows the combined fit to
  \MgI\ and \ZnII. Each bottom panel shows the value of [Cr/Zn] across
  the profile. The short-dashed line (blue) represents the solar
  depletion level whilst the dot-dashed line (red) shows the total
  depletion measured in this absorber as referenced in
  Table~\ref{tab:abundance_ratios} and Section~\ref{sec:tot_dep}. The
  black points are one set of 20$\,\rm{km}\,{\rm s}^{-1}$ bins and the
  blue crosses are a different set of similarly sized bins, offset by
  10$\,\rm{km}\,{\rm s}^{-1}$ from the first. The value of $\chi^2$
  given in the top left-hand corner of the panel is the $\chi^2$ per
  degree of freedom for the fit of the red dot-dashed line to the
  black points. That is, $\chi^2$ represents how well the profile is
  fit by a single depletion value.}
\label{fig:dep_prof}
\end{figure*}
\setcounter{subfigure}{4}
\begin{figure*}
  \centerline{
    \begin{tabular}{c}
    \subfigure[tight][The depletion profile across the $z=0.74030$ absorber towards J1107$+$0048]{
      \includegraphics[scale=0.5,trim=0 0 0 0,angle=0,clip=true]{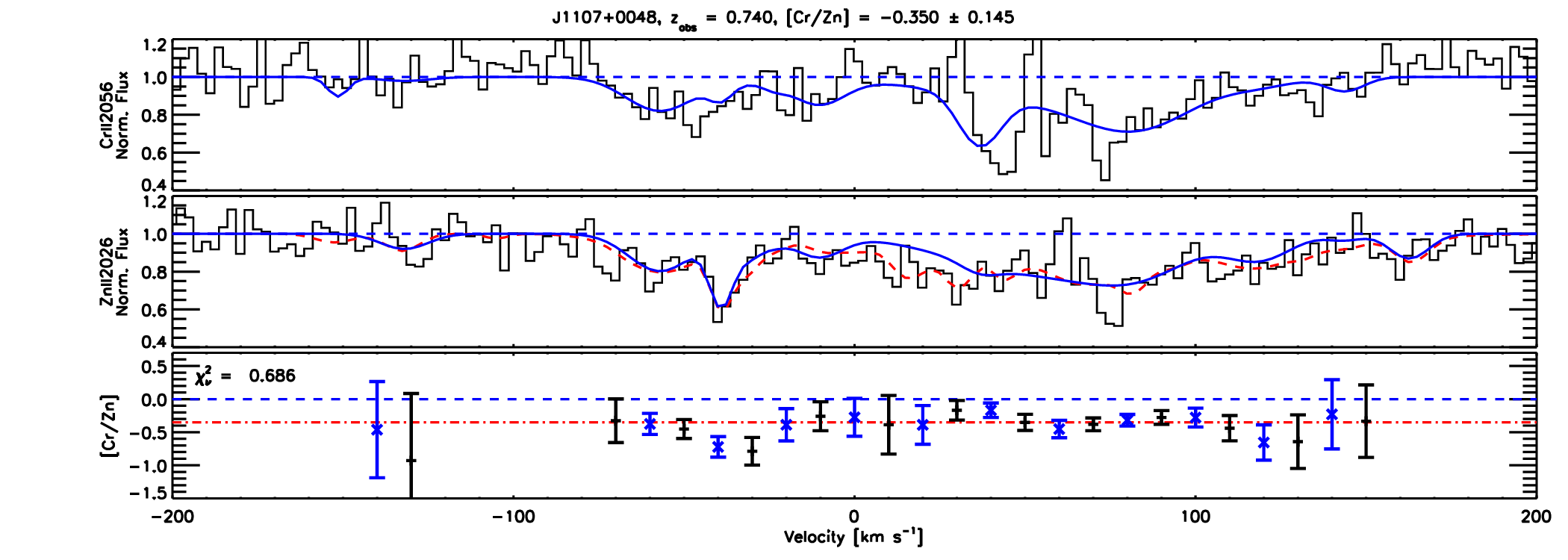}
      \label{fig:dep_prof_J1107}
    }\\
    \subfigure[tight][The depletion profile across the $z=0.74630$ absorber towards J1203$+$1028. There is a blend with a $z=1.322$ \CIV\ absorber at $-110$ and $-130\,\rm{km}\,\rm{s}^{-1}$.]{
      \includegraphics[scale=0.5,trim=0 0 0 0,angle=0,clip=true]{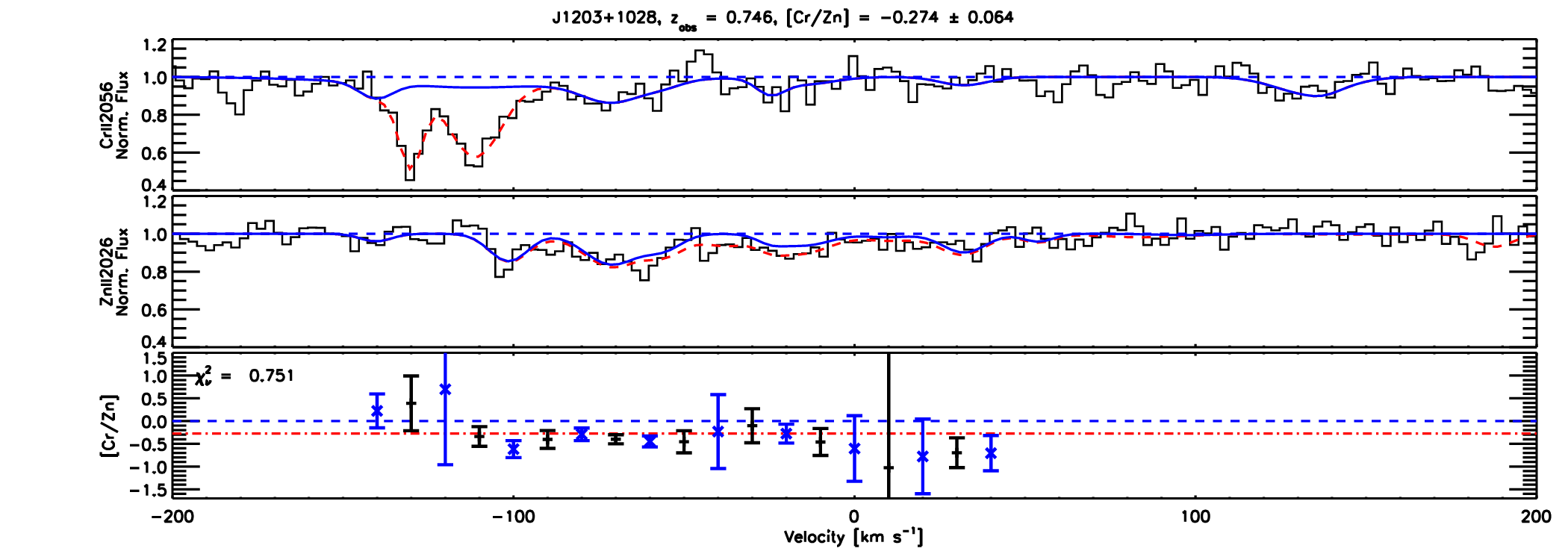}
      \label{fig:dep_prof_J1203}
    }\\
    \subfigure[tight][The depletion profile across the $z=1.24180$ absorber towards J1430$+$0149. There is a blend with a $z=1.933$ \CIV\ absorber at $-120\,\rm{km}\,\rm{s}^{-1}$.]{
      \includegraphics[scale=0.5,trim=0 0 0 0,angle=0,clip=true]{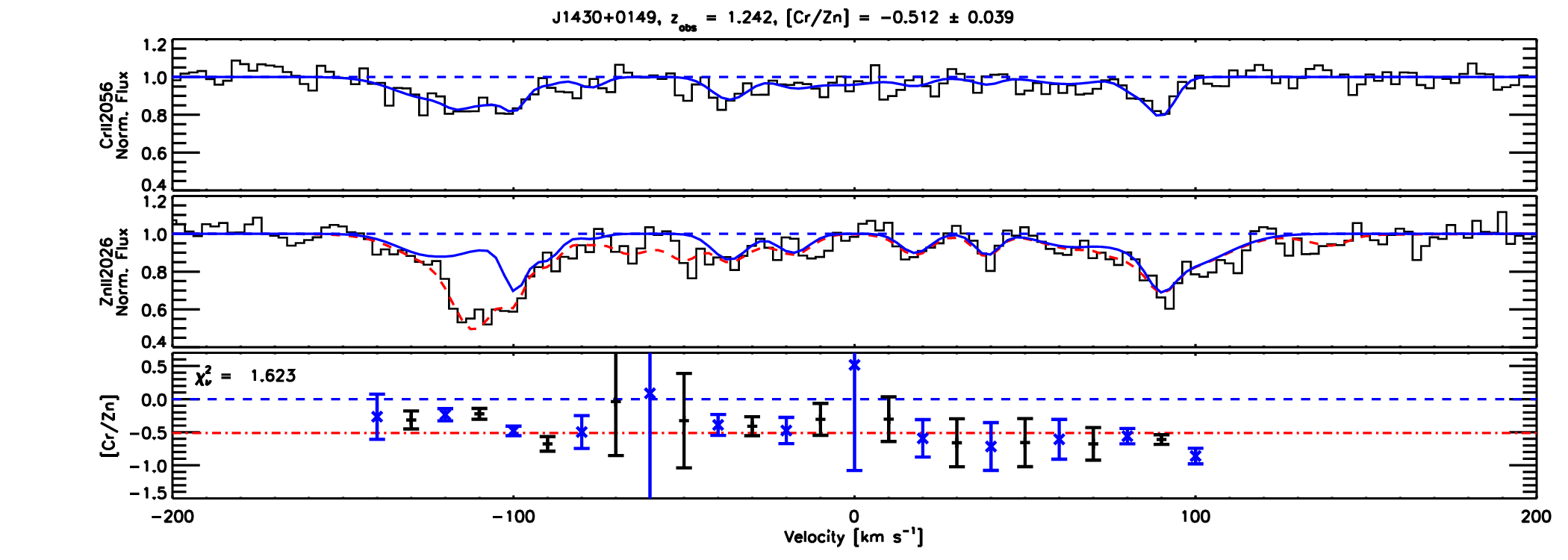}
      \label{fig:dep_prof_J1430}
    }
    \end{tabular}
  }
\contcaption{}
\end{figure*}

There is little evidence for any significant ($>0.3\,\rm{dex}$)
deviation from uniform depletion across the velocity profile in each
of the absorbers, even for those absorbers with reasonably broad
velocity profiles ($>50\,{\rm km}\,{\rm s}^{-1}$; the exception is
J0517$-$4410, to be discussed later). It is not necessarily surprising
that an absorber with complex velocity structure might appear well
mixed because velocity components do necessarily not reflect the
spatial structure of the absorbing gas. $0.3\,\rm{dex}$ variation has
$3\sigma$ significance at the SNRs of the spectra, thus the results in
Fig.~\ref{fig:dep_prof} restricts the range of dust depletion models
which may apply to these \CaII\ absorbers. In many cases it is
possible to restrict the variation to $<0.2\,\rm{dex}$ variation in
[Cr/Zn], i.e.~there is no variation with significance $>2\sigma$ for
most absorbers. See Section~\ref{sec:disc_dep} for further
discussion. We note that the velocity width of J0953$+$0801 is too
narrow for any assessment of variation across its profile with the bin
size necessary for the analysis (See
Fig.~\ref{fig:dep_prof_J0953}). In addition, the \MgI$\,\lambda2026$
contribution to the flux decrement in the region of
\ZnII$\,\lambda2026$ for J1129$+$0204 is uncertain. The uncertainty
arises because the \MgI\ contribution to the flux decrement in the
blended part of the spectrum could not be constrained from the
\MgI$\,\lambda2852$ line, which is not observed. Instead the
\MgI$\,\lambda2026$ line is constrained as the residual from fitting
the \ZnII\ and \CrII\ lines alone. As can be seen from
Fig.~\ref{fig:dep_prof_J1129}, the variation in [Cr/Zn] traces regions
with the most contamination from \MgI. Hence, the possible signature
of variation across the J1129$+$0204 profile in
Fig.~\ref{fig:dep_prof_J1129} cannot be considered robust.

Fig.~\ref{fig:dep_prof} presents two different velocity binnings, each
with the same bin size but offset from each other by
$10\,\rm{km}\,\rm{s}^{-1}$. The similarity between the two binnings
shows that the analysis is resilient to shifts in the bin
centres. Moderate changes, of order $10-20\,\rm{km}\,\rm{s}^{-1}$, to
the size of bins also has little effect on the results, other than
altering the magnitude of the associated error bars.

J0517$-$4410 has substantial variation in [Cr/Zn] across its
$\sim800\,\rm{km}\,\rm{s}^{-1}$ profile. J0517$-$4410 is also a known
sub-DLA with a high UV ionising field \citep{Quast:2008p443}, thus
much of the observed variation is likely due to differential
ionisation of the gas, rather than being purely due to differential
depletion of the gas onto dust grains. Most transitions are only
detected in the central $\sim100\,\rm{km}\,\rm{s}^{-1}$, the strongest
part of the absorber, thus we will concentrate our analysis in this
region (See Fig.~\ref{fig:dep_prof_J0517}). The advantage of
restricting the analysis is that the gas may exhibit less variation
due to ionisation on smaller scales and over this velocity range the
absorption profile is relatively uniform, except for a few velocity
bins near $-25\,\rm{km}\,\rm{s}^{-1}$, which exhibit
$\sim0.3\,\rm{dex}$ ($3\sigma$ significant) deviation.

\subsection{Chemical history}\label{sec:res_chem}

\subsubsection{Enrichment history}
We have detected both $\alpha$-capture elements and Fe-peak elements
in high resolution spectra of these \CaII\ absorbers. $\alpha$
elements such as Si, Ca and Ti\footnote{Ti is not strictly speaking an
  $\alpha$ element, but it exhibits abundance patterns similar to
  other $\alpha $-elements in Galactic stars
  \citep{Edvardsson:1995p75,Francois:2004p613}.} are created by type
II supernovae on a timescale of $< 2\times 10^{7}\,{\rm
  yr}$ after a star formation event, whereas the Fe-peak elements are
created by type Ia supernovae, which occur on a timescale of $10^8 -
10^9\,{\rm yr}$. Thus, the ratio of $\alpha$ to Fe-peak elements can
reveal something about the enrichment history and previous
star-formation in the absorber
\citep[e.g.][]{Wheeler:1989p279,DessaugesZavadsky:2007p1485}. Unfortunately
the same element ratios are also affected by dust-depletion and
disentangling the two effects is not trivial. However, if one observes
enough different transitions it is possible to begin to separate the
two effects \citep[e.g.][]{HerbertFort:2006p105}. Without a
measurement of the metallicity of the absorber, a full deconstruction
of the star-formation history of the absorber is not
possible. Therefore, we follow the example of
\cite{HerbertFort:2006p105}, presenting various abundance ratios which
are instrumental in constraining the effects of dust-depletion and
enrichment history in Table~\ref{tab:abundance_ratios}.

The abundance ratios measured in the sample of \CaII\ absorbers are
generally consistent with the DLA population
\citep{Prochaska:2003p227}. Furthermore, up to $0.33\,\rm{dex}$
variation is to be expected purely from differential dust depletion in
the sample, particularly for elements strongly affected by dust such
as Ti \citep{Savage:1996p64,DessaugesZavadsky:2006p93}.

The Metal-Strong DLAs (MSDLAs) studied by \citet{HerbertFort:2006p105}
represent an unusual subset of DLAs and so we consider whether
\CaII\ absorbers overlap at all with these rare systems, as
similarities between the populations may provide clues as to the
physical origin of the absorbers. To classify an absorber as a MSDLA,
as defined by \citet{HerbertFort:2006p105}, requires $N\left({\rm Zn
  \textsc{ii}}\right) \geq 13.15$ {\em or} $N\left({\rm Si
  \textsc{ii}}\right) \geq 15.95$. Under this definition only two of
the absorbers (J1107$+$0048 and J1129$+$0204) marginally classify as
MSDLAs. Similarly, \citet{Nestor_etal_2008a} measure five of nine
\CaII\ absorbers to be MSDLAs, although they are also all marginal
cases. Therefore, with the caveat that the highest $W_0^{3934}$
systems remain unprobed, the two populations are obviously not
identical. Rather, it seems that the tail end of the \CaII\ metal
column density distribution overlaps with the definition of MSDLAs. If
some strong \CaII\ absorbers arise, not because of sputtering of dust
grains or weak ionisation backgrounds (see introductions of
\citetalias{Wild:2006p405}; \citealp{Zych:2007p392}) but simply
because of large column densities of metals, this would be expected.

\subsubsection{Chemical uniformity}
\begin{figure*}
  \centerline{
    \begin{tabular}{cc}
      \subfigure[tight][The $\rm{[Fe/Si]}$ profile across the $z=1.02316$ absorber towards J0953$+$0801.]{
        \includegraphics[scale=0.5,trim=0 0 0 0,angle=0,clip=true]{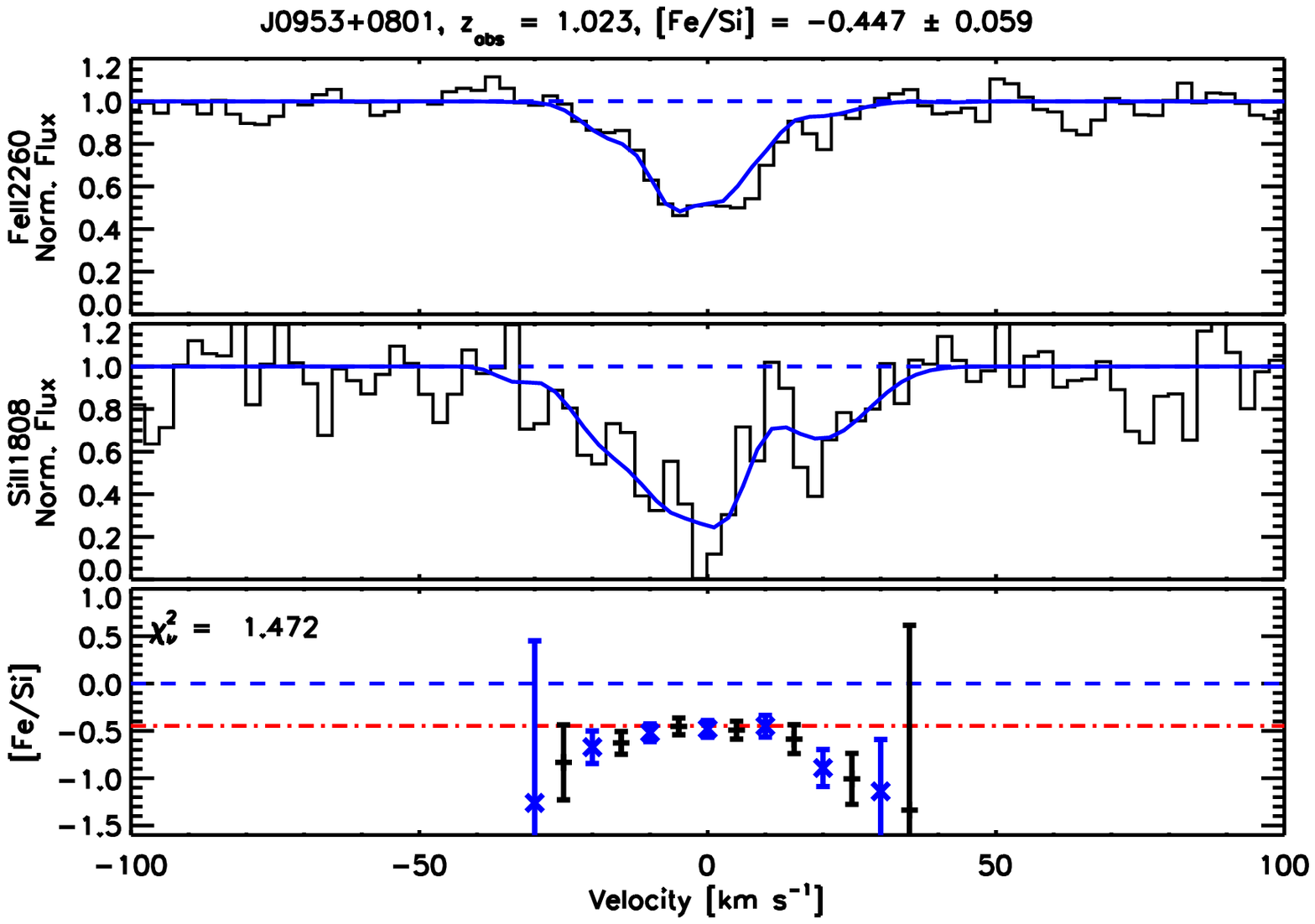}
        \label{fig:chem_prof_J0953}
      } &
      \subfigure[tight][The $\rm{[Fe/Si]}$ profile across the $z=0.74030$ absorber towards J1129$+$0204.]{
        \includegraphics[scale=0.5,trim=0 0 0 0,angle=0,clip=true]{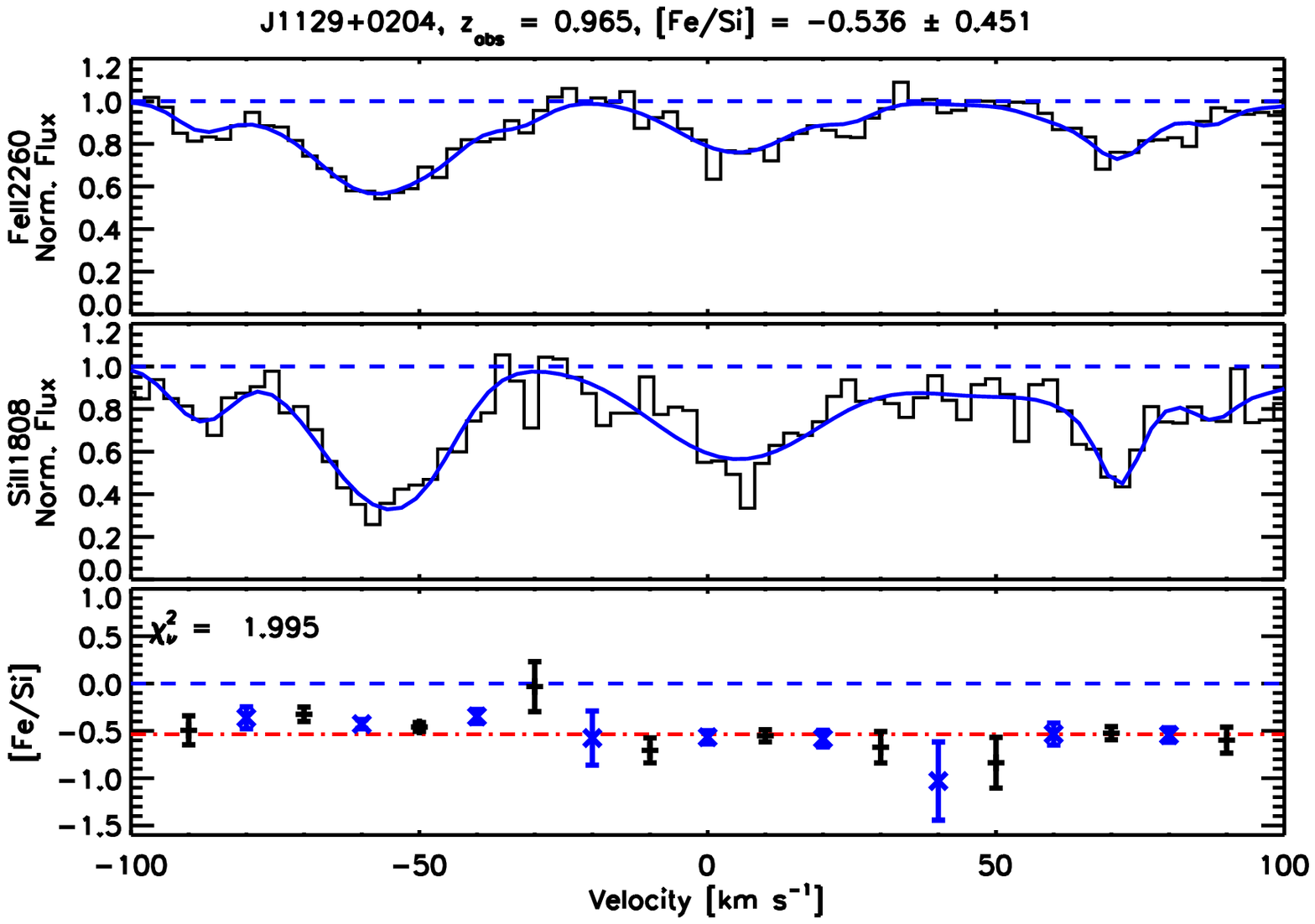}
        \label{fig:chem_prof_J1129}
      }\\
      \multicolumn{2}{c}{
      \subfigure[tight][The $\rm{[Fe/Si]}$ profile across the $z=1.14955$ absorbers towards J0517$-$4410. Note that the leftmost velocity feature seems to have a different chemical history to the rest of the profile. Given the $\rm{[Cr/Zn]}$ profile (see Fig.~\ref{fig:dep_prof_J0517}) shows similar variation, the $\rm{[Fe/Si]}$ variation is likely an effect of dust depletion, rather than differential enrichment history.]{
        \includegraphics[scale=0.5,trim=0 0 0 0,angle=0,clip=true]{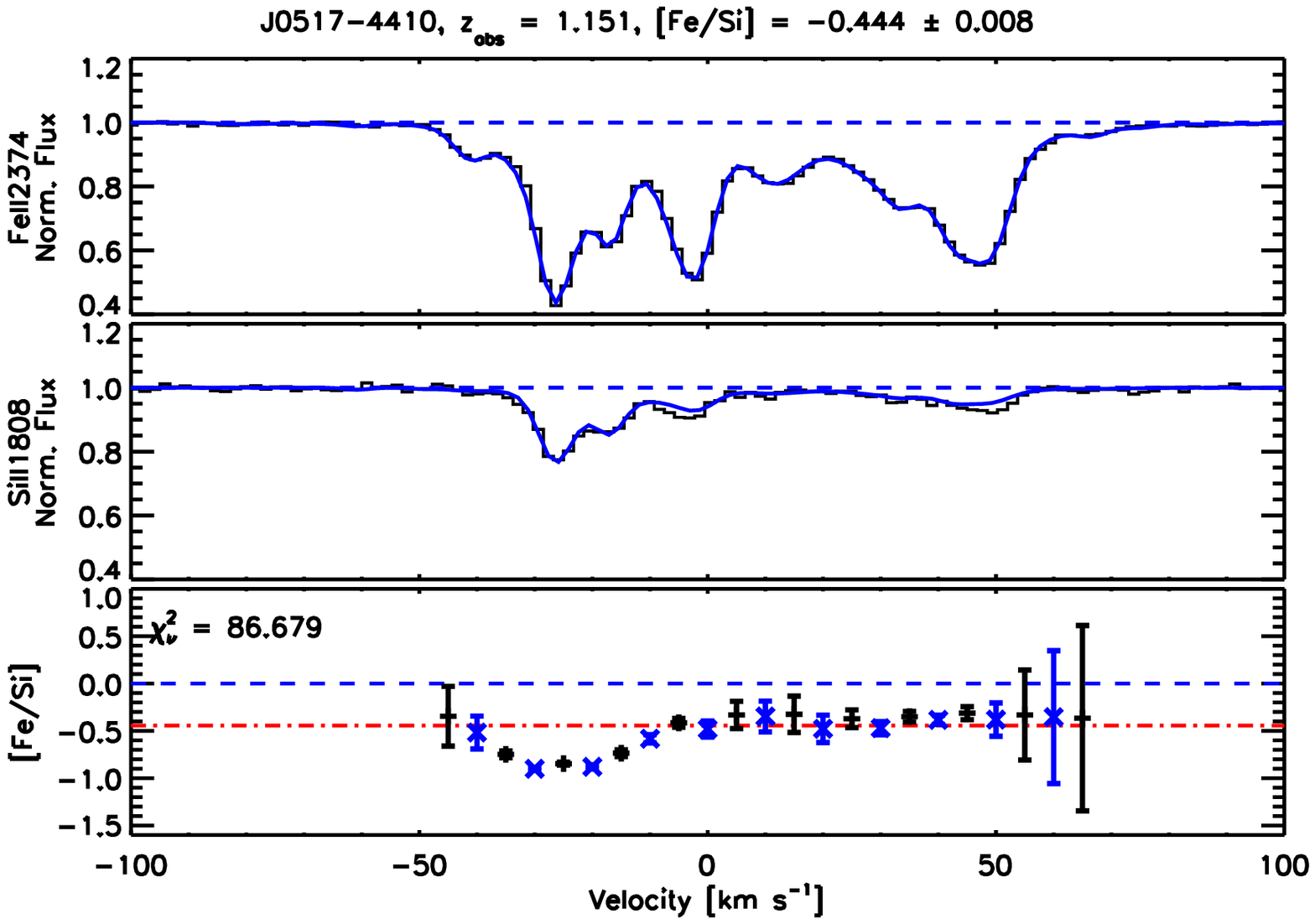}
        \label{fig:chem_prof_J0517}
      }
      }
    \end{tabular}
  }
  \caption{Each top panel shows the \FeII$\,\lambda2250$ transition,
    each middle panel shows the \SiII$\,\lambda1808$ transition. The
    flux is shown as black histograms, the continuum as a dashed line
    (blue) and the fit to the \FeII\ or \SiII\ transition is shown as
    a solid line (blue). Each bottom panel shows the value of [Fe/Si]
    across the profile. The short-dashed line (blue) represents the
    solar abundance ratio whilst the dot-dashed line (red) shows the
    total abundance ratio measured in this absorber as referenced in
    Table~\ref{tab:abundance_ratios} and
    Section~\ref{sec:res_chem}. The black points are one set of
    20$\,\rm{km}\,{\rm s}^{-1}$ bins and the blue crosses are a
    different set of similarly sized bins, offset by
    10$\,\rm{km}\,{\rm s}^{-1}$ from the first. The value of $\chi^2$
    given in the top left-hand corner of the panel is the $\chi^2$ per
    degree of freedom for the fit of the red dot-dashed line to the
    black points. That is, $\chi^2$ represents how well fit the
    profile is by a single abundance ratio value. This absorption
    profile is too narrow to assess any signature of chemical
    non-uniformity.}
  \label{fig:chem_prof}
\end{figure*}

Given the relative uniformity of the depletion profiles, as presented
in Section~\ref{sec:res_dep_prof}, any variation in the [Fe/$\alpha$]
ratio (where $\alpha$ is any $\alpha$-capture element, such as Si, Ti
or Ca) should be indicative of chemical non-uniformity in the
absorbers, rather than variation in dust depletion. For the absorbers
in our sample where we can measure [Fe/$\alpha$] the results are
presented in Fig.~\ref{fig:chem_prof}. Even without metallicity
measures it is possible to see abundance variations across the
profile, which may indicate that different components of the gas have
different enrichment histories. In the case that there is no similar
variation in the [Cr/Zn] ratio at the same velocities then the
variation is most likely due to $\alpha$-enhancement.

\setcounter{subfigure}{0}

The use of the unsaturated, unblended \FeII\ and \SiII\ lines in this
analysis also allows us to test the validity of using the fitted
absorption profile, rather than the data themselves, to calculate the
blended [Cr/Zn] depletion profiles. We observe no significant
$\ll0.1\,\rm{dex}$ difference to the derived [Fe/Si] ratios either
when analysing the Voigt profiles fits or when analysing the data
directly.

Only three absorbers have sufficient wavelength coverage to study
their [Fe/Si] ratios and we comment on each in turn. J0953$+$0801 is
not broad enough in velocity space to show significant variation
across its profile. J1129$+$0204 shows a $\sim 0.4\,\rm{dex}$
($3\sigma$) deficit of Si toward the left hand side of its profile
(See Fig.~\ref{fig:chem_prof_J1129}), however it has questionable
significance; see discussion in
Section~\ref{sec:disc_chem}. J0517$-$4410 also shows $\gtrsim
0.8\,\rm{dex}$ ($8\sigma$) variation in its [Fe/Si] profile, but this
variation is mirrored by its [Cr/Zn] profile (See
Fig.~\ref{fig:dep_prof_J0517}), which makes it likely the variation is
a further signature of differential dust depletion, rather than
$\alpha$ enhancement.

\subsection{Electron densities}\label{sec:res_ne}
Given the wavelength coverage of the spectra in our sample, there are
two sets of transitions which can be used to constrain the electron
density in each of our absorbers: \FeI$\,\lambda2484$ \&
\FeII$\,\lambda\lambda2260,2374$ and \MgI$\,\lambda2852$ \&
\MgII$\,\lambda2803$. It is not possible to measure an exact ratio for
either set of transitions, just limits. \FeI\ is not detected at the
SNR of the spectra, thus the measured $N\left({\rm Fe
  \textsc{i}}\right)$/$N\left({\rm Fe \textsc{ii}}\right)$ ratio is an
upper limit. \MgII\ is saturated, thus the measured $N\left({\rm Mg
  \textsc{i}}\right)$/$N\left({\rm Mg \textsc{ii}}\right)$ ratio is
also an upper limit. Both ratios are derived using column densities
measured via the AOD method (See Table~\ref{tab:ion_ratios} for
results).

\begin{table}
\caption{The column density ratios of neutral-to-singly ionised species in the \CaII\ absorber sample.}
\label{tab:ion_ratios}
\begin{tabular}{lcc}
\hline
Object & $N\left({\rm Fe \textsc{i}}\right)$/$N\left({\rm Fe \textsc{ii}}\right)$ & $N\left({\rm Mg \textsc{i}}\right)$/$N\left({\rm Mg \textsc{ii}}\right)$\\
\hline
J0004$-$4157 &   $<$0.0004 &  $<$0.0003\\
J0256$+$0110 &   $<$0.0008 &  $<$0.0012\\
J0334$-$0711 &   $-$              &  $<$0.0018\\
J0407$-$4410 &   $<$0.0003 &  $<$0.0008\\
J0517$-$4410 &   $<$0.0007 &  $<$0.0003\\
J0830$+$2410 &   $<$0.0010 &  $-$\\
J0846$+$0529 &   $<$0.0041  &  $<$0.0103\\
J0953$+$0801 &   $<$0.0013  &  $<$0.2063\\
J1005$+$1157 &   $-$              &  $<$0.0037\\
J1028$-$0100 &   $<$0.0018  &  $-$\\
J1107$+$0048 &   $<$0.0009 &  $<$0.0013\\
J1129$+$0204 &   $<$0.0005 &  $-$\\
J1130$-$1449 &   $<$0.0007 &  $<$0.0016\\
J1203$+$1028 &   $<$0.0159   &  $<$0.0023\\
J1211$+$1030 &   $<$0.0009 &  $-$\\
J1232$-$0224 &   $<$0.0029  &  $<$0.0017\\
J1323$-$0021 &   $<$0.0006 &  $<$0.0088\\
J1430$+$0149 &   $-$              &  $<$0.0015\\
J2328$+$0022 &   $<$0.0011  &  $<$0.0017\\
\hline

\end{tabular}

\end{table}

\begin{figure}
  \includegraphics[scale=0.5,trim=10 0 0 0,angle=0,clip=true]{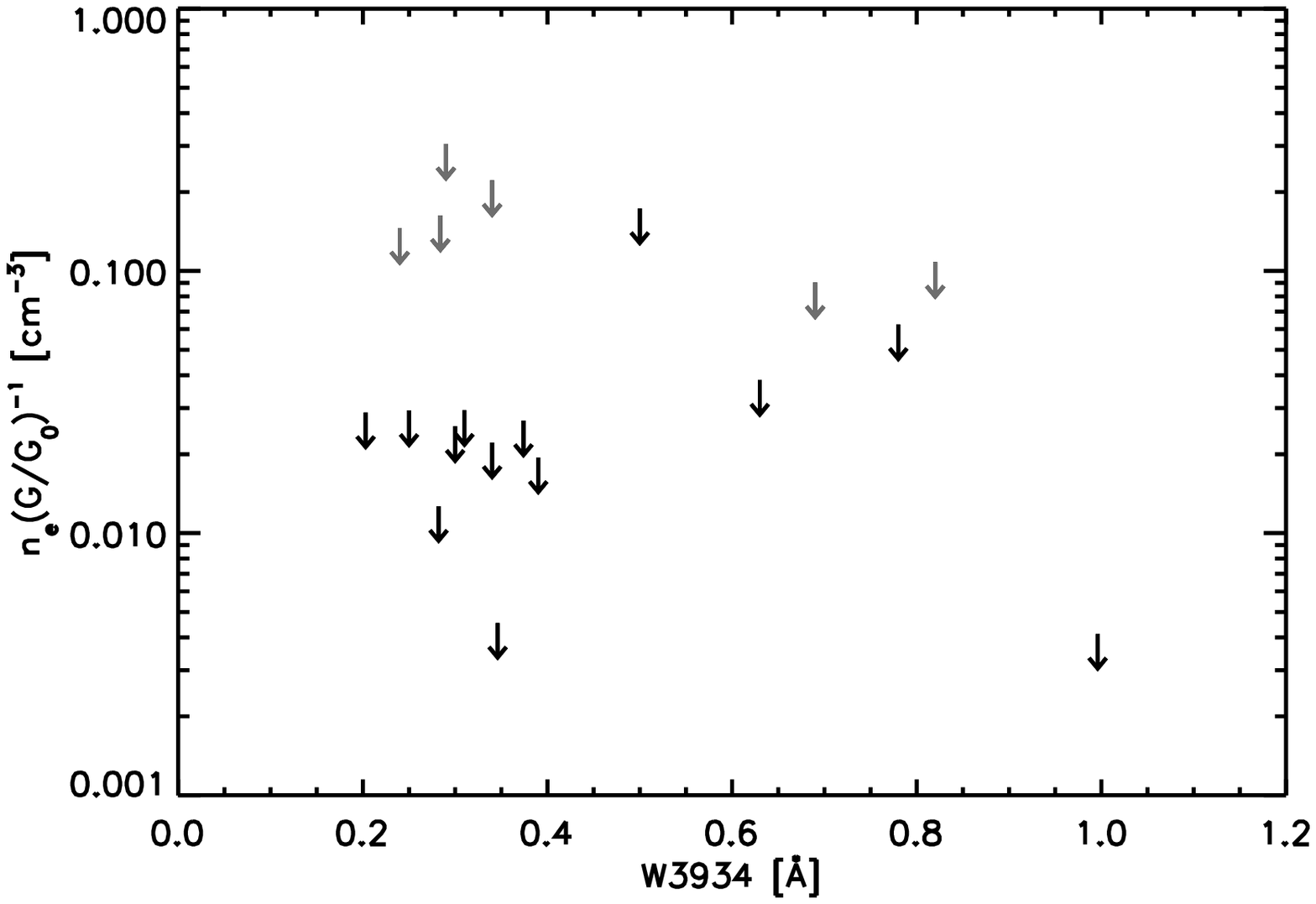}
  \caption{Upper limits on the product of the electron density $n_e$
    and the near-UV radiation field $G$ scaled to $1.7\times$ Habing's
    constant $G_0 = 2.72 \times 10^{-3} \, {\rm erg \, cm^{-2} \,
      s^{-1}} $, which roughly corresponds to the Galactic far-UV
    intensity \citep{Habing:1968p421,Gondhalekar:1980p272}.  The
    limits are established by our observed upper limits to the
    $N\left({\rm Fe \textsc{i}}\right)$/$N\left({\rm Fe
        \textsc{ii}}\right)$ and $N\left({\rm Mg
        \textsc{i}}\right)$/$N\left({\rm Mg \textsc{ii}}\right)$
    ratios under the conservative assumption that the electron
    temperature is $T_e = 8000$K. Limits based on Mg are black, whilst
    those based on Fe are grey. The results indicate the \CaII\
    absorbers are not comprised of extremely dense material.}
\label{fig:ne}
\end{figure}

Using the AOD method ensures that the column densities for each
transition are measured over the same velocity range, providing robust
limits. For saturated pixels a minimum normalised flux is assumed of
0.05 or $\sigma\left({\rm flux}\right)/5$, whichever is greater; this
provides a conservative lower limit to the optical depth. For each
absorber we derive the product $n_e\left(G/G_0\right)^{-1}$ using a
conservative estimate for the temperature, $T=8000\,\rm{K}$, which
maximises the recombination rates. Fig.~\ref{fig:ne} shows
$n_e\left(G/G_0\right)^{-1}$ versus $W_0^{3934}$ for each absorber. We
take the limit from Fe or Mg, whichever provides the greatest
constraint on $n_e$ in each case.
Fig.~\ref{fig:ne} indicates that $n_e < 0.1\,{\rm cm}^{-3}$ in
\CaII\ absorbers if we assume $G/G_0\sim1$, as in the Milky Way. Thus,
\CaII\ absorbers are not comprised of extremely dense material since
$n_{\rm H}\sim1\,{\rm atom}\,\rm{cm}^{-3}$, unless the gas is
extremely neutral ($n_{\rm H^+}/n_{\rm H}\lesssim0.01$). Here we
assume that the free electrons balance the ionized hydrogen,
$n_e/\left(n_{\rm H^+}/n_{\rm H}\right) = n_{\rm H}$.

\subsection{Absorber environment}
\begin{figure*}
\vbox{
  \hbox{
    \includegraphics[scale=0.395,trim=0 27 0 5,angle=0,clip=true]{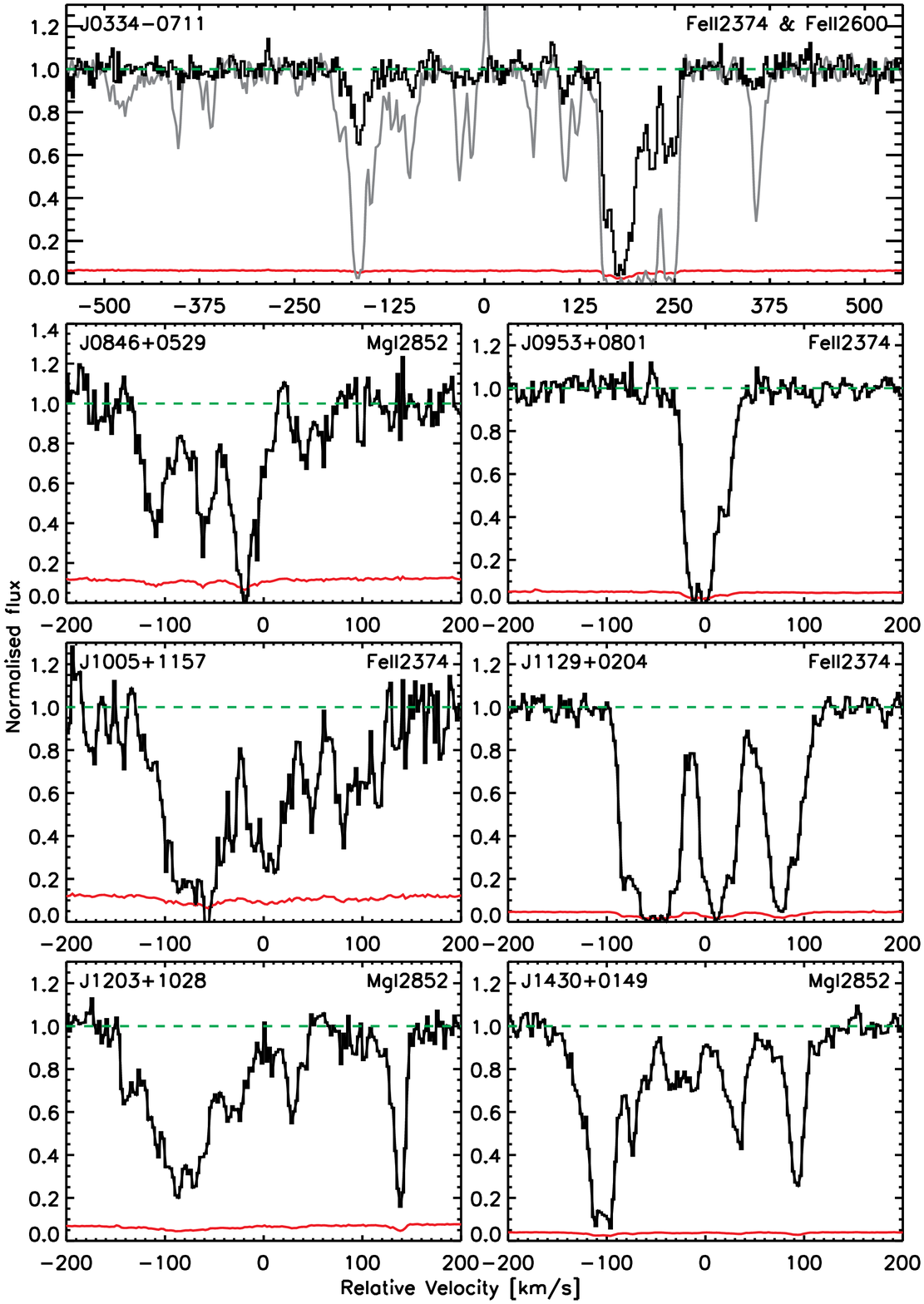}
    \includegraphics[scale=0.395,trim=0 27 0 5,angle=0,clip=true]{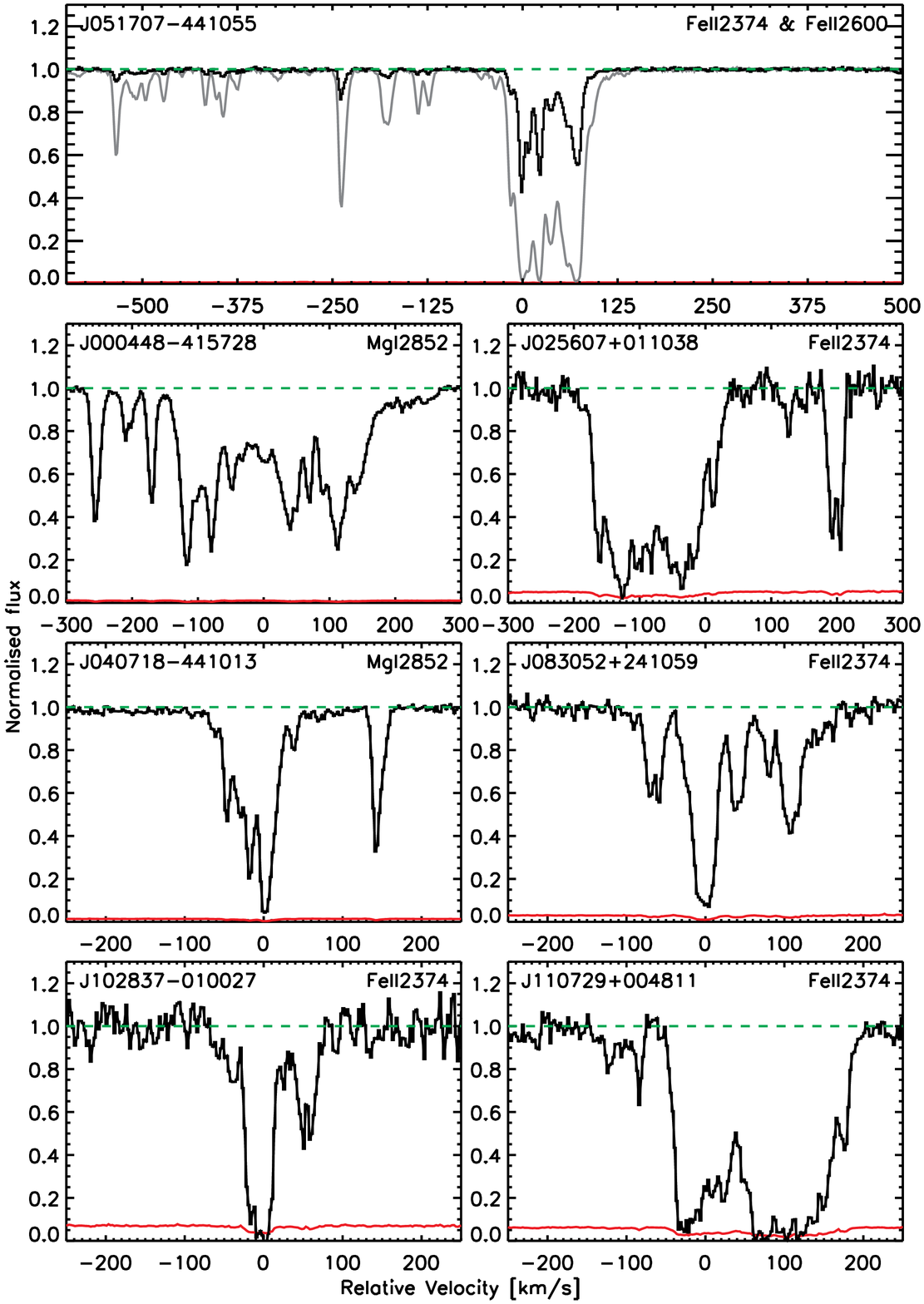}
  }
  \hbox{
    \includegraphics[scale=0.395,trim=0 200 0 200,angle=0,clip=true]{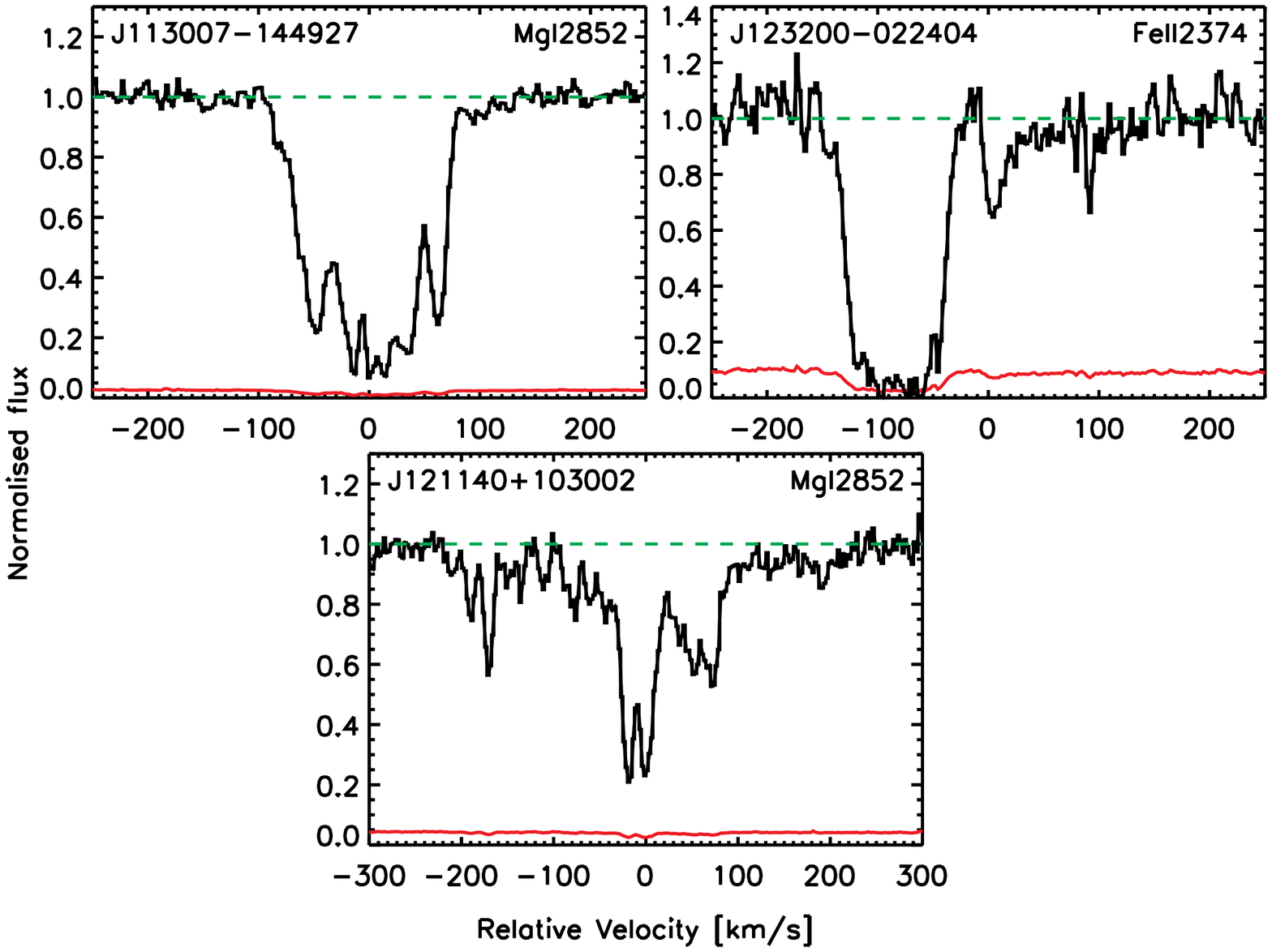}
    \includegraphics[scale=0.395,trim=0 200 0 200,angle=0,clip=true]{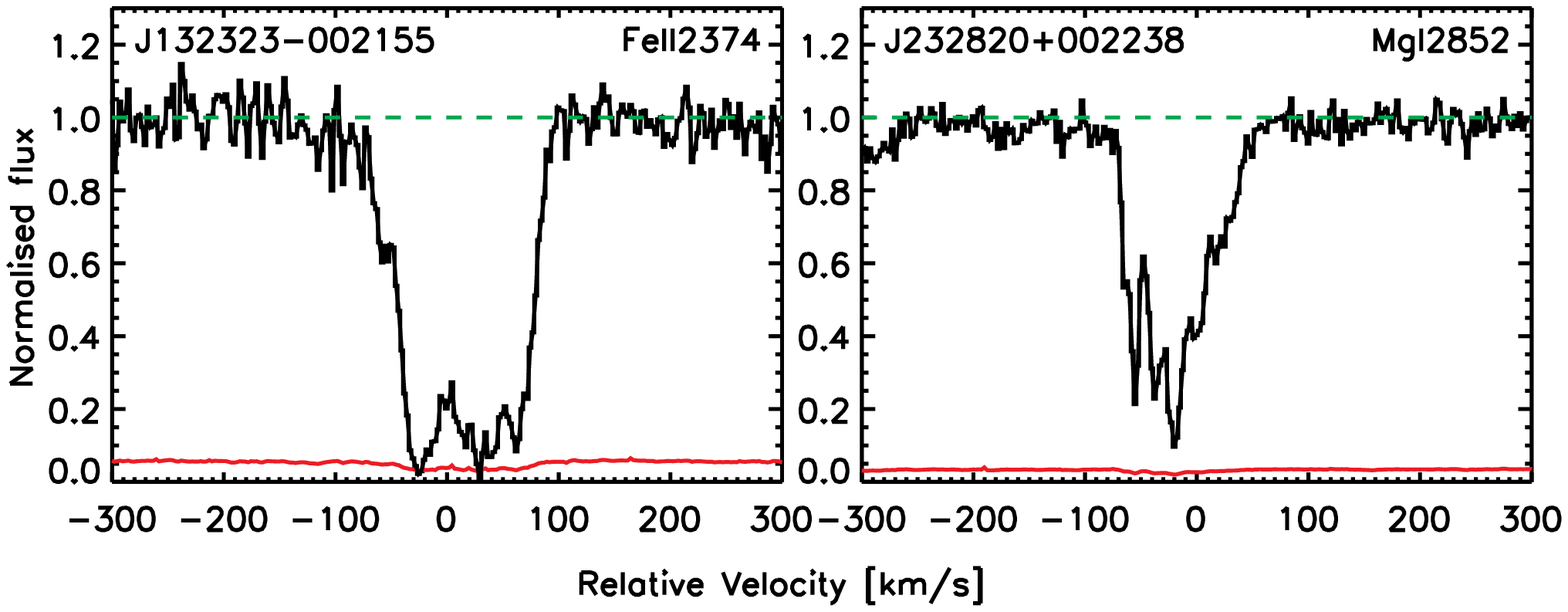}
  }
}
\caption{The velocity structure in each of the absorbers, as
  represented by an unsaturated, unblended absorption line, identified
  in the top right-hand corner of each plot. The absorber's quasar is
  identified in the top left-hand corner of each plot. The black
  histograms are the data and the red histograms are the error, whilst
  the continuum is the green short-dashed line. For the highly complex
  absorbers towards J0334$-$0711 and J0517$-$4410 we show a second
  transition with a different $f$-value in grey to reveal more of the
  structure. }
\label{fig:vel_structure}
\end{figure*}

As detailed in Section~\ref{sec:anal_env}, modelling absorbers as
rotation discs or outflows, etc.~allows us to predict the velocity
structure of such gas. Thus, it is possible to constrain the physical
environment of the absorber by ruling out mechanisms which cannot
reproduce observed velocity structures. The velocity profiles of the
\CaII\ absorbers are shown in Fig.~\ref{fig:vel_structure}. Here we
present any unsaturated transition with high SNR, such as
\MgI$\,\lambda2852$ or \FeII$\,\lambda2374$ from each absorber for
comparison. There are a wide range of different profiles associated
with \CaII\ absorbers, so it is unlikely that a single physical
process gives rise to the strong absorption; see
Section~\ref{sec:disc_env} for further discussion.

\section{Discussion}\label{sec:discussion}

\subsection{Depletions}\label{sec:disc_dep}
The \CaII\ absorbers observed appear to have very similar depletion
properties to the general DLA population (See
Fig.~\ref{fig:DLA_depletion}). Whilst this similarity may at first
seem contrary to the results of \citetalias{Wild:2006p405}, who find
\CaII\ absorbers to be amongst the most dusty absorbers known,
Fig.~\ref{fig:depletion} shows the results are consistent. Our
observations only probed the lower $W_0^{3934}$ regime, which
\citetalias{Wild:2006p405} find have much smaller levels of
depletion. The distribution of points in Fig.~\ref{fig:depletion} may
suggest that the average depletions measured by
\citetalias{Wild:2006p405} define a lower envelope to the dust
depletions of \CaII\ absorbers, rather than a simple linear
correlation. That is, low $W_0^{3934}$ absorbers may comprise both
weakly and strongly dust depleted systems, while high $W_0^{3934}$
absorbers may all be strongly dust depleted. Such a situation may
point to two distinct mechanisms producing \CaII\ absorbers.


The seven \CaII\ absorbers for which we were able to measure the dust
depletion properties across the velocity profile have variation
$\lesssim0.3\,\rm{dex}$ (corresponding to $3\sigma$) in [Cr/Zn]. Once
again, these dust depletion properties are consistent with results
from the general DLA population \citep{Prochaska:2003p110}. Given the
uniformity of the dust depletion it is unlikely that any of these
absorbers harbour molecular hydrogen since we would expect the dust
depletion to be concentrated into one or two components
\citep{Ledoux:2003p1351}.

J0517$-$4410 is exceptional in its depletion variation, evident in
both the [Cr/Zn] and [Fe/Si] profiles (See
Figs.~\ref{fig:dep_prof_J0517} \&
\ref{fig:chem_prof_J0517}). Molecular hydrogen has been detected in
this absorber at $z=1.15079$ and $1.15085$ \citep{Reimers:2003p785},
which matches with the velocity bins exhibiting increased dust
depletion. Note that narrow \CI\ absorption lines have also been
detected at the same redshifts \citep{Levshakov:2006p447}. It is
possible that ${\rm H}_2$ traces these neutral components, rather than
dust as measured using \ZnII\ and \CrII. Given that the AOD velocity
bins would smear out the depletion signature, particularly in very
narrow, cold components, the accuracy with which [Cr/Zn] traces ${\rm
  H}_2$ or \CI\ cannot be assessed more precisely here. The abundance
ratios from our analysis of J0517$-$4410 agree with the more detailed
analysis of \citet{Quast:2008p443}. The abundance ratios relative to Zn
in this absorber match very closely the SMC abundances towards Sk 155
(group A) which are consistent with MW halo cloud abundance ratios
\citep{Welty:2001p1556}, except for $\rm{[Si/Zn]} \sim
-0.36\,\rm{dex}$ and $\rm{[Ti/Zn]} < -1.9$, which are somewhat
depleted. The significance of these differences is difficult to
determine because analysis of the [Si/S] ratio along the line of sight
to Sk 155 has shown that Si is more depleted toward Sk 155 than was
apparent \citep{Sofia_etal_2006a}.

The lack of evidence for molecular hydrogen from non-uniform dust
depletion is seemingly at odds with the recent results from
\citet{Nestor_etal_2008a} who use photo-ionisation modeling to show
that \CaII\ absorbers probably have total gas densities $>1\,{\rm
  atom}\,{\rm cm}^{-3}$ -- the regime where we would expect molecular
hydrogen to be present \citep{Srianand:2005p1517}. There is tentative
evidence that stronger $W_0^{3934}$ systems should have larger total
gas densities \citep{Nestor_etal_2008a}, therefore the results
presented here may just reflect the fact we have only probed the lower
$W_0^{3934}$ regime and that it is the stronger $W_0^{3934}$ systems
which are associated with molecular hydrogen. Another possibility is
that we need to reassess some of the assumptions in the
photo-ionisation modeling of \citet{Nestor_etal_2008a}. For instance
\citet{Nestor_etal_2008a} assume that the UV radiation field in the
vicinity of their absorbers is similar to that of the solar
neighbourhood; perhaps a more general range of UV fields needs to be
considered (see Section \ref{sec:disc_ne}).

\subsection{Chemical history}\label{sec:disc_chem}
Detailed analysis of the star formation history in \CaII\ absorbers is
not appropriate given the limitations of the spectra; a wider range of
transitions, both Fe-peak and $\alpha$ elements, spanning a range of
ionisation potentials, as well as observations of the \HI\ column
density to measure metallicities are required to conduct a full
analysis of the star formation history. It is, nonetheless, possible
to use the [$\alpha$/Fe] ratio across the velocity profile to infer
whether the gas in the absorber has multiple star-formation
histories. Only two absorbers, J1129$+$0204 and J0517$-$4410, showed
any variation in [Si/Fe] $\gtrsim 0.3\,\rm{dex}$ ($3\sigma$
significance) across their profiles (see
Figs.~\ref{fig:chem_prof_J1129} \& \ref{fig:chem_prof_J0517}). Given
that dust depletion can also cause variation in [Si/Fe] as well as
differential enrichment history it is important to compare the [Si/Fe]
and [Cr/Zn] profiles to break this degeneracy.

Recall that the dust depletion profile of J1129$+$0204
(Fig.~\ref{fig:dep_prof_J1129}) has $\sim 1\sigma$ significant
variations at the velocities where the [Fe/Si] profile has $\sim
3\sigma$ significant deviation at $-25\,\rm{km}\,\rm{s}^{-1}$. It
appears that the components from $-100$ to $-25\,\rm{km}\,\rm{s}^{-1}$
may be offset in [Fe/Si] from the components from $-5$ to
$110\,\rm{km}\,\rm{s}^{-1}$. The apparent deviation is not significant
given the a posteriori nature of the statistic. Only higher SNR data
where all the individual bins between $-100$ and
$-25\,\rm{km}\,\rm{s}^{-1}$ are $>3\sigma$ from the mean can confirm
whether this deviation is real. Such a deviation would be of interest
because it is indicative of the gas having different star-formation
histories. Possible explanations would include that the absorbing gas
arises from the merging of two galaxies with different star-formation
histories. Or we could be observing chemical inhomogeneities within a
single galaxy; certainly our own Milky Way is observed to have
variations in $\alpha$-enhancement
\citep[e.g.][]{Cunha:2007p710}. Alternatively, we could be witnessing
the building blocks of a galaxy coming together at $z\sim1$. Such
assessment must wait for higher SNR data however.

The [Fe/Si] profile for J0517$-$4410 has $\gtrsim 0.8\,\rm{dex}$
($8\sigma$) deviation near $-25\,\rm{km}\,\rm{s}^{-1}$ from an
otherwise relatively uniform profile. Note, however, that a similar
$\sim 0.3\,\rm{dex}$ ($3\sigma$) deviation is seen in the [Cr/Zn]
profile. Thus it is likely that this is purely a signature of dust
depletion, rather than $\alpha$-enhancement in the absorber. Note that
J0517$-$4410 is a sub-DLA with a strong ionising background
\citep{Quast:2008p443} thus it is possible that $N_{\rm
  Fe\textsc{ii}}/N_{\rm Si\textsc{ii}} \neq N_{\rm Fe}/N_{\rm Si}$ and
some of the observed deviation in [Fe/Si] and [Cr/Zn] could be due to
variable ionisation. In addition, the trend in [Fe/Si] with
$N\left(\rm{Si\textsc{ii}}\right)$ follows what one would expect if
photo-ionisation were causing the variation. Nonetheless, across the
region we have analysed here, the fraction of highly ionised to lowly
ionised gas appears relatively constant \citep[][on-line
  material]{Quast:2008p443}, thus it is likely the variation is due to
dust depletion, rather than ionisation. This argument is strengthened
by the fact that molecular hydrogen and \CI\ are detected at the same
velocities (See Section~\ref{sec:disc_dep}).

\subsection{Electron densities}\label{sec:disc_ne}
In Section \ref{sec:res_ne} we show that the \CaII\ absorbers in our
sample have $n_e<0.1\,\rm{cm}^{-3}$. This relies on the assumption
that the radiation field in these absorbers is similar to or weaker
than the radiation field of the Milky Way. Whilst in a few cases the
radiation field may be greater than this, it seems unlikely to be the
case in most absorbers as a stronger radiation field would lead to
more highly ionized gas, thus this remains a robust upper limit. Note
that the extragalactic UV background is about $10\times$ weaker in the
far UV than the Galactic value \citep{Sujatha:2008p189} thus absorbers
experiencing strong UV radiation fields must be associated with
galaxies emitting a lot of energy in the far UV.

The electron densities and gas densities derived in Section
\ref{sec:res_ne} are within the range predicted by
\citet{Nestor_etal_2008a}. \citet{Srianand:2005p1517} find molecular
hydrogen to be associated with absorbers whose gas densities are
$n_{\rm H}\gtrsim 1\,\rm{atom}\,\rm{cm}^{-3}$, which is similar to the
probable gas densities in the \CaII\ absorber sample derived in
Section \ref{sec:res_ne}, $n_{\rm H}\sim
1\,\rm{atom}\,\rm{cm}^{-3}$. Fig.~\ref{fig:ne} provides little
evidence for greater gas densities at stronger $W_0^{3934}$, as found
by \citet{Nestor_etal_2008a}. However, we can only place limits on
$n_e$, rather than direct measurements so a detailed comparison is not
possible. For instance, if stronger $W_0^{3934}$ absorbers
preferentially arise in stronger radiation fields for a given
ionisation fraction, then Fig.~\ref{fig:ne} would imply greater
$n_{\rm H}$ at higher $W_0^{3934}$.

Three absorbers in our sample, those towards J0256$+$0110,
J1107$+$0048 and J1323$-$0021, overlap with the absorbers modelled by
\citet{Nestor_etal_2008a}. The {\sc cloudy} models predict specific
electron densities for each absorber: 0.050, 0.035, and
$0.058\,\rm{cm}^{-3}$ for J0256$+$0110, J1107$+$0048 and J1323$-$0021,
respectively (D.~B.~Nestor, private communication). Combining the
electron densities measured by \citeauthor{Nestor_etal_2008a} with the
limits on $n_e\left(G/G_0\right)^{-1}$ derived in this paper we place
lower limits on the radiation field strengths of $2.6\times$,
$1.6\times$ and $0.5\times$ $\left(G/G_0\right)$ in the absorbers
towards J0256$+$0110, J1107$+$0048 and J1323$-$0021,
respectively. Further measurements would be required to comment on
whether there is a trend with $W_0^{3934}$ and radiation field
strength, though it is interesting to note that the weakest field is
associated with the absorber with the highest $W_0^{3934}$ of the
three. Note that these are only rough estimates of the field strength
as the electron densities measured by \citeauthor{Nestor_etal_2008a}
depend on the strength of the radiation field used in the {\sc cloudy}
modelling. That is, running the {\sc cloudy} models again with a
different value for $G/G_0$ will result in a different value of $n_e$
returned from the model, which in turn will lead to a new
determination of $G/G_0$.

Generally Mg places stronger constraints on $n_e$ than Fe in the
spectra and, in any case, all the constraints are limits, so it is
difficult to compare them with the measurements of
\citet{Nestor_etal_2008a}. Constraining $n_e$ and hence $n_{\rm H}$
more accurately might be achieved by spectral observations of the
\CIIf\ and \SiIIf\ fine structure lines. However, note that these
transitions constrain $n_e$ more tightly if the gas is predominantly
ionised \citep{Prochaska_1999a}. Higher ionisation state lines such as
\CIV, may arise in distinct regions and therefore not bear too
directly on the \CaII\ absorbing gas.

\subsection{Environment}\label{sec:disc_env}
\begin{figure}
  \includegraphics[scale=0.5,trim=30 10 0 0,angle=0,clip=true]{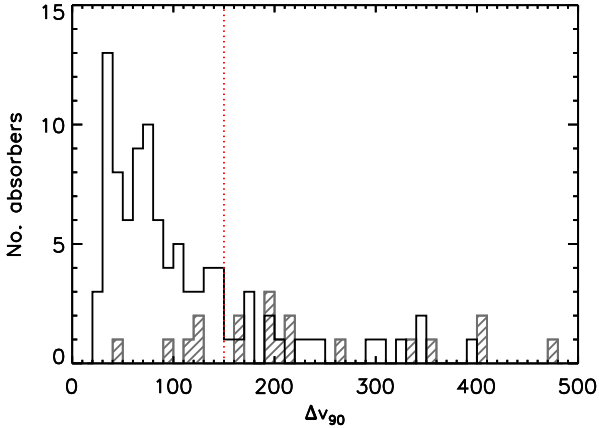}
  \caption{The measured $\Delta v_{90}$ distribution for the DLA
    sample of \citet{Prochaska:2003p227}, shown as the black
    histogram. The $\Delta v_{90}$ distribution for our sample of
    \CaII\ absorbers is also shown as the hashed-grey histogram. The
    number of observed \CaII\ absorbers is too small for a meaningful
    statistical comparison of the two populations. Given the
    $W_0^{3934}$ detection limit for \CaII\ absorbers implies a cut in
    $\Delta v_{90}$, the two absorber populations could have the same
    intrinsic $\Delta v_{90}$ distributions. The approximate cut
    implied in $\Delta v_{90}$ at $150\,\rm{km}\,\rm{s}^{-1}$ is shown
    as a dotted vertical line (See discussion in Section
    \ref{sec:disc_env}).}
  \label{fig:v90_dist}
\end{figure}

Velocity structure by itself gives us limited insight into the true
physical environments hosting \CaII\ absorption as we can reproduce
the velocity profiles of all absorbers via various models in the
literature (Recall Section~\ref{sec:anal_env}). A further consequence
of such models is that it is possible for multiple physical processes
to produce similar velocity profiles. It is even possible to take a
pathological line of sight through an outflow or inflow to reproduce
what looks like a simple disc profile
\citep[e.g.][]{Charlton:1998p1325}. However, it is not possible for a
pure disc model to reproduce an outflow-like profile because not
enough high-velocity components are recovered from the model
\citep[e.g.][]{Kacprzak:2008p922}. At a cursory glance it seems that
the \CaII\ absorber sample has much greater velocity widths than the
general DLA population; that is, they are less disk-like than general
DLAs. Taking a sample of 95 DLAs with $z>1.6$ based on observations
from ESI\footnote{Echelle Spectrograph and Imager on the Keck-II
  telescope in Hawaii}, HIRES and UVES as described in
\cite{Prochaska:2003p227} we can measure their velocity widths using
$\Delta v_{90}$, as described by \cite{Prochaska:1997p73}. We find
that the median velocity width of such as sample is $\left<\Delta
v\right>^{\rm DLA} \simeq 75\,\rm{km}\,\rm{s}^{-1}$, whereas the
\CaII\ absorber sample presented here has a median velocity width,
$\left<\Delta v\right>^{\rm Ca\textsc{ii}} \simeq
200\,\rm{km}\,\rm{s}^{-1}$ (See Table~\ref{tab:EWs} for $\Delta
v_{90}$ measurements). Although we made no conscious decision to
select high velocity width absorbers the \CaII\ sample is, in fact,
biased due to the limit of detecting a $0.2\,\rm{\AA}$ equivalent
width feature at $4\sigma$ significance in SDSS spectra. The exact
velocity cut this implies depends on the depth of the
\CaII\ absorption and because we do not have this information for the
DLA sample it is non-trivial to apply a cut to those data. For
instance, a $0.6\,{\rm \AA}$ equivalent width feature in a spectrum
with SNR$\sim 20$, would require $\Delta v_{90} \sim
120\,\rm{km}\,{s}^{-1}$ to be detectable. Furthermore, if one compares
the distribution of $\Delta v_{90}$ measurements of the two samples
(See Fig.~\ref{fig:v90_dist}) it is evident that the number of DLAs
dramatically increases at $\Delta v_{90} \lesssim
120\,\rm{km}\,\rm{s}^{-1}$. Given the $W_0^{3934}$ detection limit for
\CaII\ absorbers implies a cut in $\Delta v_{90}$ and the small sample
size, the two absorber populations could have the same intrinsic
$\Delta v_{90}$ distributions. Above $150\,\rm{km}\,\rm{s}^{-1}$,
where velocity-selection effects are less important, there is a 78 per
cent chance the two $\Delta v_{90}$ distributions are drawn from the
same underlying distribution, based on a KS-test.

Note that the selection criterion, $W_0^{2796} \geq 0.5\,\rm{\AA}$
implies a velocity cut on the \CaII\ sample. If we conservatively
assume that the \MgII\ absorption is completely saturated, then the
minimum velocity width is $\Delta v_{90}\sim50\,\rm{km}\,\rm{s}^{-1}$,
for our SDSS selected sample. Our \CaII\ absorber sample includes
systems down to this limit. This selection effect will not be as
important as the selection effect due to the $W_0^{3934}$ limit of any
larger sample studied in the future because the bulk of DLAs observed
have $\Delta v_{90}$ greater than $50\,\rm{km}\,\rm{s}^{-1}$.

Irrespective of velocity-selection effects, it is clear there are a
mixture of narrow and wide velocity profiles in the \CaII\ absorber
sample (See Fig~\ref{fig:vel_structure}). The above result is
suggestive that strong \CaII\ absorbers consist of two populations,
one which is due to a line-of-sight passing at low impact parameter
through a quiescent galaxy disc and the other due to a line-of-sight
passing through a complex outflow, merger or galaxy cluster. As a
further point, \CaII\ absorption has been detected in many High
Velocity Clouds (HVCs) in the Milky Way
\citep[e.g.][]{Welty:1996p533,Bekhti:2008p3204}. If one modelled a
galaxy as a disc $+$ HVCs, one would still not be able to reproduce
the high velocity components of the extragalactic \CaII\ absorbers
because the $W_0^{3934}$ of HVCs are too small.

Simulations of DLAs in a cosmological setting
\citep[e.g.][]{Pontzen:2008p1540} significantly under-predict the
relative number of complex absorber profiles compared to simple
disc-like profiles, suggesting that we do not fully understand the
relative importance of different physical mechanisms for producing
absorbers. Nonetheless, most, if not all, \CaII\ absorbers studied
here are {\em not} simple disc-like systems as their \FeII\ absorption
line profiles are too complex (See Fig.~\ref{fig:vel_structure}).

\section{Conclusions}\label{sec:conclusions}
In this work we present the first high resolution study of
\CaII-selected absorption systems, using spectra from the UVES and
HIRES echelle spectrographs. We divided our analysis into three parts,
examining the properties of dust, $\alpha$-enhancement and environment
of the absorbers in turn, finding the following:
\begin{enumerate}
  \item We demonstrate for the first time the feasibility of using AOD
    methods on blended absorption profiles, such as
    \ZnII$\,\lambda2062$ and \CrII$\,\lambda2062$, by utilising the
    Voigt profile fits, rather than the flux, which enables the
    assessment of velocity profile uniformity. We find, as previous
    authors \citep[e.g.][]{Prochaska:2003p110}, that the AOD method is
    robust against binning size and positioning. Furthermore, we show
    there is little difference in applying the AOD to the fluxes
    compared to an absorption profile fit to the spectra, thus
    allowing us to use the method on blended lines. The Voigt profile
    AOD method allows us to combine the power of the more traditional
    Voigt profile fitting and the AOD methods when studying the
    absorber.
  \item From a study of dust depletion across the absorber profile
    using the \ZnII\ and \CrII\ transitions we conclude that the
    depletion in these absorbers is uniform at the $0.3\,\rm{dex}$
    level, which corresponds to deviations $<3\sigma$ in relative
    abundances at the SNR of the spectra. It is, therefore, unlikely
    that these absorbers harbour molecular hydrogen, as when ${\rm
      H}_2$ is present the depletion is concentrated in one or two
    velocity components. The absorbers studied here have low
    $W_0^{3934} \left(<0.7\,\rm{\AA}\right)$, thus higher $W_0^{3934}$
    absorbers may still trace ${\rm H}_2$. These results are
    consistent with \citet{Nestor_etal_2008a}, whose photo-ionisation
    modeling implied that only high $W_0^{3934}$ \CaII\ absorbers
    should be associated with molecular hydrogen. If echelle
    observations of higher $W_0^{3934}$ systems fail to show evidence
    for molecular hydrogen then it may be necessary to reevaluate the
    assumptions behind the modeling of \citet{Nestor_etal_2008a}, such
    as the strength of the ionising UV background and the assumption
    of solar abundance ratios.
  \item J0517$-$4410 stands out from the remainder of the absorber
    sample, exhibiting variation in dust depletion which may map to
    the detection of molecular hydrogen or \CI\ in this absorber. The
    disadvantage of the AOD method is that binning in velocity space
    smears out the signal of dust depletion, precluding identification
    of the velocity components responsible. Thus it may be that the
    narrow \CI\ and ${\rm H}_2$ lines occur at the same velocities,
    whilst the broader \ZnII\ and \CrII\ lines do not correspond to
    the same gas. Narrow components of \ZnII\ or \CrII\ would be
    hidden by the broader absorption profile, so even a detailed Voigt
    profile analysis of this absorber would not illuminate the
    situation further. Furthermore, ionization effects on the measured
    variation in dust depletion across the profile may be important
    \citep{Quast:2008p443}.
  \item There are only a few systems where we can
    study the chemical uniformity of the absorbers by examining their
    [Fe/Si] ratios. J1129$+$0204 shows $\sim0.3\,\rm{dex}$ ($3\sigma$
    significance) variation in its [Fe/Si] profile that is unlikely be
    explained by variable dust depletion, though at the SNR of the
    data this cannot be certain.
  \item The \CaII\ absorbers studied here are not comprised of
    extremely dense material. Assuming a Milky Way-like radiation
    field there electron densities are, $n_e<0.1\,\rm{cm}^{-3}$. Thus,
    unless the gas is very neutral ($n_{\rm H^+}/n_{\rm
      H}\lesssim0.01$), $n_{\rm H}\sim
    1\,\rm{atom}\,\rm{cm}^{-3}$. Comparing the three absorbers which
    overlap between our sample and that of \citet{Nestor_etal_2008a},
    we find a range of UV radiation field strengths are likely in
    these \CaII\ absorbers, from $0.5\times$ to $2.6\times$ the mean
    Galactic field strength.
  \item We conclude that most, if not all, \CaII\ absorbers studied
    here are {\em not} simple disc-like systems because they have
    complex, broad velocity profiles. Simulations only partially
    constrain the astrophysics of the observed velocity profiles of
    absorbers as multiple mechanisms can produce similar velocity
    profiles.  
\end{enumerate}

Whilst this work offers the first opportunity to examine
\CaII\ absorbers at high resolution, many questions as to the origin
of these rare absorbers and the environments of their host galaxies
remain unanswered. For further ground to be made in our understanding
of \CaII\ absorption systems, future studies will require echelle
observations of larger samples, including the, as yet unprobed, higher
$W_0^{3934}$ systems as well as host galaxy Integral Field Unit (IFU)
observations.

\section*{Acknowledgments}
BJZ is supported by the UK Science and Technology Facilities Council
(STFC, formerly PPARC). BJZ thanks the Centre for Astrophysics \&
Supercomputing, Swinburne University of Technology, Melbourne for
their support during his visit. MTM thanks the Australian Research
Council for a QEII Research Fellowship (DP0877998). PCH acknowledges
support from the STFC-funded Galaxy Formation and Evolution programme
at the Institute of Astronomy. JXP acknowledges funding through an NSF
CAREER grant (AST-0548180) and NSF grant (AST-0709235). Thanks to Bob
Carswell (RFC), Gary Ferland, Daniel Nestor and Max Pettini (MP) for
useful discussion. Thanks also to RFC for aid with \VPFIT\ and MP for
the DLA depletions data for Fig.~\ref{fig:DLA_depletion}.

We thank our anonymous referee for helpful feedback, which improved
the flow and consistency of the paper.

This work was based on observations collected at the European Southern
Observatory, Chile, as part of program 79.A-0656 and from data in the
ESO Science Archive.

Some of the data presented herein were obtained at the W.M. Keck
Observatory, which is operated as a scientific partnership among the
California Institute of Technology, the University of California and
the National Aeronautics and Space Administration. The Observatory was
made possible by the generous financial support of the W.M. Keck
Foundation. The authors wish to recognize and acknowledge the very
significant cultural role and reverence that the summit of Mauna Kea
has always had within the indigenous Hawaiian community.  We are most
fortunate to have the opportunity to conduct observations from this
mountain.

Funding for the SDSS and SDSS-II has been provided by the Alfred P.
Sloan Foundation, the Participating Institutions, the National Science
Foundation, the U.S. Department of Energy, the National Aeronautics
and Space Administration, the Japanese Monbukagakusho, the Max Planck
Society, and the Higher Education Funding Council for England. The
SDSS Web Site is \href{http://www.sdss.org/}{http://www.sdss.org/}.



\appendix

\section{Absorber Voigt profile fits}\label{adx:fits}
Here we present the Voigt profile fits to our absorption profiles for
the absorbers and species listed in Table \ref{tab:abundances}.
\begin{figure*}
\vbox{
  \hbox{
    \includegraphics[width=0.33\textwidth,trim=0 15 0 0,angle=0,clip=false]{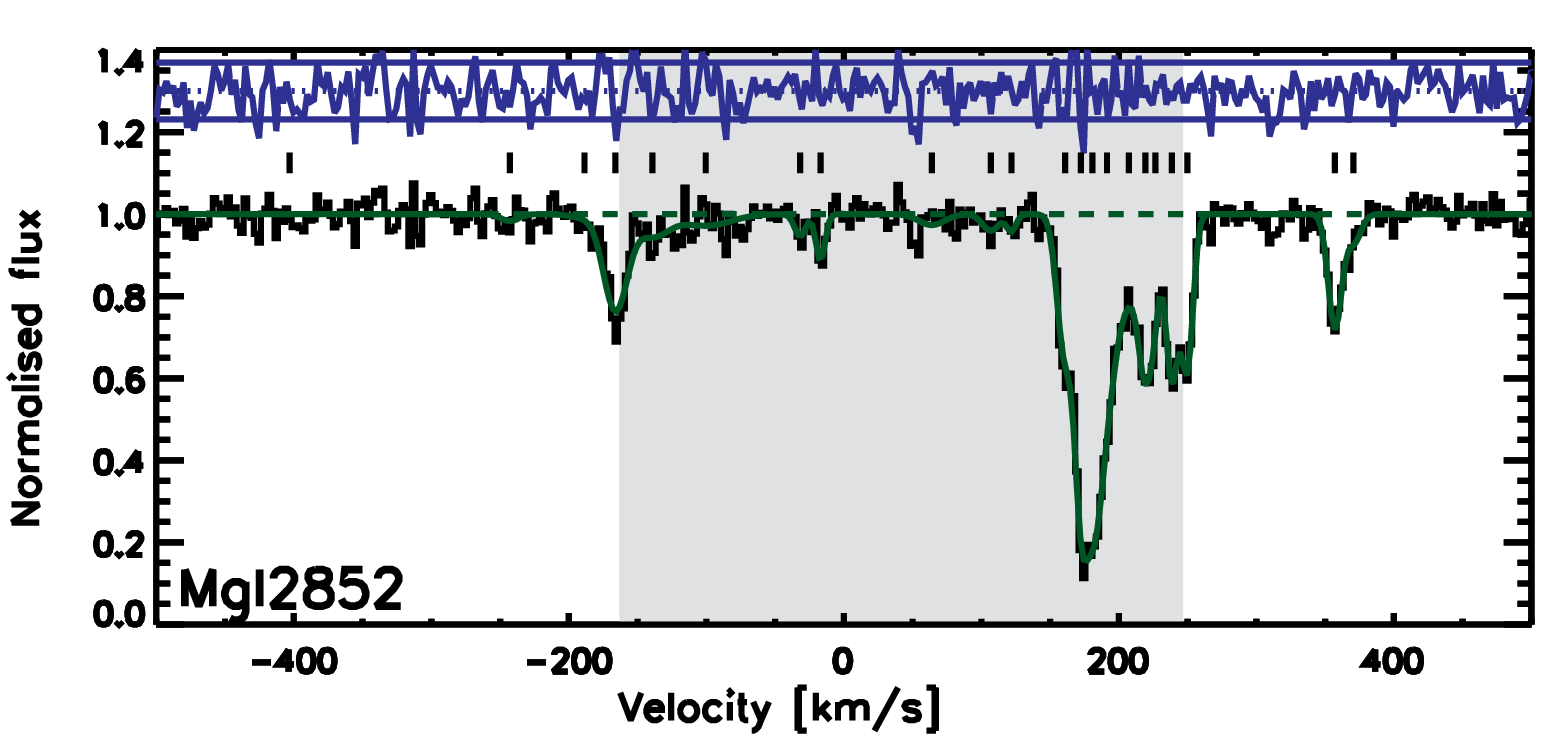}
    \includegraphics[width=0.33\textwidth,trim=0 15 0 0,angle=0,clip=false]{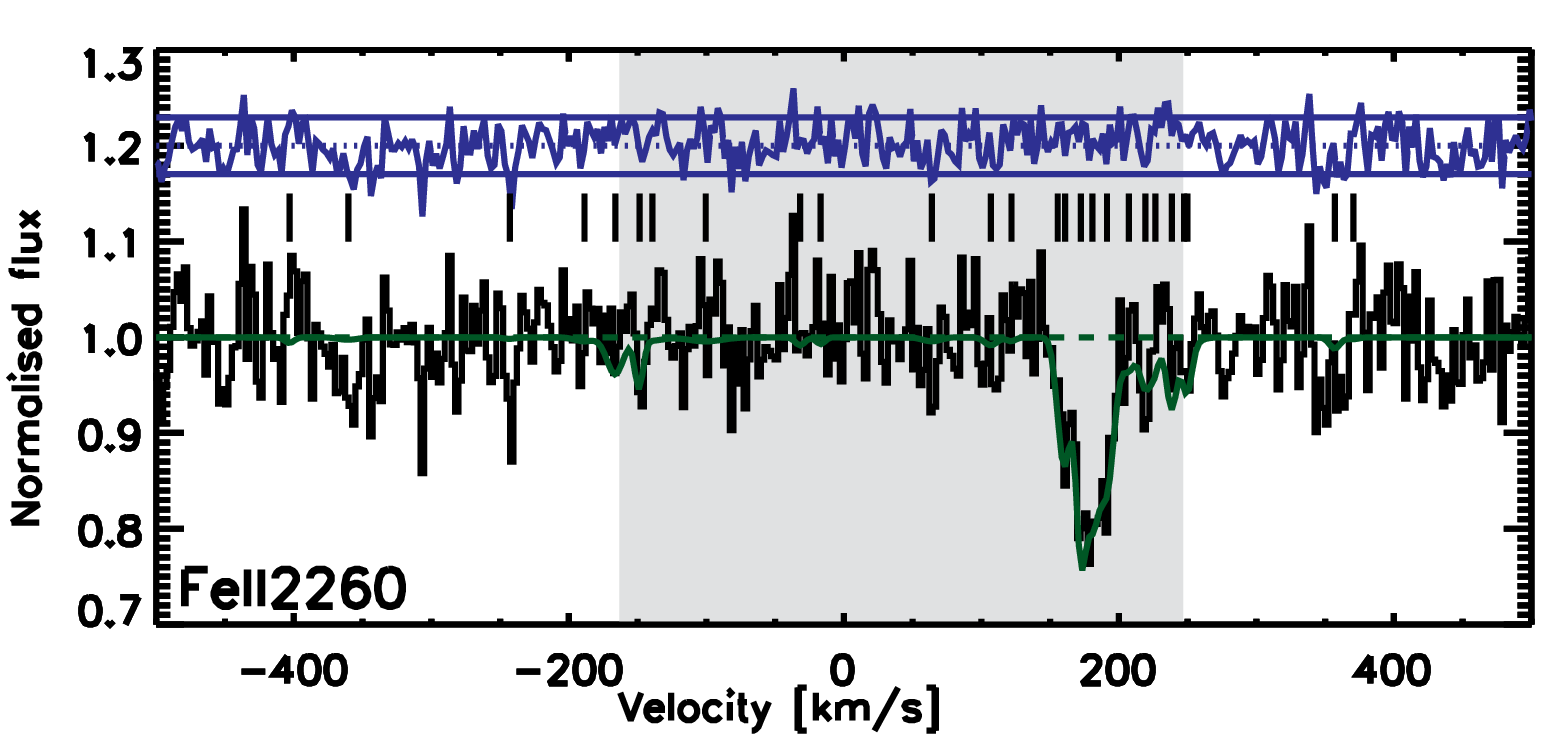}
    \includegraphics[width=0.33\textwidth,trim=0 15 0 0,angle=0,clip=false]{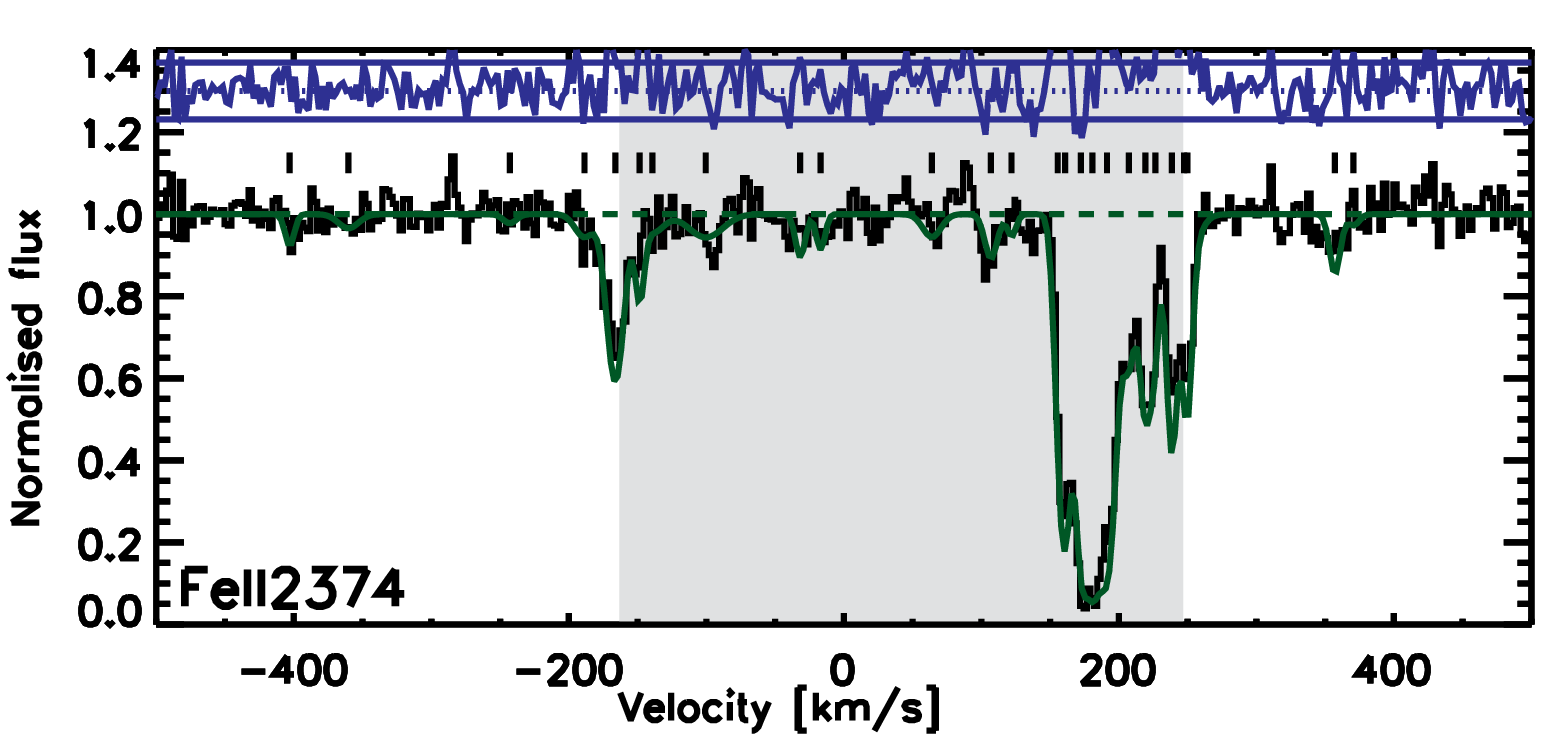}
  }
  \hbox{
    \includegraphics[width=0.33\textwidth,trim=0 15 0 0,angle=0,clip=false]{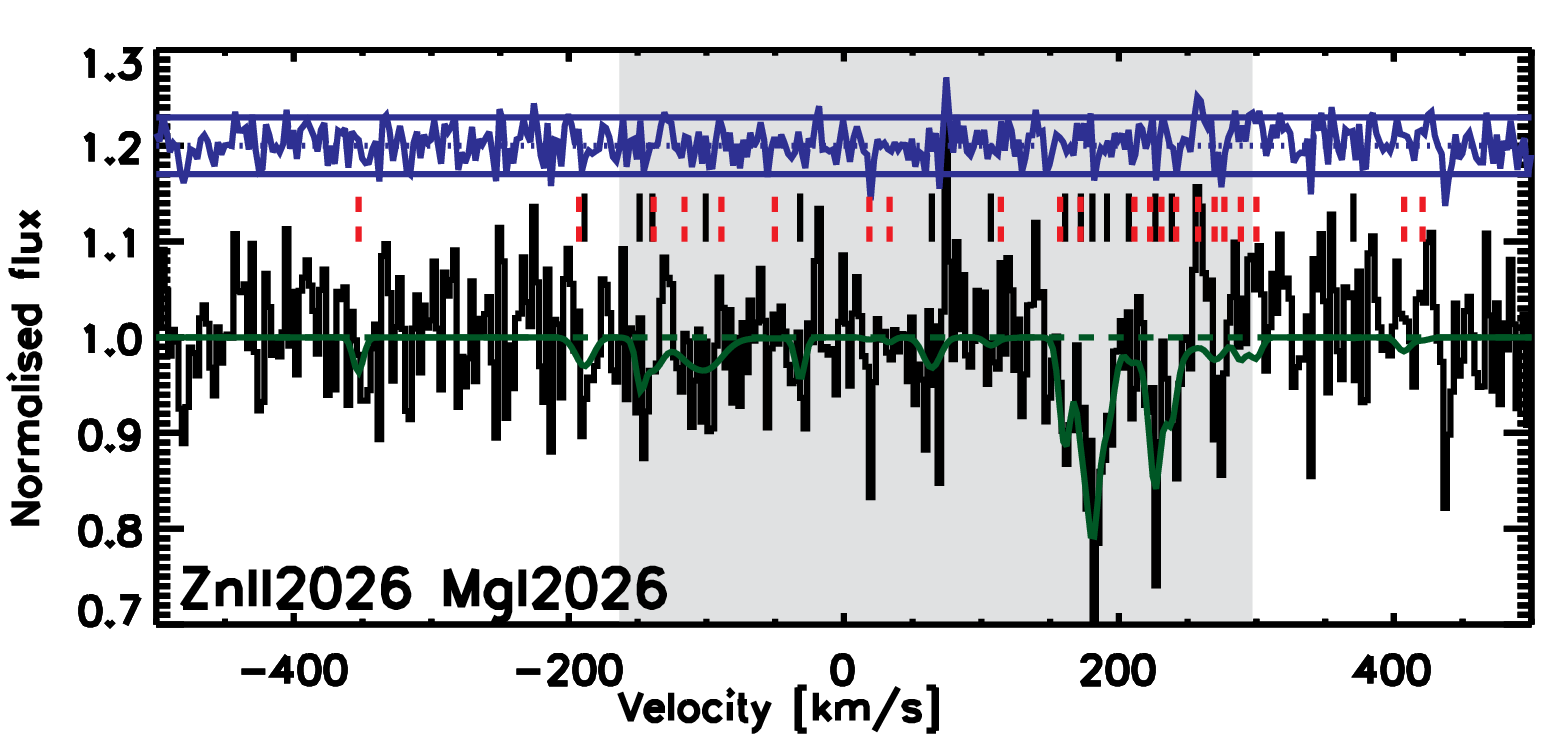}
    \includegraphics[width=0.33\textwidth,trim=0 15 0 0,angle=0,clip=false]{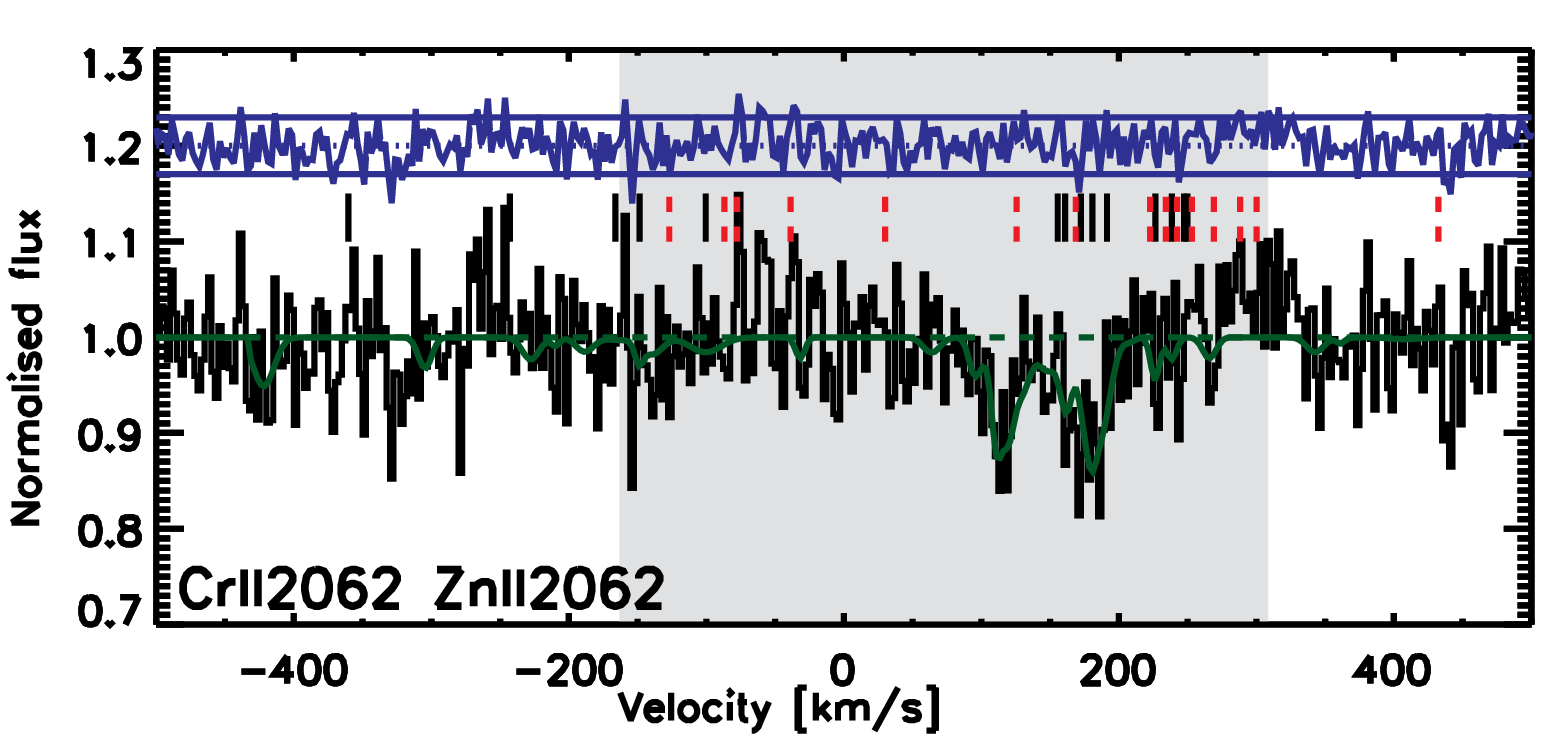}
    \includegraphics[width=0.33\textwidth,trim=0 15 0 0,angle=0,clip=false]{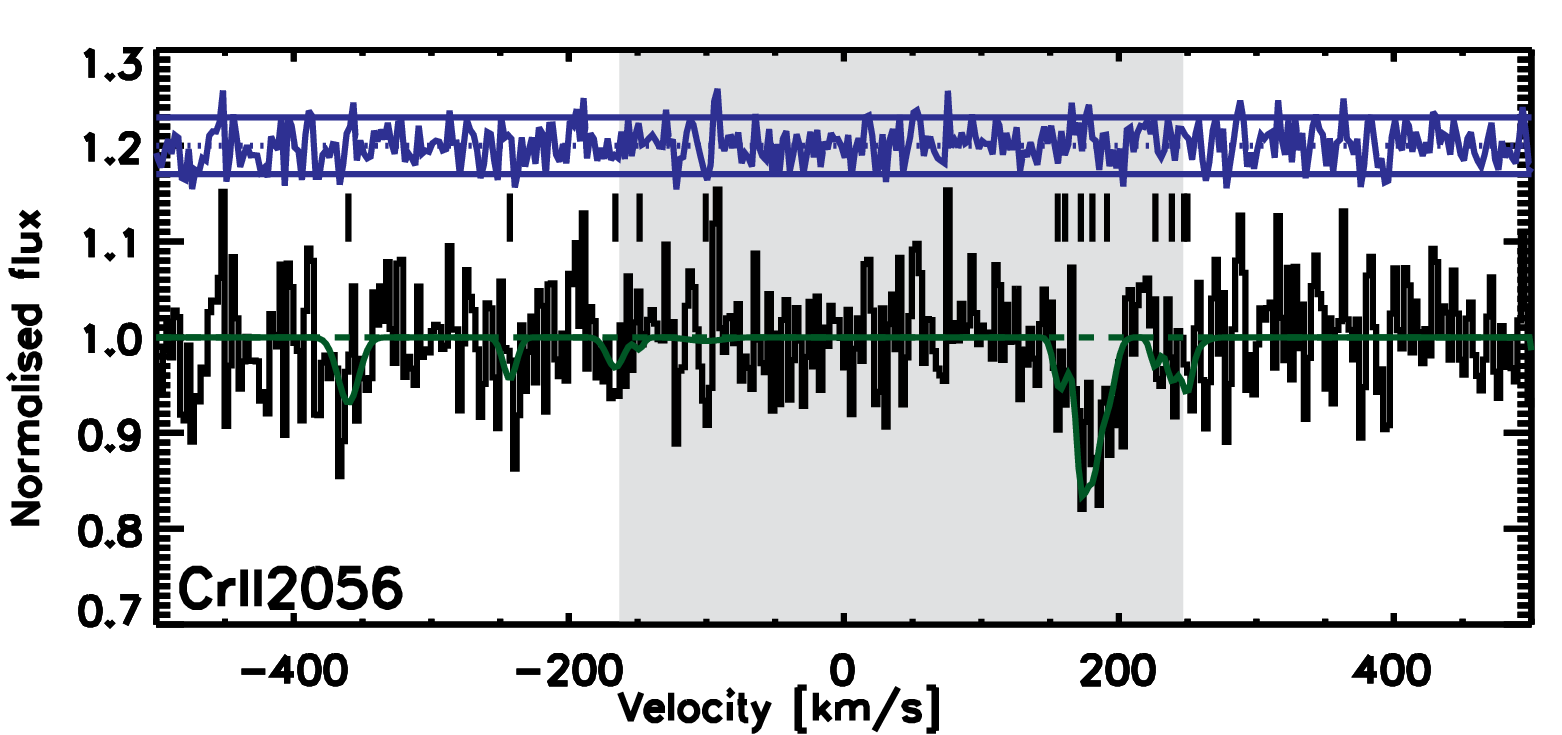}
  }
  \hbox{
    \includegraphics[width=0.33\textwidth,trim=0 15 0 0,angle=0,clip=false]{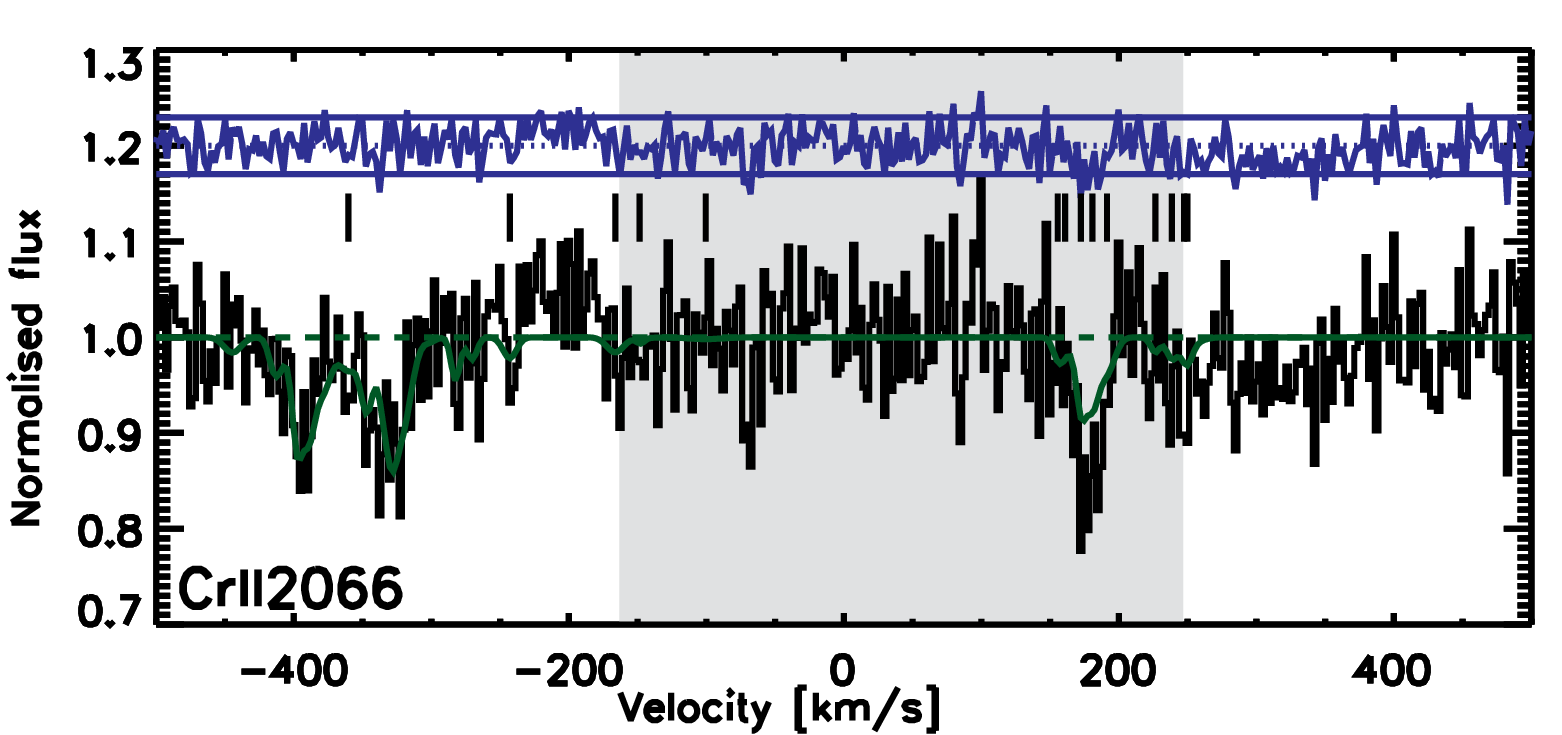}
    \includegraphics[width=0.33\textwidth,trim=0 15 0 0,angle=0,clip=false]{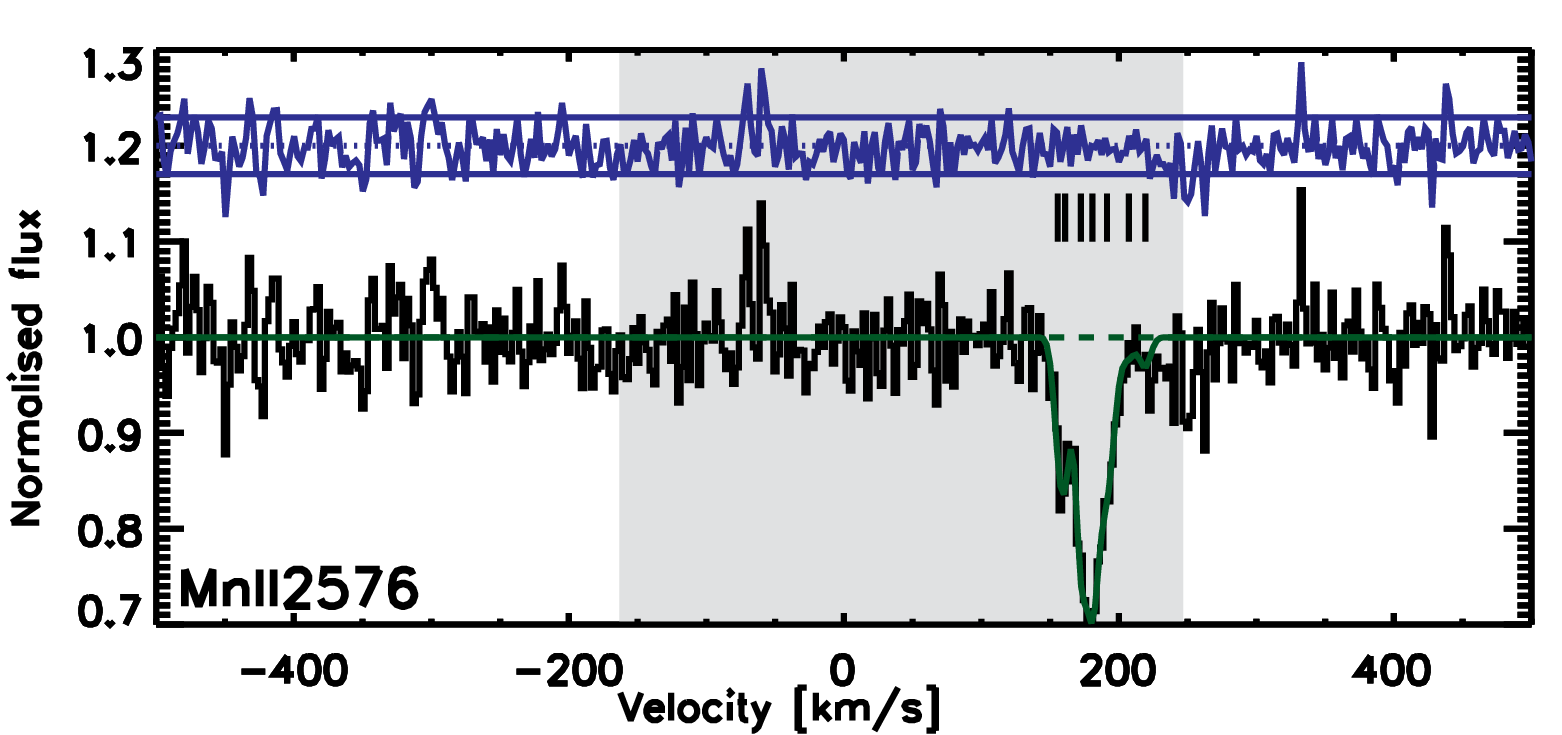}
    \includegraphics[width=0.33\textwidth,trim=0 15 0 0,angle=0,clip=false]{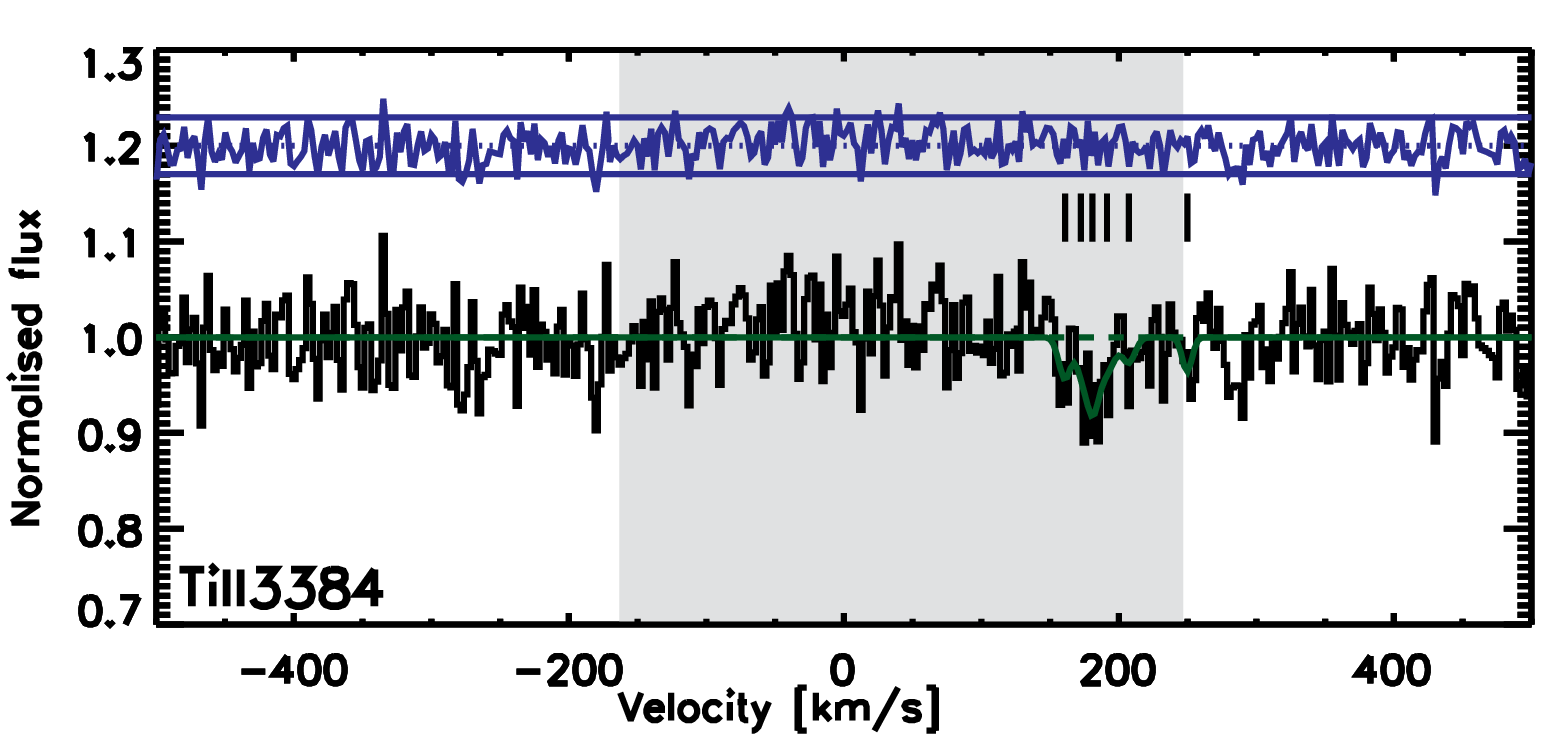}
  }
}
\caption{Voigt profile fits to the $z=0.59760$ absorber towards
  J0334$-$0711. A label (bottom-left) in each panel specifies the
  transitions incorporated in the profile fits to the spectral segment
  shown. The data are represented by the black histograms, whilst the
  continuum is the dashed line. The fit is shown as a solid line
  through the data. The residuals between the spectra and the fits,
  normalised by the 1$\sigma$ error, are plotted above the
  spectra. Individual Voigt profile components are marked with
  ticks. In regions with blends the higher wavelength transition is
  marked with dashed ticks. The region representing $\Delta v_{90}$
  is lightly shaded behind the data.}
\label{fig:fit_J0334m0711}
\end{figure*}
\begin{figure*}
\vbox{
  \hbox{
    \includegraphics[width=0.33\textwidth,trim=0 15 0 0,angle=0,clip=false]{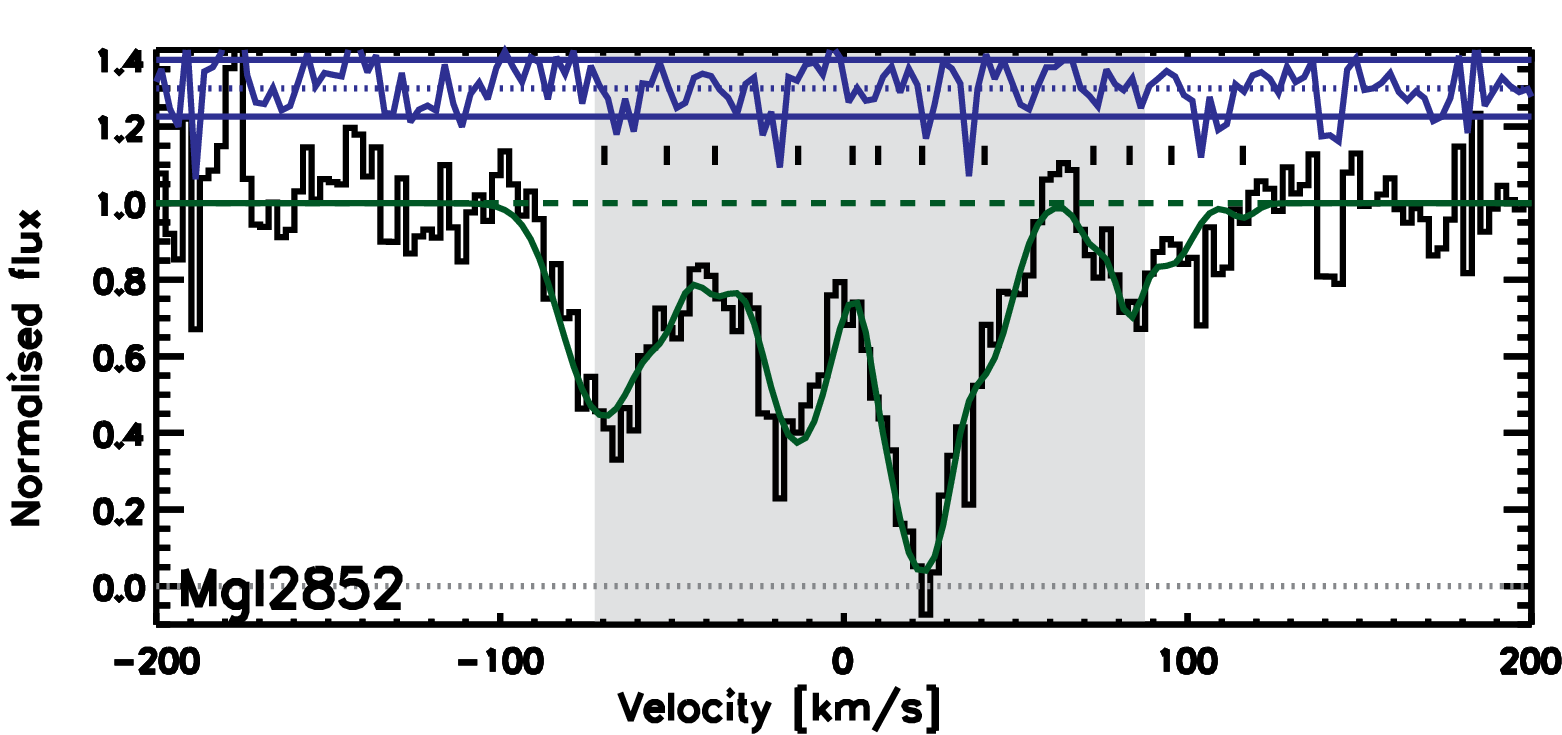}
    \includegraphics[width=0.33\textwidth,trim=0 15 0 0,angle=0,clip=false]{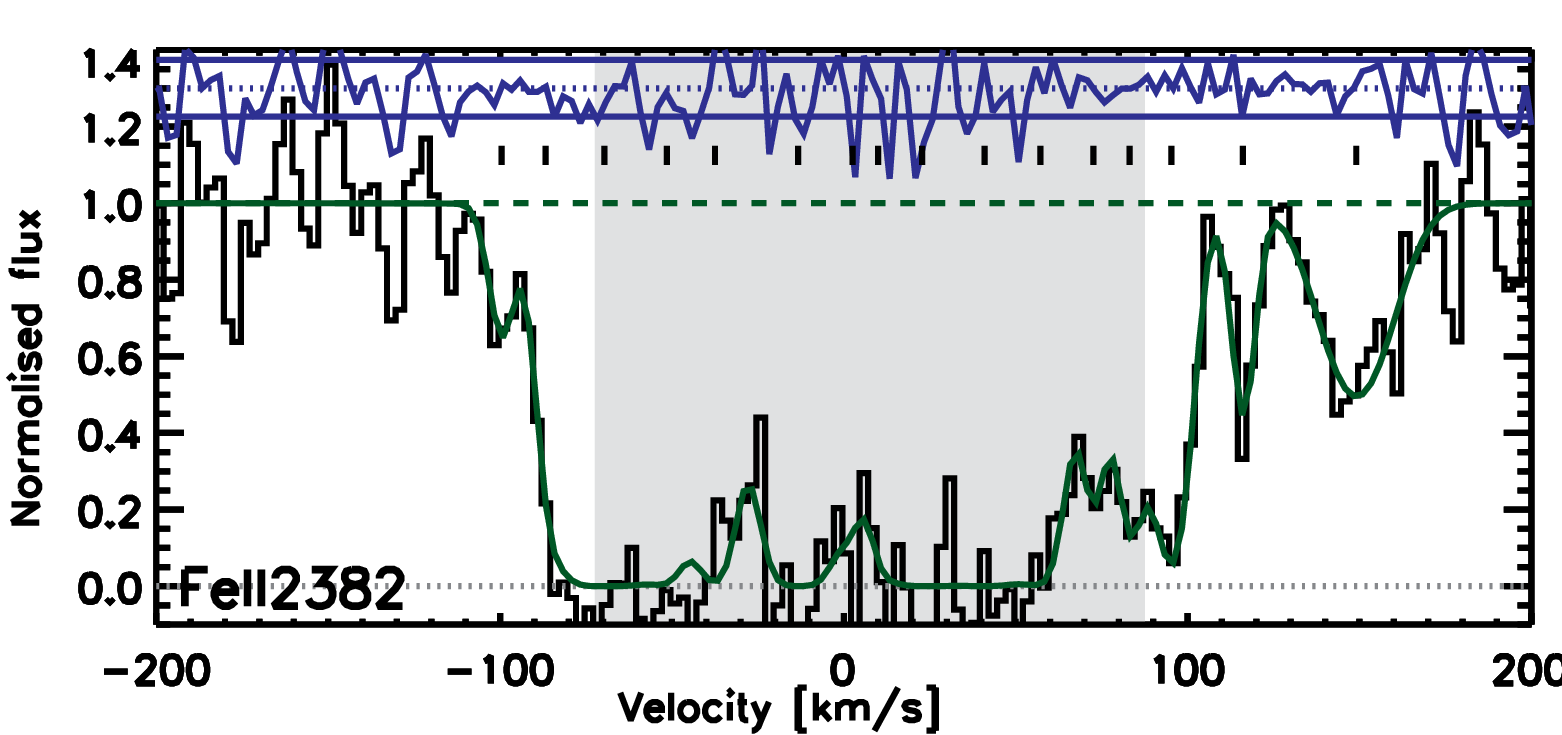}
    \includegraphics[width=0.33\textwidth,trim=0 15 0 0,angle=0,clip=false]{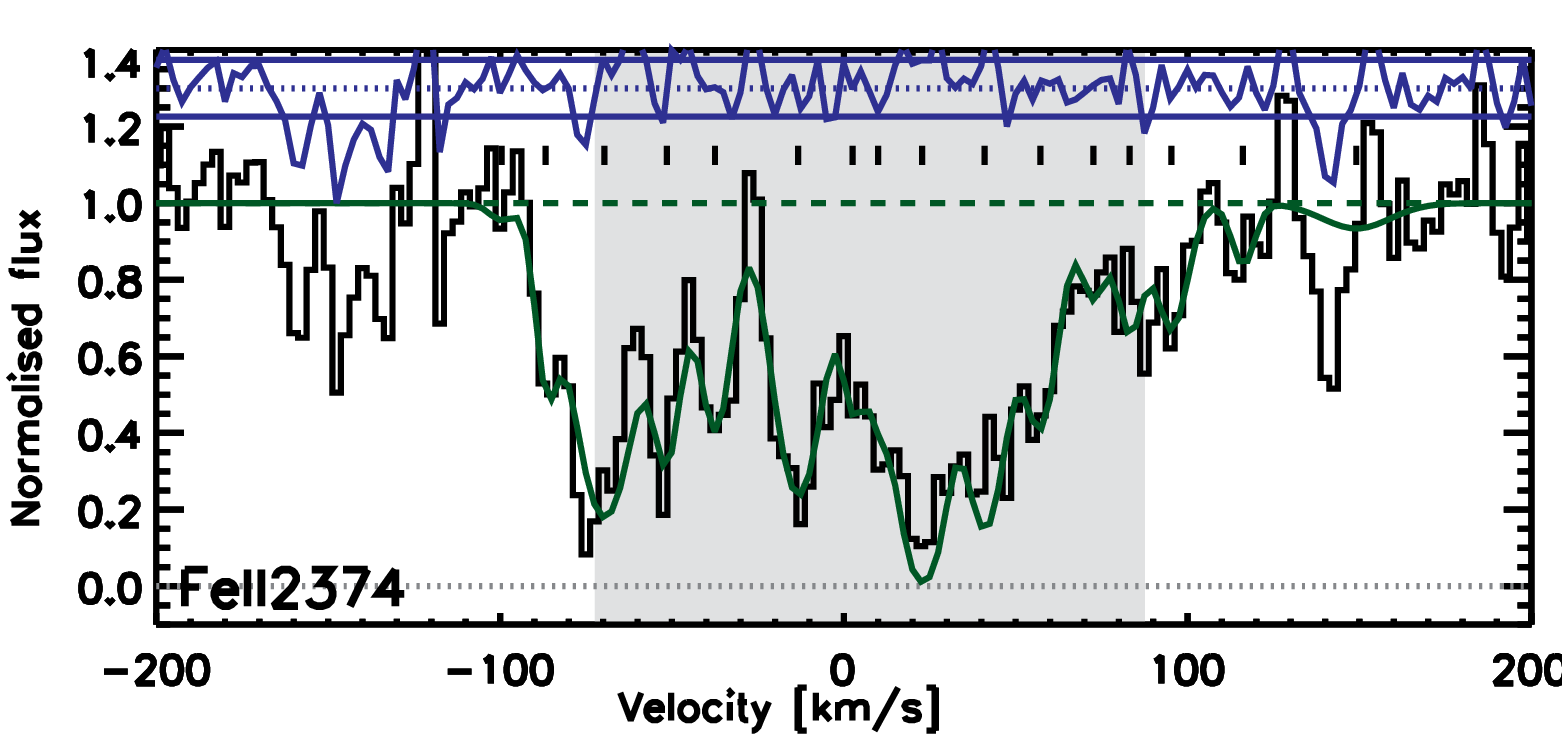}
    }
  \hbox{
    \includegraphics[width=0.33\textwidth,trim=0 15 0 0,angle=0,clip=false]{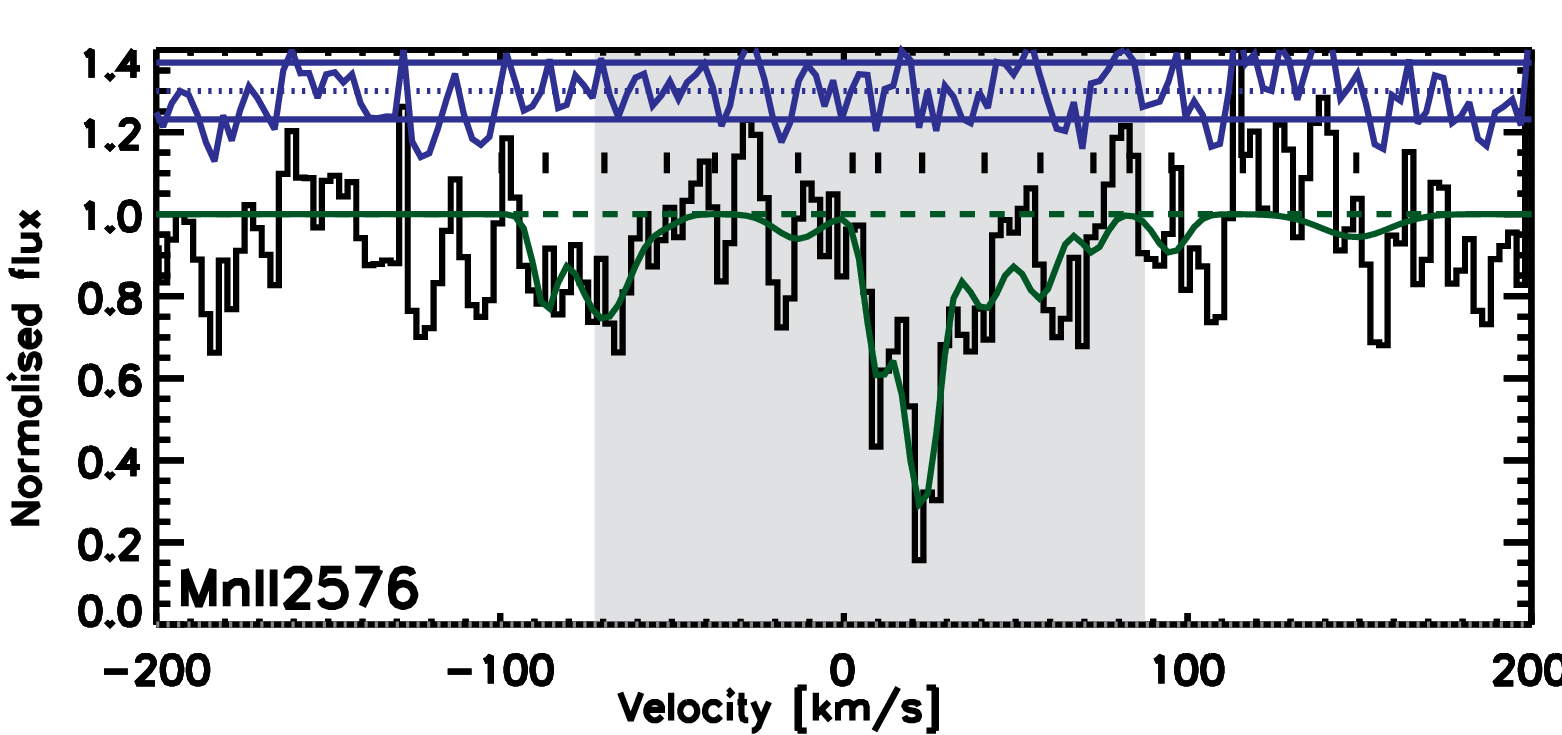}
    \includegraphics[width=0.33\textwidth,trim=0 15 0 0,angle=0,clip=false]{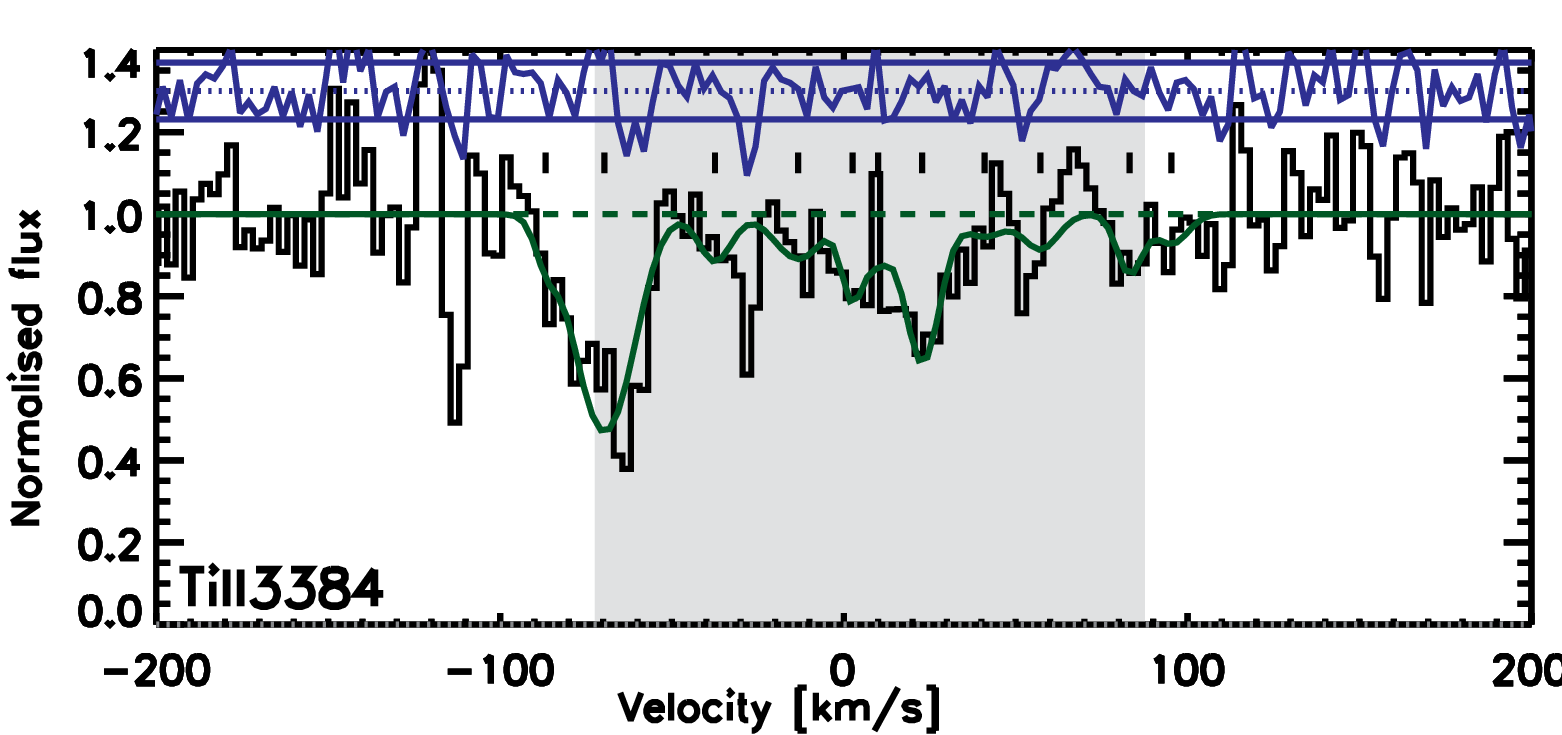}
    \includegraphics[width=0.33\textwidth,trim=0 15 0 0,angle=0,clip=false]{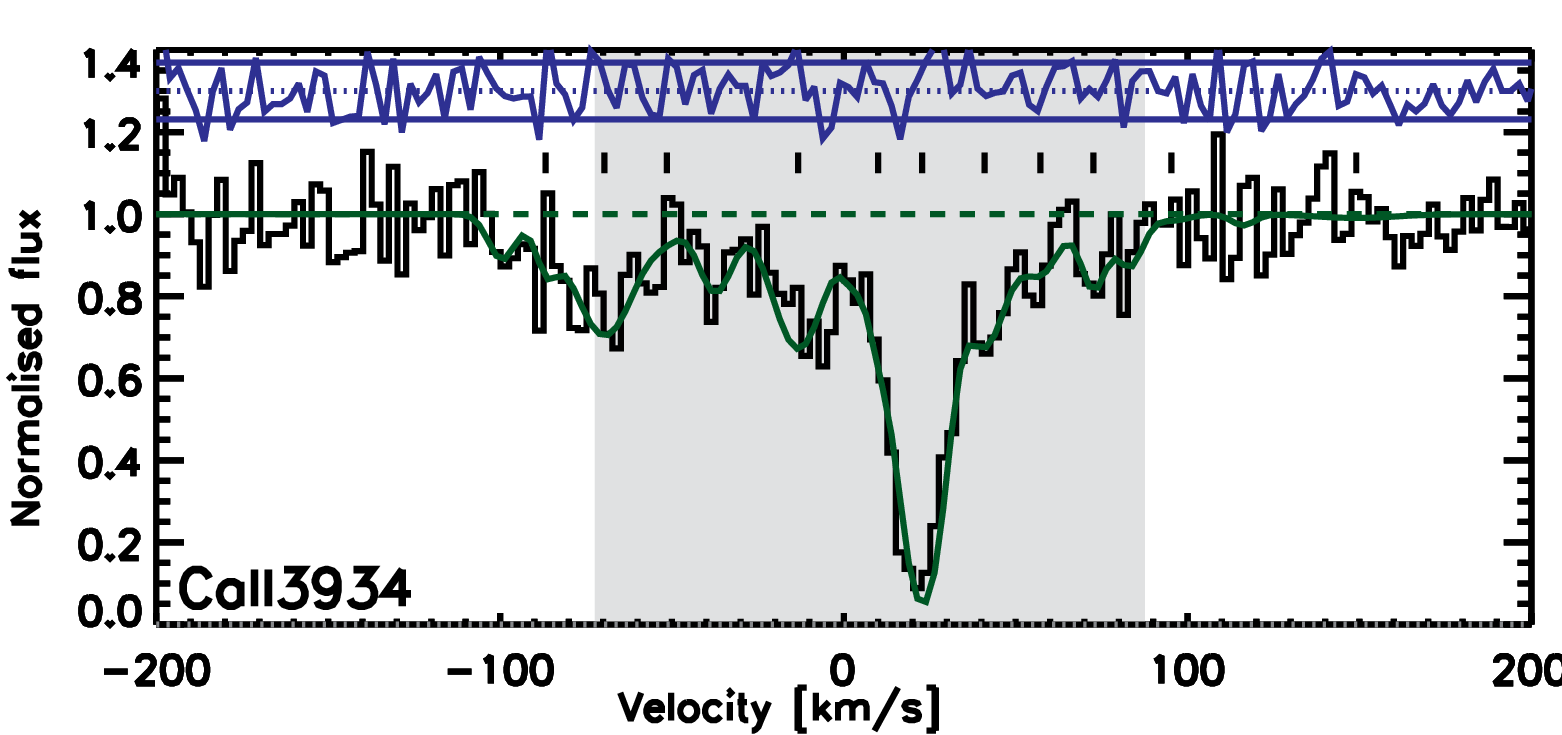}
  }
}
\caption{Voigt profile fits to the $z=0.74294$ absorber towards
  J0846$+$0529, see Fig.~\ref{fig:fit_J0334m0711} for description. }
\label{fig:fit_J0846p0529}
\end{figure*}
\begin{figure*}
\vbox{
  \hbox{
    \includegraphics[width=0.33\textwidth,trim=0 15 0 0,angle=0,clip=false]{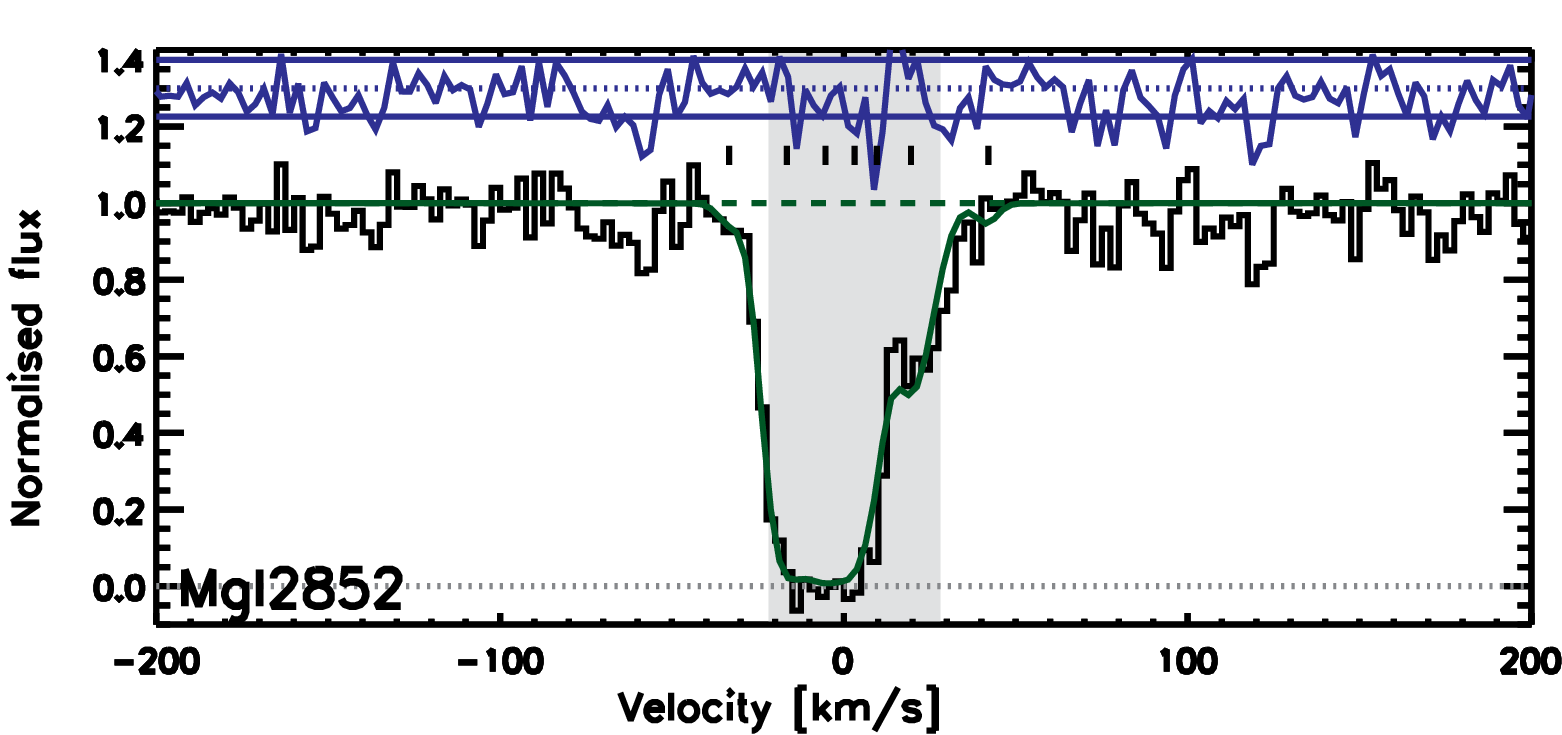}
    \includegraphics[width=0.33\textwidth,trim=0 15 0 0,angle=0,clip=false]{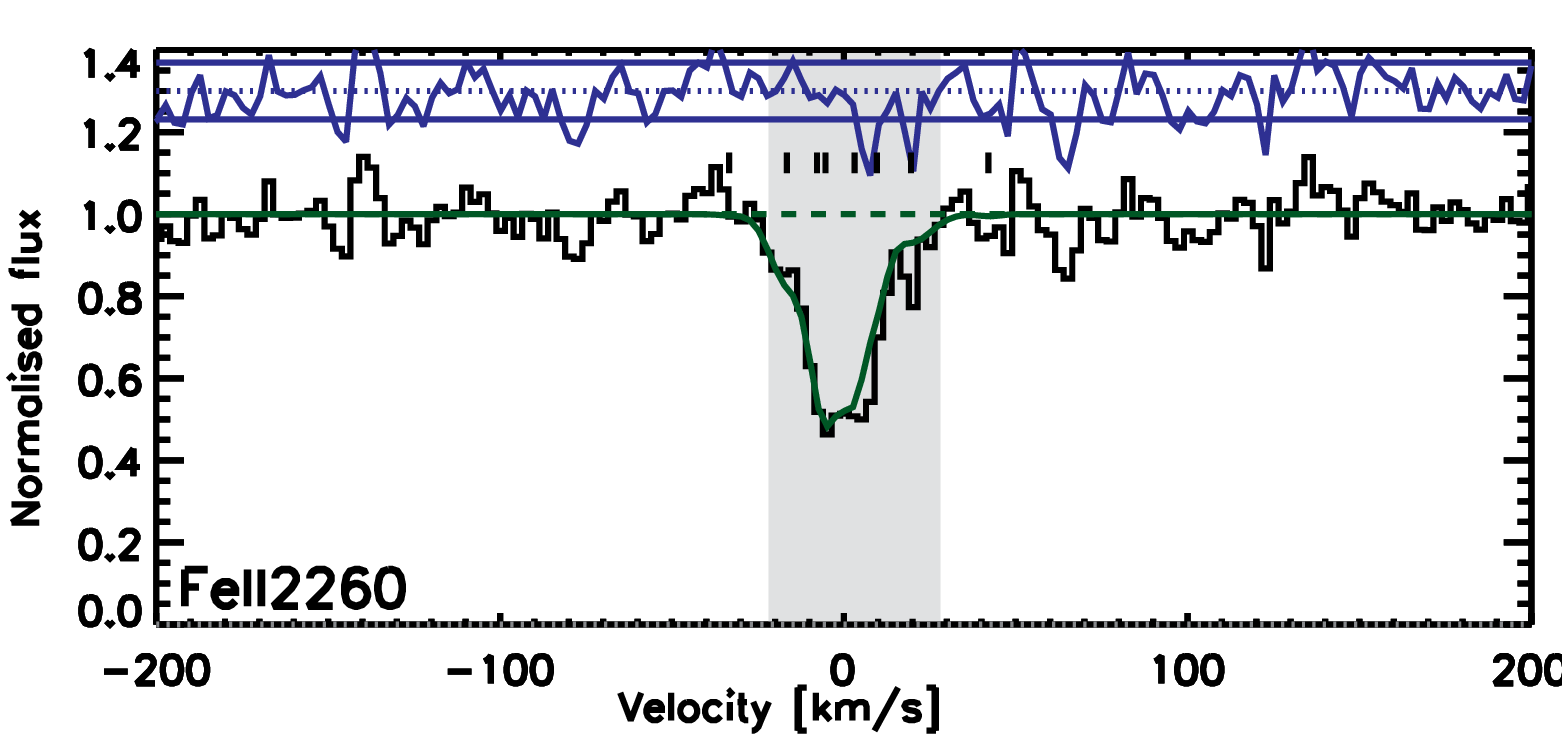}
    \includegraphics[width=0.33\textwidth,trim=0 15 0 0,angle=0,clip=false]{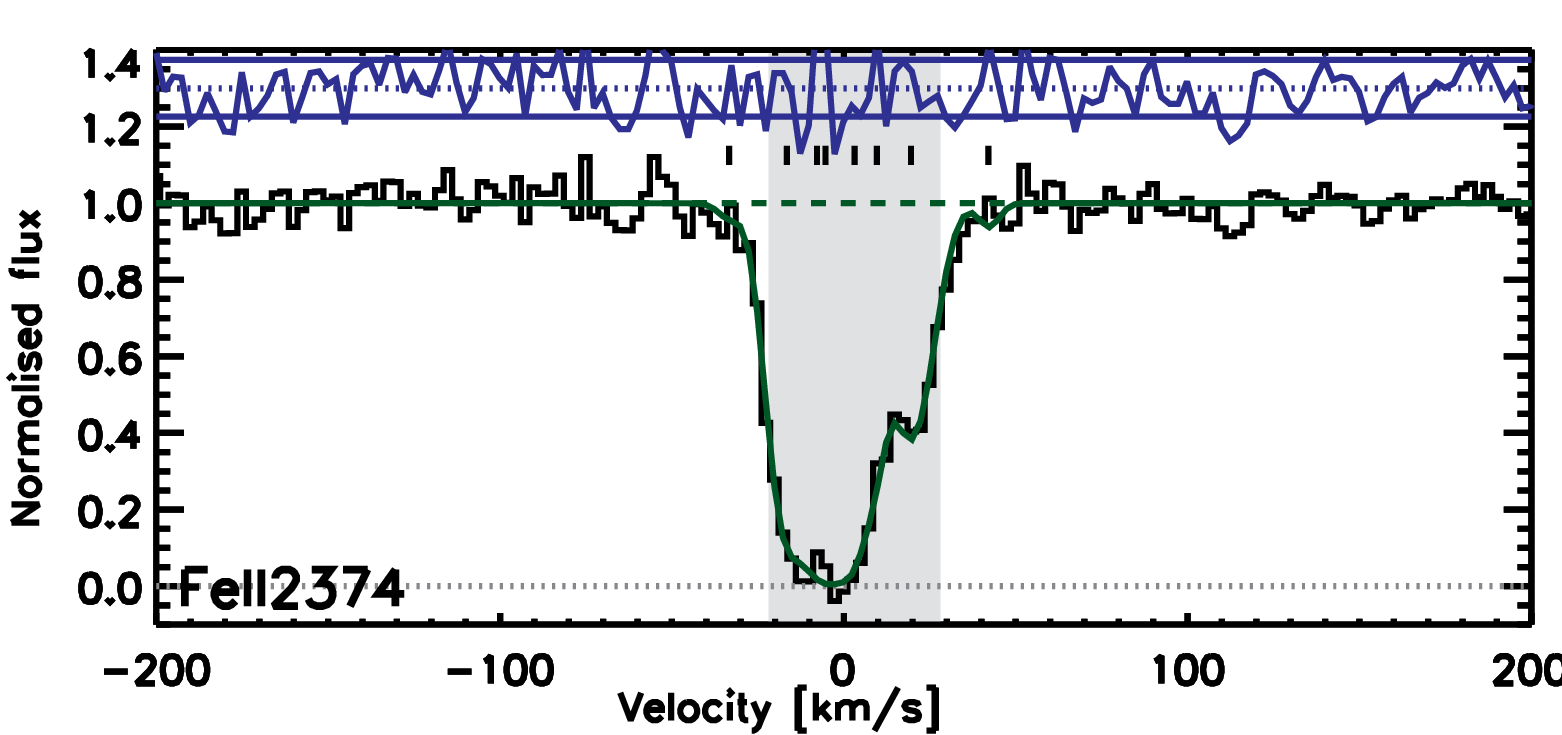}
  }
  \hbox{
    \includegraphics[width=0.33\textwidth,trim=0 15 0 0,angle=0,clip=false]{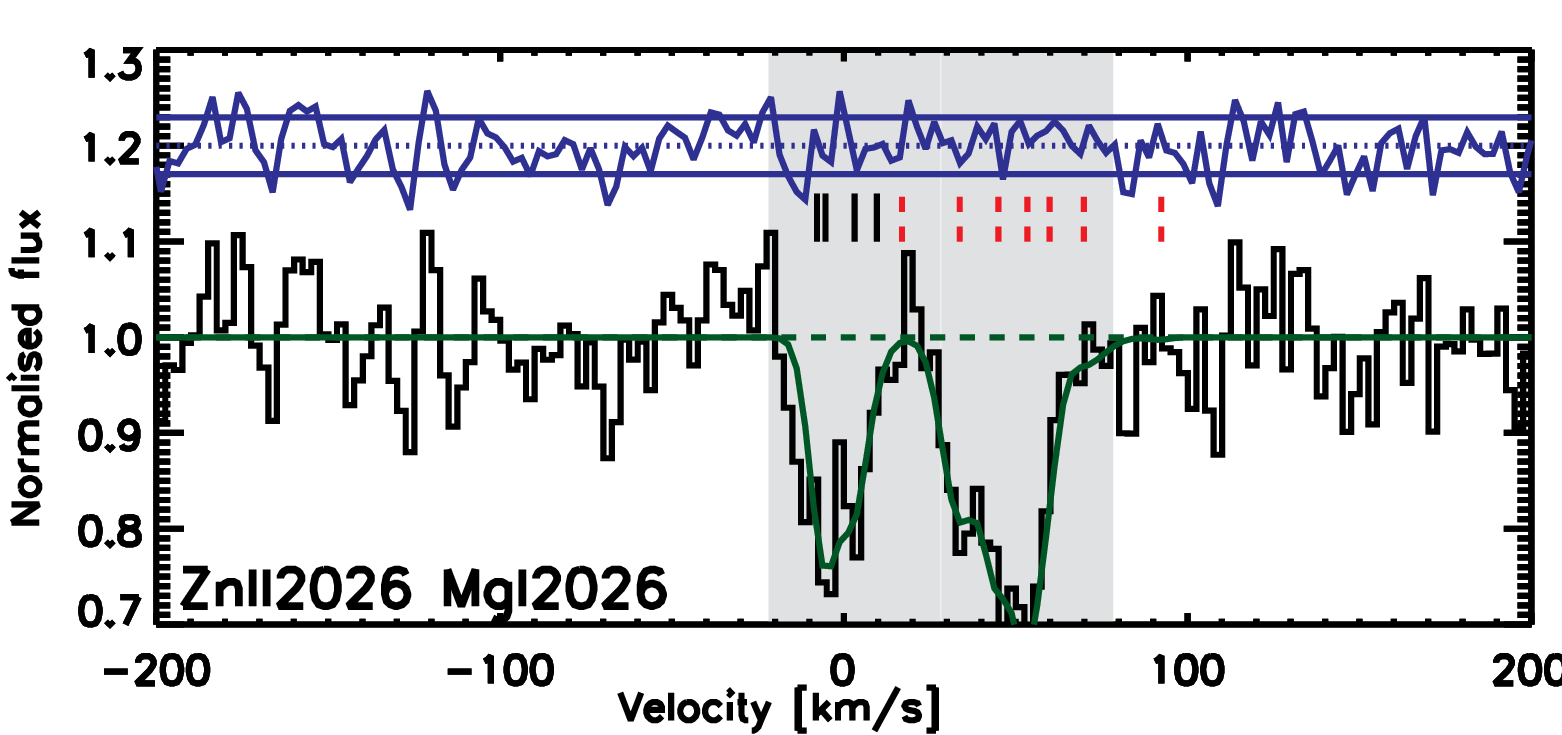}
    \includegraphics[width=0.33\textwidth,trim=0 15 0 0,angle=0,clip=false]{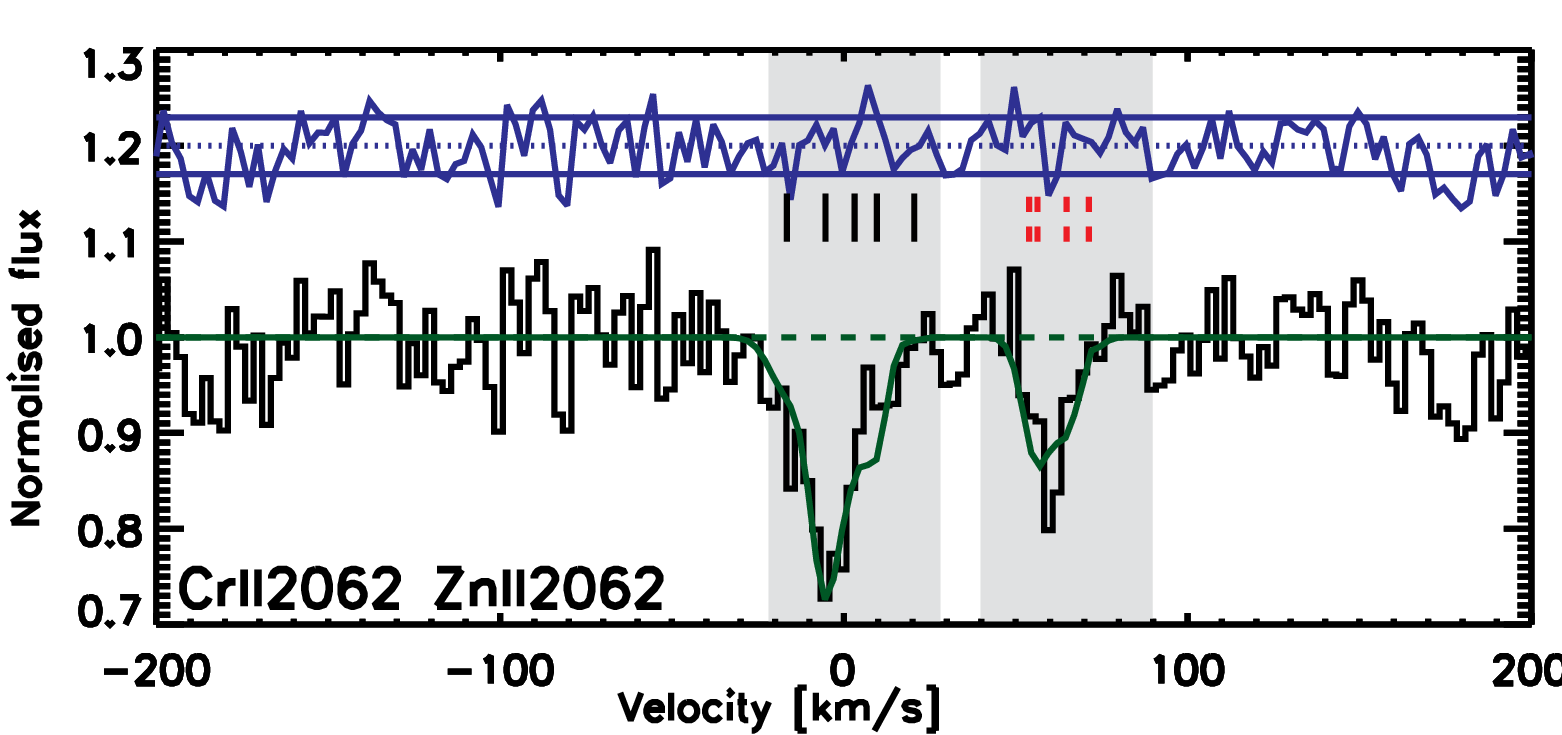}
    \includegraphics[width=0.33\textwidth,trim=0 15 0 0,angle=0,clip=false]{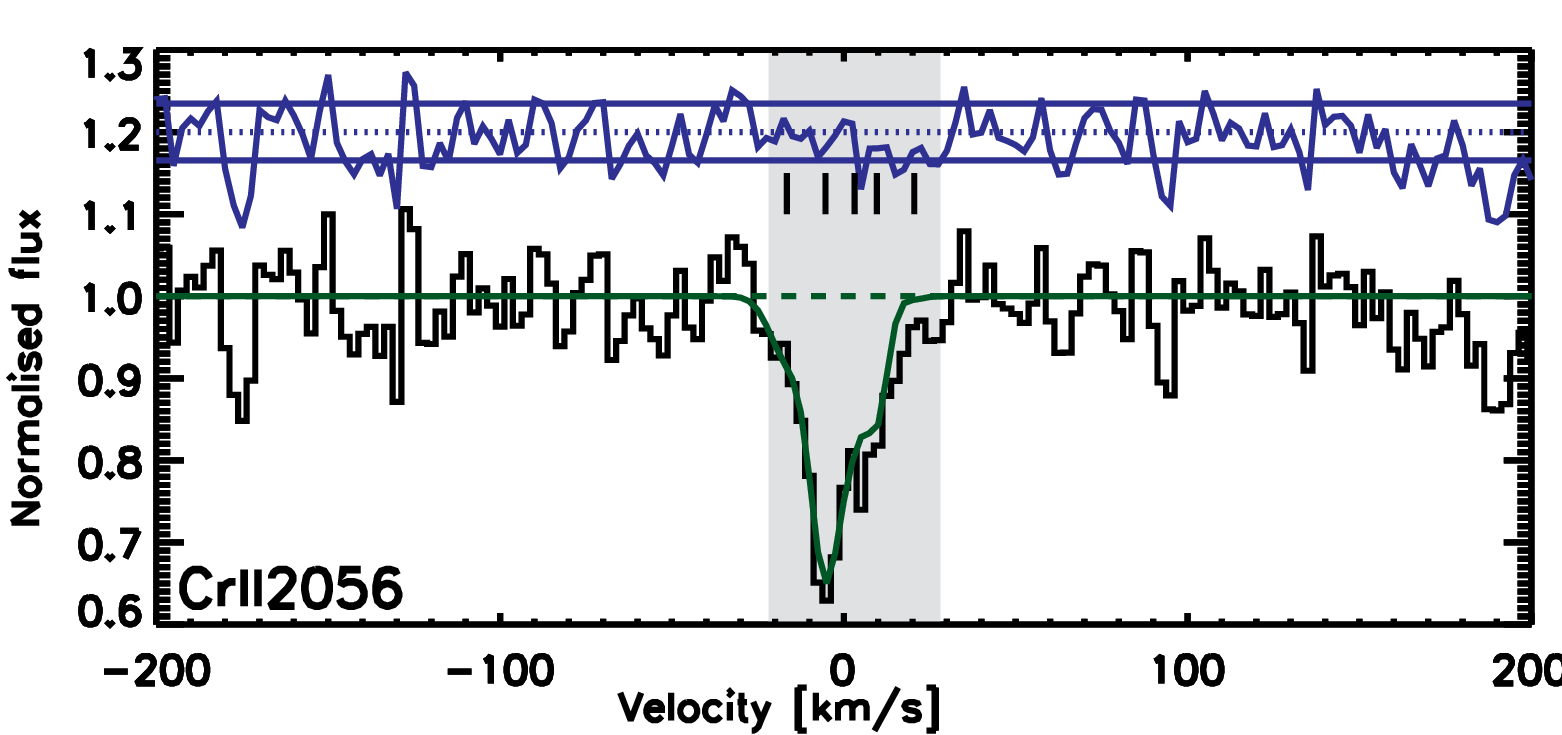}
  }
  \hbox{
    \includegraphics[width=0.33\textwidth,trim=0 15 0 0,angle=0,clip=false]{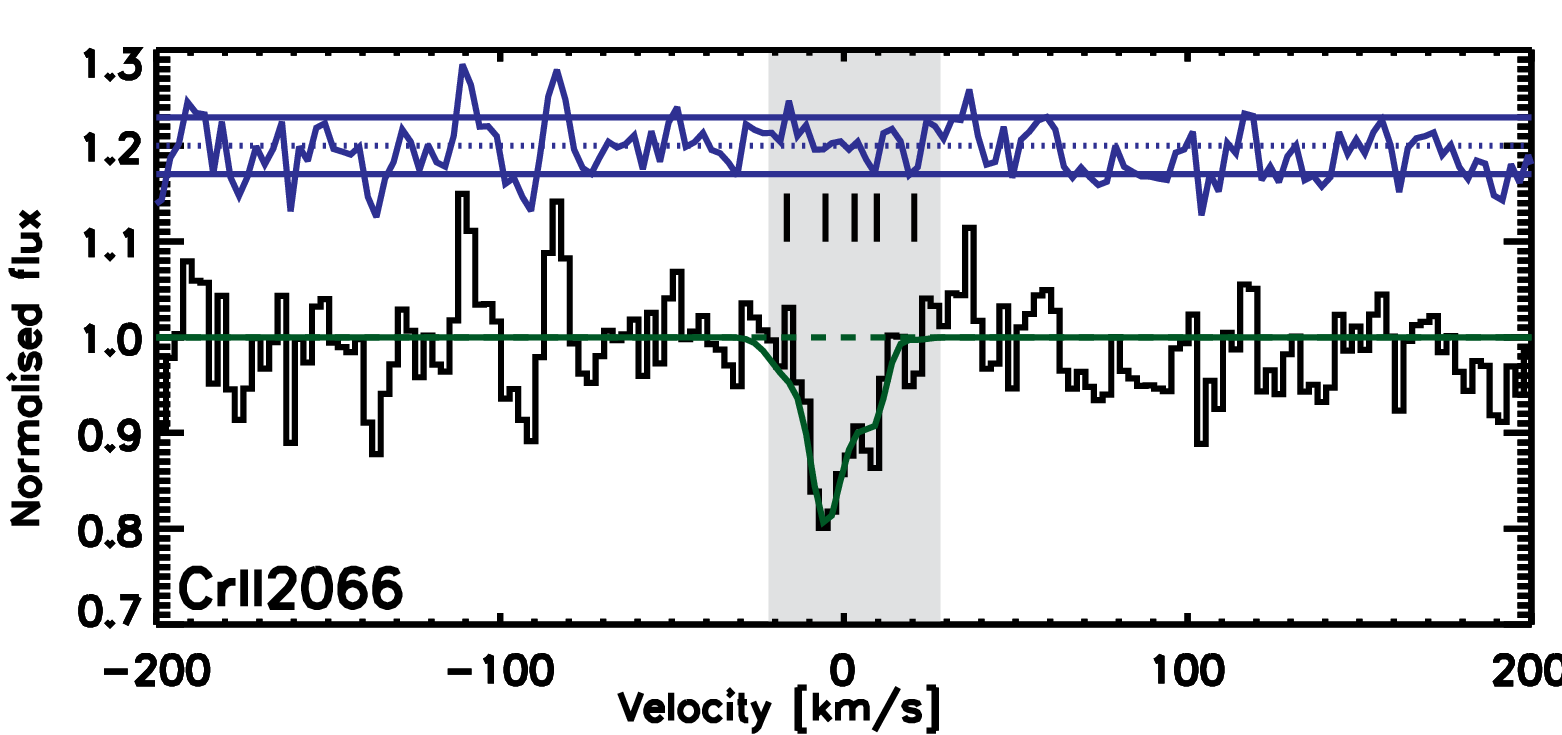}
    \includegraphics[width=0.33\textwidth,trim=0 15 0 0,angle=0,clip=false]{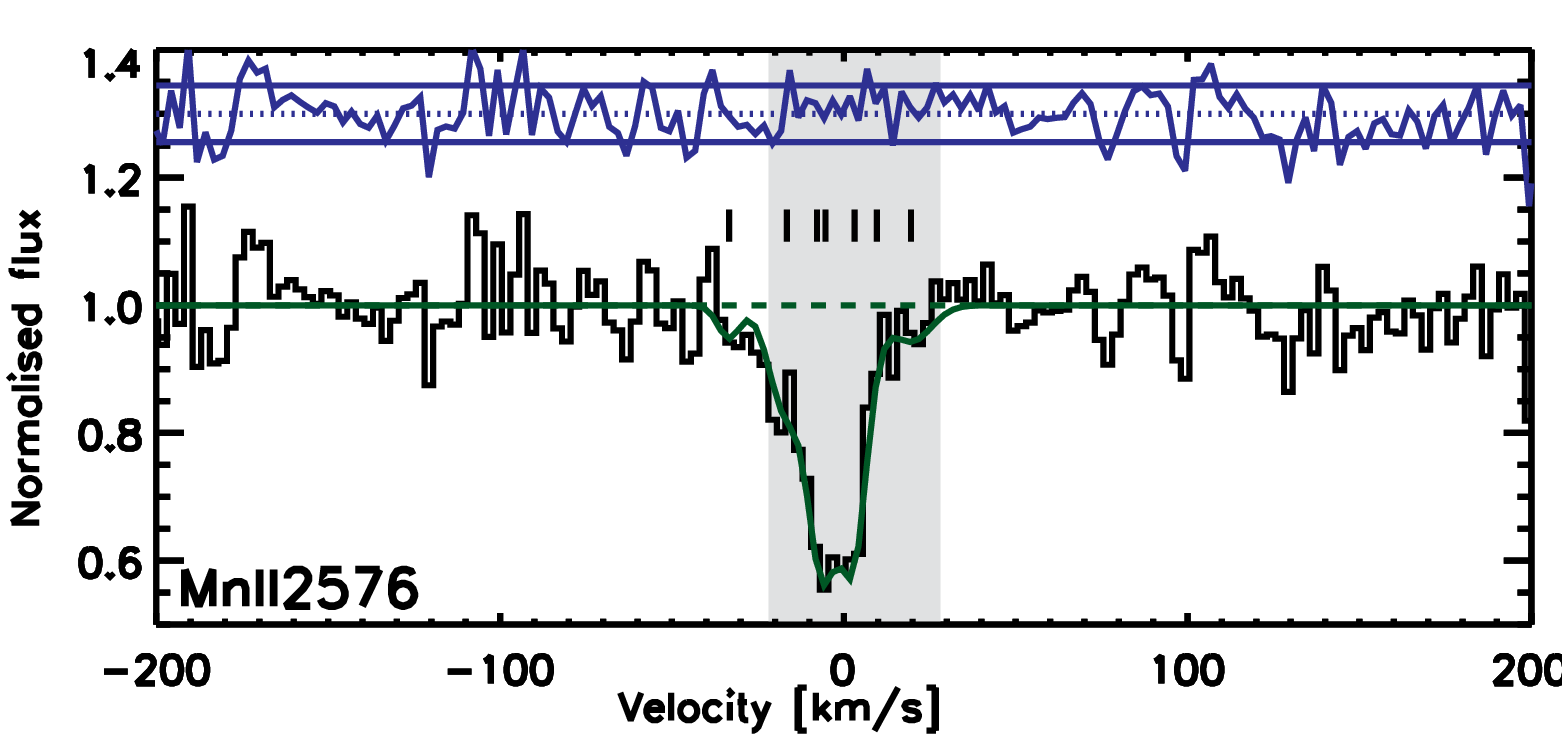}
    \includegraphics[width=0.33\textwidth,trim=0 15 0 0,angle=0,clip=false]{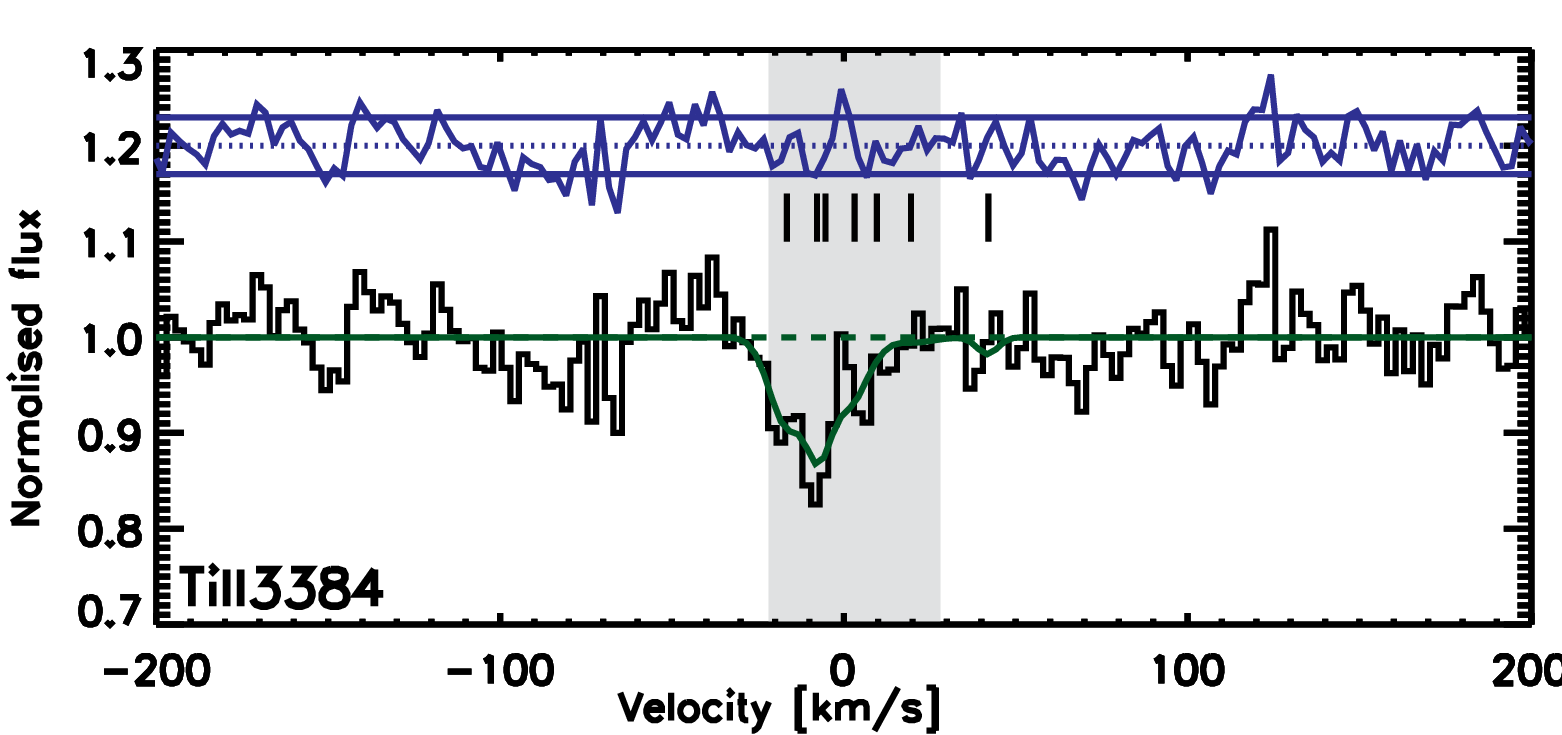}
  }
  \hbox{
    \includegraphics[width=0.33\textwidth,trim=0 15 0 0,angle=0,clip=false]{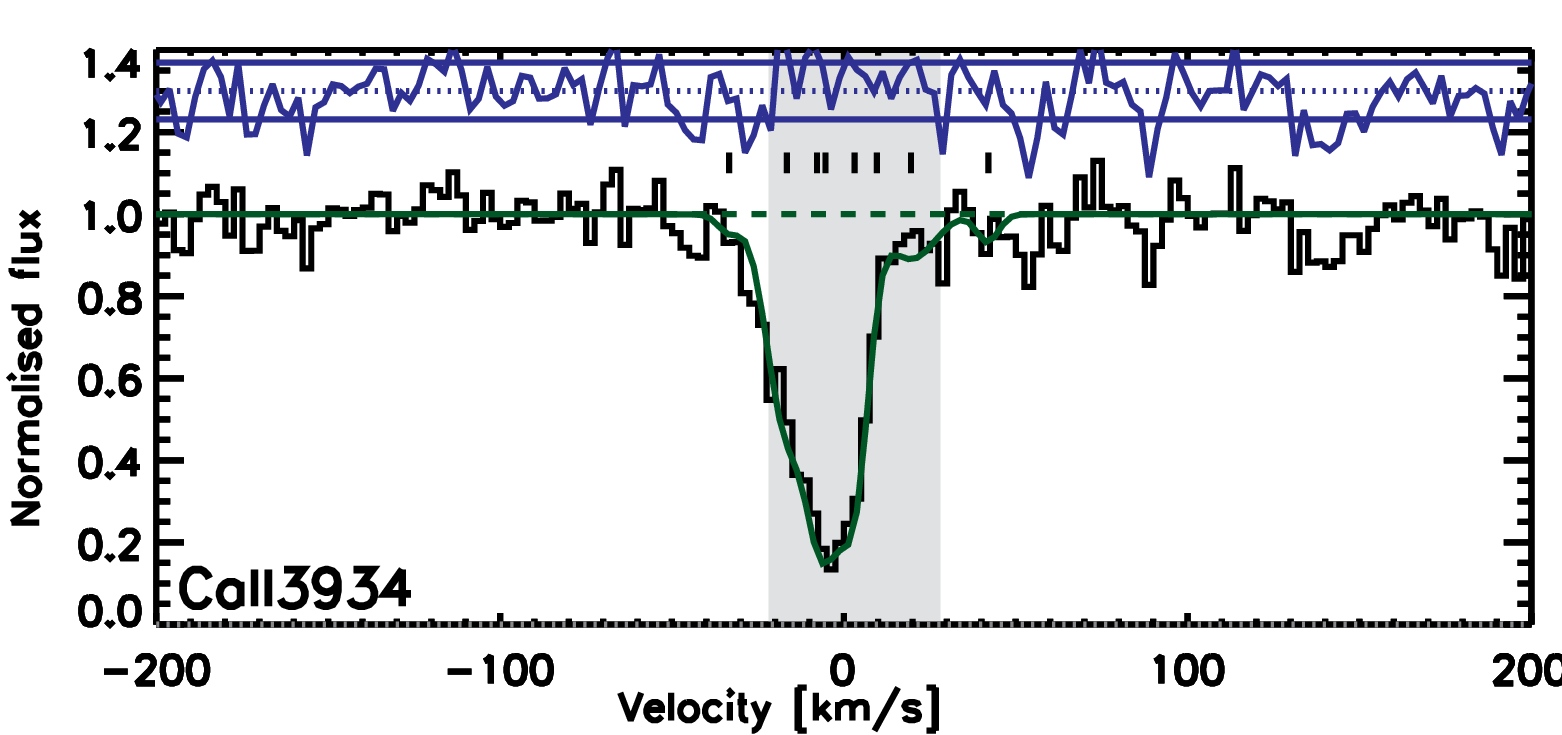}
    \includegraphics[width=0.33\textwidth,trim=0 15 0 0,angle=0,clip=false]{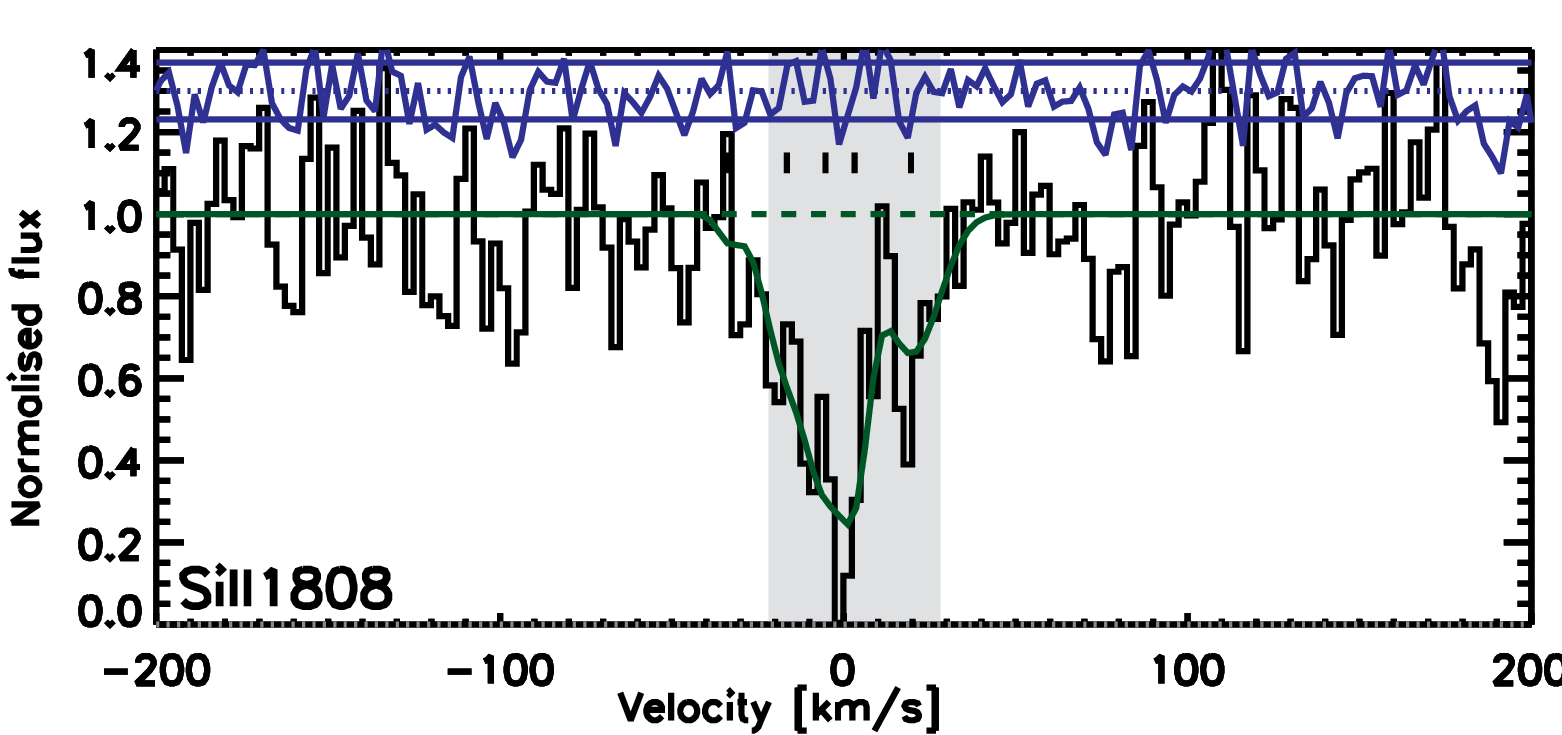}
    \includegraphics[width=0.33\textwidth,trim=0 15 0 0,angle=0,clip=false]{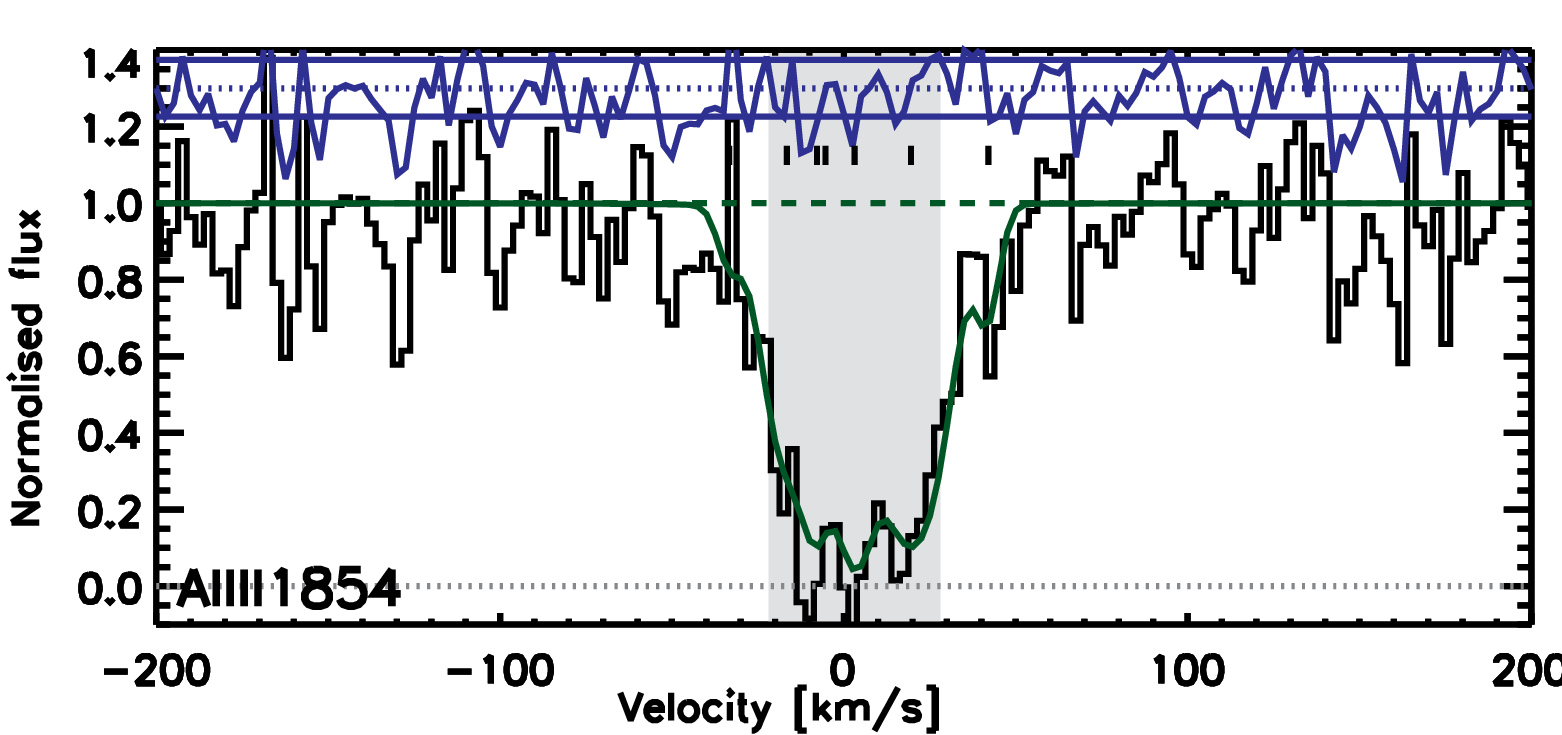}
  }
}
 \caption{Voigt profile fits to the $z=1.02316$ absorber towards J0953$+$0801, see Fig.~\ref{fig:fit_J0334m0711} for description. }
 \label{fig:fit_J0953p0801}
 \end{figure*}
\begin{figure*}
\vbox{
  \hbox{
    \includegraphics[width=0.33\textwidth,trim=0 15 0 0,angle=0,clip=false]{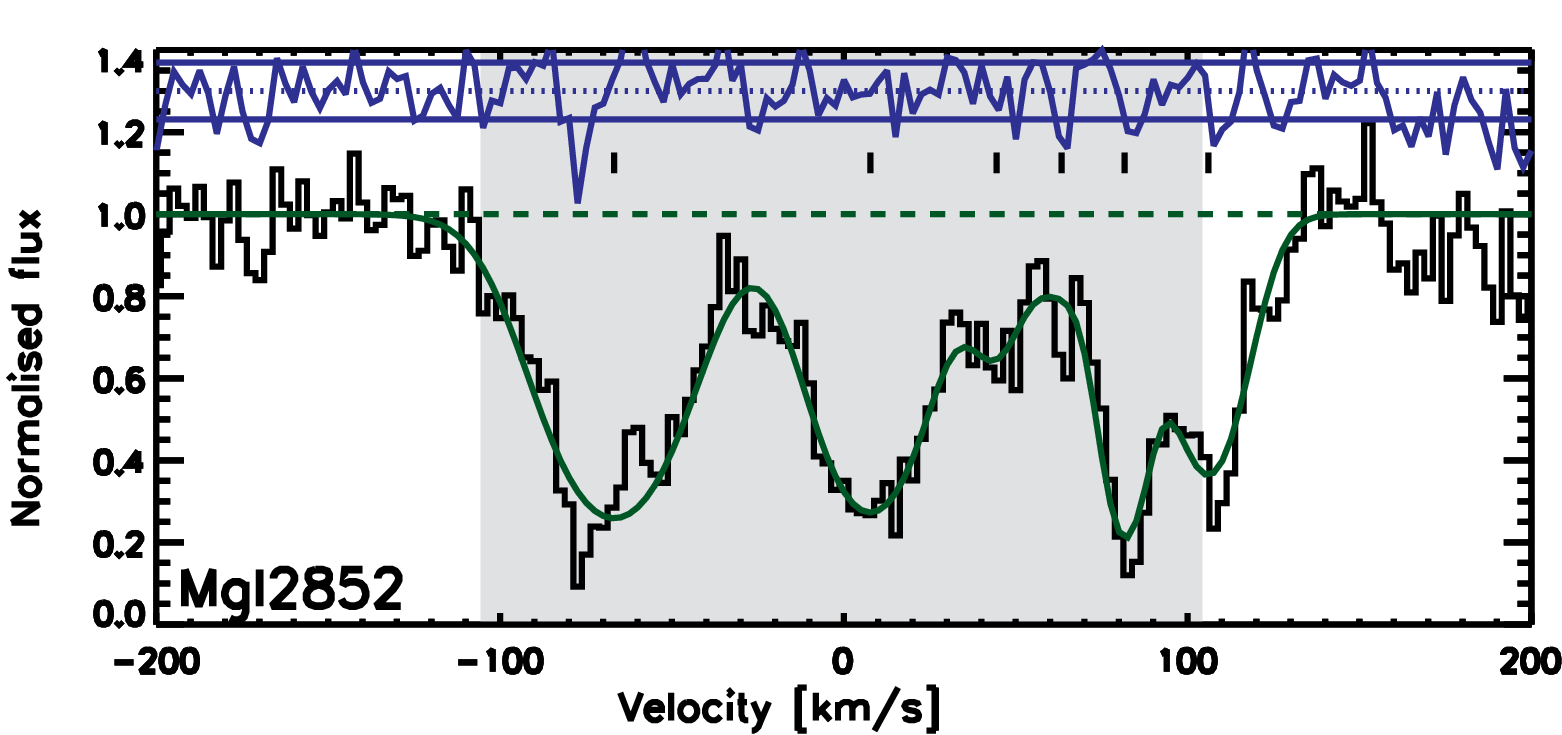}
    \includegraphics[width=0.33\textwidth,trim=0 15 0 0,angle=0,clip=false]{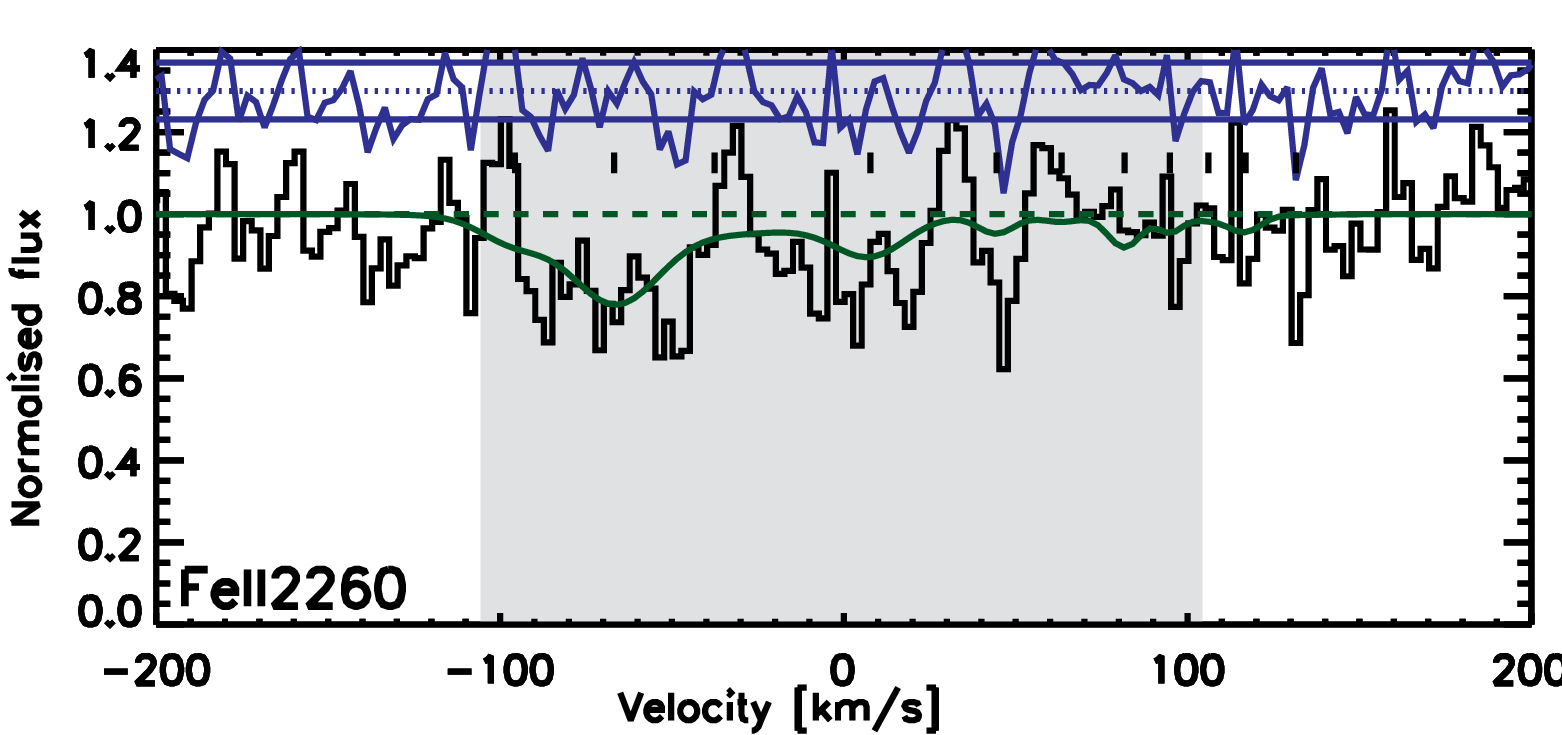}
    \includegraphics[width=0.33\textwidth,trim=0 15 0 0,angle=0,clip=false]{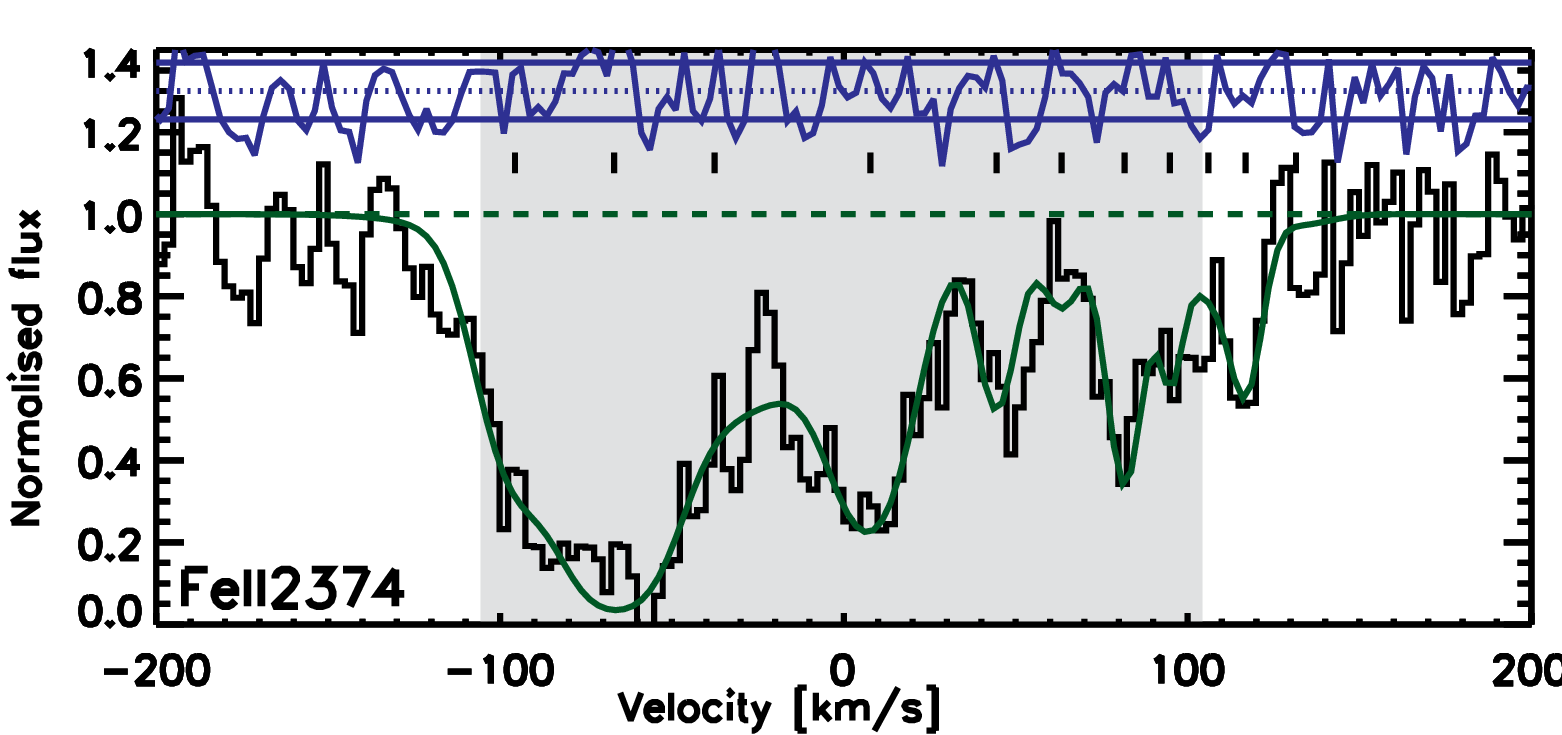}
  }
  \hbox{
    \includegraphics[width=0.33\textwidth,trim=0 15 0 0,angle=0,clip=false]{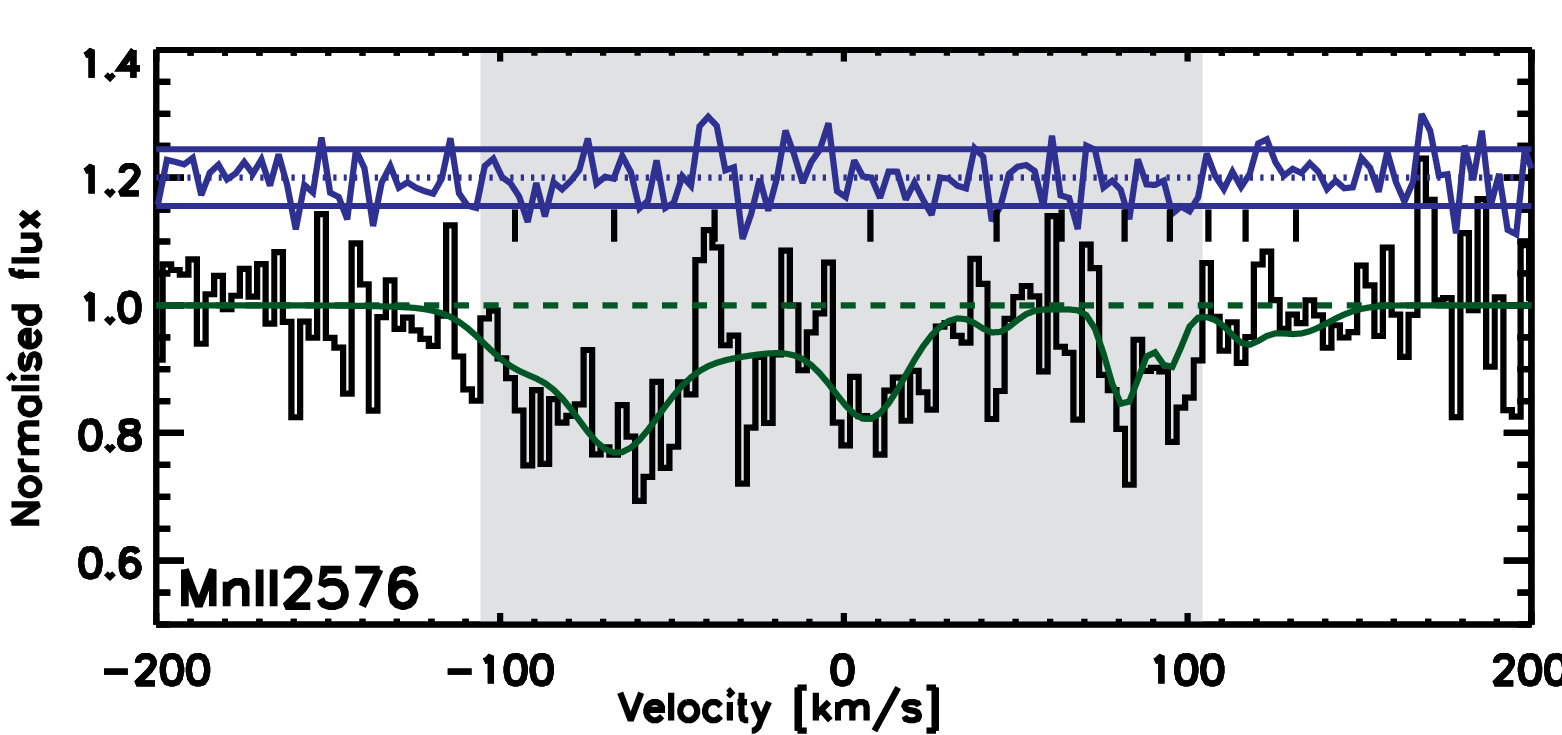}
  }
}
\caption{Voigt profile fits to the $z=0.83460$ absorber towards J1005$+$1157, see Fig.~\ref{fig:fit_J0334m0711} for description. }
\label{fig:fit_J1005p1157}
\end{figure*}
\begin{figure*}
\vbox{
  \hbox{
    \includegraphics[width=0.33\textwidth,trim=0 15 0 0,angle=0,clip=false]{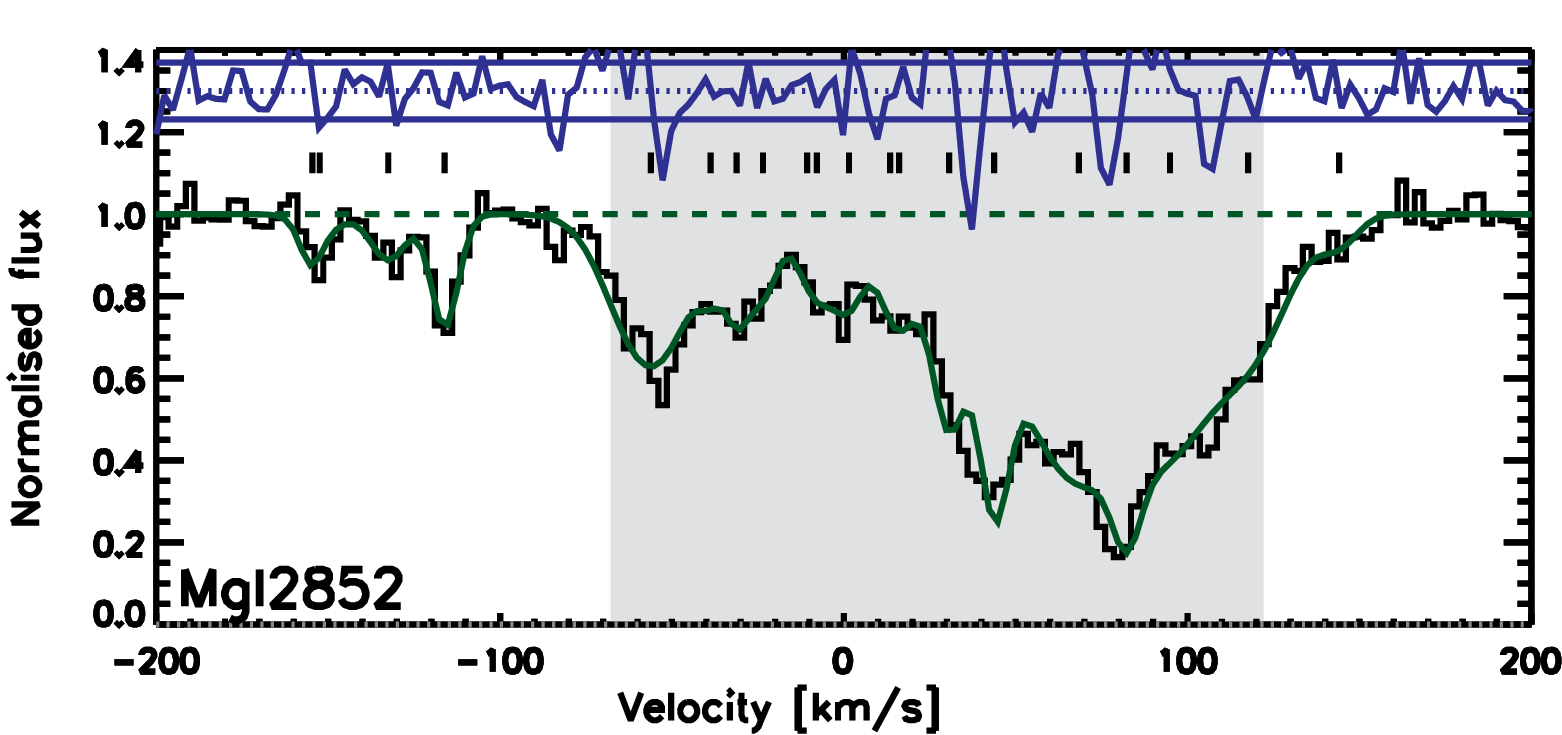}
    \includegraphics[width=0.33\textwidth,trim=0 15 0 0,angle=0,clip=false]{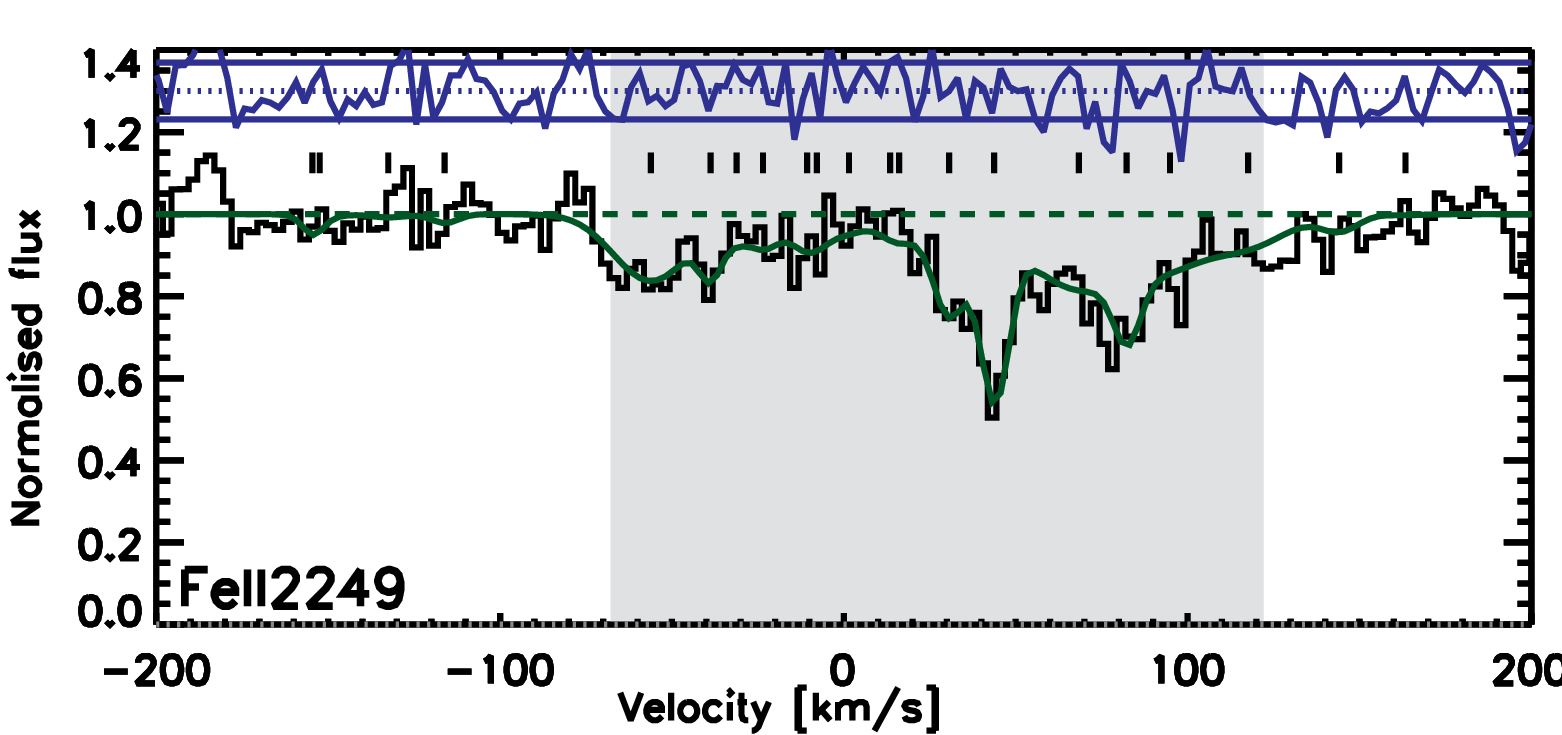}
    \includegraphics[width=0.33\textwidth,trim=0 15 0 0,angle=0,clip=false]{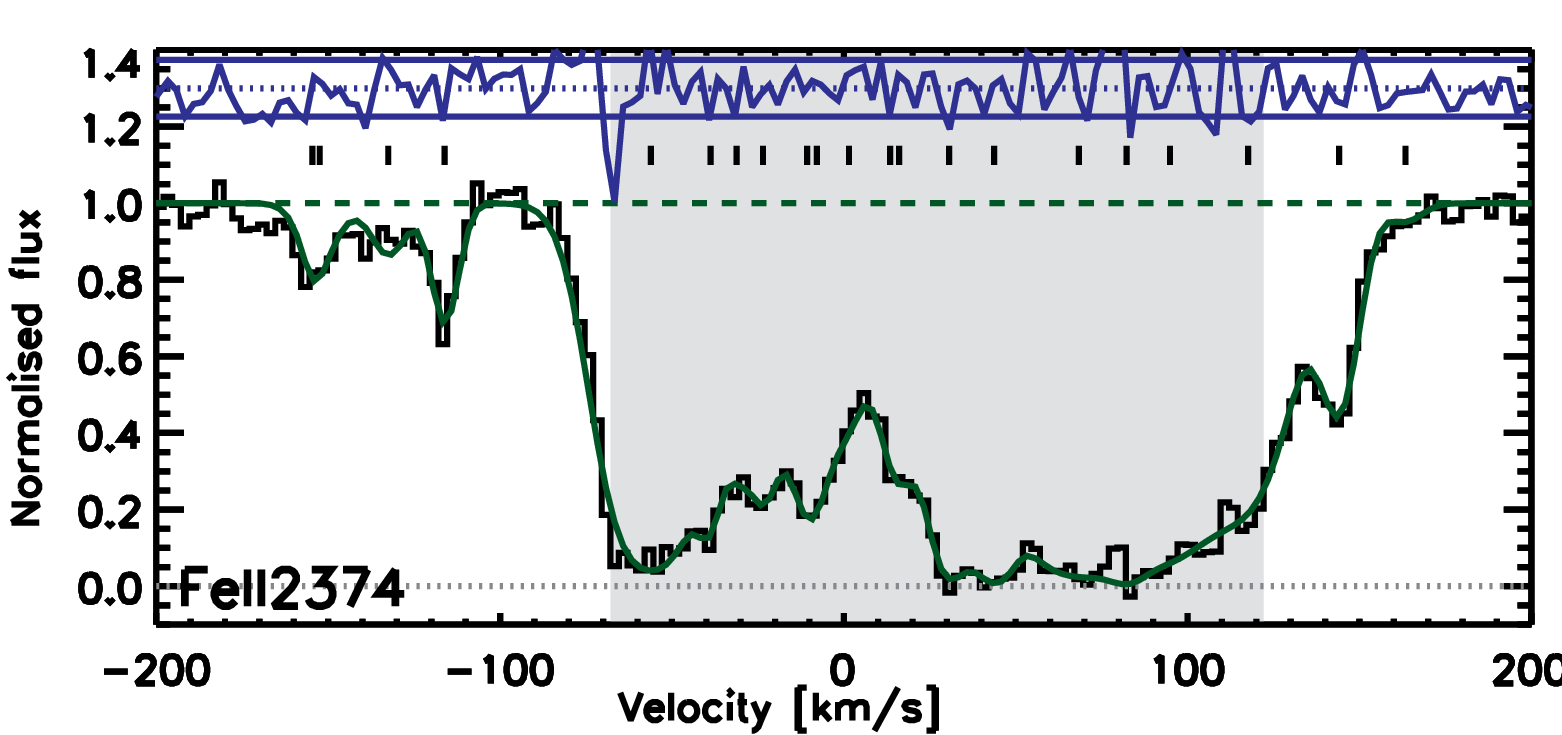}
  }
  \hbox{
    \includegraphics[width=0.33\textwidth,trim=0 15 0 0,angle=0,clip=false]{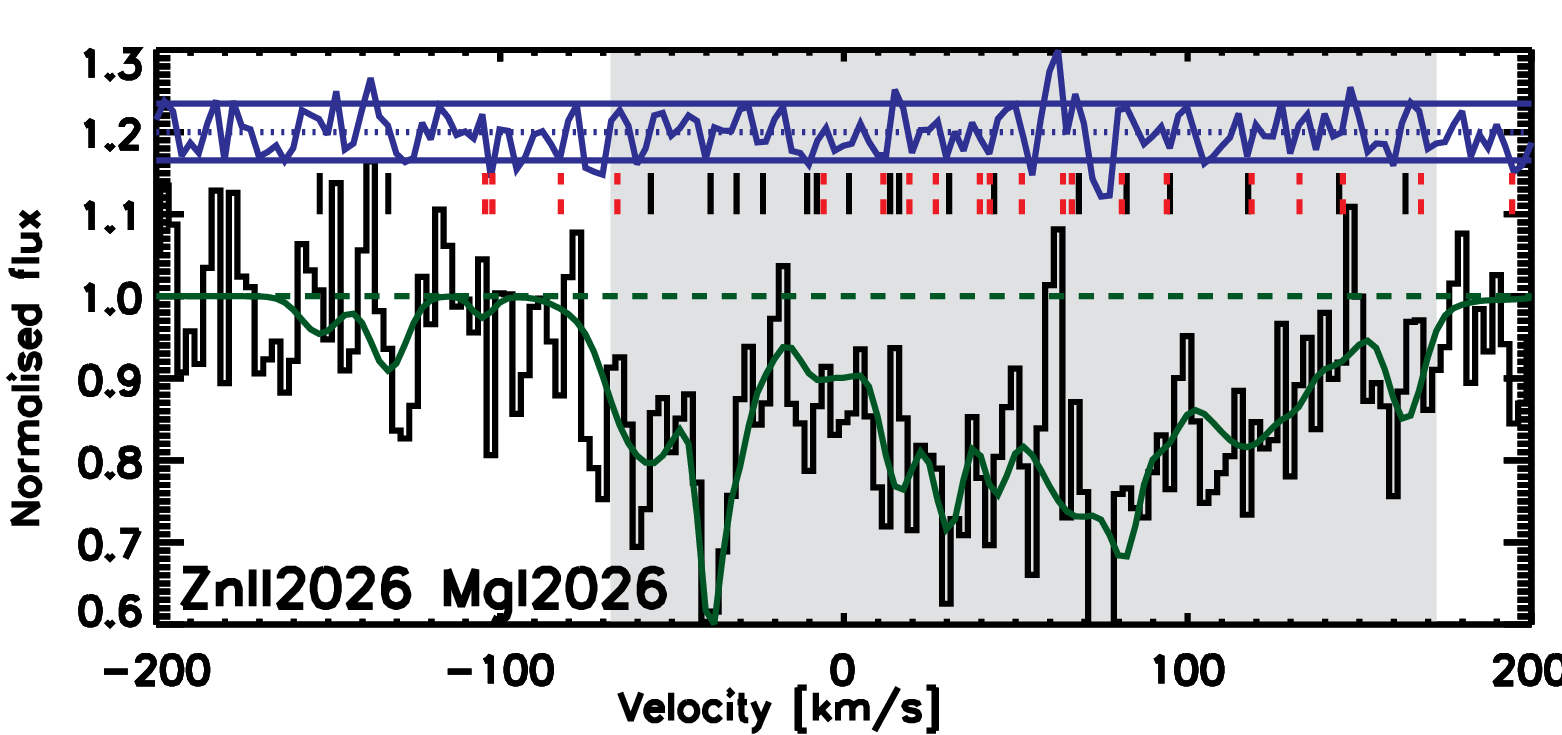}
    \includegraphics[width=0.33\textwidth,trim=0 15 0 0,angle=0,clip=false]{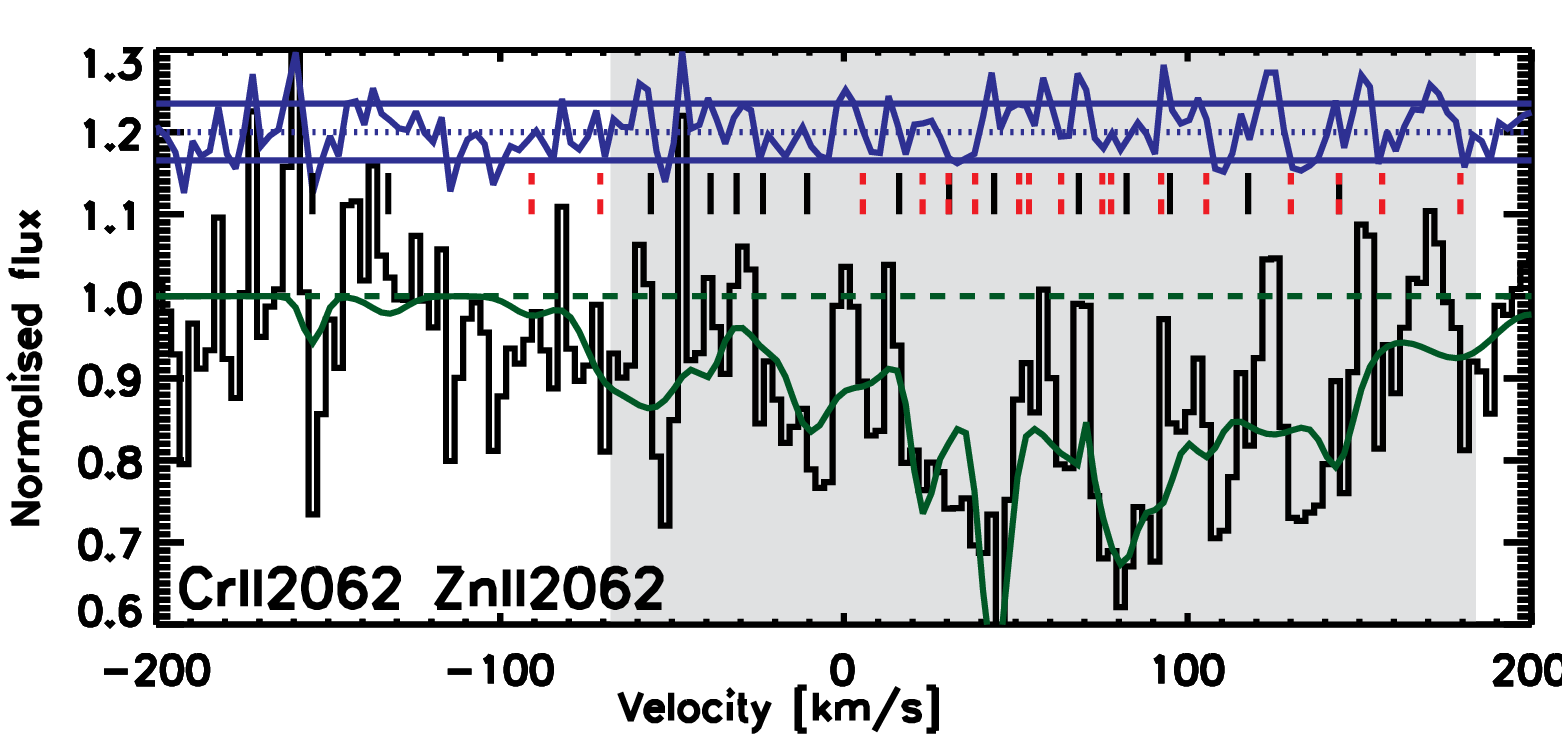}
    \includegraphics[width=0.33\textwidth,trim=0 15 0 0,angle=0,clip=false]{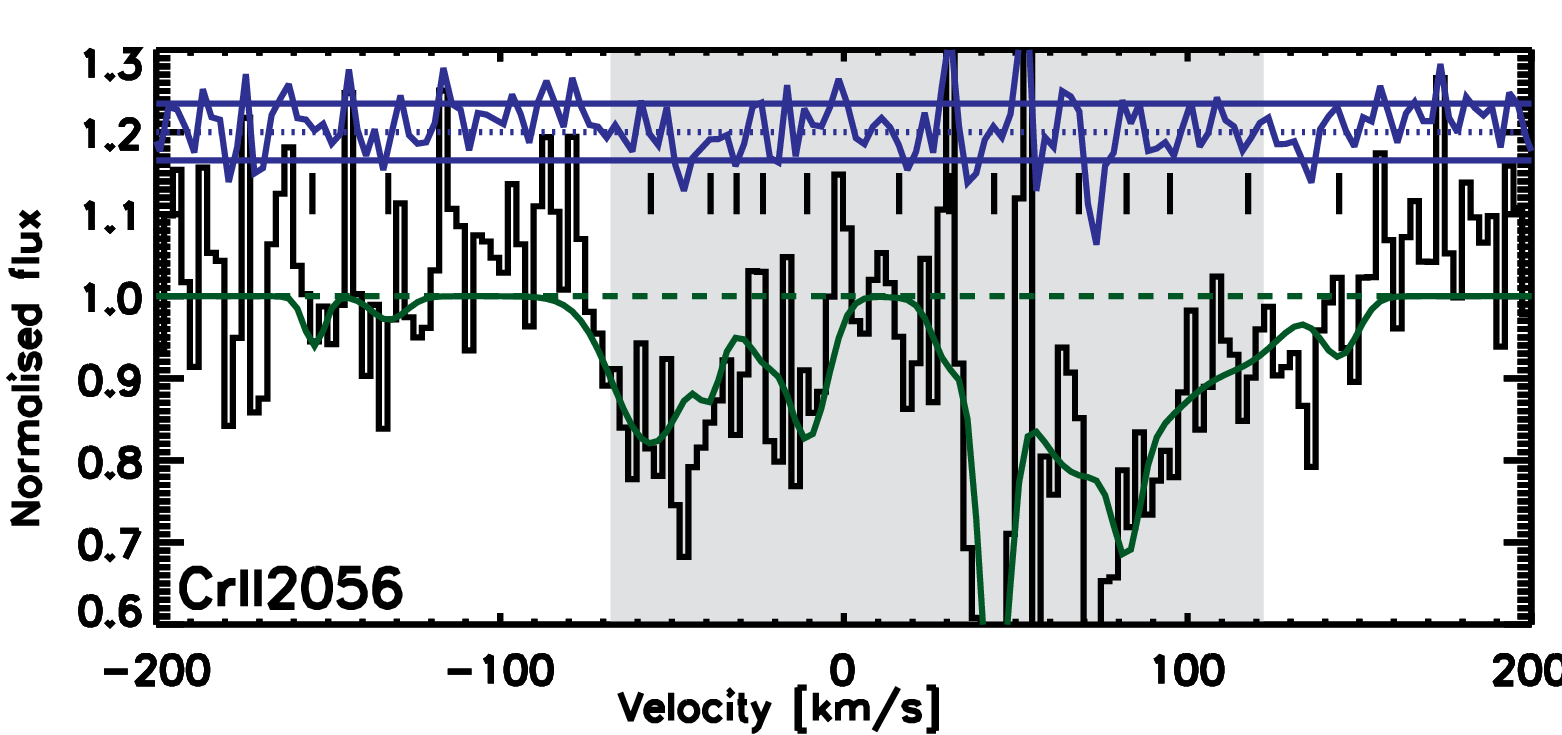}
  }
  \hbox{
    \includegraphics[width=0.33\textwidth,trim=0 15 0 0,angle=0,clip=false]{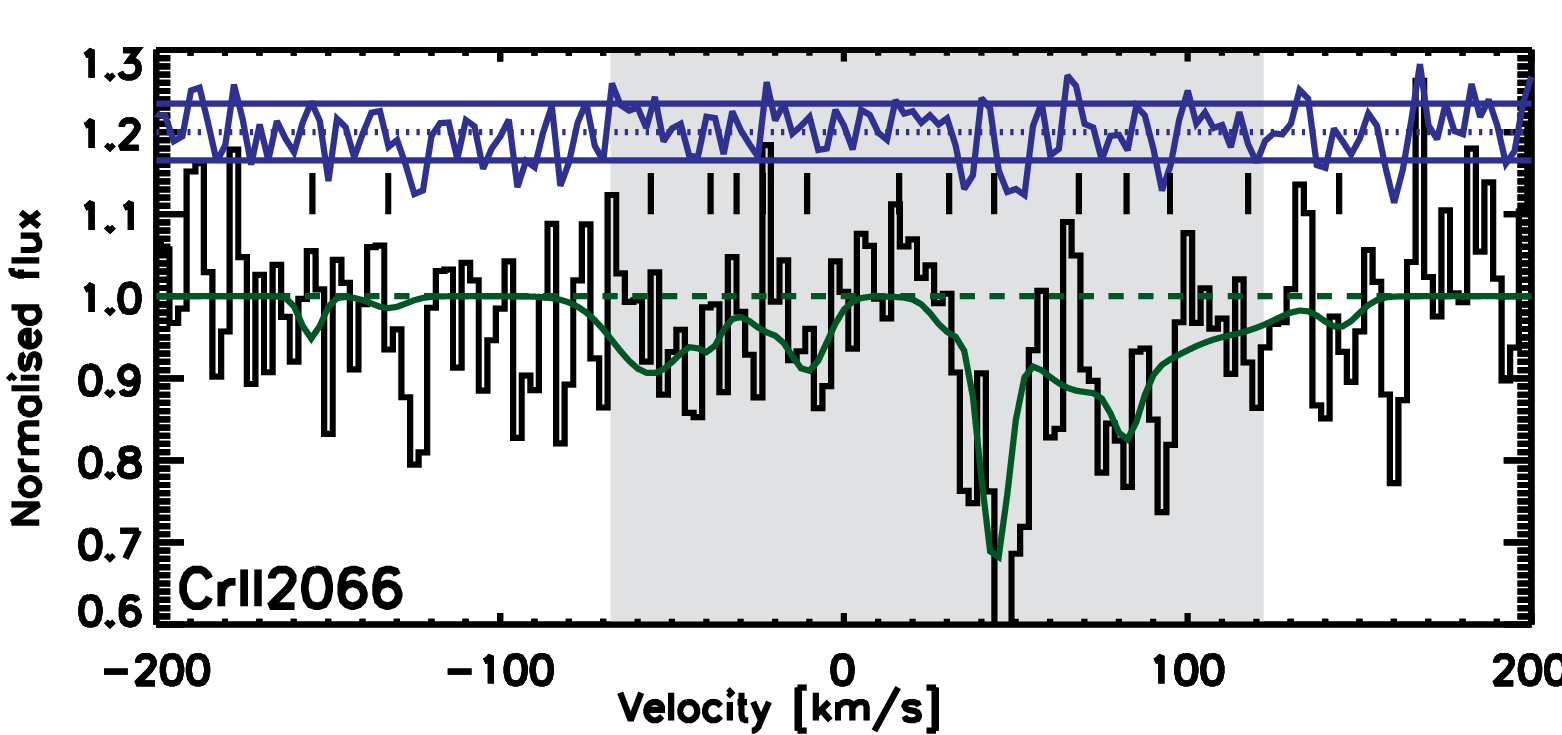}
    \includegraphics[width=0.33\textwidth,trim=0 15 0 0,angle=0,clip=false]{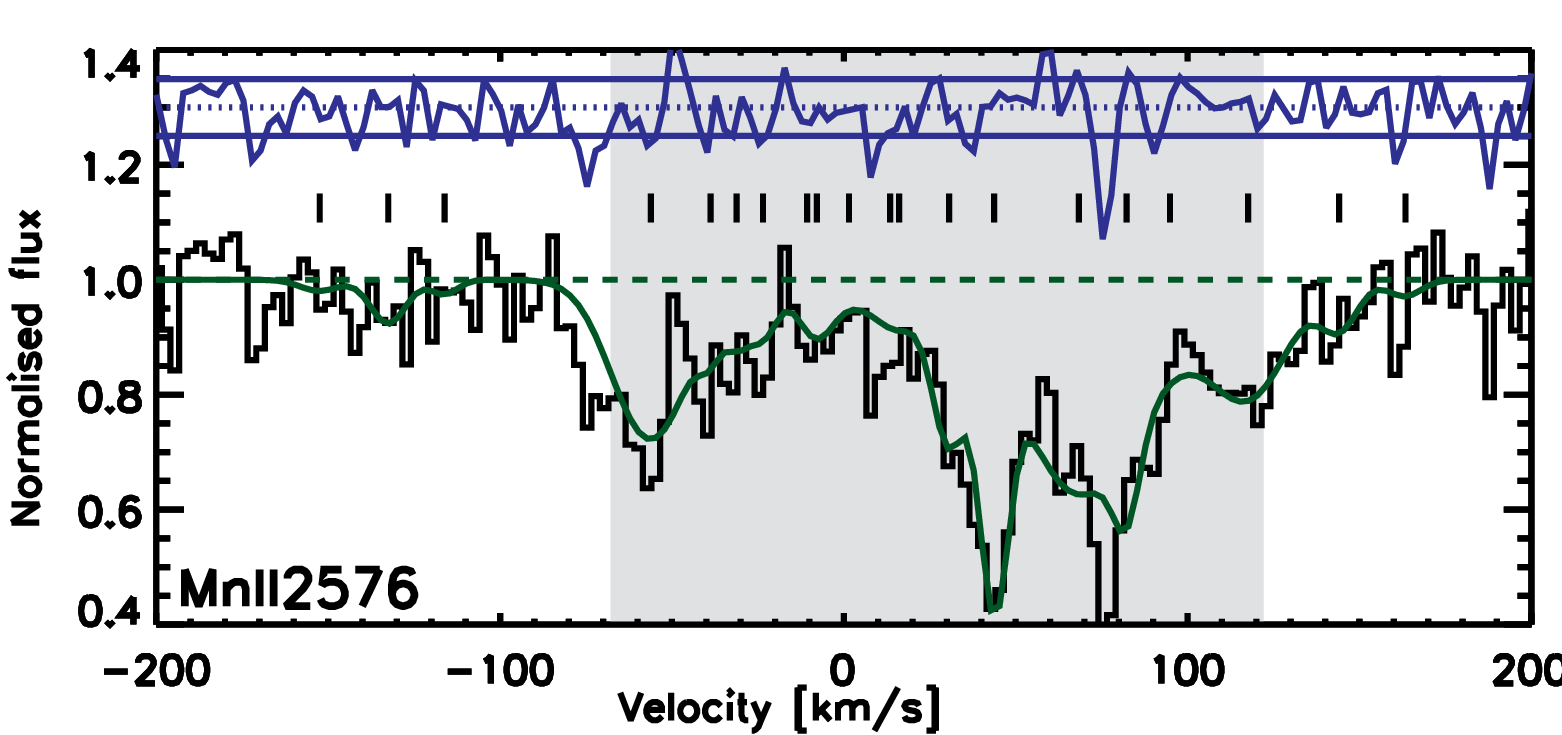}
    \includegraphics[width=0.33\textwidth,trim=0 15 0 0,angle=0,clip=false]{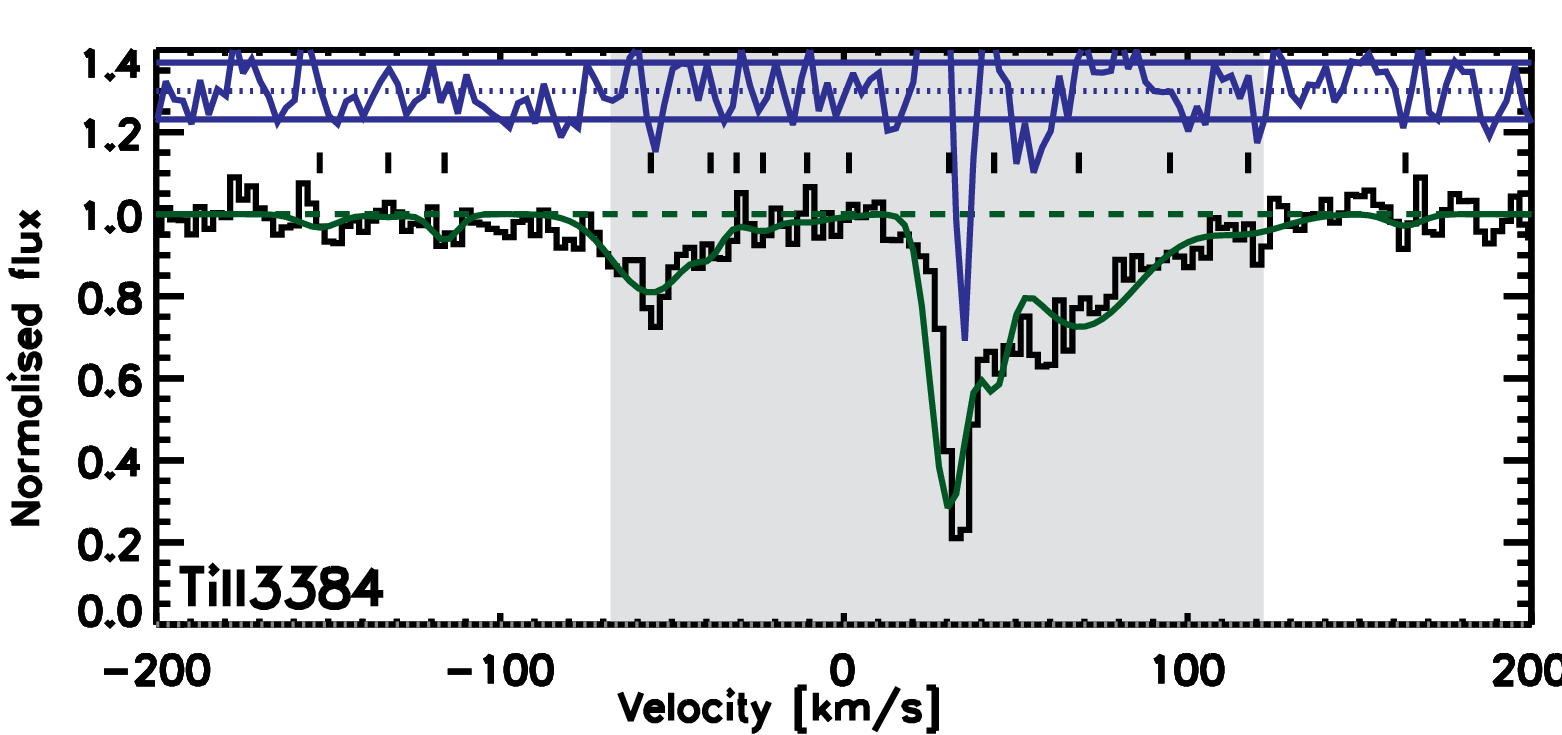}
  }
}
 \caption{Voigt profile fits to the $z=0.74030$ absorber towards
   J1107$+$0048, see Fig.~\ref{fig:fit_J0334m0711} for description. }
 \label{fig:fit_J1107p0048}
 \end{figure*}
\begin{figure*}
\vbox{
  \hbox{
    \includegraphics[width=0.33\textwidth,trim=0 15 0 0,angle=0,clip=false]{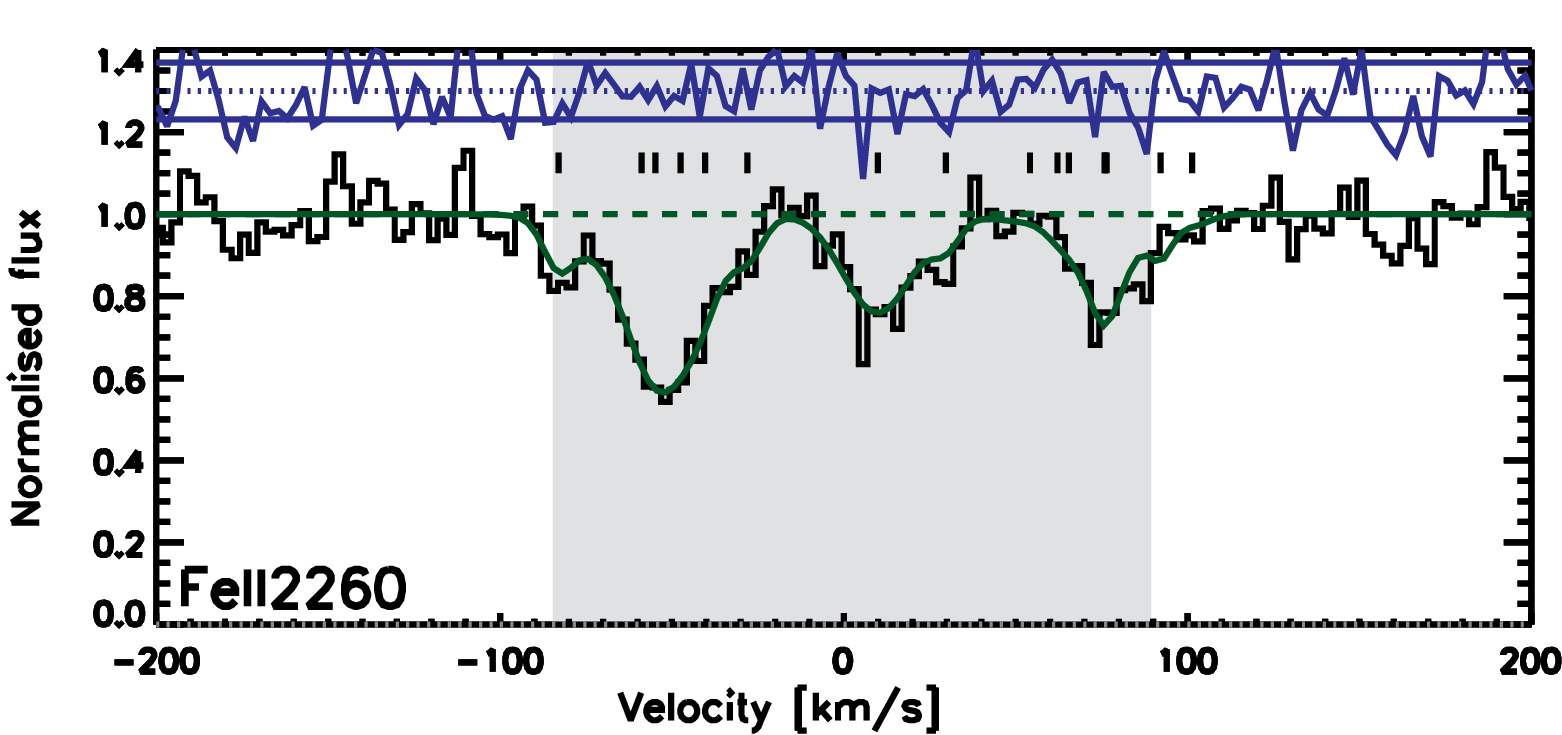}
    \includegraphics[width=0.33\textwidth,trim=0 15 0 0,angle=0,clip=false]{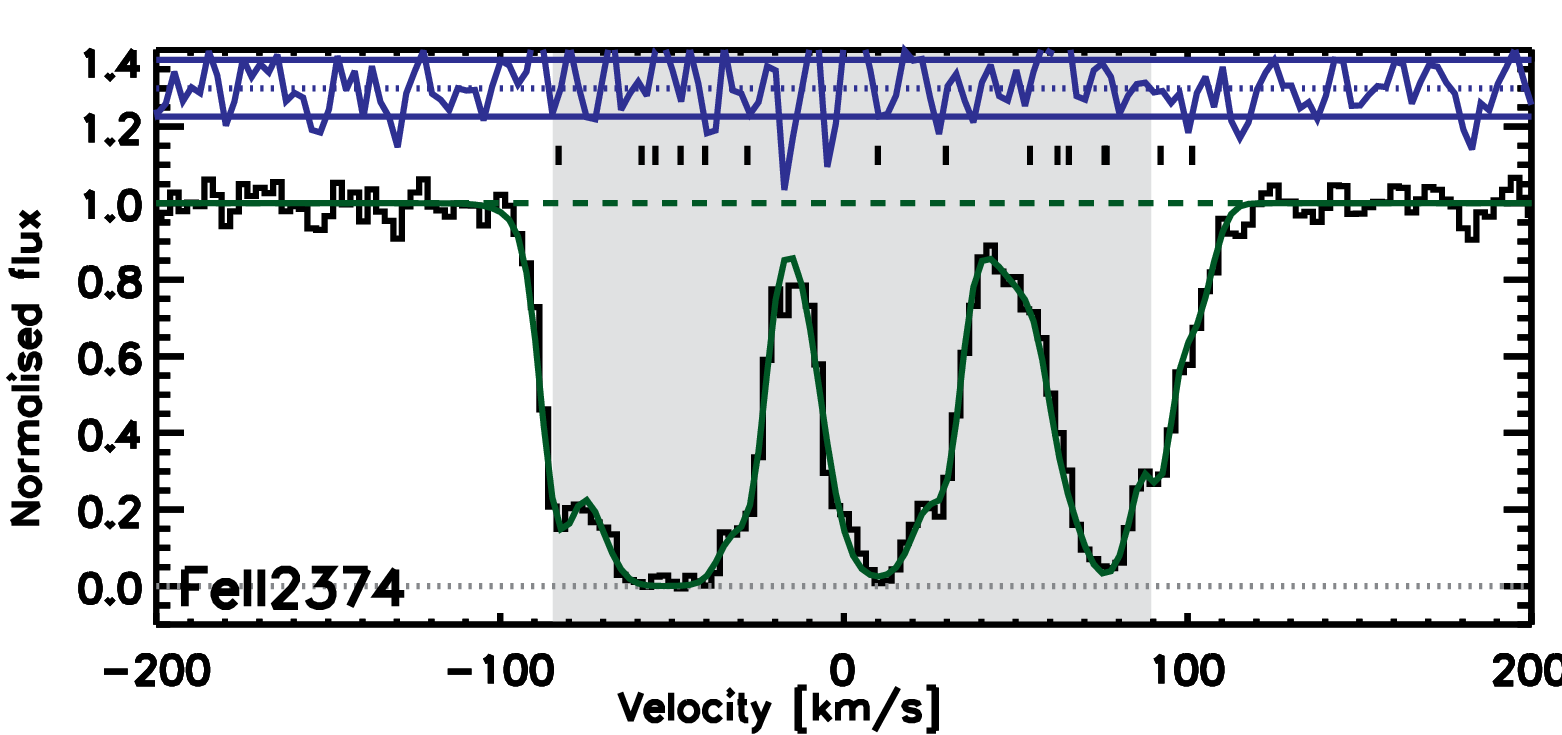}
    \includegraphics[width=0.33\textwidth,trim=0 15 0 0,angle=0,clip=false]{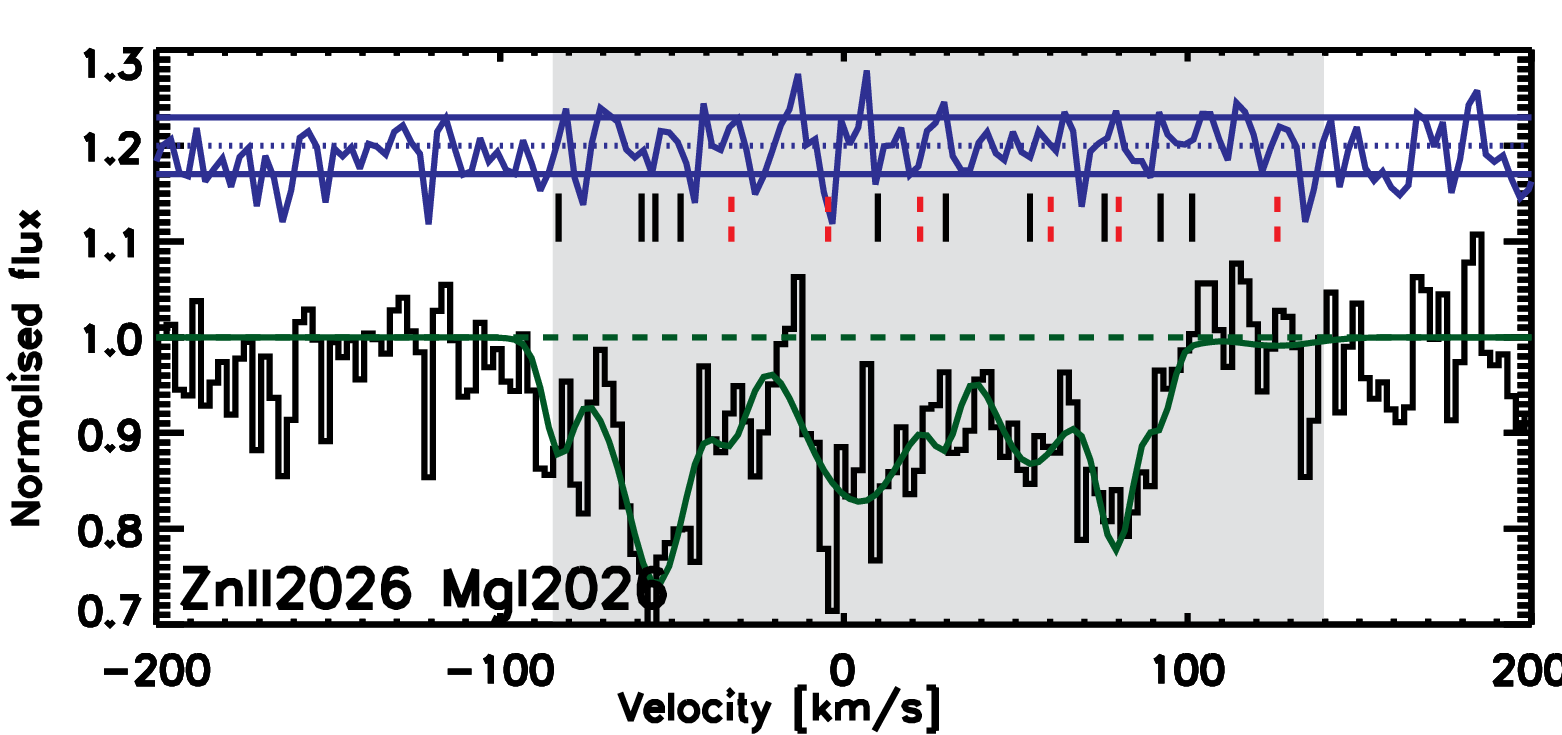}
  }
  \hbox{
    \includegraphics[width=0.33\textwidth,trim=0 15 0 0,angle=0,clip=false]{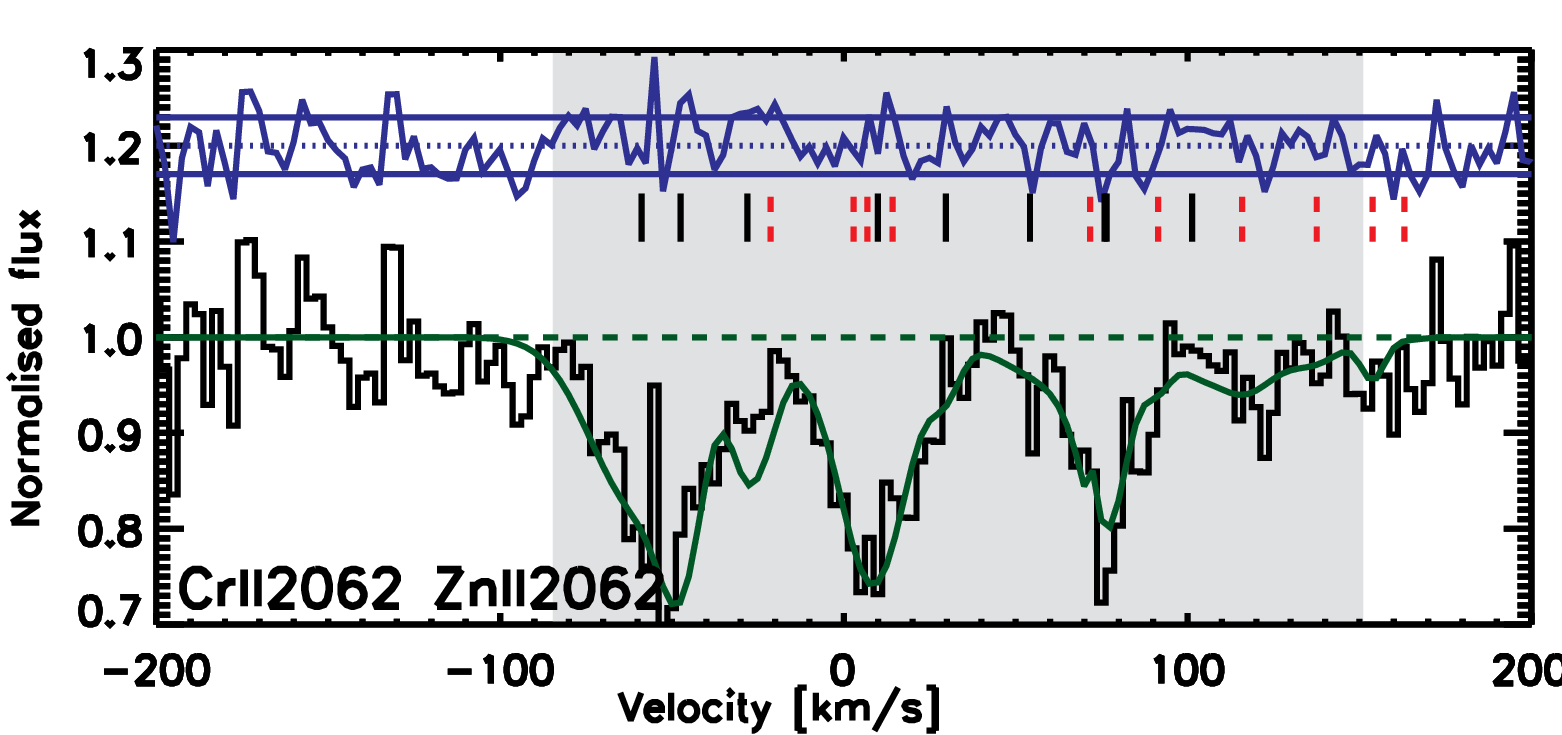}
    \includegraphics[width=0.33\textwidth,trim=0 15 0 0,angle=0,clip=false]{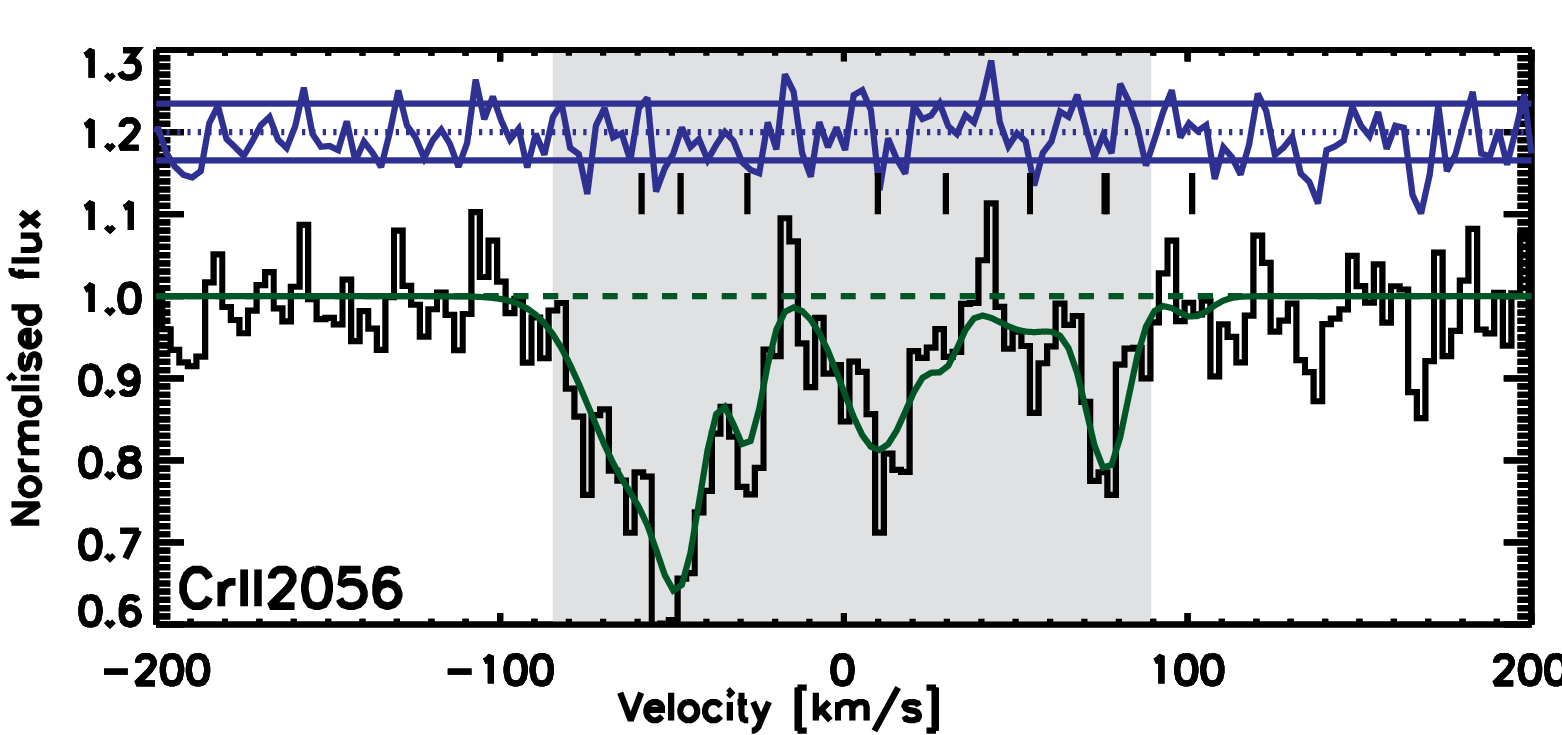}
    \includegraphics[width=0.33\textwidth,trim=0 15 0 0,angle=0,clip=false]{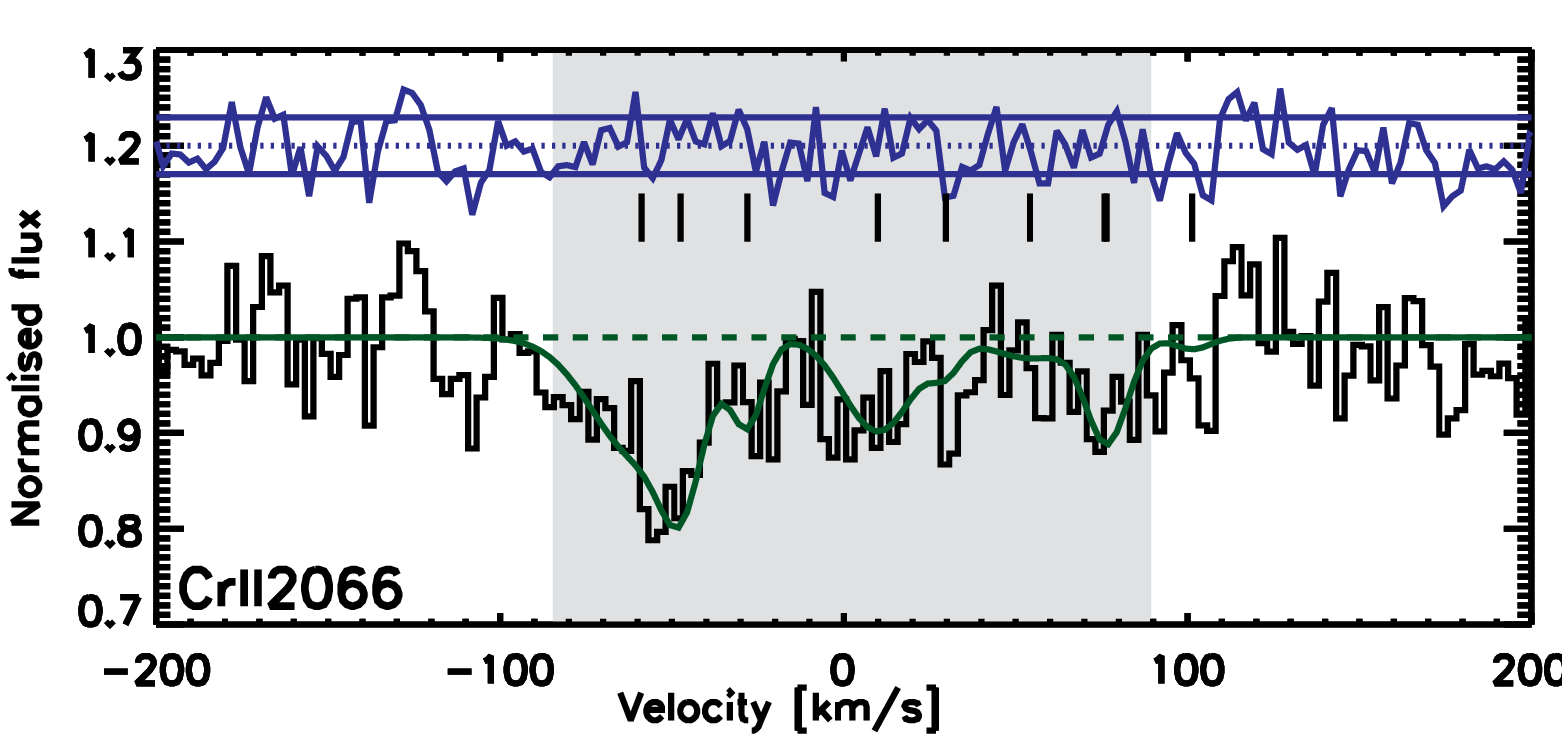}
  }
  \hbox{
    \includegraphics[width=0.33\textwidth,trim=0 15 0 0,angle=0,clip=false]{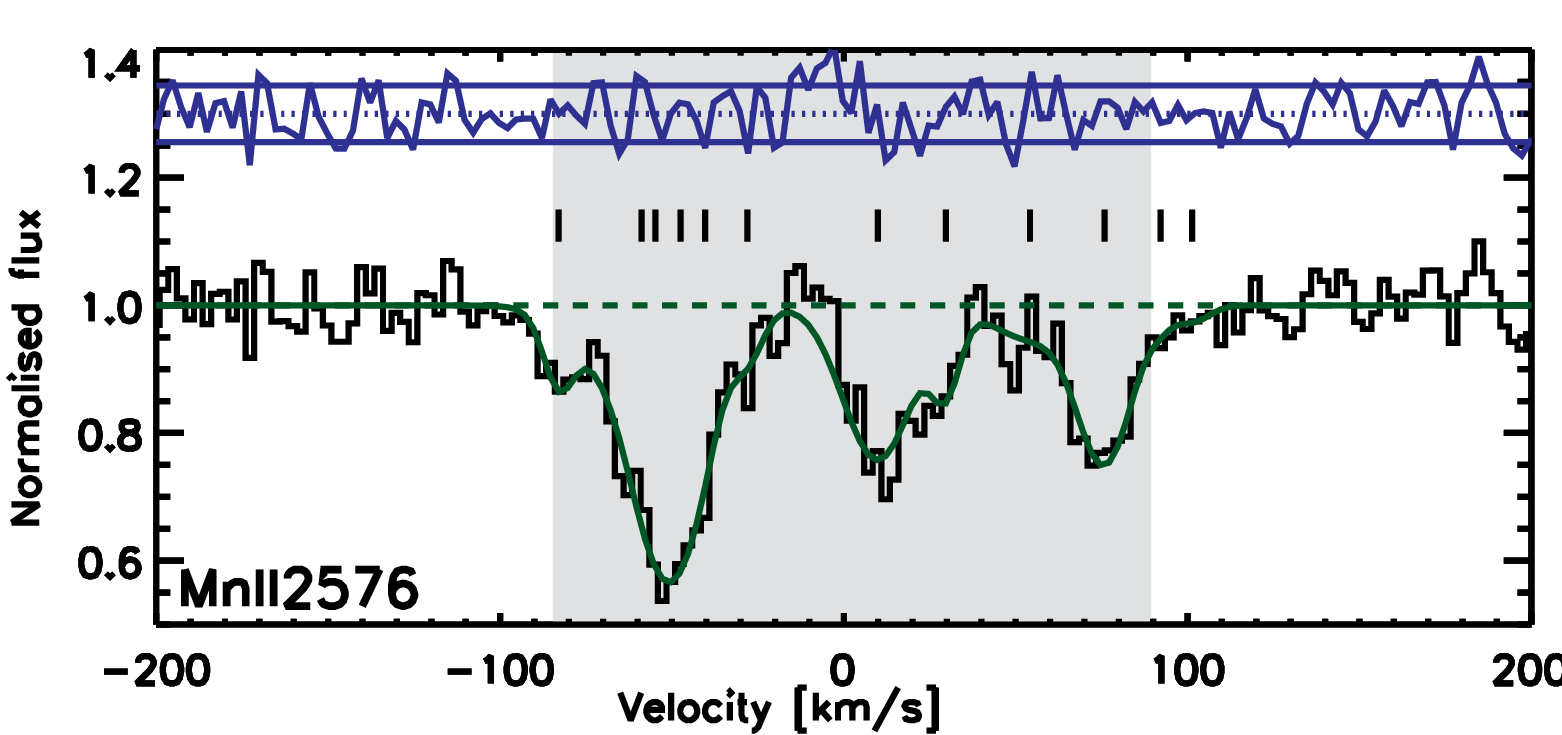}
    \includegraphics[width=0.33\textwidth,trim=0 15 0 0,angle=0,clip=false]{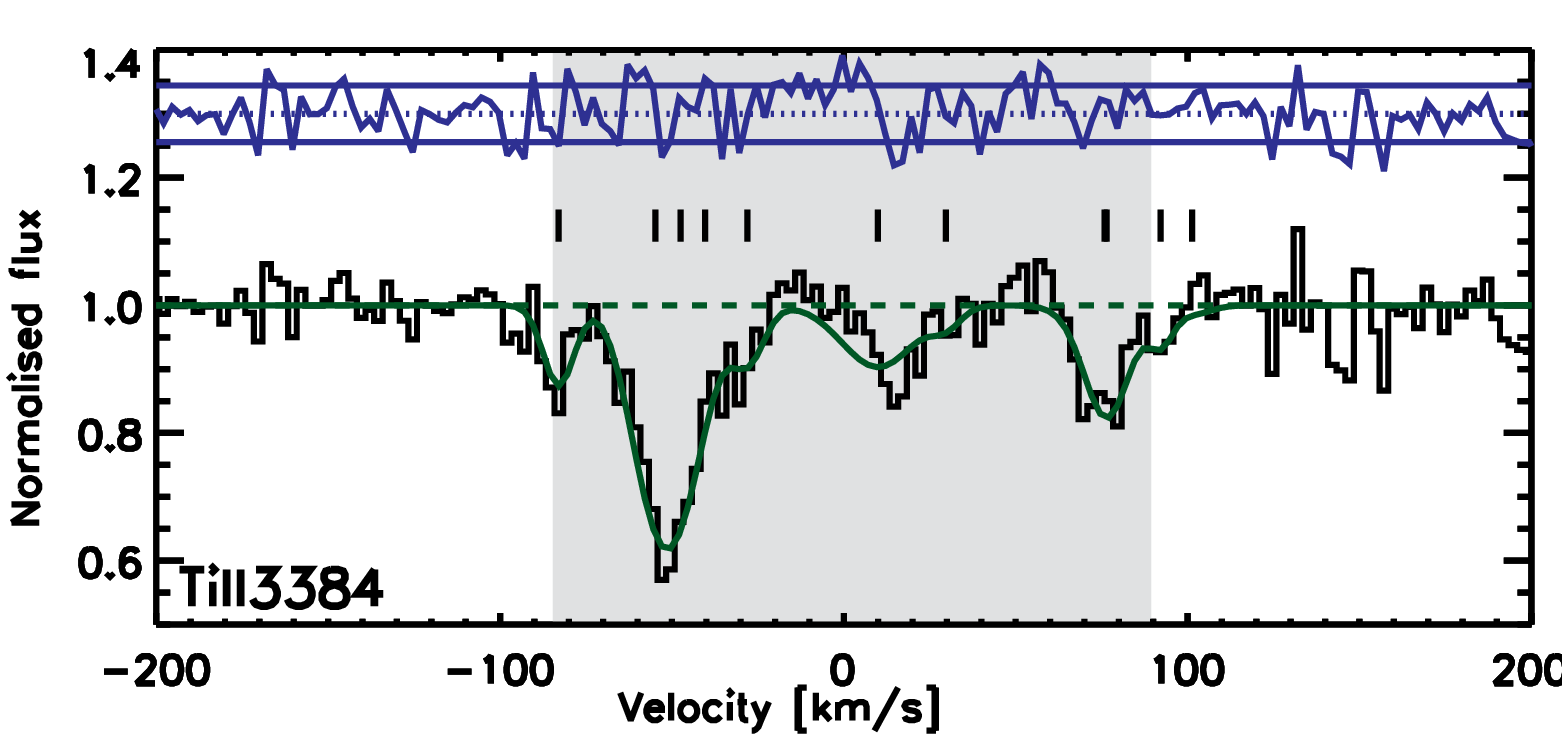}
    \includegraphics[width=0.33\textwidth,trim=0 15 0 0,angle=0,clip=false]{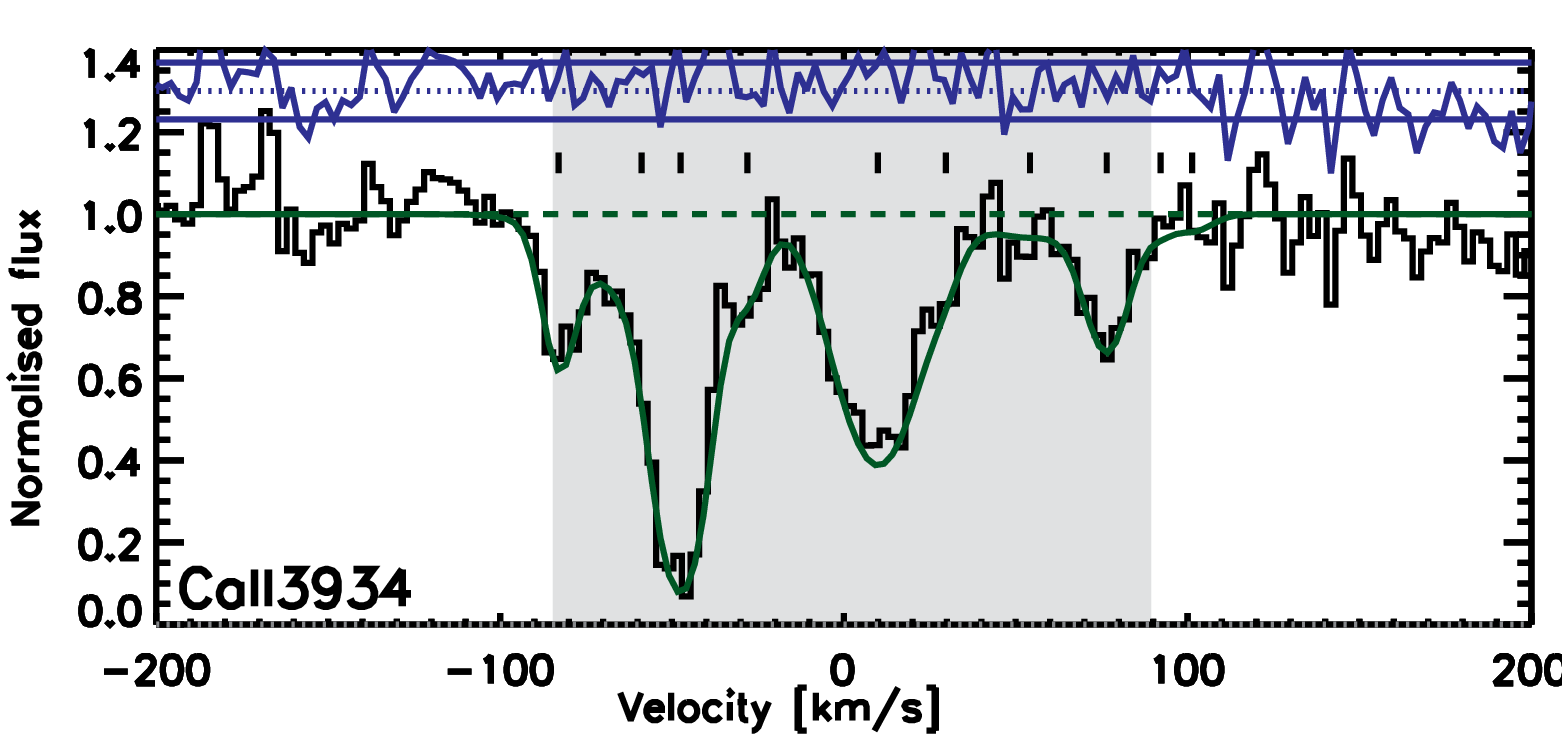}
  }
  \hbox{
    \includegraphics[width=0.33\textwidth,trim=0 15 0 0,angle=0,clip=false]{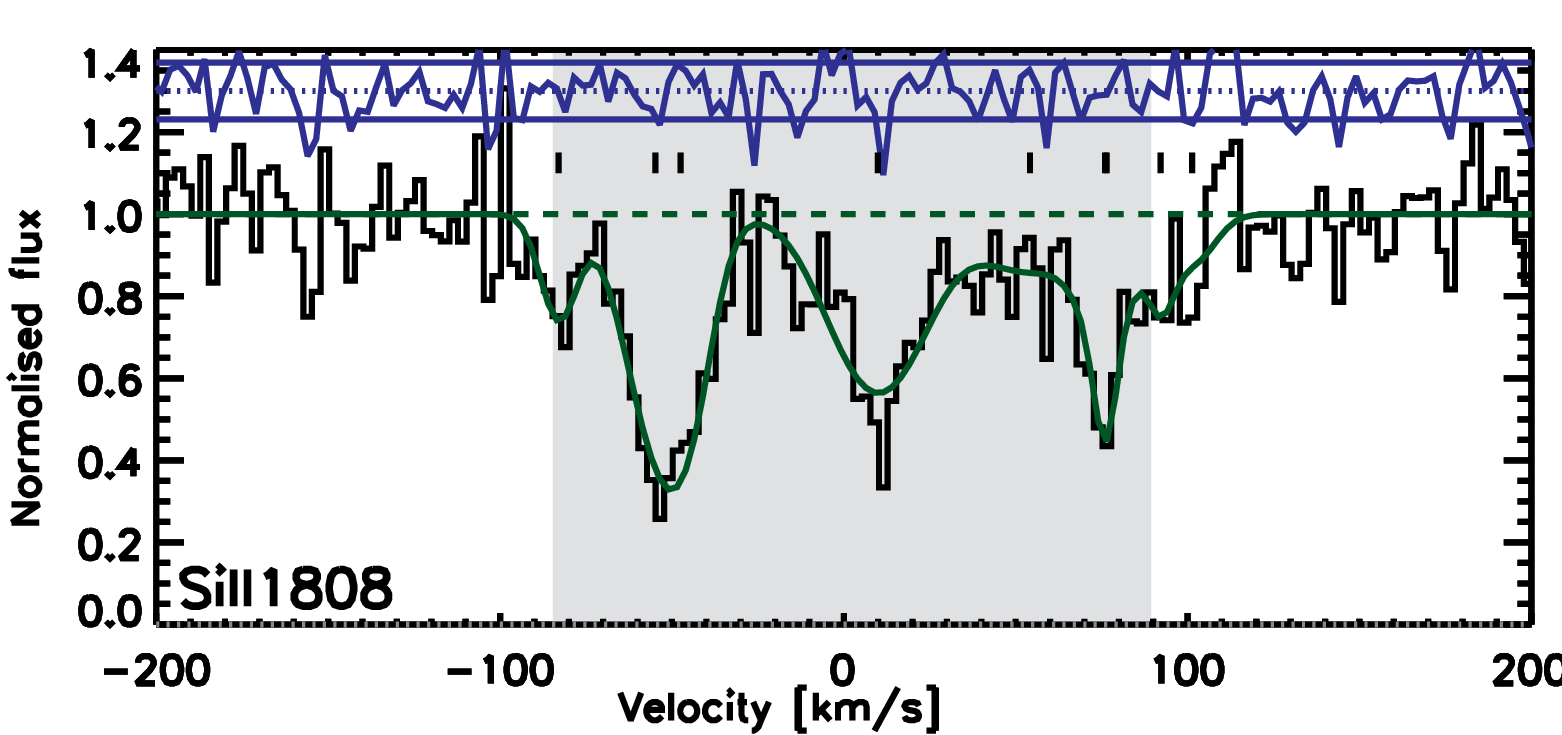}
    \includegraphics[width=0.33\textwidth,trim=0 15 0 0,angle=0,clip=false]{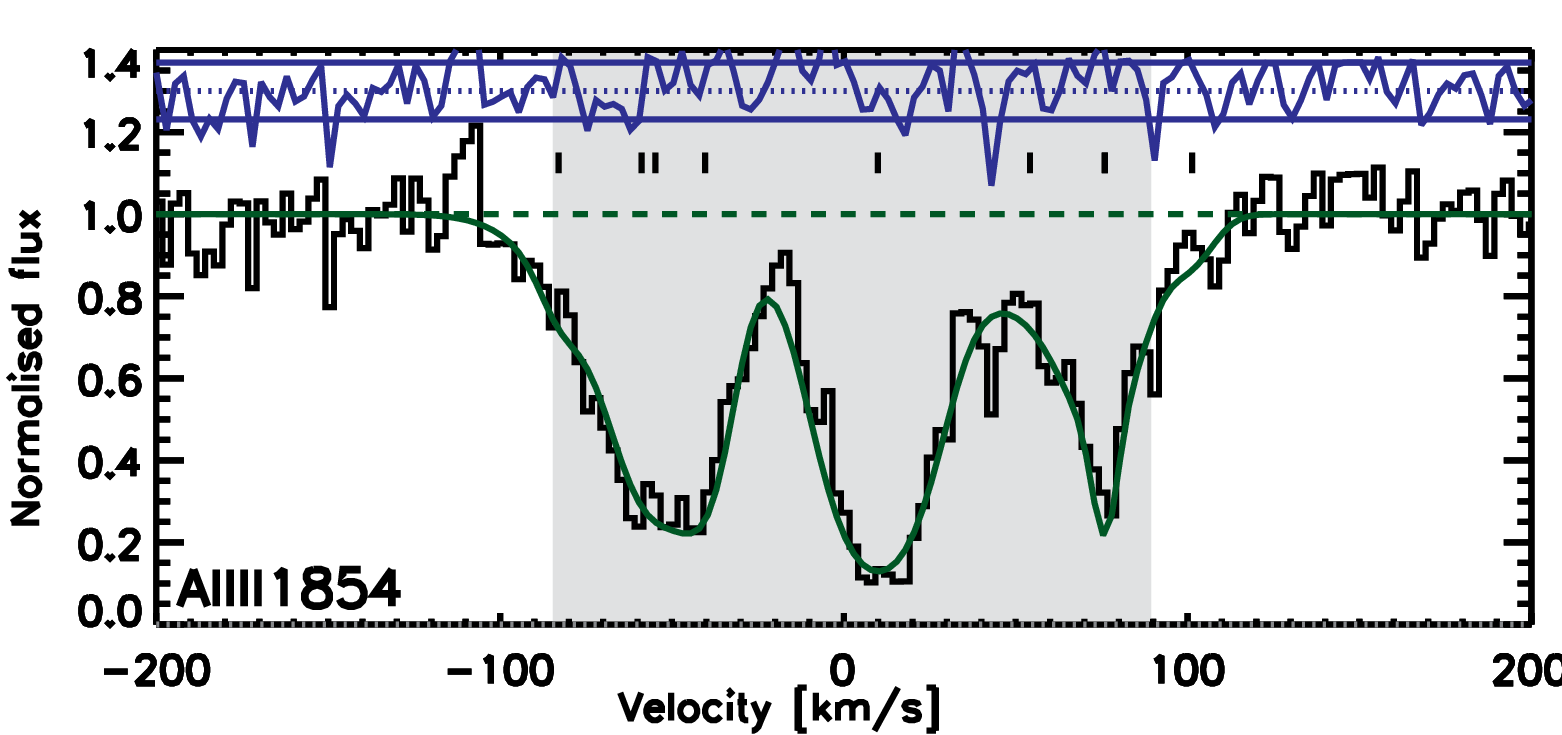}
  }
}
\caption{Voigt profile fits to the $z=0.96497$ absorber towards
  J1129$+$0204, see Fig.~\ref{fig:fit_J0334m0711} for description. }
\label{fig:fit_J1129p0204}
\end{figure*}
\begin{figure*}
\vbox{
  \hbox{
    \includegraphics[width=0.33\textwidth,trim=0 15 0 0,angle=0,clip=false]{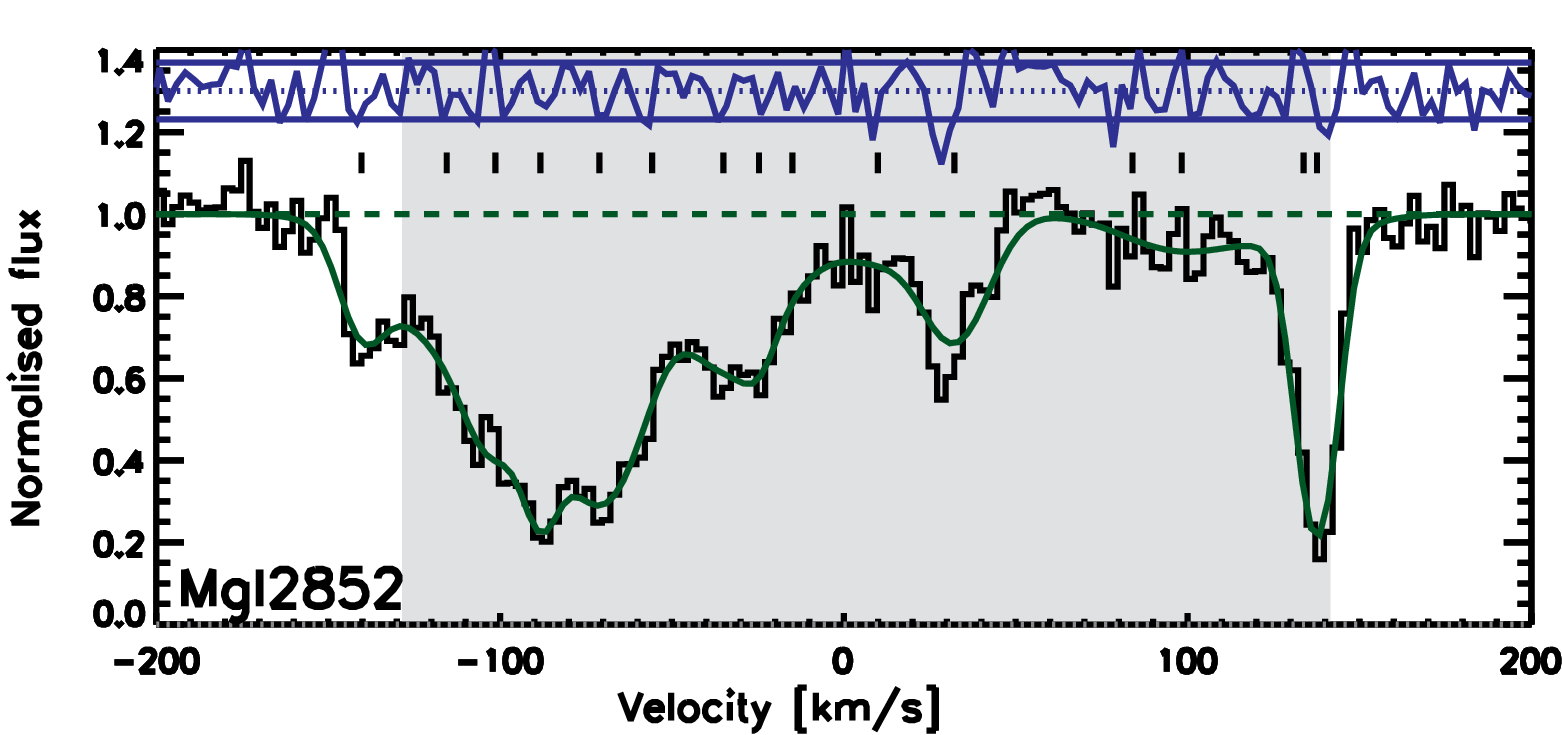}
    \includegraphics[width=0.33\textwidth,trim=0 15 0 0,angle=0,clip=false]{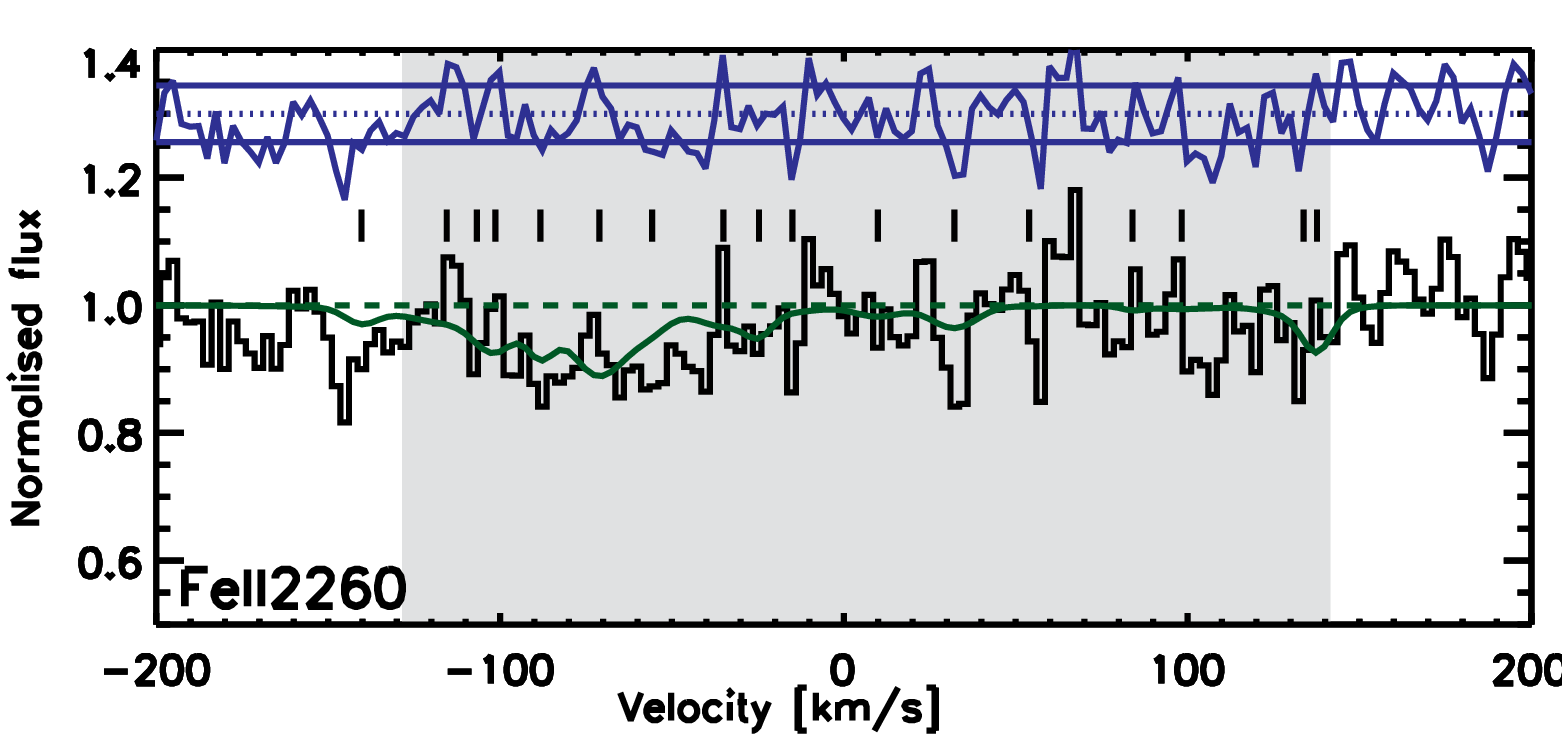}
    \includegraphics[width=0.33\textwidth,trim=0 15 0 0,angle=0,clip=false]{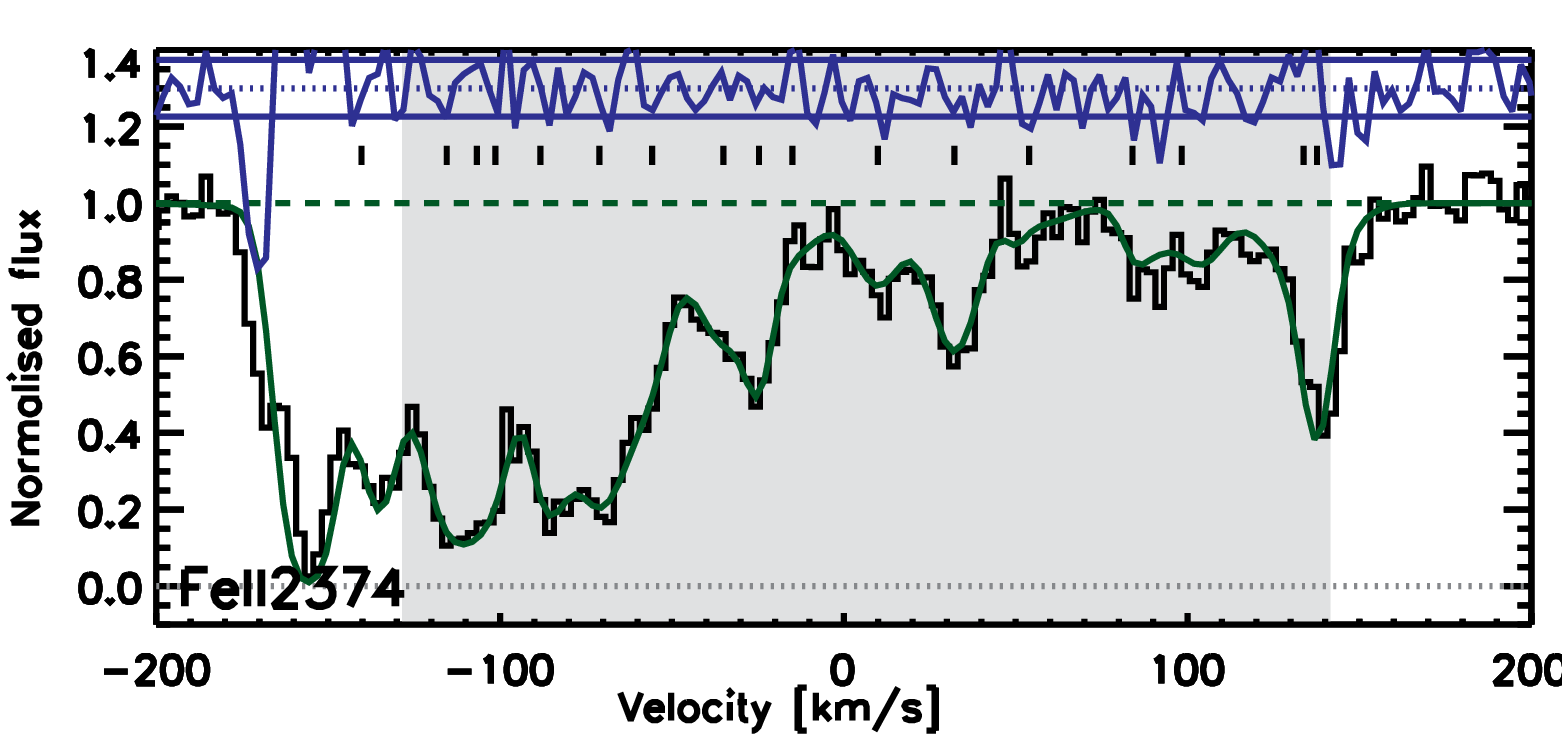}
  }
  \hbox{
    \includegraphics[width=0.33\textwidth,trim=0 15 0 0,angle=0,clip=false]{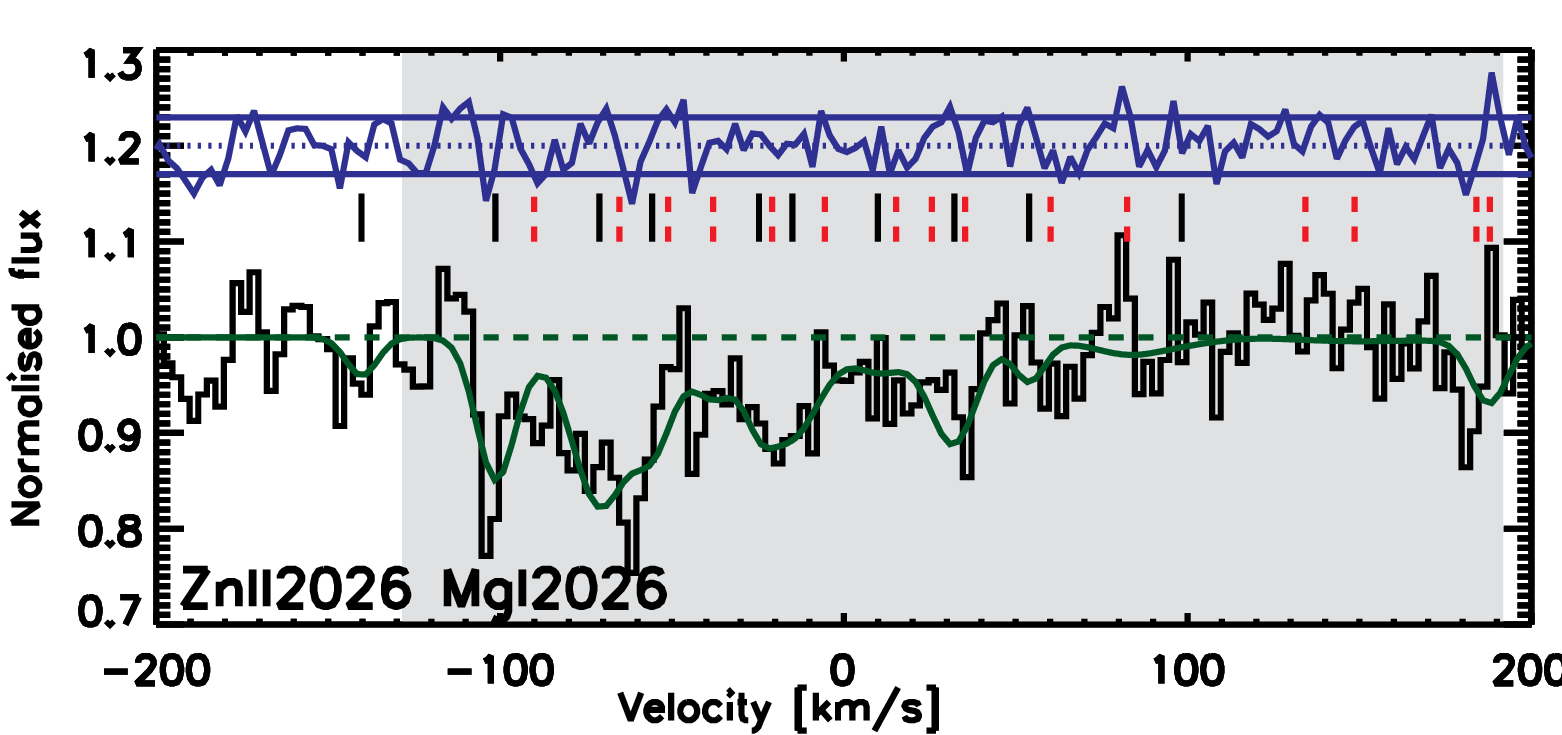}
    \includegraphics[width=0.33\textwidth,trim=0 15 0 0,angle=0,clip=false]{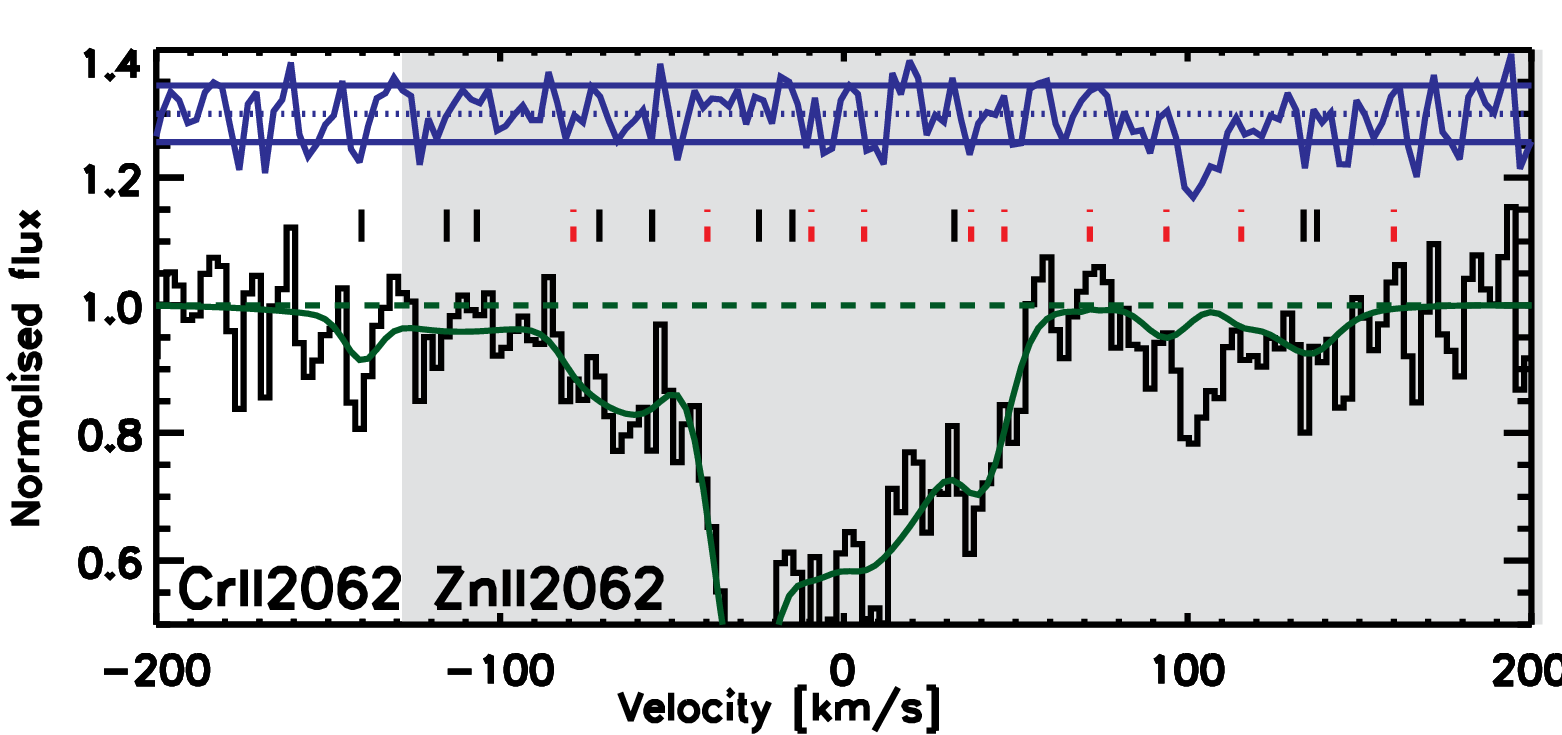}
    \includegraphics[width=0.33\textwidth,trim=0 15 0 0,angle=0,clip=false]{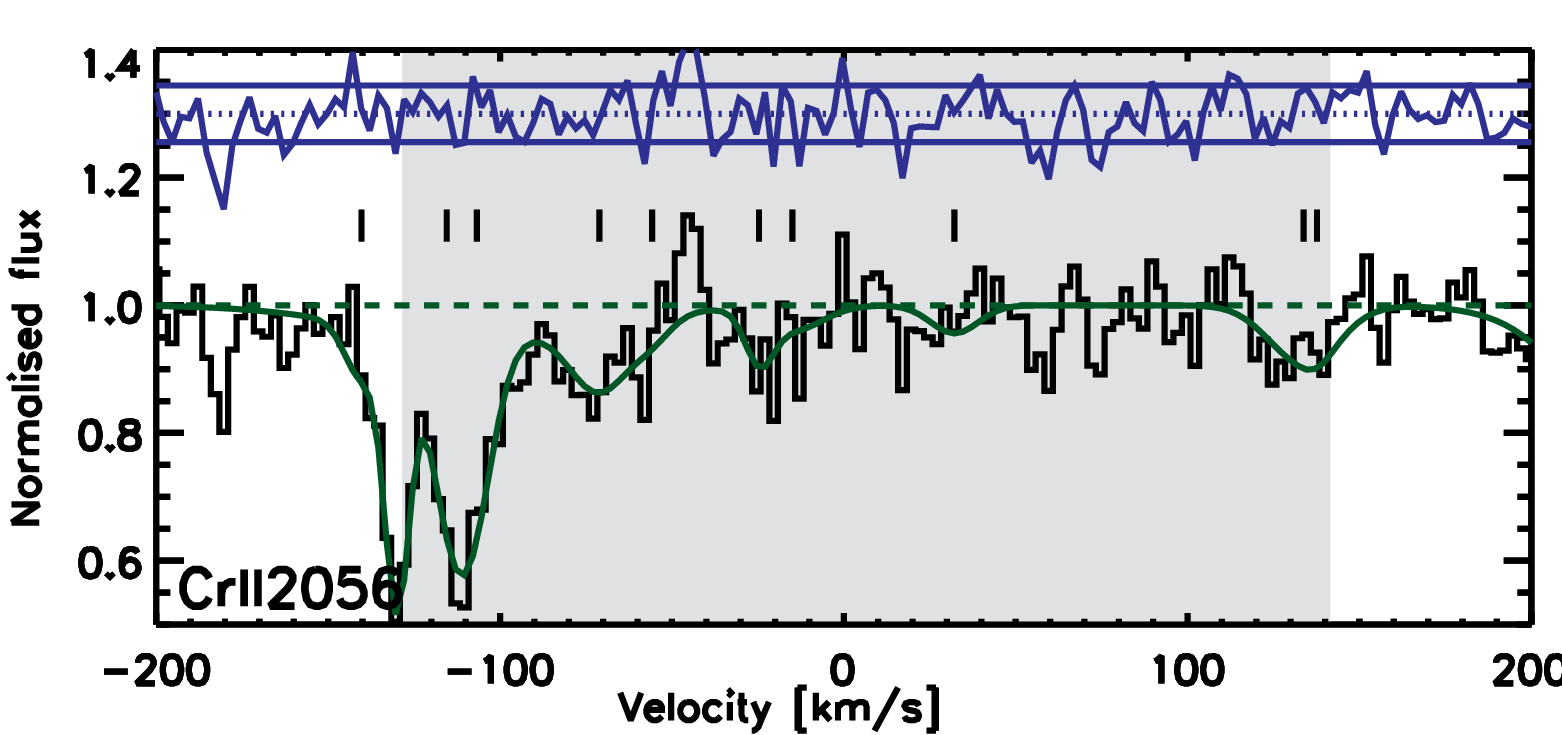}
  }
  \hbox{
    \includegraphics[width=0.33\textwidth,trim=0 15 0 0,angle=0,clip=false]{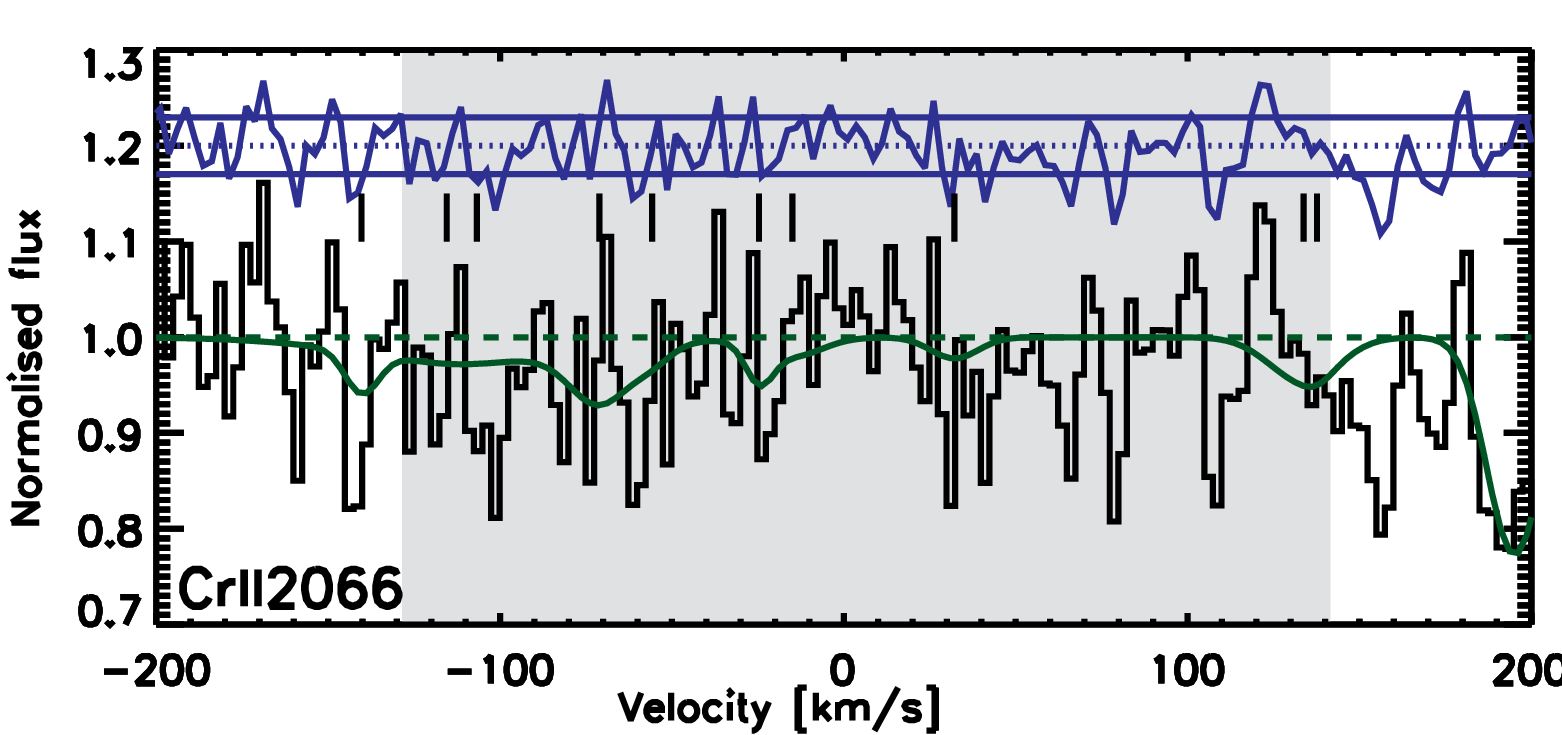}
    \includegraphics[width=0.33\textwidth,trim=0 15 0 0,angle=0,clip=false]{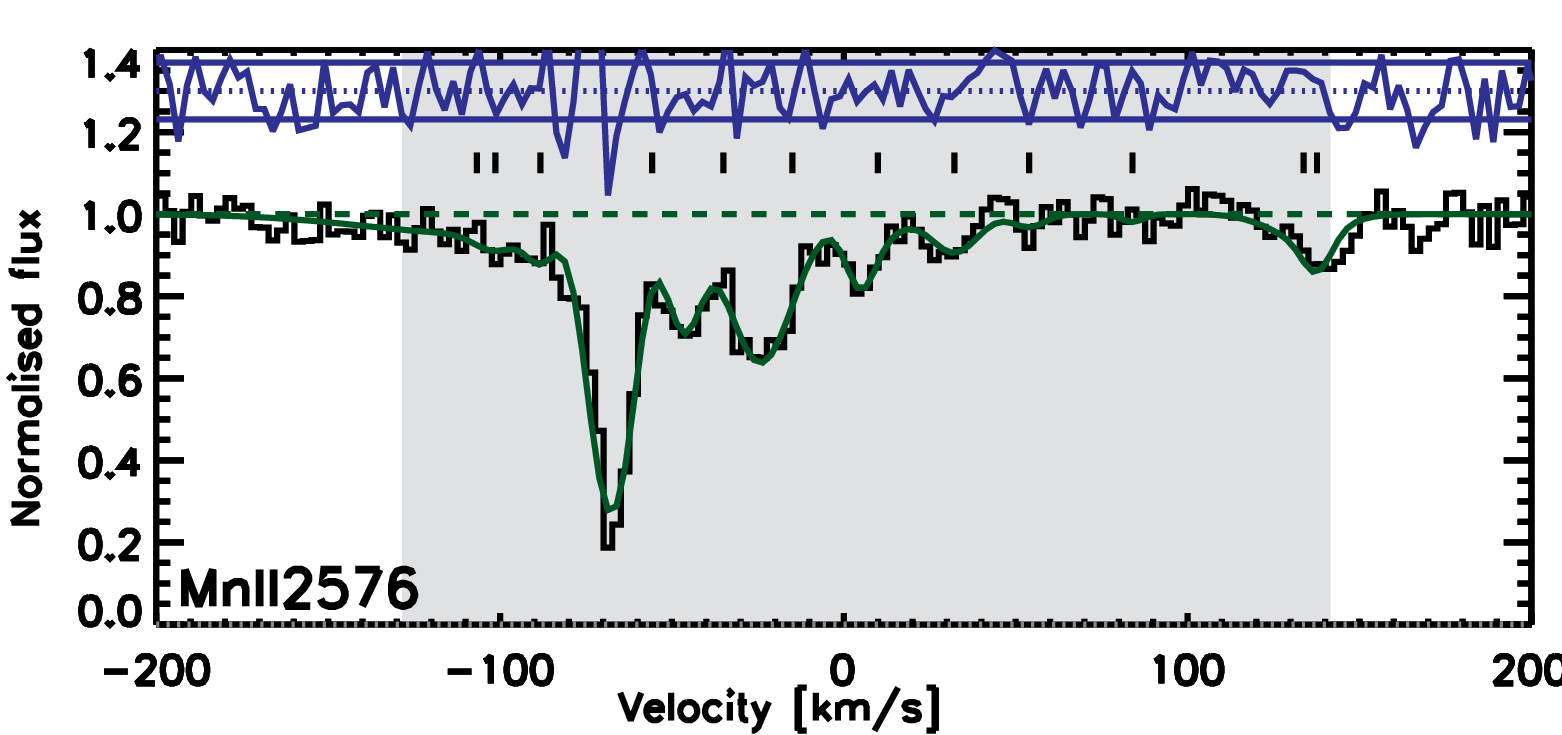}
    \includegraphics[width=0.33\textwidth,trim=0 15 0 0,angle=0,clip=false]{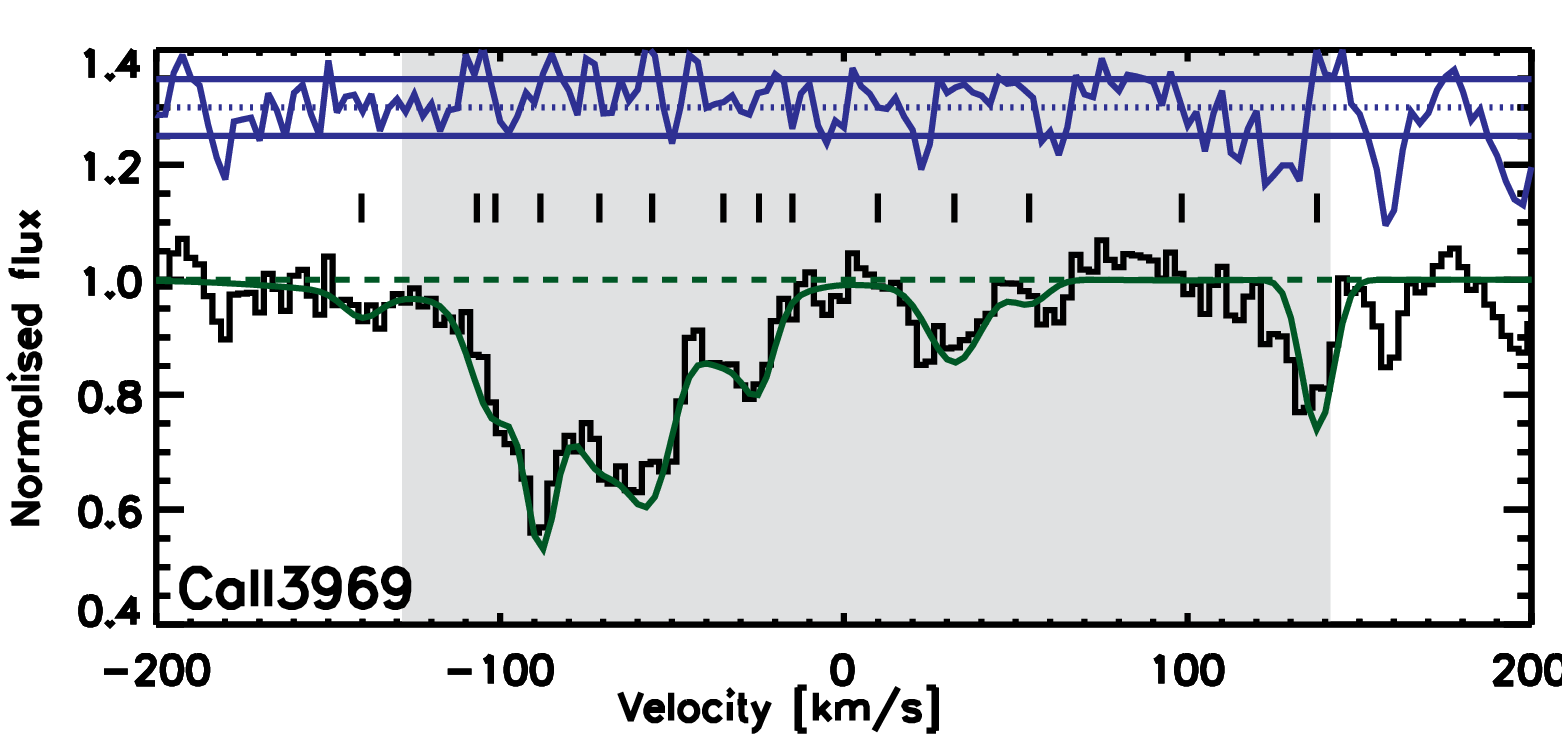}
  }
}
\caption{Voigt profile fits to the $z=0.74630$ absorber towards
  J1203$+$1028, see Fig.~\ref{fig:fit_J0334m0711} for
  description. Note there is a blend from an absorber at $z=0.739$ in
  the \FeII$\,\lambda2374$ profile and blends from absorbers at
  $z=1.32$ and $1.57$ in the \CrII\ and \ZnII$\,\lambda2062$ profiles.}
\label{fig:fit_J1203p1028}
\end{figure*}
\begin{figure*}
\vbox{
  \hbox{
    \includegraphics[width=0.33\textwidth,trim=0 15 0 0,angle=0,clip=false]{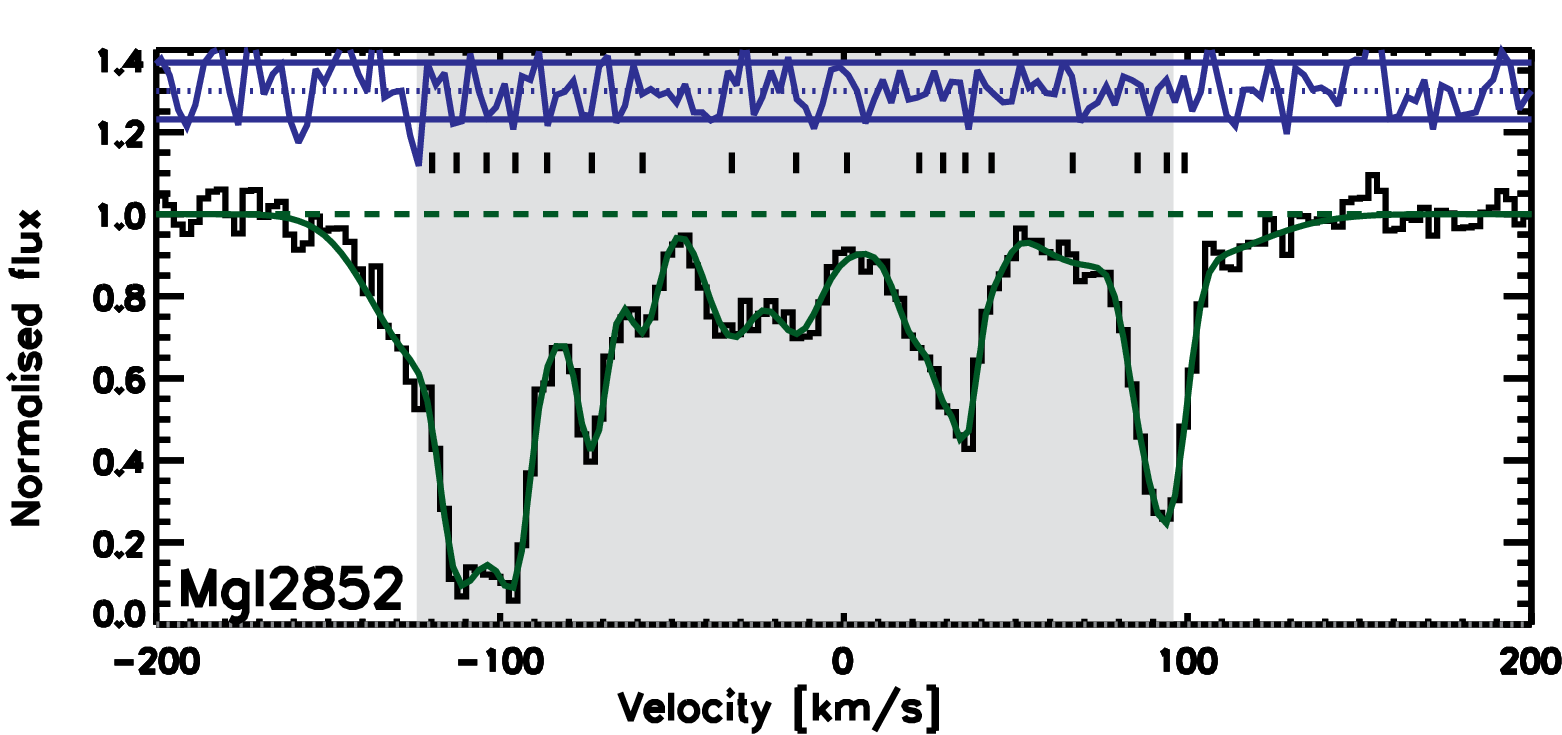}
    \includegraphics[width=0.33\textwidth,trim=0 15 0 0,angle=0,clip=false]{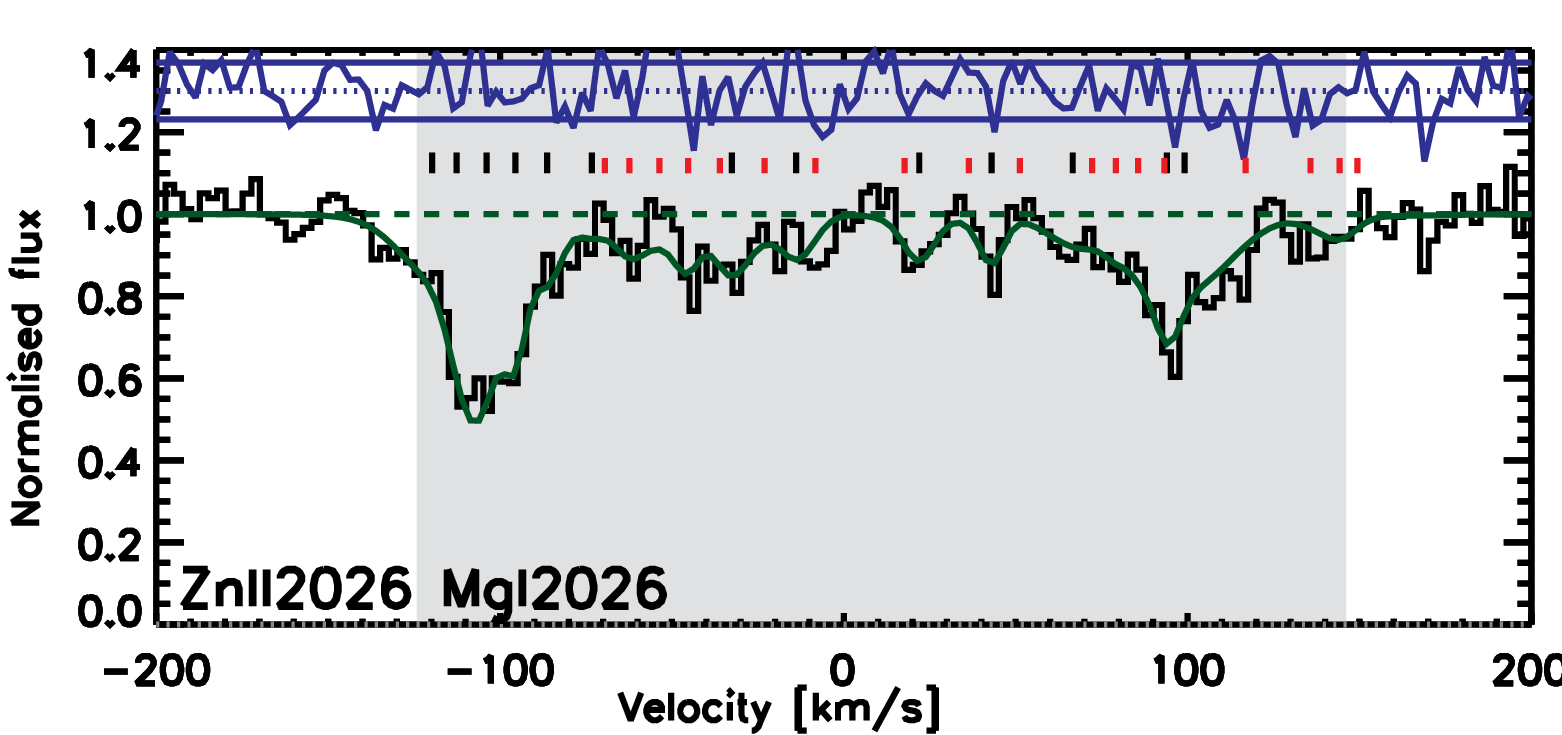}
    \includegraphics[width=0.33\textwidth,trim=0 15 0 0,angle=0,clip=false]{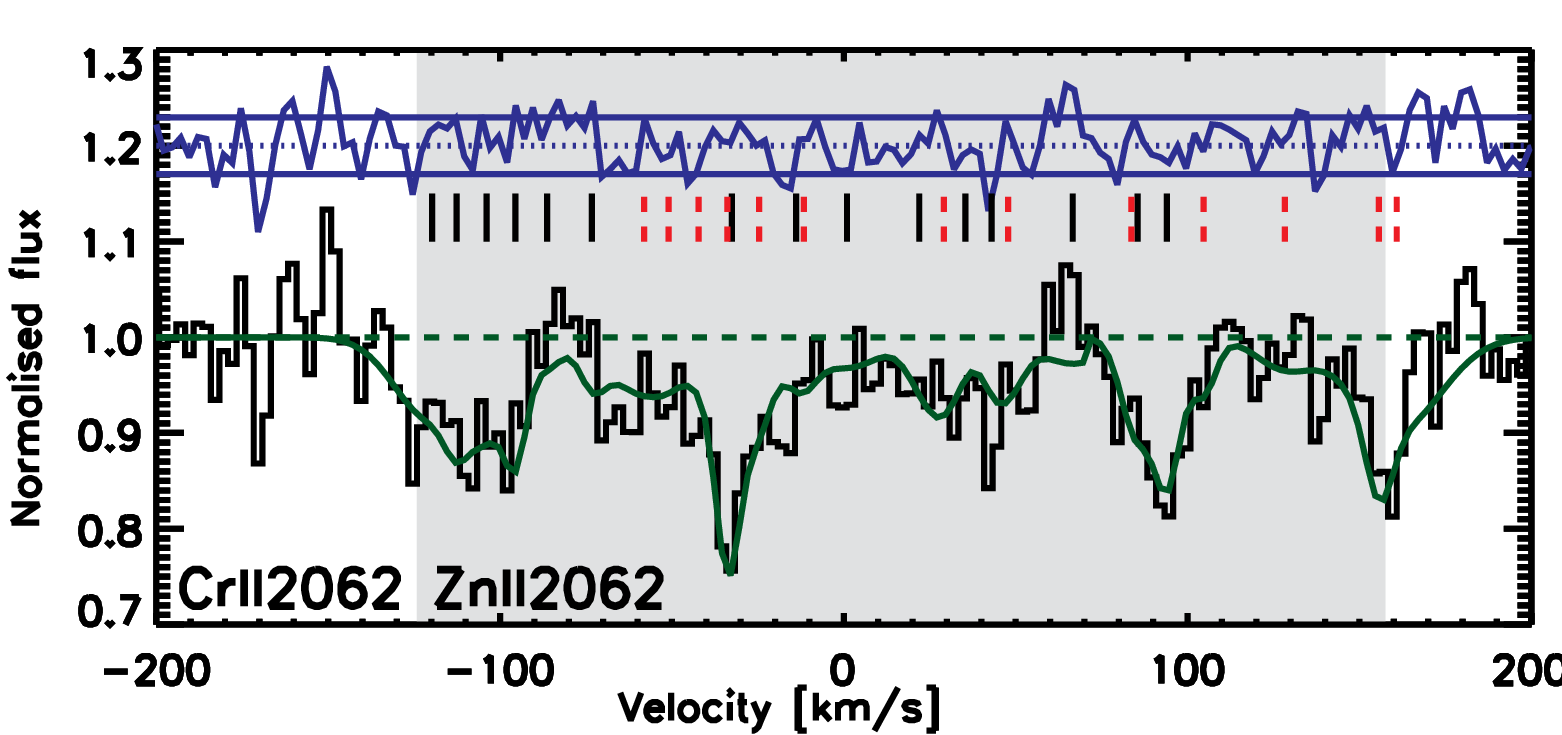}
  }
  \hbox{
    \includegraphics[width=0.33\textwidth,trim=0 15 0 0,angle=0,clip=false]{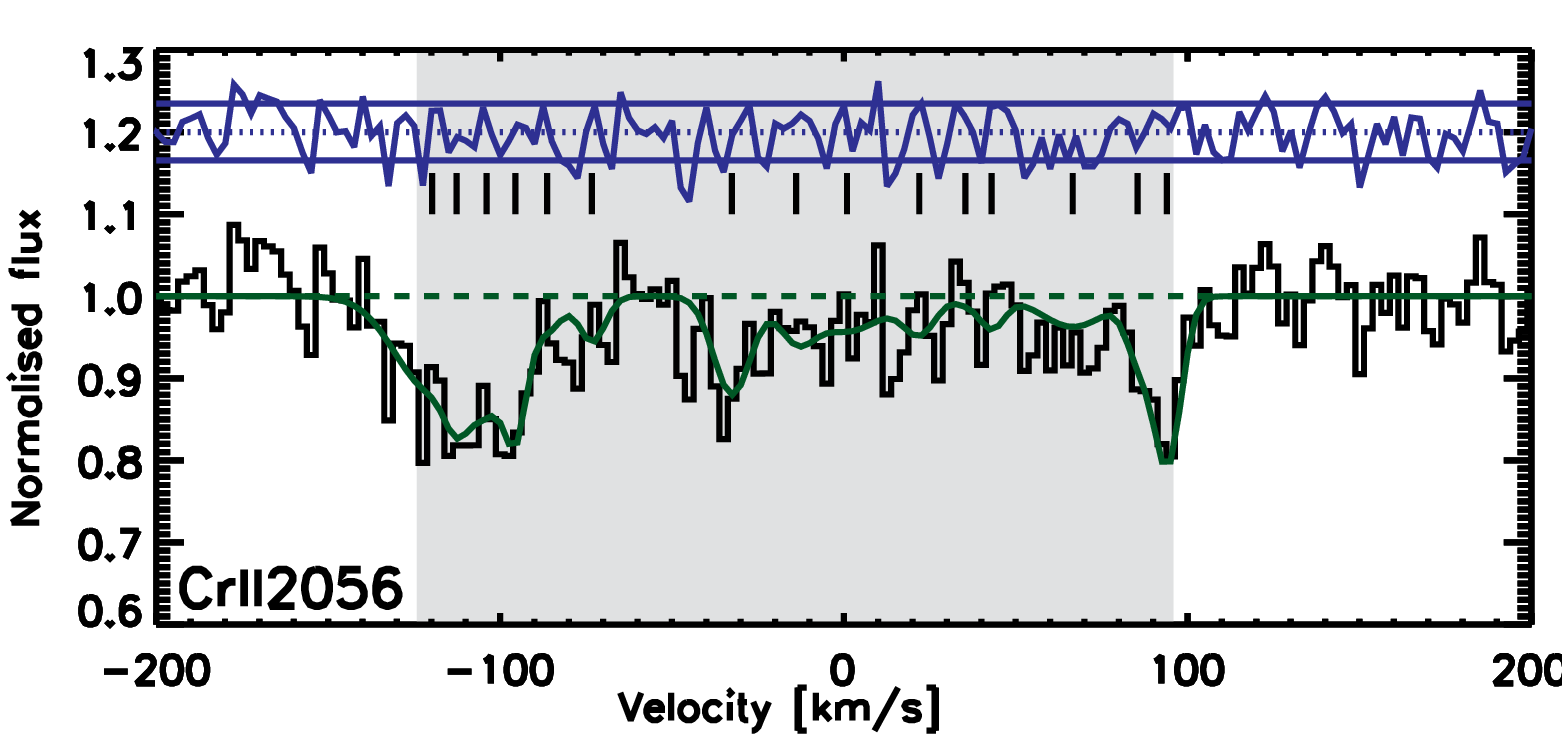}
    \includegraphics[width=0.33\textwidth,trim=0 15 0 0,angle=0,clip=false]{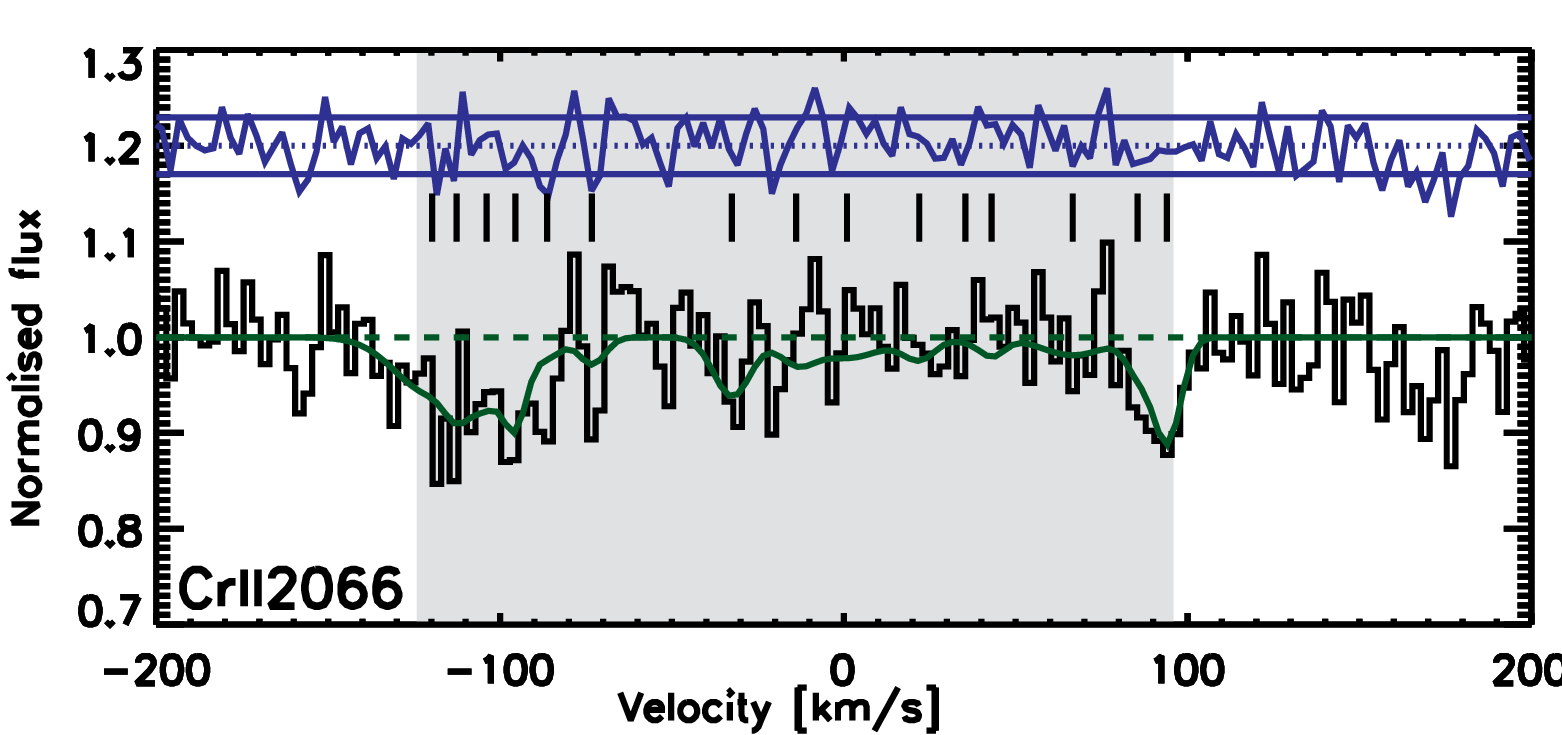}
    \includegraphics[width=0.33\textwidth,trim=0 15 0 0,angle=0,clip=false]{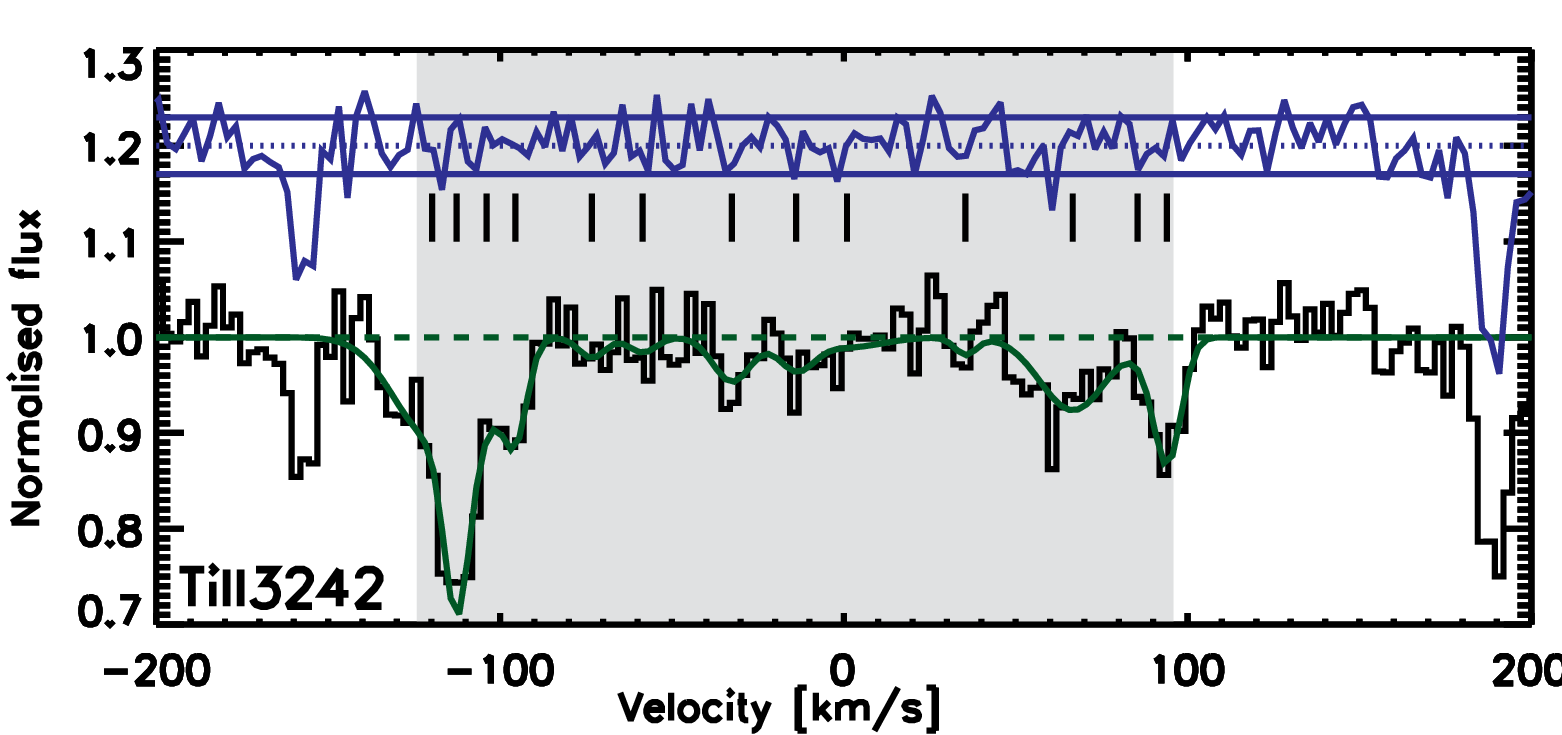}
  }
  \hbox{
    \includegraphics[width=0.33\textwidth,trim=0 15 0 0,angle=0,clip=false]{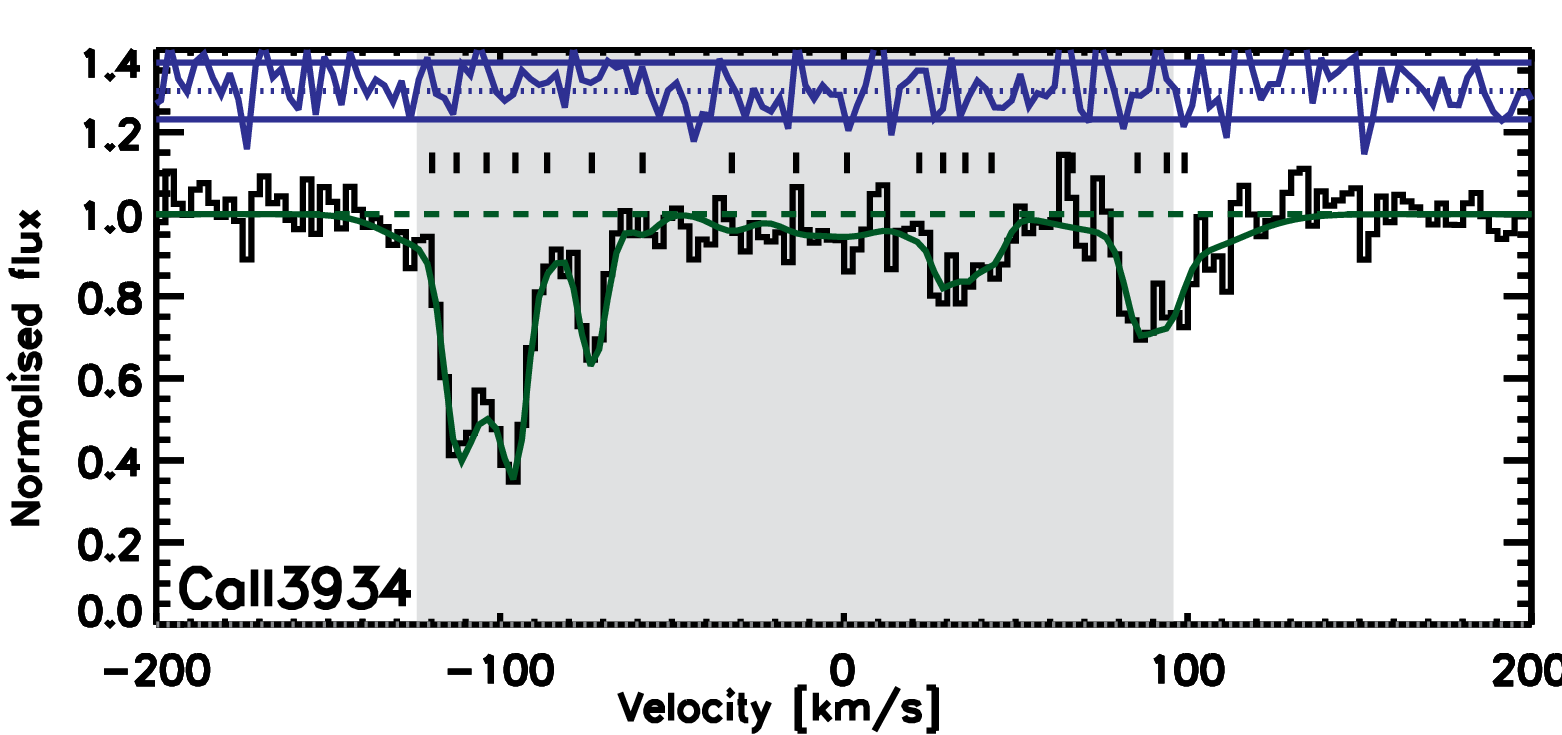}
    \includegraphics[width=0.33\textwidth,trim=0 15 0 0,angle=0,clip=false]{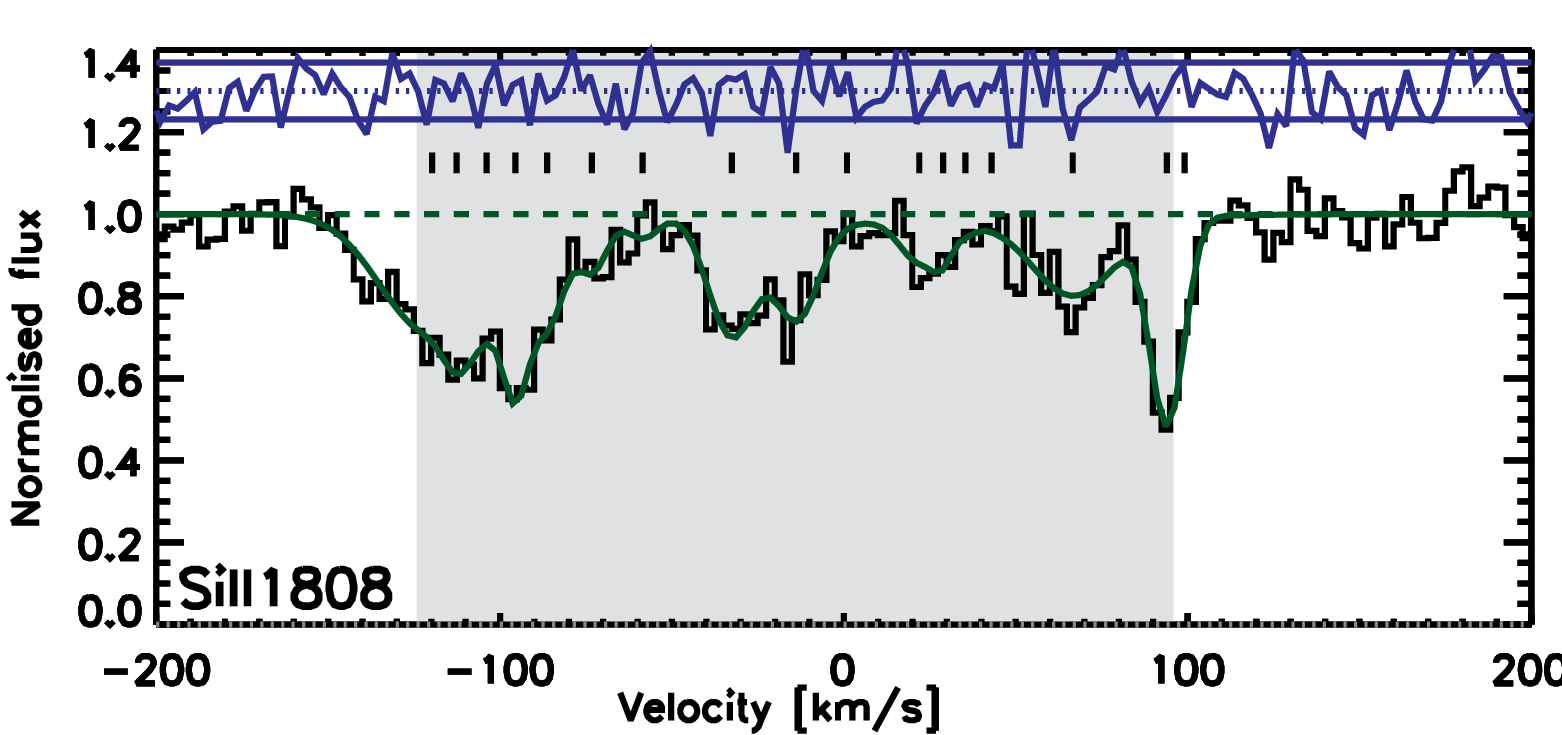}
    \includegraphics[width=0.33\textwidth,trim=0 15 0 0,angle=0,clip=false]{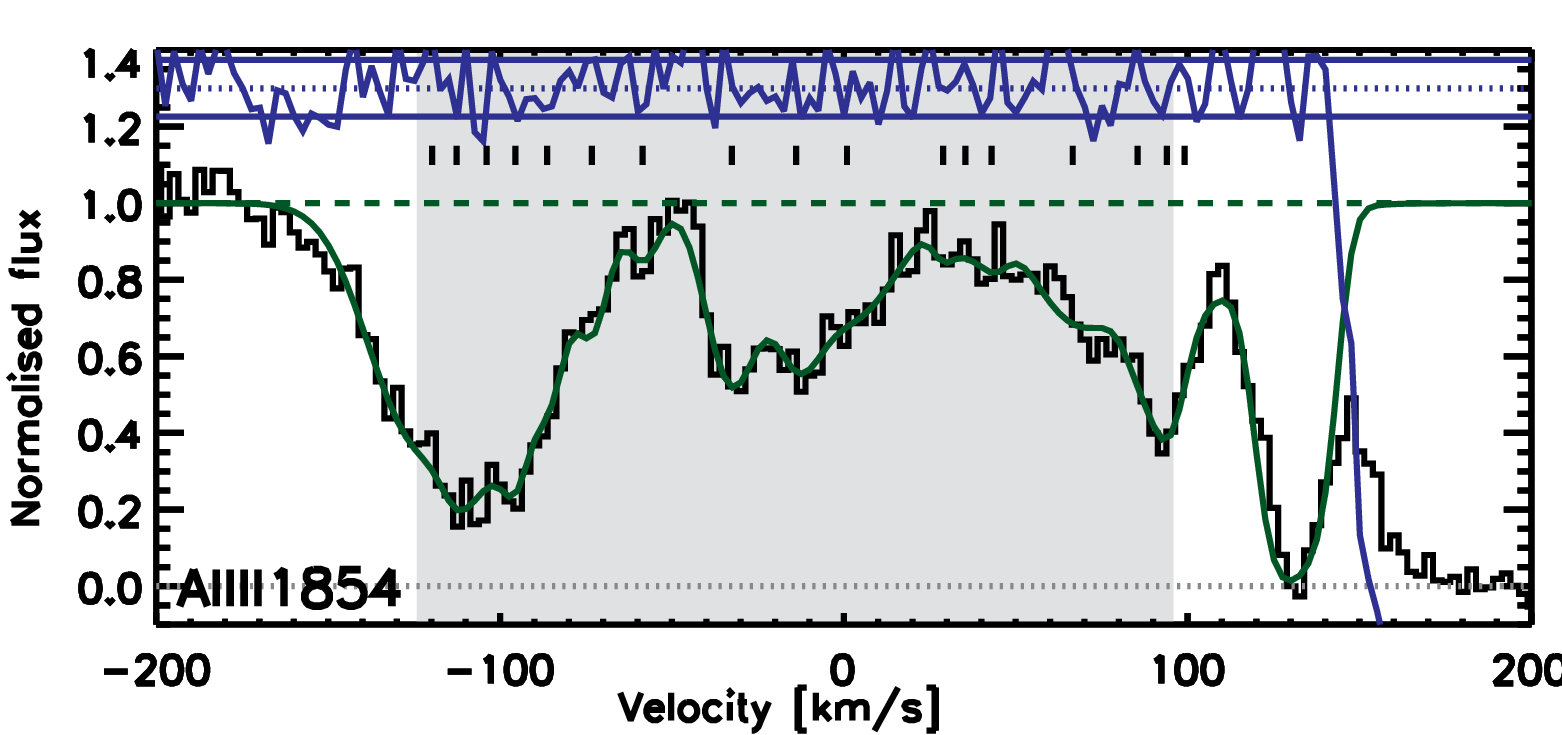}
  }
}
\caption{Voigt profile fits to the $z=1.24180$ absorber towards
  J1430$+$0149, see Fig.~\ref{fig:fit_J0334m0711} for description. }
\label{fig:fit_J1430p0149}
\end{figure*}
\begin{figure*}
\vbox{
  \hbox{
    \includegraphics[width=0.33\textwidth,trim=0 15 0 0,angle=0,clip=false]{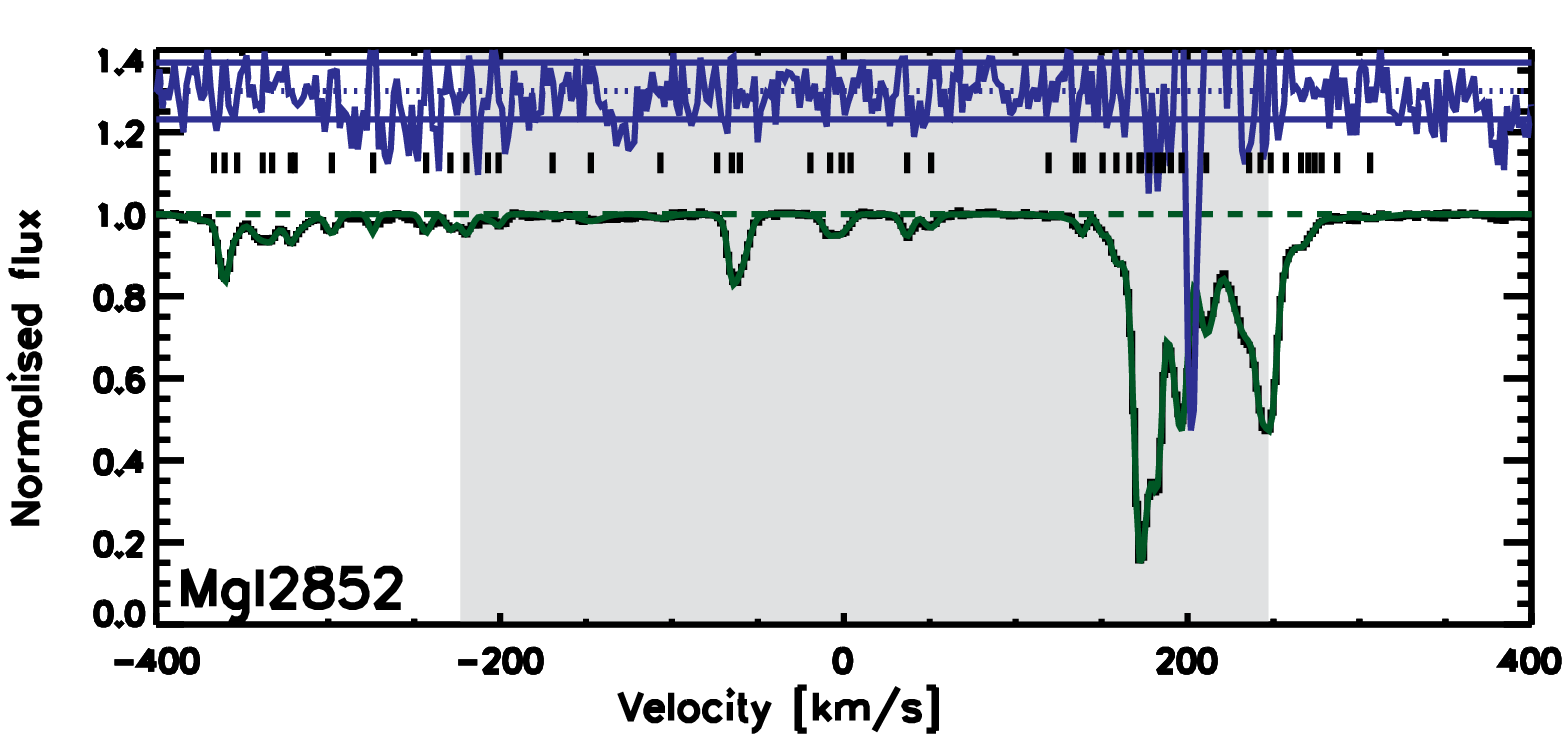}
    \includegraphics[width=0.33\textwidth,trim=0 15 0 0,angle=0,clip=false]{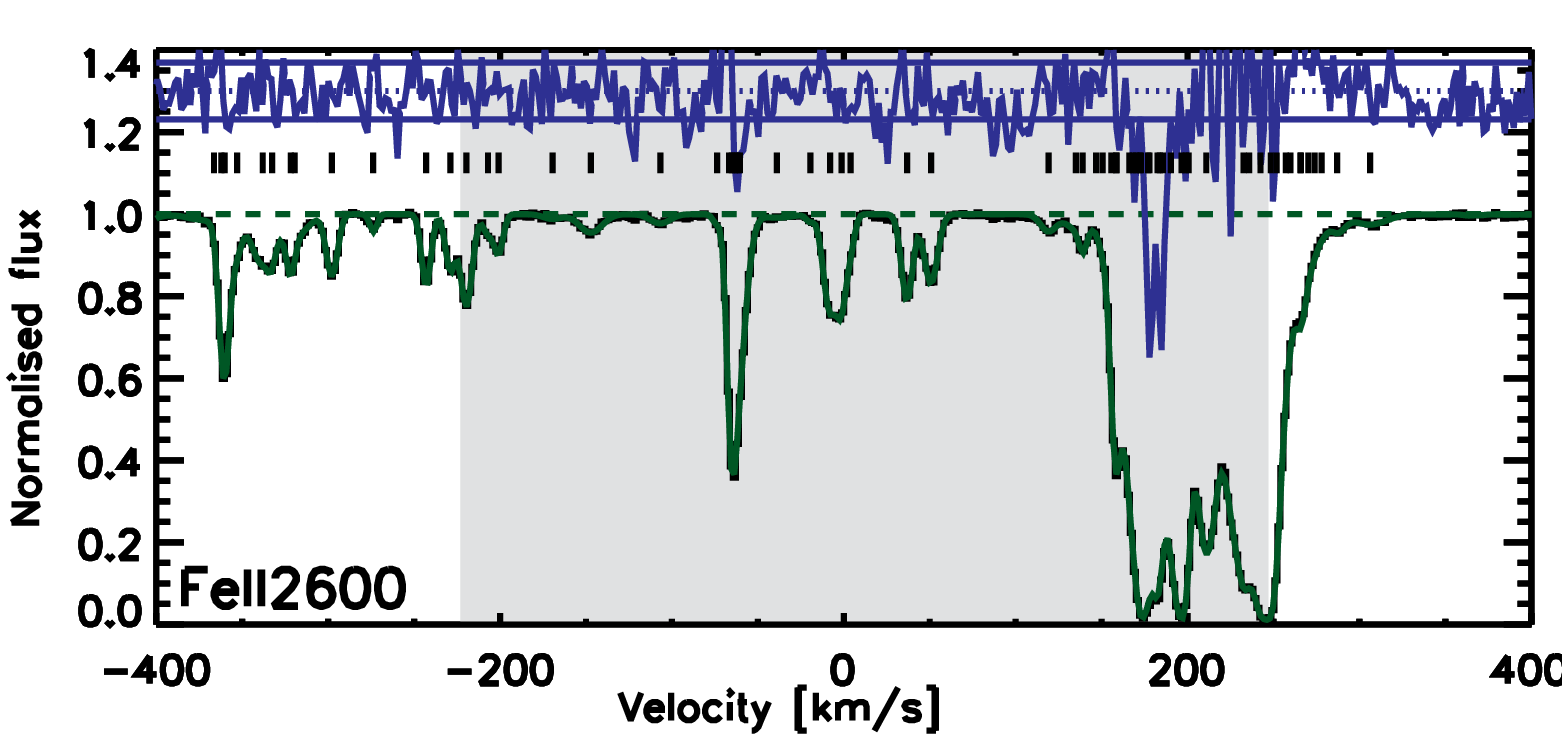}
    \includegraphics[width=0.33\textwidth,trim=0 15 0 0,angle=0,clip=false]{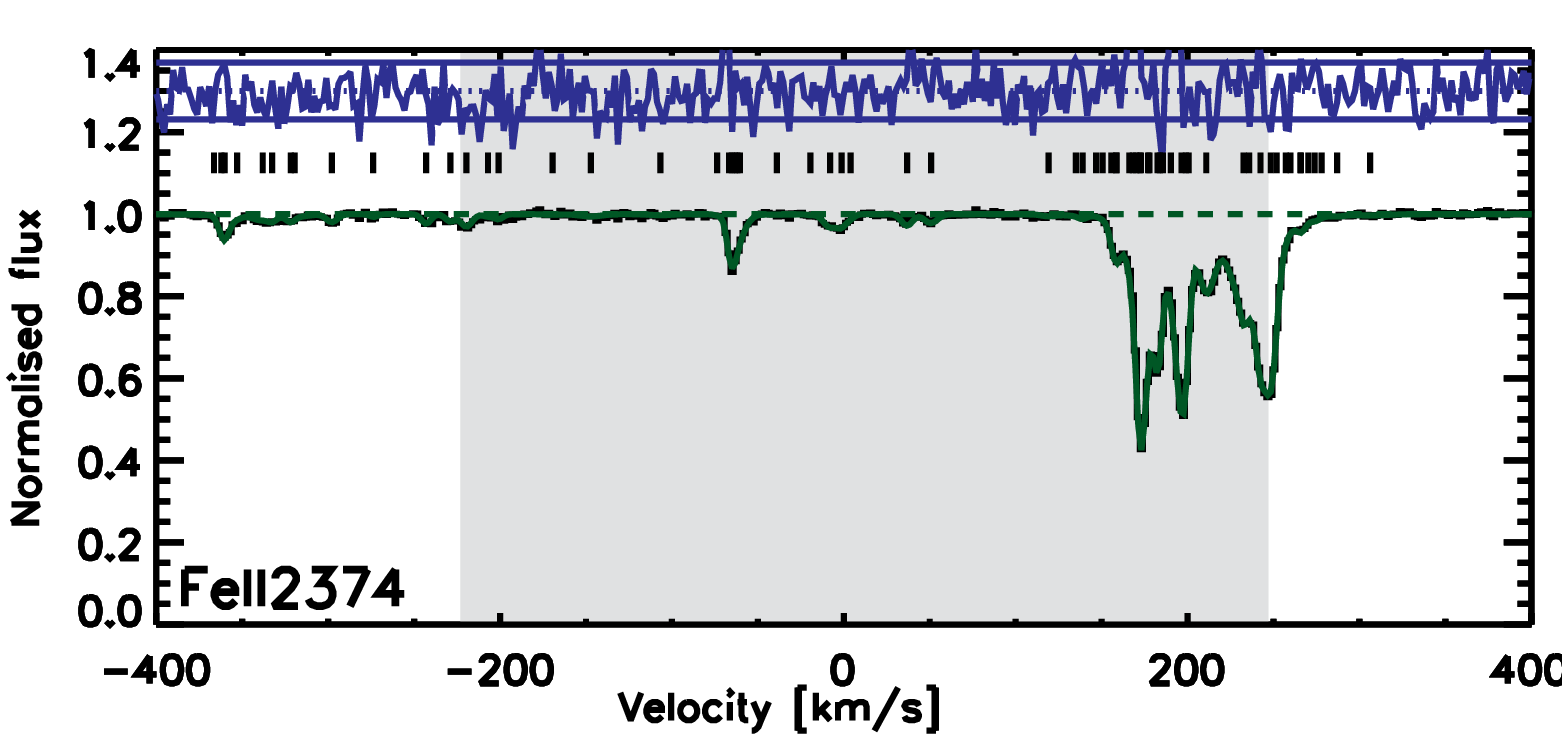}
  }
  \hbox{
    \includegraphics[width=0.33\textwidth,trim=0 15 0 0,angle=0,clip=false]{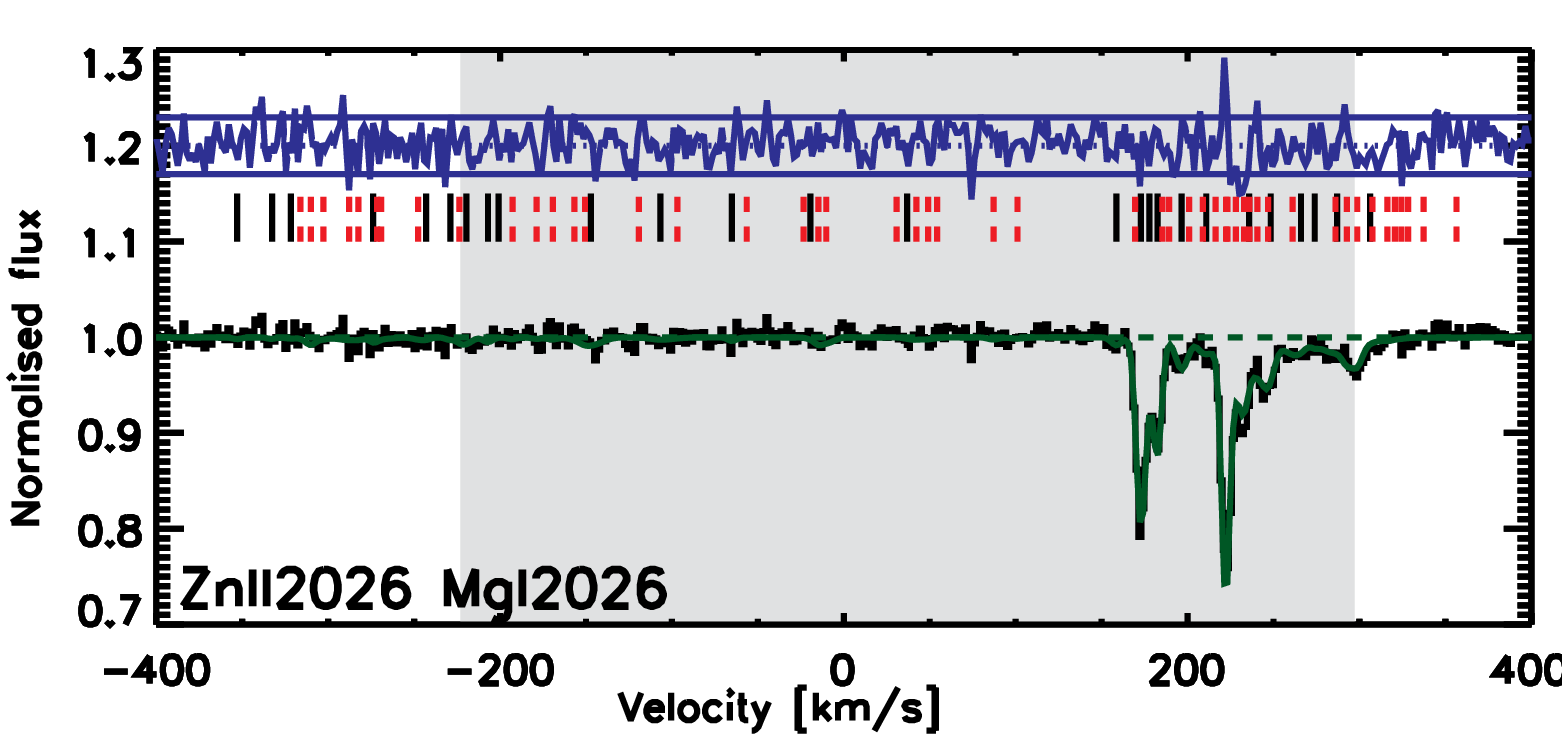}
    \includegraphics[width=0.33\textwidth,trim=0 15 0 0,angle=0,clip=false]{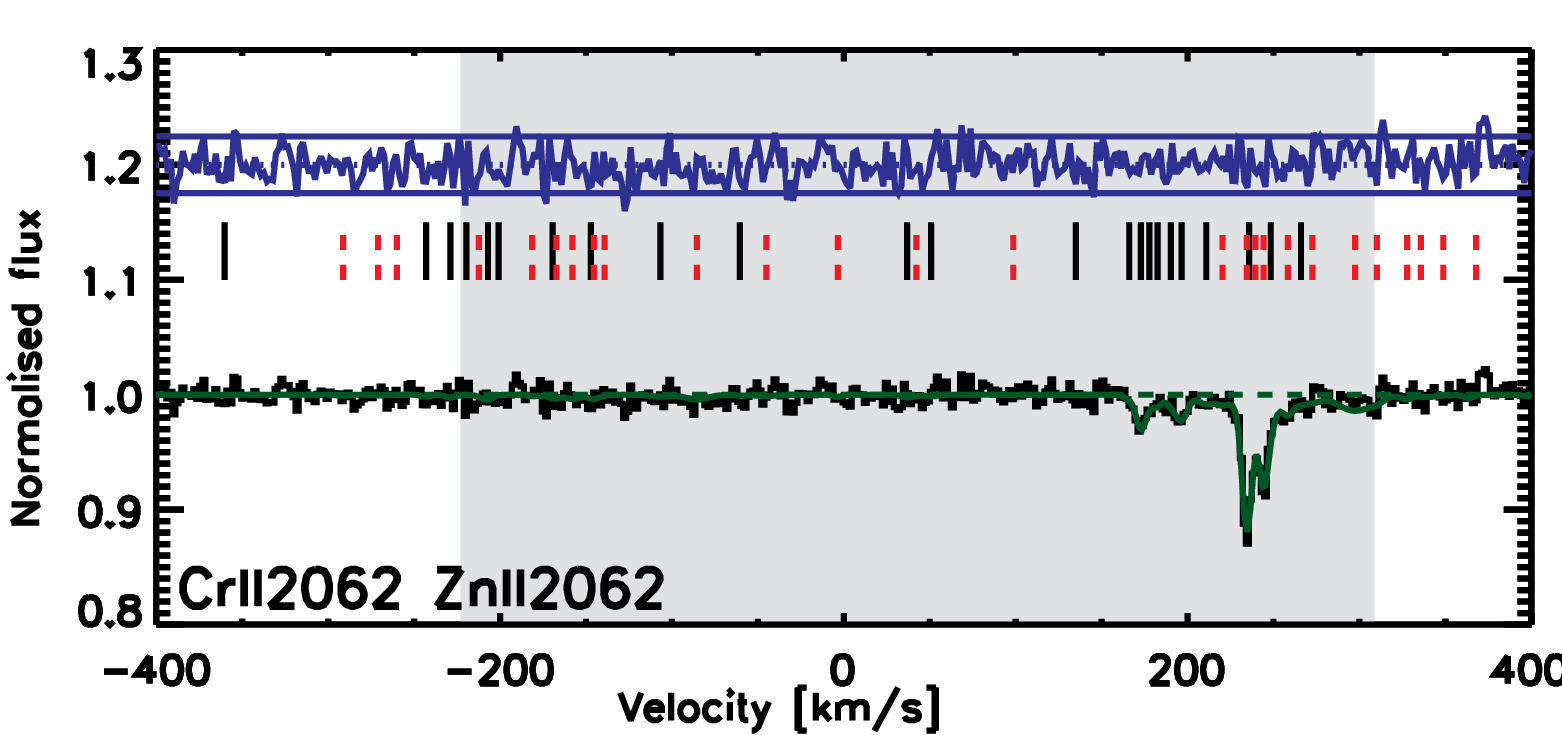}
    \includegraphics[width=0.33\textwidth,trim=0 15 0 0,angle=0,clip=false]{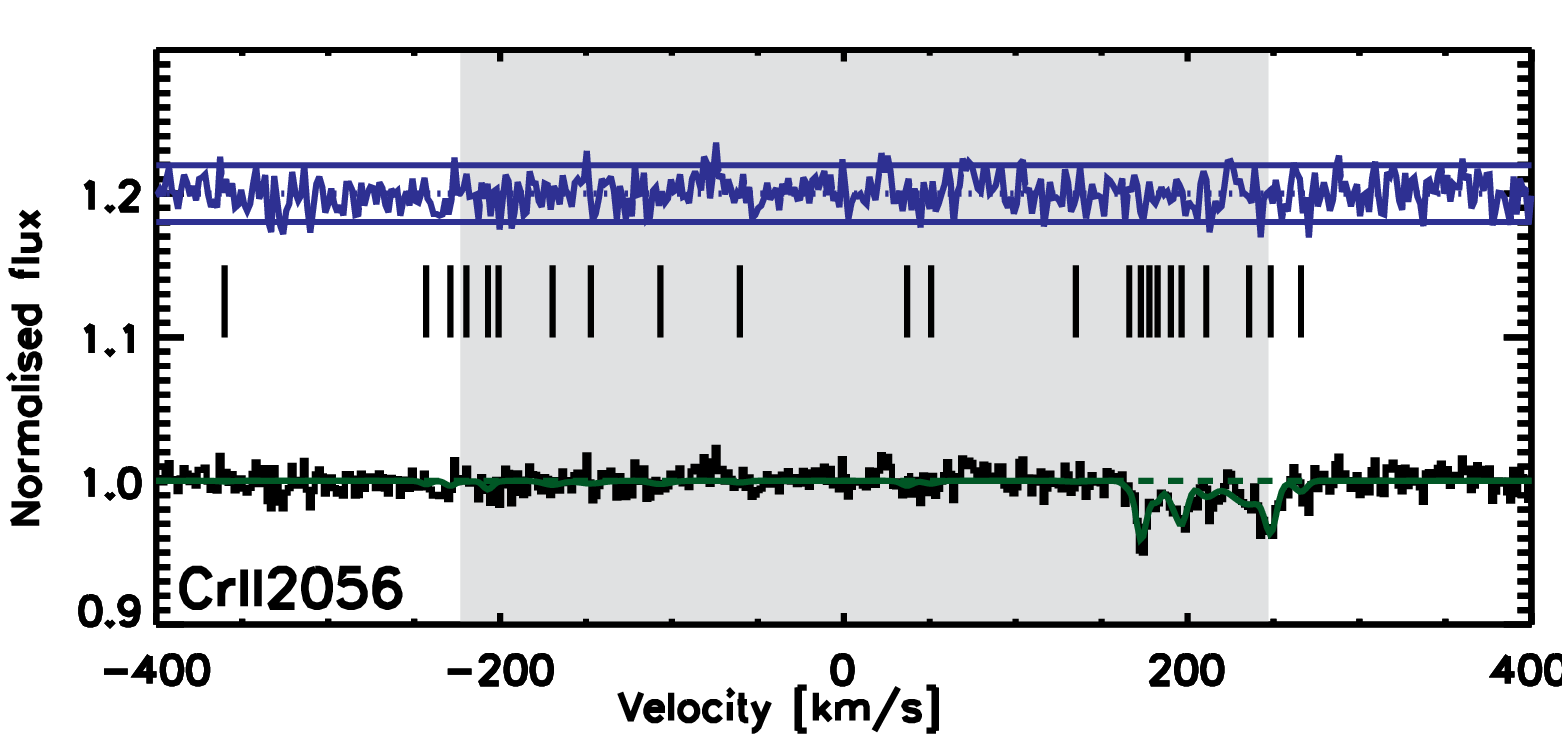}
  }
  \hbox{
    \includegraphics[width=0.33\textwidth,trim=0 15 0 0,angle=0,clip=false]{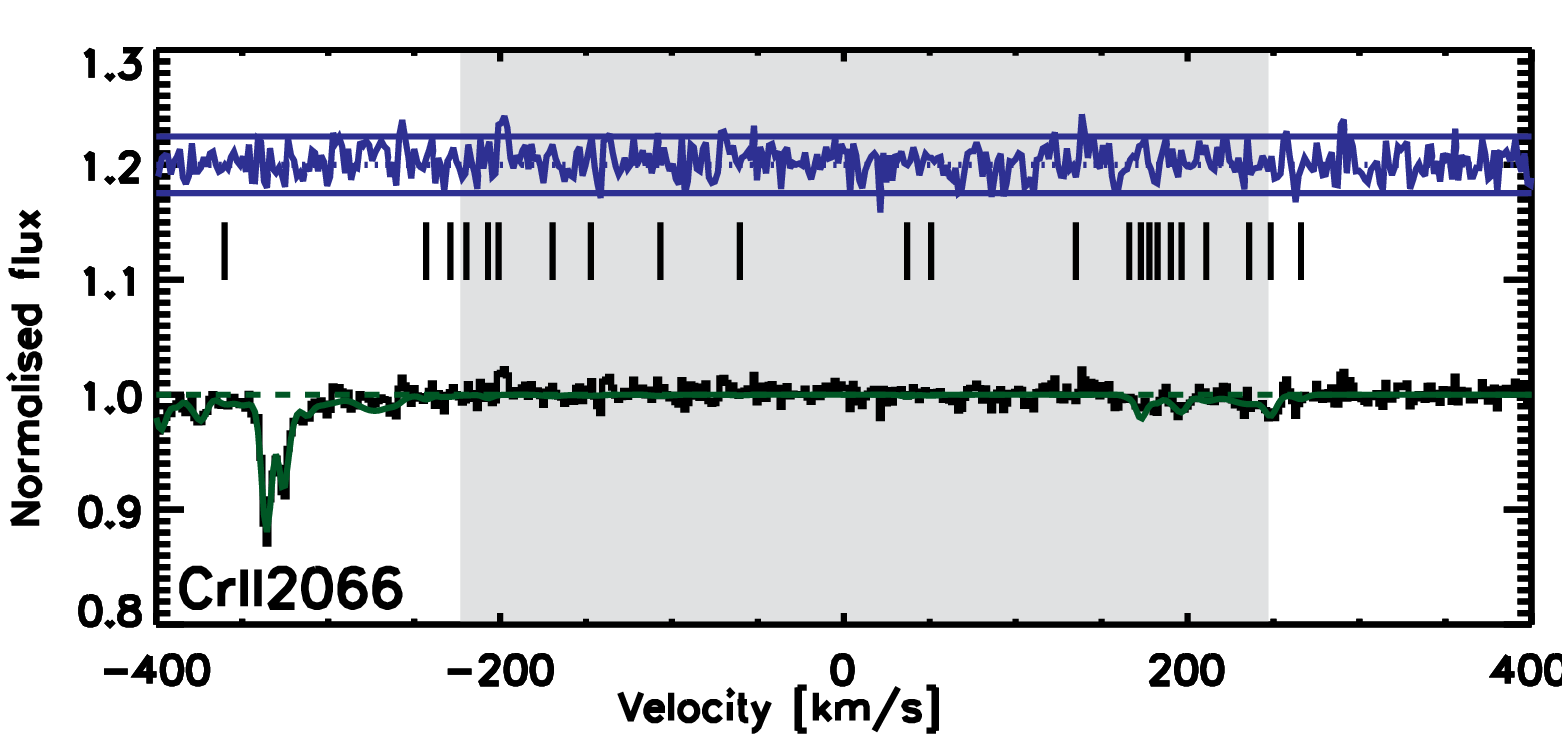}
    \includegraphics[width=0.33\textwidth,trim=0 15 0 0,angle=0,clip=false]{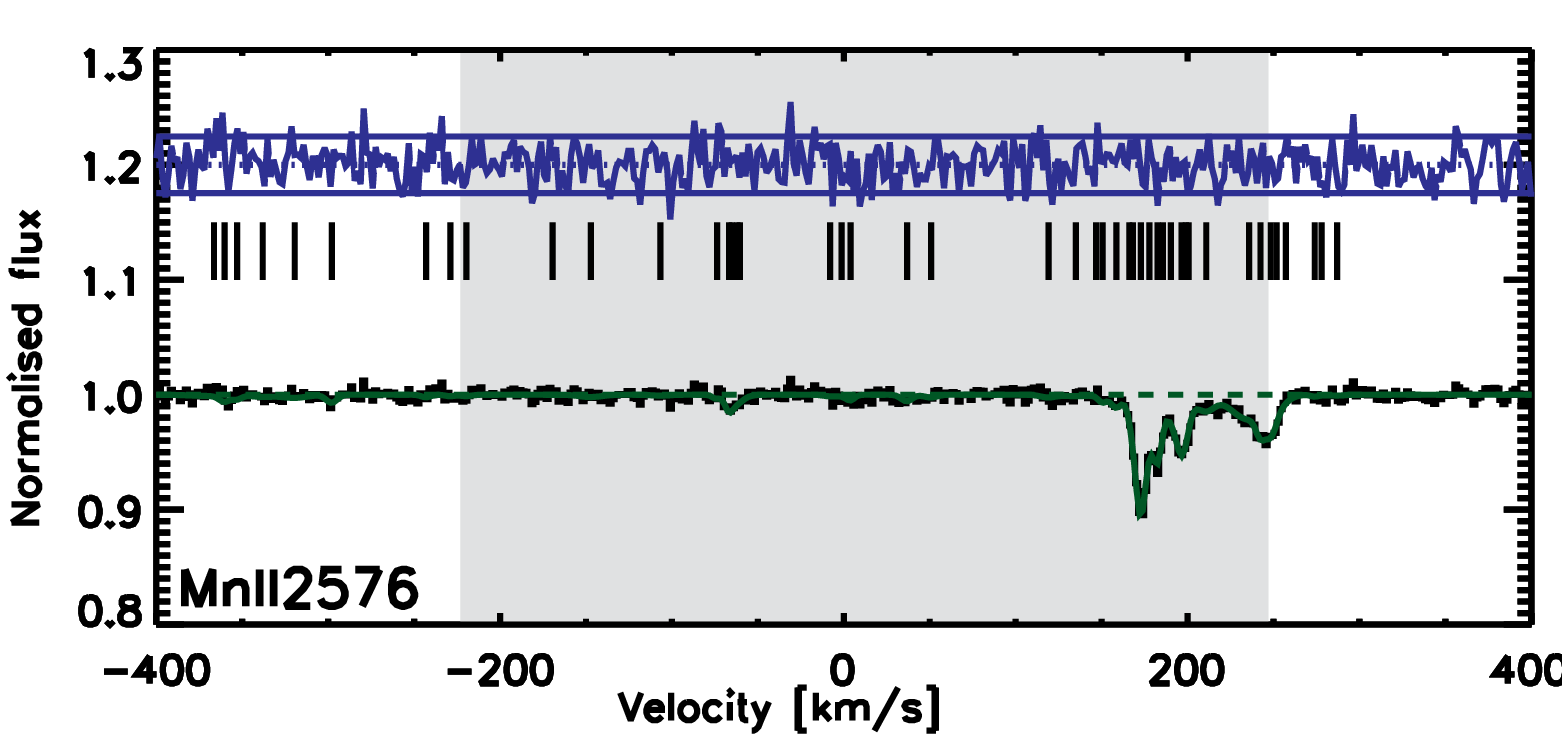}
    \includegraphics[width=0.33\textwidth,trim=0 15 0 0,angle=0,clip=false]{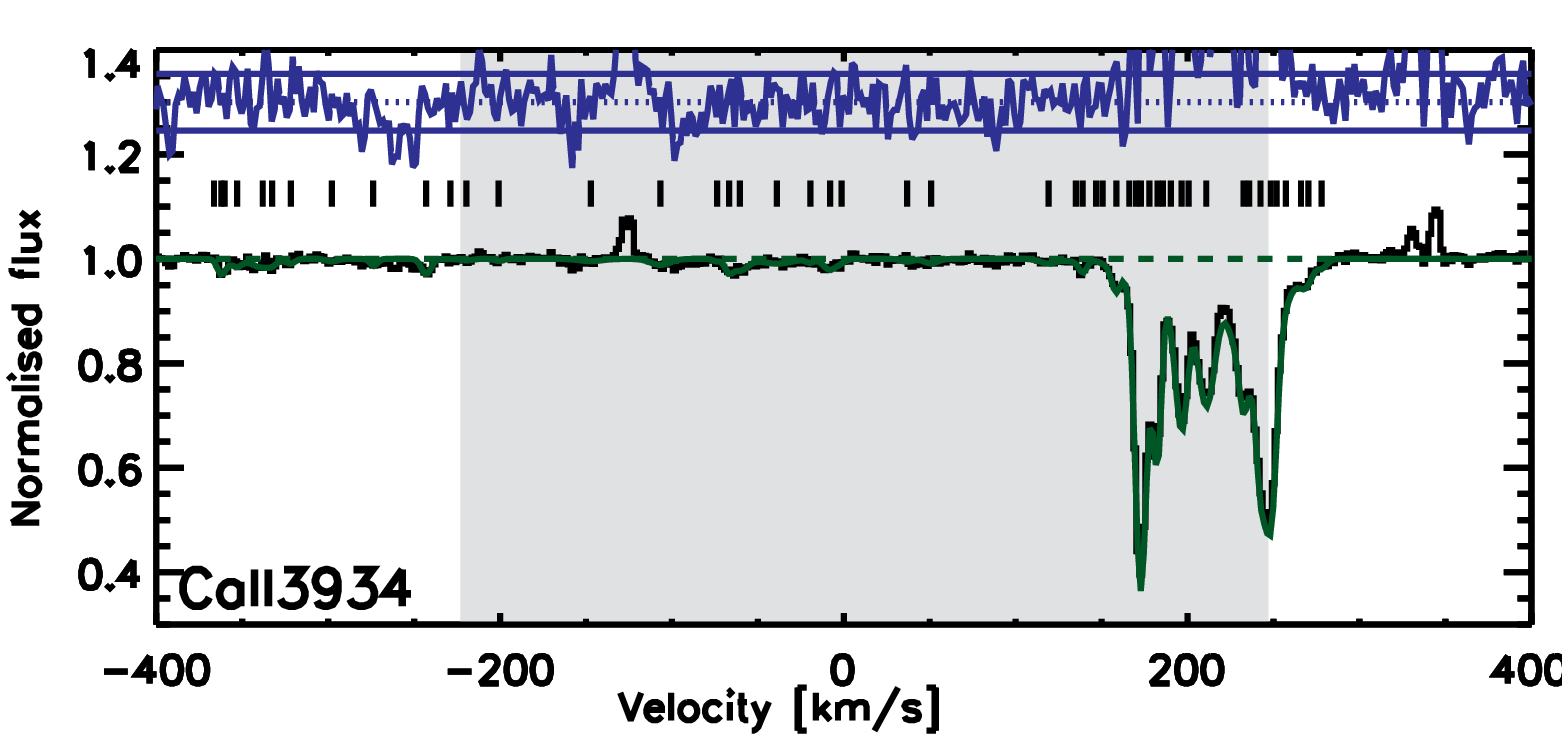}
  }
  \hbox{
    \includegraphics[width=0.33\textwidth,trim=0 15 0 0,angle=0,clip=false]{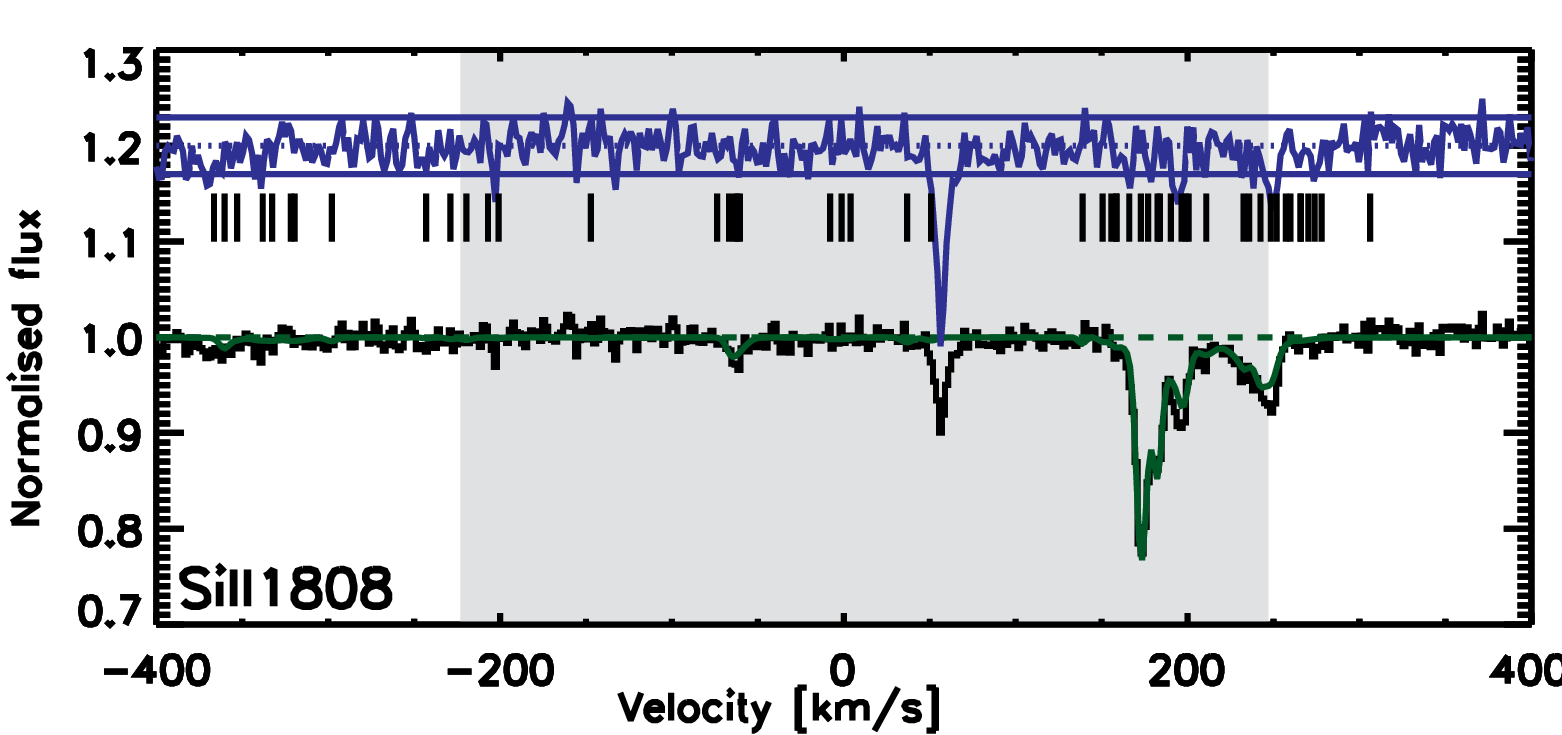}
    \includegraphics[width=0.33\textwidth,trim=0 15 0 0,angle=0,clip=false]{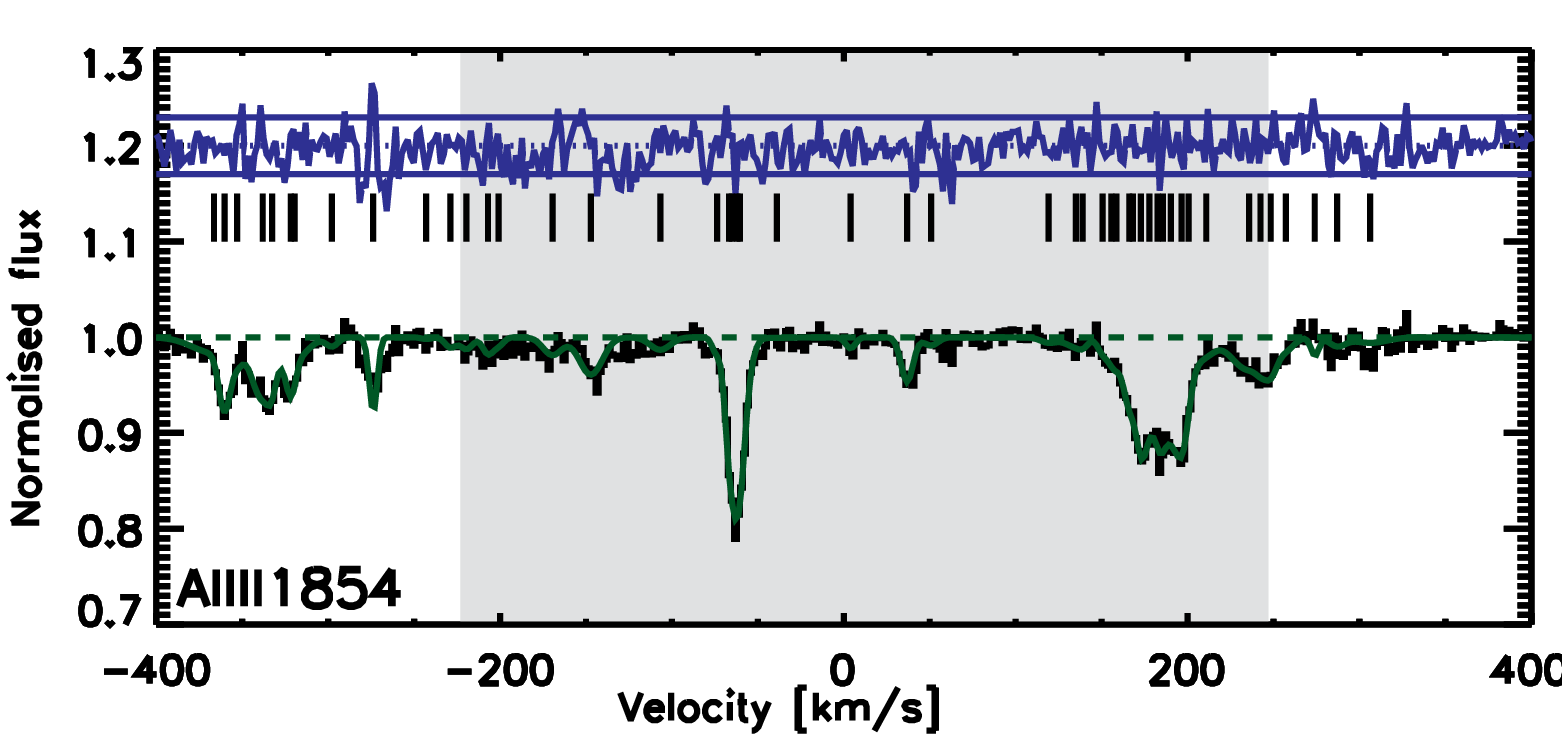}
  }
}
\caption{Voigt profile fits to the $z=1.14955$ absorber towards J0517$-$4410, see Fig.~\ref{fig:fit_J0334m0711} for description. }
\label{fig:fit_J0517m4410}
\end{figure*}

\section{Unfitted data}\label{adx:MgII}
\begin{figure*}
\vbox{
  \hbox{
    \includegraphics[scale=0.395,trim=0 15 0 5,angle=0,clip=true]{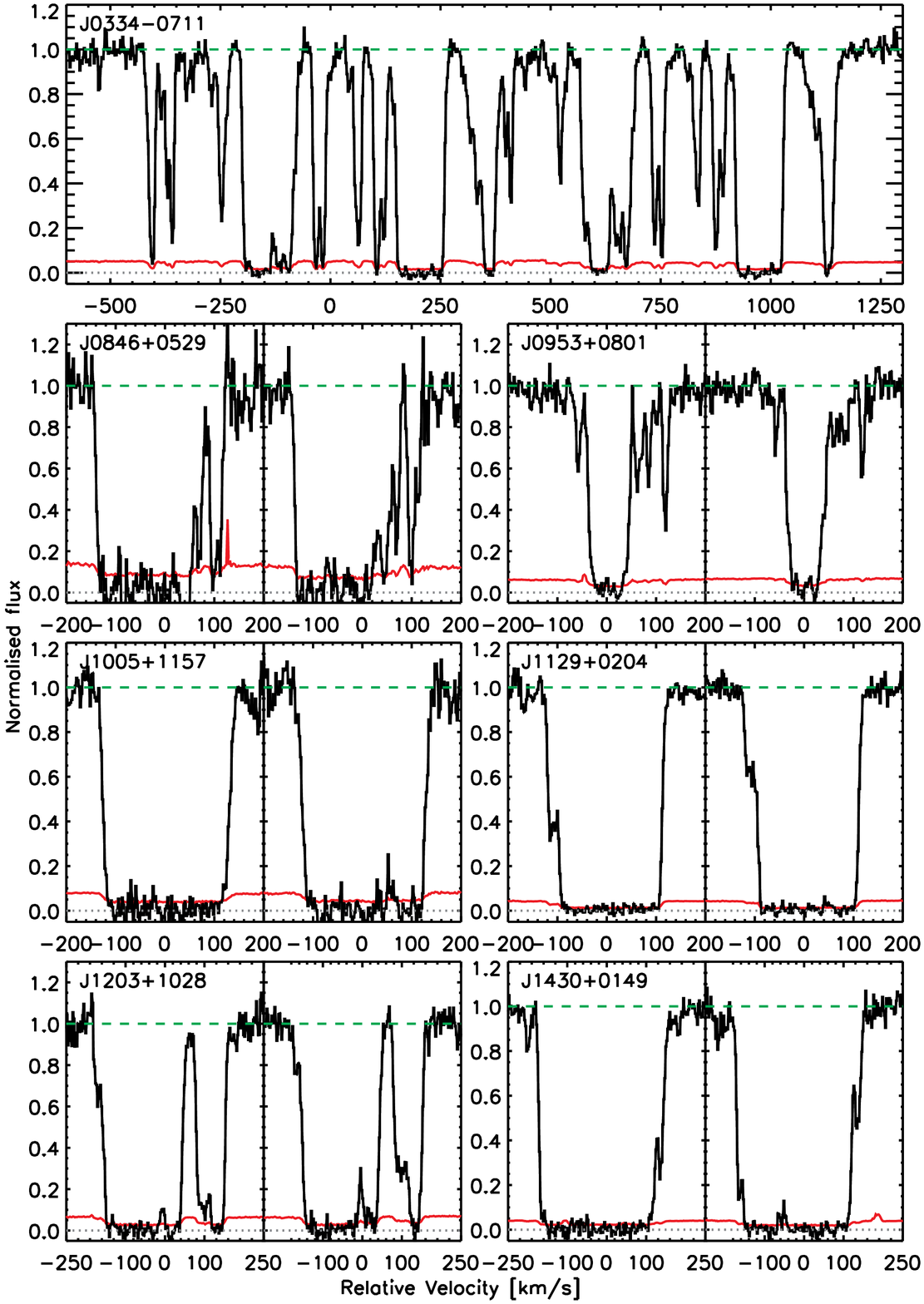}
    \includegraphics[scale=0.395,trim=0 15 0 5,angle=0,clip=true]{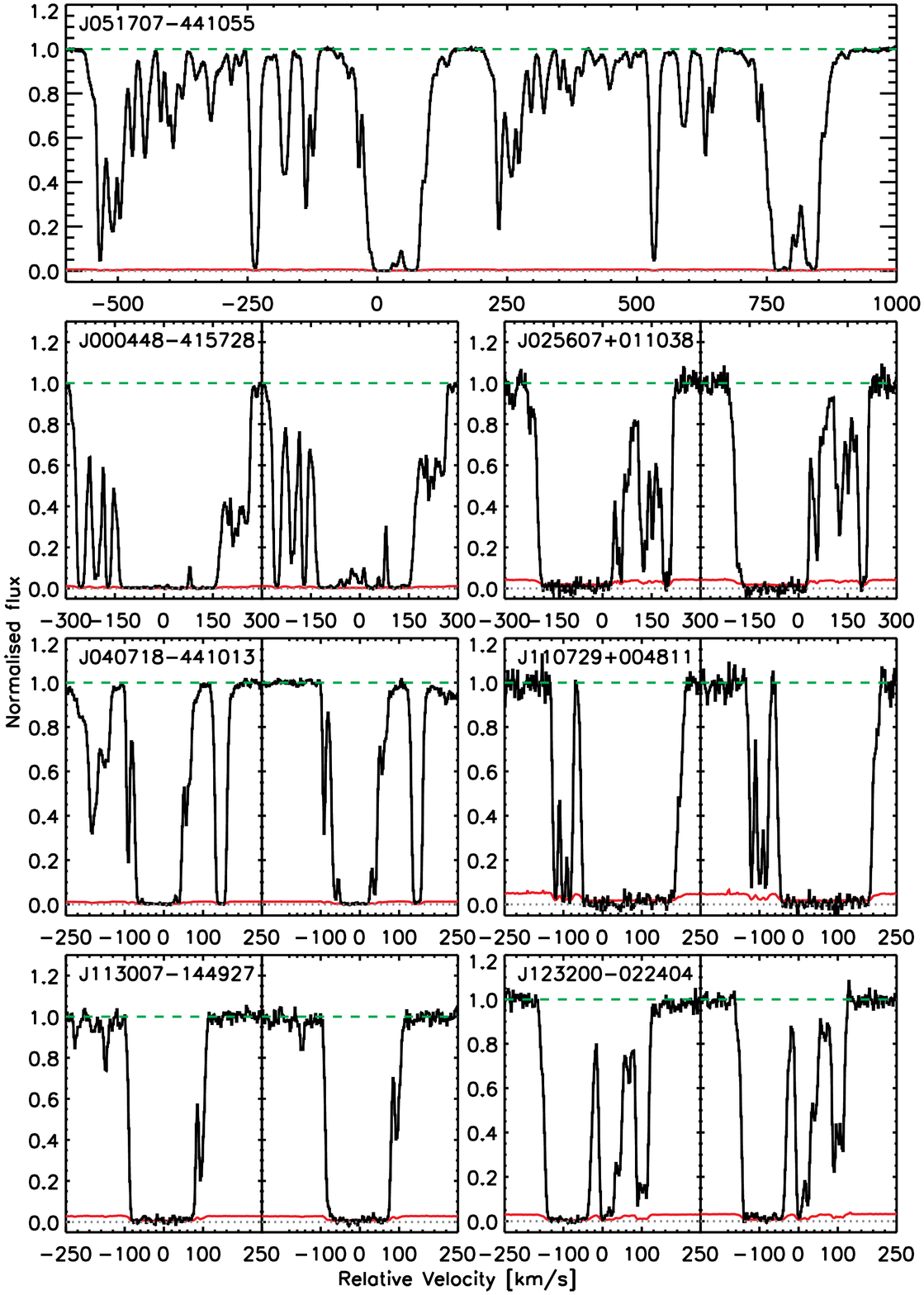}
  }
  \hbox{
    \includegraphics[scale=0.395,trim=0 410 0 200,angle=0,clip=true]{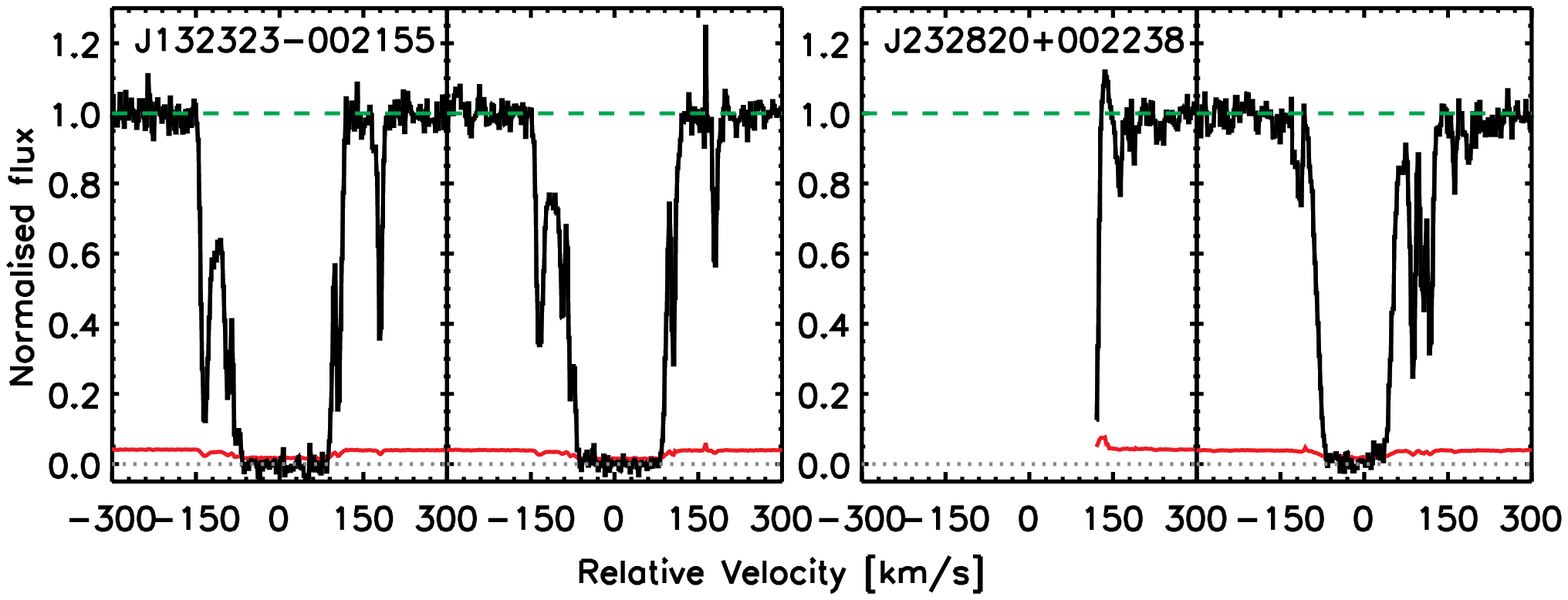}
  }
}
\caption{The \MgII\ absorption line structure in each of our absorbers
  with UVES or HIRES data with sufficient spectral coverage. The
  absorber's quasar is identified in the top left-hand corner of each
  panel. The black histograms are the data and the red histograms
  are the associated error, whilst the adopted continuum is shown by
  the dashed line. }
\label{fig:mgii_structure}
\end{figure*}

\begin{figure*}
\vbox{
  \hbox{
    \includegraphics[scale=0.385,trim=0 5 0 0,angle=0,clip=true]{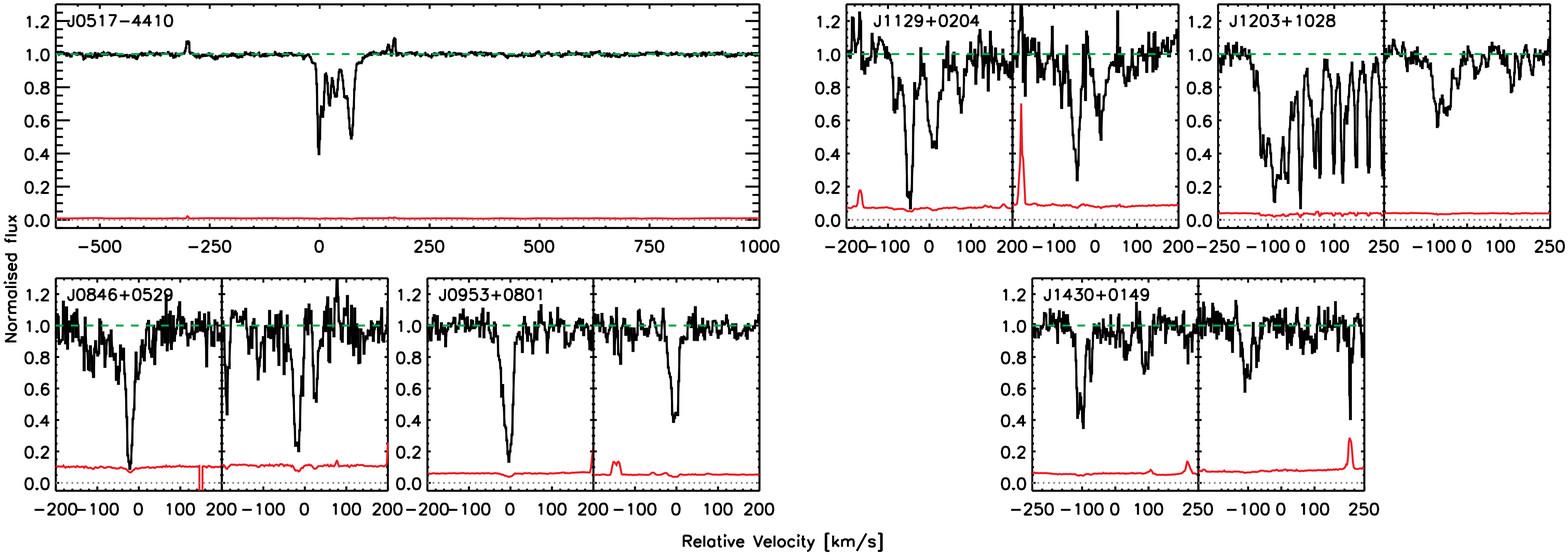}
  }
}
\caption{The \CaII\ absorption line structure in each of our absorbers
  with UVES or HIRES data with sufficient spectral coverage. The
  absorber's quasar is identified in the top left-hand corner of each
  panel. The black histograms are the data and the red histograms
  are the associated error, whilst the adopted continuum is shown by
  the dashed line. }
\label{fig:caii_structure}
\end{figure*}

Here we present the \MgII\ absorption line profiles which were not
fitted (See Fig.~\ref{fig:mgii_structure}). We also collate the
observed \CaII\ profiles into one figure as a comparative aid (See
Fig.~\ref{fig:caii_structure}).

\bspsmall

\label{lastpage}

\end{document}